\documentclass[fleqn,usenatbib]{mnras}
\pdfoutput=1

\usepackage[T1]{fontenc}

\DeclareRobustCommand{\VAN}[3]{#2}
\let\VANthebibliography\thebibliography
\def\thebibliography{\DeclareRobustCommand{\VAN}[3]{##3}\VANthebibliography}

\usepackage{graphicx}	
\usepackage{amsmath}	
\usepackage{amssymb}	
\usepackage{commath}    
\usepackage{enumitem}   
\usepackage{subcaption} 
\usepackage{mathtools}  
\usepackage{threeparttable}  
\usepackage{booktabs}   
\usepackage{newtxtext,newtxmath}

\newcommand{\nonthermal}[1][]{non-thermal#1 }
\newcommand{\lc}[1][]{light crossing#1 }
\newcommand{\ts}[1][]{time-scale#1 }
\newcommand{\cutoff}[1][]{cut-off#1 }

\newcommand{\quoted}[2][]{`#2'#1 }

\newcommand{\organiz}[1][]{organiz#1 }
\newcommand{\generaliz}[1][]{generalis#1 }
\newcommand{\normaliz}[1][]{normalis#1 }
\newcommand{\analyz}[1][]{analys#1 }
\newcommand{\crosssection}[1][]{cross-section#1 }
\newcommand{\characteriz}[1][]{characteriz#1 }

\newcommand{\summariz}[1][]{summariz#1 }
\newcommand{\parameteriz}[1][]{parametriz#1 }
\newcommand{\ling}[1][]{lling#1 }
\newcommand{\led}[1][]{lled#1 }
\newcommand{\spellor}[1][]{our#1 }

\newcommand{\radzone}[1][]{radiation zone#1 }

\newcommand{\me}{\ensuremath{m_e}}
\newcommand{\eph}{\ensuremath{\epsilon_{\rm ph}}}
\newcommand{\uph}{\ensuremath{U_{\rm ph}}}
\newcommand{\gmax}{\ensuremath{\gamma_{\rm max}}}
\newcommand{\gx}{\ensuremath{\gamma_{\rm X}}}
\newcommand{\gcool}{\ensuremath{\gamma_{\rm cool}}}
\newcommand{\gcooli}[1]{\ensuremath{\gamma_{\mathrm{cool,}#1}}}
\newcommand{\gct}{\ensuremath{\tilde{\gamma}_{\rm cool}}}
\newcommand{\gcti}[1]{\ensuremath{\tilde{\gamma}_{\mathrm{cool,}#1}}}
\newcommand{\tcoolt}{\ensuremath{t_{\rm cool,T}}}
\newcommand{\tcoolk}{\ensuremath{t_{\rm cool,IC}}}
\newcommand{\gradt}{\ensuremath{\gamma_{\rm rad,T}}}
\newcommand{\gradk}{\ensuremath{\gamma_{\rm rad,IC}}}
\newcommand{\gkn}{\ensuremath{\gamma_{\rm KN}}}
\newcommand{\knp}{\ensuremath{q}}
\newcommand{\gth}{\ensuremath{\gamma_{\rm pp}}}
\newcommand{\sigc}{\ensuremath{\sigma_{\rm c,0}}}
\newcommand{\sigh}{\ensuremath{\sigma_{\rm h,0}}}
\newcommand{\sigcgen}{\ensuremath{\sigma_{\rm c}}}
\newcommand{\sighgen}{\ensuremath{\sigma_{\rm h}}}
\newcommand{\sighn}[1]{\ensuremath{\sigma_{\mathrm{h,}#1}}}
\newcommand{\lmfp}{\ensuremath{\lambda_{\rm mfp}}}
\newcommand{\tread}{\ensuremath{t_{\rm ra}}}
\newcommand{\go}{\ensuremath{\gamma_1}}
\newcommand{\gf}{\ensuremath{\gamma_2}}
\newcommand{\gon}[1][]{\ensuremath{\go^{(#1)}}}
\newcommand{\gfn}[1][]{\ensuremath{\gf^{(#1)}}}
\newcommand{\pt}{\ensuremath{p_{\rm T}}}
\newcommand{\pk}{\ensuremath{p_{\rm KN}}}
\newcommand{\ngen}{\ensuremath{N}}

\newcommand{\ptclsym}{N}
\newcommand{\Ngen}[1][]{\ensuremath{\ptclsym_{\gamma\gamma}^{(#1)}}}

\newcommand{\fnc}{f_{\rm nocool}}
\newcommand{\fncn}[1][]{\ensuremath{\fnc^{(#1)}}}
\newcommand{\fncmin}{f_{\rm nocool,min}}
\newcommand{\avgen}[1][]{\bar{\gamma}_{N#1}}
\newcommand{\avgeninj}{\bar{\gamma}_Q}
\newcommand{\avgenninj}[1][]{\ensuremath{\avgeninj^{(#1)}}}
\newcommand{\avgenn}[1][]{\ensuremath{\avgen[]^{(#1)}}}

\newcommand{\Ngg}[1][]{\ensuremath{\ptclsym_{\gamma\gamma}^{(#1)}}}
\newcommand{\Nph}[1][]{\ensuremath{N_{\rm ph}^{(#1)}}}
\newcommand{\injsym}{Q}
\newcommand{\Qinj}[1][]{\ensuremath{\injsym_{\gamma\gamma}^{(#1)}}}
\newcommand{\Qgg}[1][]{\ensuremath{\injsym_{\gamma\gamma}^{(#1)}}}
\newcommand{\newbornsym}{n_{\gamma\gamma}}
\newcommand{\newbornn}[1][]{\ensuremath{\newbornsym^{(#1)}}}
\newcommand{\avgy}[1][]{\ensuremath{\bar{y}^{(#1)}}}
\newcommand{\pgg}[1][]{\ensuremath{\Gamma_{#1}}}
\newcommand{\Qnorm}[1][]{\ensuremath{B_{#1}}}
\newcommand{\zmax}{\ensuremath{1100}}
\newcommand{\thickness}{\ensuremath{\Delta}}

\newcommand{\corlab}{\ensuremath{disc}}

\newcommand{\fne}{f_{\rm noesc}}
\newcommand{\xeqh}{x_{n+1} = h(x_n)}

\DeclareMathOperator{\Li}{Li}
\newcommand{\argmin}[1][]{\ensuremath{\underset{#1}{\mathrm{arg \, min}}}}
\DeclareMathOperator{\sign}{sign}
\DeclarePairedDelimiter\ceil{\lceil}{\rceil}

\ifdefined \machineIsMaxwell
    
\else
    
\fi

\ifdefined \compilingRevision
    \newcommand{\revtext}[1]{\textcolor{blue}{#1}}
\else
    \newcommand{\revtext}[1]{}
\fi

\title[Klein-Nishina Reconnection]{Pair-Regulated Klein-Nishina Relativistic Magnetic Reconnection with Applications to Blazars and Accreting Black Holes}

\author[J. M. Mehlhaff et al.]{
J. M. Mehlhaff,$^{1}$\thanks{E-mail: john.mehlhaff@colorado.edu}
G. R. Werner,$^{1}$
D. A. Uzdensky,$^{1}$
M. C. Begelman$^{2, 3}$
\\
$^{1}$Center for Integrated Plasma Studies, Physics Department, 390 UCB, University of Colorado, Boulder, CO 80309, USA\\
$^{2}$JILA, University of Colorado and National Institute of Standards and Technology, 440 UCB, Boulder, CO 80309, USA\\
$^{3}$Department of Astrophysical and Planetary Sciences, 391 UCB, Boulder, CO 80309, USA
}

\date{Accepted XXX. Received YYY; in original form ZZZ}

\pubyear{2021}

\begin{document}
\label{firstpage}
\pagerange{\pageref{firstpage}--\pageref{lastpage}}
\maketitle

\begin{abstract}
    Relativistic magnetic reconnection is a powerful agent through which magnetic energy can be tapped in astrophysics, energizing particles that then produce observed radiation. In some systems, the highest energy photons come from particles Comptonizing an ambient radiation bath supplied by an external source. If the emitting particle energies are high enough, this inverse Compton (IC) scattering enters the Klein-Nishina regime, which differs from the low-energy Thomson IC limit in two significant ways. First, radiative losses become inherently discrete, with particles delivering an order-unity fraction of their energies to single photons. Second, Comptonized photons may pair-produce with the ambient radiation, opening up another channel for radiative feedback on magnetic reconnection. We analytically study externally illuminated highly magnetized reconnecting systems for which both of these effects are important. We identify a universal (initial magnetization-independent) quasi-steady state in which gamma-rays emitted from the reconnection layer are absorbed in the upstream region, and the resulting hot pairs dominate the energy density of the inflow plasma. However, a true pair cascade is unlikely, and the number density of created pairs remains subdominant to that of the original plasma for a wide parameter range. Future particle-in-cell simulation studies may test various aspects. Pair-regulated Klein-Nishina reconnection may explain steep spectra (quiescent and flaring) from flat-spectrum radio quasars and black hole accretion disc coronae. 
\end{abstract}


\section{Introduction}
Many accreting and jet-launching compact objects host tenuous, highly magnetized plasmas that are prone to dissipation through collisionless relativistic magnetic reconnection \citep{bf94, lu03, l05}. Reconnection represents an important pathway that transfers free magnetic energy to plasma internal and kinetic energy, two forms that may then be channeled into observable radiation. In some circumstances, the two steps in this energy conversion process -- from the magnetic field to plasma through reconnection, and from plasma to light through radiative processes -- can be imagined as happening separately. This is true, for example, when the plasma radiative cooling time is longer than the characteristic \ts[]on which reconnection occurs.

If, however, the emitting particles cool on \ts[s]comparable to -- or even much shorter than -- the reconnection dynamical time, then radiative and reconnection physics are inseparable. The coupling between them may, furthermore, be facilitated by much more than just optically thin radiative cooling, where the emitting particles suffer radiative drag but the produced photons passively escape the system. In some situations, the optical depths to various processes, including Thomson scattering and pair-creation, may exceed unity, affording the emitted photons further opportunity to impact the ongoing reconnection process. Even when these optical depths are small -- but especially when they are not -- a self-consistent approach that models radiation and reconnection simultaneously is required to capture the modifications that the various radiative interactions may make to the reconnection-powered photon spectrum. In this paper, we term any regime of this kind, in which reconnection and photon processes are inextricably coupled, as a \quoted{radiative}regime of magnetic reconnection \citep{u16}.

Due to high magnetic and radiation energy densities, reconnection in relativistic compact object environments is likely to be highly radiative, and much recent work on radiative reconnection is related to these systems, including studies of:
pulsar winds \citep{p12, cp17, cpd20}, pulsar wind nebulae \citep{ucb11, cub12, cwu13, cwu14b, cwu14a, ynz16}, pulsar magnetospheres \citep{l96, us14, psc15, cps16, ps18_philippov, hps19}, magnetar magnetospheres \citep{sgu19}, gamma-ray bursts \citep{mu12}, accreting black holes \citep{b17, wpu19, sb20}, blazars \citep{ngb11, nbc12, nyc18, on20, mwu20}, and black hole magnetospheres \citep{ppc19, ccd20}.
Some studies have not focused on a single object class, but have still been motivated by some combination of the above \citep[e.g.][]{jh09, um11, u11, u16, hps20, nb20}.
 
In many astrophysical contexts, the reconnection region is expected to be illuminated by an external source of soft photons -- with energies much lower than the electron temperature -- and inverse Compton (IC) scattering of these photons dominates the emitted light. For some such systems, observations further suggest that particles emitting at the highest photon energies do so in the Klein-Nishina regime. For example, TeV observations of the flat-spectrum radio quasar (FSRQ) PKS~1222+21 \citep{magic11} indicate that the observed TeV photons, if Comptonized from radiation impinging on the jet from a hot dust region, are produced in the marginal Klein-Nishina regime of the IC process \citep{mwu20}. Thus, if the radiating particles are accelerated by relativistic reconnection (for which a case has been made by \citealt{nbc12} and \citealt{mwu20}), then it is likely that the collective reconnection dynamics are significantly impacted by Klein-Nishina IC effects. In an entirely different type of system, an X-ray binary, recent \textit{Fermi} observations reveal that the spectral \cutoff[]in the quiescent high-luminosity state of Cyg X-1 lies in the $40$-$80$ MeV range \citep{zmc17}. As in the case of FSRQs, this hints that the most energetic particles Comptonize ambient photons (sourced, in this case, by the accretion disc) in the Klein-Nishina regime.

In addition to qualitatively modifying the radiative cooling experienced by particles, Klein-Nishina physics also has important consequences for the scattered photons. When Comptonized deep in the Klein-Nishina limit, these may go on to pair produce with their parent population of ambient seed photons. Therefore, an astrophysically relevant treatment of reconnection with Klein-Nishina Compton cooling must also account for pair production.

\begin{figure}
    \centering
    \includegraphics[width=\columnwidth, trim=0 0 105 0, clip]{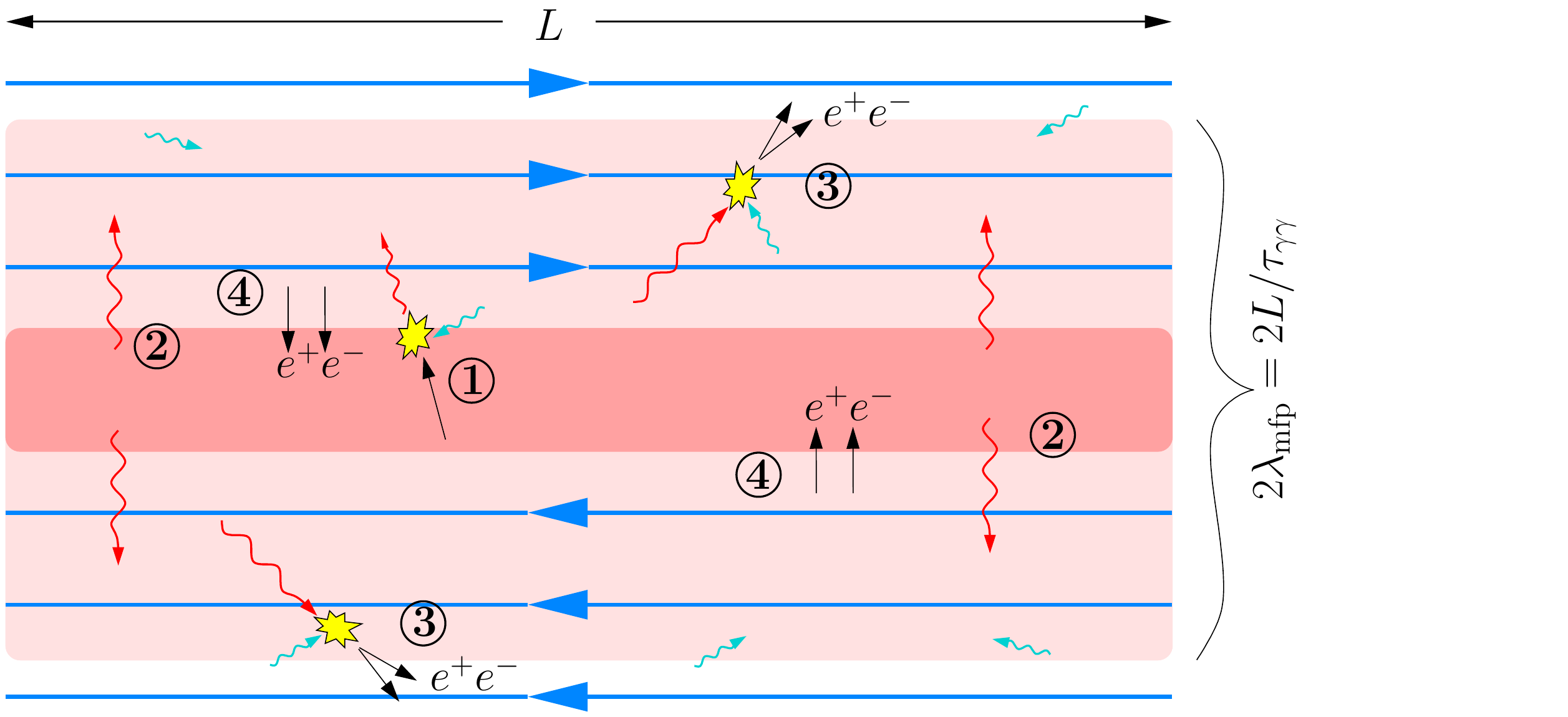}
    \caption{Schematic of Klein-Nishina radiative magnetic reconnection when the system is optically thick~$\tau_{\gamma\gamma} \gg 1$ to pair-production. In this regime, the mean free path~$\lmfp$ of a high-energy photon is less than the system size~$L$:~$\lmfp = L / \tau_{\gamma\gamma} \ll L$. Thus, every high-energy photon (red wiggled arrows) annihilates against an ambient seed photon (cyan wiggled arrows) before escaping the system. Newborn pairs are denoted by black arrows and magnetic field lines by blue arrows. The reconnection layer is opaque pink; the region penetrated by high-energy photons (one mean free path away from the layer) is transparent pink. See text for a description of stages~$1-4$.}
    \label{fig:knppdiagram}
\end{figure}

So motivated, we provide, in this study, an analytic model for a relatively unexplored regime of radiative reconnection: the \textit{pair-regulated Klein-Nishina regime}. Our model hinges on a self-regulation mechanism that we diagram in Fig.~\ref{fig:knppdiagram} and describe below. In that description as throughout this text, the term \quoted{reconnection layer}(sometimes just \quoted[)]{layer}refers to the region permeated by reconnected magnetic flux; the term \quoted{upstream}(sometimes \quoted[)]{inflow}refers to the region filled with unreconnected flux.

First (step~$1$ in Fig.~\ref{fig:knppdiagram}), particles accelerated in the reconnection layer Comptonize ambient seed photons to gamma-ray energies. Second (step~$2$ in Fig.~\ref{fig:knppdiagram}), IC-produced gamma-rays penetrate into the upstream region about one pair-production mean free path from the layer. While propagating, these photons are immune to secondary IC scattering because the Thomson optical depth~$\tau_{\rm T}$ is very small (even though the pair-production optical depth~$\tau_{\gamma\gamma}$ exceeds unity). In step~$3$, high-energy photons are absorbed by the background radiation, producing pairs in the upstream plasma. Newborn pairs are then advected toward the layer. While en route, they radiatively cool, and thus some of their initial energy never returns to the layer. Nevertheless, the created pairs remain hot enough that their energy density dominates that of the originally present colder upstream particles. Thus, the plasma feeding the layer in step~$4$ possesses a reduced magnetization -- the ratio of magnetic energy density to total (original~$+$ hot pairs) matter enthalpy density. This inhibits particle acceleration and subsequent photon emission in the reconnection layer, closing the negative feedback loop. 

Our model predicts two types of pair-regulated Klein-Nishina reconnection dynamics. If only a small fraction of the energy radiated away from the layer is recaptured as hot pairs, reconnection enters a steady state \characteriz[ed]by a \textit{universal} (independent of the initial value) pair-regulated magnetization. However, if the nearly all of the radiated energy gets swept back into the layer as pairs, the steady state is never realized. Instead, the system overshoots its theoretical fixed point solution, getting caught between two extreme magnetization states. In these \quoted[,]{swing cycles}the high magnetization state yields efficient above-threshold photon emission from the layer and subsequent injection of hot pairs into the upstream region. This initiates a very low/pair-loaded magnetization. Here, pair production is quenched until the created pairs vacate the inflow plasma by entering the layer, restoring the high magnetization. 

In both a steady state and a swing cycle, the created particles, when present, dominate the upstream pressure, but a prolific pair cascade is not expected. The (power-law) distribution of pairs injected into the inflow region, though potentially quite broad, is too steep for later pair generations -- born from photons emitted by earlier upstream generations -- to outnumber the first generation. Furthermore, for a wide range of parameters, the newborn pairs are also few in number relative to the original plasma particles.

This last aspect of our model qualitatively departs from earlier treatments of radiative reconnection with pair-production \citep[e.g.][]{l96, u11, hps19}. Rather than being dressed in a coat of pairs that dominates both the upstream matter energy and number densities, the reconnection layer in this regime is self-consistently fed by a few high-energy newborn particles that control only the energy, and hence the magnetization, of inflowing material.

Our model also predicts the power-law index of the particle energy distribution yielded by Klein-Nishina reconnection. However, this index is sensitive to our free parameters and simplifying assumptions. Therefore, although the present analysis holds promise for making contact with observed features (e.g.\ the photon power-law index) of FSRQs and black hole accretion disc coronae (ADCe), future numerical studies, for which this work lays the foundation, are necessary to refine the model and make testable predictions.

Our discussion is \organiz[ed]{}as follows. In section~\ref{sec:thomradreconn}, we review some of the salient features of radiative magnetic reconnection when low-energy Thomson IC scattering is the dominant radiative process. We focus here on \organiz[ing]{}the relevant energy scales in the reconnection problem, classifying different radiative regimes from the hierarchy of these scales. This provides a jumping-off point from which to \generaliz[e]{}our discussion to Klein-Nishina IC losses in section~\ref{sec:knradreconn}. After introducing scales and classifying regimes of Klein-Nishina reconnection in that section, we present our model, which pertains to a subset of the available scale hierarchies, in section~\ref{sec:knradreconn_mod}. We comment on the model's observable features and its implications for FSRQ jets and ADCe in the high-luminosity states of black hole X-ray binaries in section~\ref{sec:applications}. We conclude in section~\ref{sec:conclusions}.

\section{Review of Thomson radiative reconnection}
\label{sec:thomradreconn}
Although a variety of radiation mechanisms (e.g. synchrotron emission) may impact the dynamics of relativistic collisionless magnetic reconnection, in this work, we specialize to the case where external IC scattering is the dominant and most dynamically consequential radiative channel. In the present section, we further restrict our discussion to the case where the IC process occurs in the low-energy Thomson limit (defined below). This allows us both to review some of the principal results applicable in this regime and to assemble a set of energy scales that will anchor our exploration of Klein-Nishina radiative reconnection in subsequent sections.

\subsection{Single-particle Thomson IC cooling}
To begin, we review Thomson inverse Compton radiative cooling at the level of individual particles. Afterwards, we extend the discussion to include collective effects governed by the interplay between radiation and reconnection.

We consider a static, homogeneous, isotropic bath of ambient radiation with spectral energy density~$u_{\rm ph}(\epsilon)$. An ultrarelativistic electron or positron with energy~$E = \gamma \me c^2 \gg \me c^2$ traversing this radiation field preferentially IC-scatters photons that appear strongly blueshifted in its rest frame: the lab-frame photon energy~$\epsilon$ transforms to the rest frame energy~$\epsilon' \sim \gamma \epsilon$ (except for a negligible few photons that travel within an angle~$1 / \gamma \ll 1$ of the particle's velocity and so transform to much smaller energies). If~$\epsilon' \ll \me c^2$, then the encounter reduces to Thomson scattering, which is approximately elastic, yielding final (scattered) rest-frame photon energy~$\epsilon'_{\rm f} \simeq \epsilon'$. Moreover, the Thomson differential cross section lacks a strong angular dependence. Thus, very few photons are emitted to within angle~$1 / \gamma$ opposite the scattering particle's velocity, and the vast majority of scatterings yield typical final lab-frame photon energy~$\epsilon_{\rm f} \sim \gamma \epsilon'_{\rm f} \sim \gamma^2 \epsilon$.

The same assumptions that place us in the Thomson regime also imply that the particle loses a very small fraction of its energy in one collision:~$\epsilon_{\rm f} / \gamma \me c^2 \sim \epsilon' / \me c^2 \sim \gamma \epsilon / \me c^2 \ll 1$. This allows IC radiative losses to be treated classically -- as a continuous drag force,~$\pmb{f}_{\rm T} = -(4/3) \sigma_{\rm T} \gamma^2 \uph \pmb{\beta}$, where~$\sigma_{\rm T} = 8 \pi e^4 / 3 \me^2 c^4$ is the Thomson cross section,~$\uph \equiv \int \dif \epsilon \, u_{\rm ph}(\epsilon)$ is the total background radiation energy density, and~$c \pmb{\beta}$ is the scattering particle's velocity vector \citep[cf.][]{bg70, rl79, p82, pss83, u16, wpu19, sb20, mwu20}. Importantly,~$\pmb{f}_{\rm T}$ depends only on~$\uph$ and not on the spectral distribution~$u_{\rm ph}(\epsilon)$ of IC seed photons. Thus, provided a plasma radiates purely in the Thomson IC regime and is also optically thin to Thomson scattering, the collective dynamics are insensitive to the incident spectrum (though the \textit{Comptonized} spectrum is not). The Thomson IC radiated power per particle is
\begin{align}
    P_{\rm T}(\gamma) = \abs{c \pmb{\beta} \cdot \pmb{f}_{\rm T}} = \frac{4}{3} \sigma_{\rm T} c \beta^2 \gamma^2 \uph \, .
    \label{eq:pthom}
\end{align}

All of the features discussed here -- continuous emission,~$\epsilon_{\rm f} \propto \gamma^2$, and incident-spectrum independence -- change significantly when we later allow particles to experience general Klein-Nishina Compton losses. However, before we get there, let us move on from this single-particle picture of IC cooling and chart out how Thomson radiative cooling impacts the collective reconnection dynamics.

\subsection{Thomson IC effects on collective plasma behavi\spellor}
\label{sec:collicthom}
We first introduce a set of parameters, cast as particle energy scales (Lorentz factors), that \characteriz[e]Thomson IC scattering in the context of magnetic reconnection. We then examine the radiative regimes represented by the possible scale hierarchies. 

\subsubsection{Reconnection energy scales}
The reconnecting magnetic field,~$B_0$, can be recast in terms of a length scale: the nominal relativistic gyroradius~$\rho_0 \equiv \me c^2 / e B_0$. This is a useful form for comparing~$B_0$ with other length scales in the problem. For example, introducing the length of the reconnection layer~$L$ (see Fig.~\ref{fig:knppdiagram}), one may define a \quoted[:]{Hillas criterion}the energy,
\begin{align}
    \gamma_{\rm Hillas} \equiv \frac{e B_0 L}{\me c^2} = \frac{L}{\rho_0} \, ,
    \label{eq:ghillas}
\end{align}
of a particle with Larmor radius~$\gamma_{\rm Hillas} \rho_0$ equal to the system size \citep{h84}.\footnote{For simplicity, we assume that~$L$ \characteriz[es]the size of the system in all spatial dimensions in addition to its horizontal length (Fig.~\ref{fig:knppdiagram}).} Equivalently,~$\gamma_{\rm Hillas}$ is the energy imparted to a particle accelerated across the system by an electric field of strength~$B_0$.

While~$\gamma_{\rm Hillas}$ gives a firm upper bound on the achievable particle energies, a more practical scale for reconnection problems is
\begin{align}
    \gmax \equiv 0.1 \gamma_{\rm Hillas} \, ,
    \label{eq:gmax}
\end{align}
corresponding to extreme acceleration \citep[][]{abd02, ucb11, cub12}.
We use~$\gmax$ instead of~$\gamma_{\rm Hillas}$ because the reconnection electric field,~$E_{\rm rec}$, is not quite as strong as~$B_0$, but about equal to~$\beta_{\rm rec} B_0$ where~$\beta_{\rm rec}$ is the dimensionless reconnection rate. For collisionless reconnection,~$\beta_{\rm rec}$ can be expressed in terms of the Alfv\'en speed~$v_{\rm A}$ as~$\beta_{\rm rec} \simeq 0.1 v_{\rm A} / c$. In the relativistic limit treated in this paper,~$v_{\rm A} \simeq c$ and hence~$\beta_{\rm rec} \simeq 0.1$.

It is sometimes helpful to equivalently define~$\gmax$ by equating \ts[s.]We therefore introduce the acceleration time for a particle being linearly accelerated by the electric field~$E_{\rm rec} \simeq 0.1 B_0$:
\begin{align}
    t_{\rm X}(\gamma) \equiv \frac{\gamma \me c^2}{e c E_{\rm rec}} = \frac{\gamma \me c^2}{\beta_{\rm rec} e c B_0} \simeq \frac{10 \gamma \rho_0}{c} \simeq 1.6 \frac{2 \pi \gamma \rho_0}{c} \, ,
    \label{eq:tx}
\end{align}
equal to~$10 / 2 \pi \simeq 1.6$ cyclotron periods in the magnetic field~$B_0$.
Here, subscript \quoted{X}denotes that the spatial regions where this type of linear acceleration is active usually surround magnetic X-points -- locations where the magnetic field reconnects and near which particles become unmagnetized \citep[e.g.][]{ucb11, cub12}. Equivalently to~(\ref{eq:gmax}), one may define~$\gmax$ such that a particle is X-point-accelerated over one system \lc[]time:
\begin{align}
    t_{\rm X}(\gmax) \equiv \frac{L}{c} \, .
    \label{eq:gmaxalt}
\end{align}

Another important reconnecting system parameter is the (combined electron~$+$ positron) upstream plasma density~$n_0$. With~$B_0$ fixed,~$n_0$ can be cast in terms of the (dimensionless) upstream cold magnetization
\begin{align}
    \sigc \equiv \frac{B_0^2}{4 \pi n_0 \me c^2} \, ,
    \label{eq:sigcup}
\end{align}
Physically,~$\sigc$ represents the initial magnetic energy per particle:~$\sigc \me c^2 / 2$. Since reconnection delivers an appreciable fraction of the magnetic energy to the plasma,~$\sigc$ also \characteriz[es]the average Lorentz factor~$\langle \gamma \rangle$ of reconnection-energized particles (absent radiative cooling). Assuming that half of the initial magnetic energy is dissipated, we have~$\langle \gamma \rangle \sim \sigc / 4$ \citep[cf.][]{spg15, wuc16, sb20}. In this way, just like~$\gmax$ furnishes a characteristic particle energy scale to stand in for the system size~$L$, the cold magnetization provides an energy scale that acts as a proxy for the upstream number density~$n_0$.

We report here, for reference, one more important dimensionless parameter, the hot magnetization \citep[cf.][]{mwf14a, wub18}
\begin{align}
    \sigh \equiv \frac{B_0^2}{4 \pi w_0} \, .
    \label{eq:sighdef}
\end{align}
Here,~$w_0 = p_0 + u_0$ is the relativistic plasma enthalpy density with~$p_0$ the upstream pressure and~$u_0$ the internal energy density. For a relativistically hot upstream plasma with temperature~$\theta_0 = T_0 / \me c^2 \gg 1$ the enthalpy density is~$w_0 \simeq 4 \theta_0 n_0 \me c^2$~($p_0 = \theta_0 n_0 \me c^2 = u_0 / 3$); for a cold~$\theta_0 \ll 1$ plasma,~$w_0 \simeq u_0 \simeq n_0 \me c^2$ is dominated by rest-mass energy. Thus, in the relativistically hot case,~$\sigh \simeq 1 / (2 \beta_{\rm pl}) = \sigc / (4 \theta_0) \ll \sigc$ where~$\beta_{\rm pl} = 8 \pi p_0 / B_0^2$ is the plasma beta parameter, but in the opposite limit,~$\sigh \simeq \sigc$. Physically, the hot magnetization determines whether the energy flux into the reconnection region is dominated by the magnetic field~($\sigh \gg 1$) or by the matter~($\sigh \ll 1$). In the former~$\sigh \gg 1$ limit, the Alfv\'en speed,~$v_{\rm A} = c \sqrt{\sigh/(1 + \sigh)}$, approaches~$c$ and, as a result, magnetic reconnection may drive not only relativistic individual particle motion (which merely requires~$\sigc \gg 1$) but also relativistic bulk flows. Thus,~$\sigh \gg 1$ is really the defining feature of \quoted{relativistic}reconnection. In the spirit of our present discussion, one may regard~$\sigh$ as a proxy for the upstream temperature~$\theta_0$, especially when~$\theta_0 \gg 1$. However, we do not make explicit use of~$\sigh$ for some time (until section~\ref{sec:knradreconn_mod}). For now, we simply assume that we are in the relativistic limit of reconnection,~$\sigh \gg 1$, and, within that context, scope out the possible regimes by ordering our other important energy scales (which so far include~$\gmax$ and~$\sigc$). 

Armed with~$\gmax$ and~$\sigc$ (and also assuming~$\sigh \gg 1$), we can describe the size of the relativistic reconnection system. A \quoted{large}system satisfies~$\gmax \gg \sigc$. In terms of the global geometry, this implies a high aspect ratio: the system is much longer than the microscopic current sheet thickness, of order~$\langle \gamma \rangle \rho_0$ -- the typical Larmor radius of reconnection-energized particles (specifically,~$\gmax \gg \sigc \Rightarrow L \gg 10 \sigc \rho_0$; cf.\ \citealt{wuc16}). This renders the layer plasmoid-unstable and initiates plasmoid-dominated reconnection \citep[e.g.][]{jd11}.

In terms of individual particles,~$\gmax \gg \sigc$ also alleviates system-size constraints on particle energization, at least up to the mean Lorentz factor~$\langle \gamma \rangle \sim \sigc/4$. In addition, when~$1 \ll \theta_0 \ll \langle \gamma \rangle \sim \sigc / 4$ and, hence,~$\sigh \gg 1$, previous non-radiative 2D particle-in-cell (PIC) simulations have found that direct/fast acceleration by the reconnection electric field saturates at~$\gx \sim (\mathrm{several}) \sigc$ \citep[e.g.~$4 \sigc$;][]{wuc16, knp18}. Thus, when~$\gmax \gg \sigc$, both the physics governing average-energy particle acceleration and the direct acceleration mechanism are unencumbered by the system size.

However, we remark that there are, in addition to primary X-point acceleration, other energization channels in the large-system plasmoid-dominated regime: most of them take place on slower \ts[s]but are not limited to Lorentz factors of order several~$\sigc$. We term these \quoted{secondary}acceleration processes because they typically operate on plasma that has already been processed into the layer (the region of reconnected magnetic flux). One example of such a process is adiabatic heating by slowly compressing reconnected magnetic fields inside plasmoids \citep{ps18, hps20}. A different, but related, example is a Fermi-type mechanism \citep{dsc06, dds14, gld14, gld15, gll16, gld19, gld20} in which particles bounce from end to end across contracting plasmoids. (See \citealt{u20} for a review of secondary acceleration mechanisms in~2D reconnection.) The existence of such acceleration channels may, without radiative cooling, allow the highest particle energies to grow well beyond~$\gx$ -- even if the dynamics of average particles and X-point acceleration top out at much lower energies. This is one aspect where even weakly radiative reconnection differs from its non-radiative counterpart. Radiative losses can impose a system-size-independent high-energy \cutoff[]even on acceleration processes with no intrinsic upper energy limit \citep[e.g.][]{hps20, mwu20}, potentially allowing the spectrum of accelerated particles to become independent of~$L$. With this in mind, we now turn to quantifying the impact of Thomson IC radiative cooling on reconnection. 

As discussed above, particles emitting in the Thomson limit suffer a drag force~$\pmb{f}_{\rm IC}$ determined by the total background radiation energy density~$\uph$. Thus, Thomson radiation introduces just one extra parameter,~$\uph$, into the reconnection problem. To cast~$\uph$ as an energy scale, we define the Lorentz factor,~$\gradt$, at which radiative drag matches the acceleration force from the reconnection electric field,~$e E_{\rm rec} \simeq 0.1 e B_0$, or (equivalently) such that the X-point acceleration time,~$t_{\rm X}(\gamma)$, equals the Thomson cooling time,~$\tcoolt(\gamma) \equiv \gamma \me c^2 / P_{\rm T}(\gamma)$, over which a particle radiates a significant fraction of its energy~$\gamma$. Putting~$f_{\rm IC}(\gradt) \equiv 0.1 e B_0$ [or~$t_{\rm X}(\gradt) \equiv \tcoolt(\gradt)$] yields \citep[cf.][]{u16, n16, wpu19, sb20, mwu20}
\begin{align}
    \gradt \equiv \sqrt{\frac{0.3 e B_0}{4 \sigma_{\rm T} \uph}} \, .
    \label{eq:gradt}
\end{align}
There is also a third way to define~$\gradt$: it is the Lorentz factor of a particle that cools in~$\sim 1$ cyclotron period in the magnetic field~$B_0$ \citep[cf.][]{u16}.

Despite the fact that it is defined only by balancing radiative cooling against X-point energization, the energy~$\gradt$ firmly radiatively caps the achievable particle energies \citep{wpu19, mwu20}. This is because acceleration by the coherent reconnection electric field may be the fastest significant particle energization channel in magnetic reconnection.\footnote{It is unclear whether motional electric fields~$\pmb{E}_{\rm motion} = -\pmb{v} \times \pmb{B} / c$~($v \simeq v_{\rm A} \simeq c$) -- e.g.\ due to rapidly moving compact plasmoid cores where~$B > B_0$ and, hence,~$E_{\rm motion} \sim v_{\rm A} B / c > B_0$ -- yield an overall faster effective acceleration than X-point regions. A similar remark applies to momentary pulses of high electric fields associated with waves launched at plasmoid mergers \citep{pus19}. In both cases, the issue is not just one of field strength but also of spatio-temporal coherence.} Other channels \citep[such as those discussed by][]{ps18, gld19, hps20, gld20}, while not saturating at~$\gx$ like X-point acceleration, are much slower and hence radiatively stall at Lorentz factors less than~$\gradt$ \citep[e.g.][]{mwu20}.

To sum up, we now have three energy scales, each \characteriz[ing]different physical parameters in relativistic reconnection. Two of these,~$\gmax$ and~$\sigc$ -- representing, respectively, the system size~$L$ and the upstream particle density~$n_0$ -- are non-radiative, common to all reconnection problems. The third scale,~$\gradt$, encodes the energy density of ambient radiation and is unique to reconnection with IC cooling. Furthermore, one of these energies,~$\sigc$, splits into two: it represents both the average energy of reconnection-energized particles~$\langle \gamma \rangle \sim \sigc / 4$ and the intrinsic maximum energy deliverable by the X-point acceleration mechanism~$\gx \sim 4 \sigc$ in the large-system, plasmoid-mediated regime. We later argue (section~\ref{sec:kbprospects}) that, under some circumstances,~$\gx$ may exceed the nominal value~$\gx = 4 \sigc$ found from 2D PIC simulations \citep{wuc16, knp18}, but in the case where~$\gx$ truly is of order~$4 \sigc$,~$\gx$ and~$\langle \gamma \rangle$ are only offset by about a factor of~$16$.

\subsubsection{Regimes of Thomson radiative reconnection}
\label{sec:thomregimes}
Next, we enumerate the various orderings of the scales discussed above and examine the physical regimes each ordering represents. To simplify this program, we concentrate solely on the large-system regime with a relativistic amount of magnetic energy per particle:~$\gmax \gg \sigc \gg 1$ (or, if we split the scale~$\sigc$ into~$\gx \sim 4 \sigc$ and~$\langle \gamma \rangle \sim \sigc / 4$, the regime~$\gmax \gg \gx \gg \langle \gamma \rangle \gg 1$). Different scale hierarchies are then realized by inserting~$\gradt$ into various positions of the base ordering~$\gmax \gg \gx \gg \langle \gamma \rangle \gg 1$. The possible orderings are summarized in Table~\ref{table:tregimes}.
\begin{table*}
\centering
\begin{threeparttable}
\caption{Scale hierarchies and associated radiative regimes of relativistic reconnection subject to Thomson IC losses. A relativistic large-system ordering~$\gmax > \gx > \langle \gamma \rangle > 1$ is assumed throughout. Inserting~$\gradt$ and~$\gcool$ at different locations in this base ordering corresponds to different regimes. To guide the eye, the \quoted{independent}parameters (i.e.\ independent of~$\gmax$,~$\gx$, and~$\langle \gamma \rangle$ but not of each other)~$\gradt$ and~$\gcool$ are typeset in red. The regimes are semantically distinguished based on which particles -- either those in the bulk of the particle energy distribution near the average energy~$\langle \gamma \rangle$ or those in the high-energy tail -- cool on \ts[s]shorter than~$L/c$ (strong cooling) versus those that cool in one gyroperiod (saturated cooling). For Thomson IC radiation reaction (but not in the more general Klein-Nishina limit) saturated cooling is more efficient than, and therefore implies, strong cooling.}
\label{table:tregimes}
\begin{tabular}{llcccc}
    \toprule
    & & \multicolumn{2}{c}{Bulk particles} & \multicolumn{2}{c}{High-energy particles} \\
    \cmidrule(lr){3-4} \cmidrule(lr){5-6}
    & & Strong & Saturated & Strong & Saturated \\
    Scale hierarchy & Regime name & cooling (Y/N) & cooling (Y/N) & cooling (Y/N) & cooling (Y/N) \\
    \midrule
    $\textcolor{red}{\gcool} > \textcolor{red}{\gradt} > \gmax > \gx > \langle \gamma \rangle$ & Non-radiative & N & N & N & N \\
    $\gmax > \textcolor{red}{\gradt} > \textcolor{red}{\gcool} > \gx > \langle \gamma \rangle$ & Quasi non-radiative & N & N & Y\tnote{*} & N \\
    \midrule
    $\gmax > \textcolor{red}{\gradt} > \gx > \textcolor{red}{\gcool} > \langle \gamma \rangle$ & & N & N & Y & N \\
    $\gmax > \textcolor{red}{\gradt} > \gx > \langle \gamma \rangle > \textcolor{red}{\gcool}$ & & Y & N & Y & N \\
    $\gmax > \gx > \textcolor{red}{\gradt} > \textcolor{red}{\gcool} > \langle \gamma \rangle$ & & N & N & Y & Y \\
    $\gmax > \gx > \textcolor{red}{\gradt} > \langle \gamma \rangle > \textcolor{red}{\gcool}$ & & Y & N & Y & Y \\
    $\gmax > \gx > \langle \gamma \rangle > \textcolor{red}{\gradt} > \textcolor{red}{\gcool}$ & Extremely radiative & Y & Y & Y & Y \\
    \bottomrule
\end{tabular}
\begin{tablenotes}
\item[*]Highest-energy particles may or may not achieve strong cooling. If they do, it is not through impulsive X-point acceleration (see text).
\end{tablenotes}
\end{threeparttable}
\end{table*}

This procedure is made more conceptually transparent if we introduce the derived scale~$\gcool$, the Lorentz factor of a particle that cools in one dynamical time~$L/c$ of the system. Writing~$\tcoolt(\gcool) \equiv L / c$ yields
\begin{align}
    \gcool \equiv \gradt^2 / \gmax \, .
    \label{eq:gcool}
\end{align}
Interestingly, the radiatively limited Lorentz factor~$\gradt$ is always intermediate between~$\gcool$ and~$\gmax$, equal to the geometric mean of those two scales.
Note that one may have~$\gcool < 1$, in which case~$\gcool$ does not correspond to a physical Lorentz factor. In that case, all particles cool to non-relativistic energies in~$\tcoolt(1) < L / c$. Related to~$\gcool$ is the compactness of the system
\begin{align}
    \ell \equiv \frac{\uph \sigma_{\rm T} L}{\me c^2} = \frac{3}{4 \gcool} \, .
    \label{eq:compact}
\end{align}
The time for a particle to cool from any initial~$\gamma$ to~$\gamma = 1$ is~$\sim L / c \ell$.

We begin our exploration of the various radiative regimes by quantifying the \textit{non-radiative} limit~$\gcool > \gradt > \gmax \gg \sigc \gg 1$. Hereafter, we do not list~$\sigc \gg 1$ explicitly. We also only use~\quoted{$>$}(not~\quoted[)]{$\gg$}symbols, with the understanding that all regimes become more distinct when the corresponding scales are well-separated. The regime~$\gcool > \gradt > \gmax > \sigc$ corresponds to the limit~$\uph \to 0$ and, hence,~$\gradt \to \infty$. Here, no particle radiates a significant fraction of its energy within one dynamical time~$L/c$, effectively decoupling radiation from reconnection. This is the regime mentioned in the Introduction where magnetic reconnection can, in principle, be studied on its own and the radiative signatures calculated independently.

The first step up in radiative efficiency might be called the \textit{quasi non-radiative regime}~$\gmax > \gradt > \gcool > \gx > \langle \gamma \rangle$. Here, primary X-point acceleration does not impart enough energy to particles so that they significantly radiate on one dynamical time~$L/c$, let alone so that they achieve radiative saturation~$\gradt$. Secondary acceleration channels, on the other hand, might be able to deliver particles to energies~$\geq \gcool$ so that those particles radiate faster than the global \ts[.]However, this depends on the detailed nature of each secondary acceleration process -- whether any of them radiatively stall above~$\gcool$ is not guaranteed. 

As an example of a secondary acceleration mechanism relevant to the quasi non-radiative regime, one may consider particles slowly energized inside of adiabatically compressing magnetic islands (also \quoted[),]{plasmoids}as detailed by \citet{ps18} and \citet{hps20}. The energy where radiative losses shut this process down is determined by matching the plasmoid compression time to the particle cooling \ts[.]As reported by \citet{hps20}, this upper-limit energy is~$\gamma_{\rm sec} \sim \me c^2 \beta_{\rm rec} / w \sigma_{\rm T} \uph$, where~$w \simeq 0.1 L$ is the size of the largest plasmoids formed by reconnection. [We ignore that smaller plasmoids compress faster and so may yield, for the smaller number of particles they contain, higher~$\gamma_{\rm sec}$ \citep{hps20}.] In effect,~$\gamma_{\rm sec} \sim \ell^{-1} = (4/3)\gcool$. Thus, this secondary energization channel only barely accelerates some particles up to~$\gcool$, and most reach energies much less than this. Hence, considering only this secondary mechanism, the reconnection process can be regarded as marginally non-radiative, with only very few highest-energy particles cooling in less than one dynamical time.

Increasing the cooling efficiency once more brings us to the first of several truly \textit{radiative} regimes where radiation is dynamically important for at least some of the particles. Aptly naming these regimes is cumbersome because different particles can experience varying degrees of radiative efficiency. The high-energy particles, for example, may be rapidly cooled and the particles at the average energy~$\langle \gamma \rangle$ cooled quite slowly. Therefore, we do not classify these regimes globally, calling the entire system weakly or strongly radiative, but we refer to them based on which populations of particles cool on various \ts[s.]

We call particles \textit{strongly cooled} if radiation reaction causes them to lose an appreciable fraction of energy in less than one dynamical time~$L/c$ -- i.e.\ if their Lorentz factors exceed~$\gcool$. This agrees with typical notions of strong cooling in astrophysics, which indicate that radiative cooling occurs faster than some macroscopic system \ts[.]Correspondingly, we call particles with~$\gamma < \gcool$ \textit{weakly cooled} (because they are not strongly cooled) and sometimes \textit{non-radiative} (because they do not radiate appreciably in a dynamical time). The particles with much higher energies, close to~$\gradt > \gcool$, we say exhibit \textit{saturated cooling}: their Lorentz factors are radiatively saturated because intense emission prevents further energization~($\tcoolt = t_{\rm X}$). Although, in the Thomson IC limit, particles undergoing saturated cooling radiate much more efficiently than just strongly cooled particles, this is not always true once Klein-Nishina effects come into play (see section~\ref{sec:collickn} and Fig.~\ref{fig:tcoolic}). Thus, we wish to avoid associating the saturated cooling regime with a term connoting excessively efficient or fast cooling (e.g.\ \quoted[).]{very strong cooling} 

Using these terms, we see that the scale hierarchy~$\gmax > \gradt > \gx > \gcool > \langle \gamma \rangle$ indicates that average (or \quoted[)]{bulk}particles are weakly radiative; most of them cool slower than~$L/c$~($\langle \gamma \rangle < \gcool$). Meanwhile, at least some of the high-energy particles accelerated by the primary X-point channel are strongly cooled~($\gx > \gcool$). Even so, radiative losses are not so fast as to hinder direct X-point acceleration, with~$t_{\rm X}(\gx)$ faster than~$\tcoolt(\gx)$ because~$\gx < \gradt$. Thus, X-point acceleration (because it is intrinsically capped to below~$\gradt$) -- and, hence, all other (known) secondary energization channels (because they are slow) -- cannot deliver particles up to the radiative saturation limit~$\gradt$.

Permuting scales again by swapping the positions of~$\gcool$ and~$\langle \gamma \rangle$, we arrive in the regime~$\gmax > \gradt > \gx > \langle \gamma \rangle > \gcool$. Here, most particles radiate strongly because~$\gcool < \langle \gamma \rangle < \gx$. However, like in the previous regime, virtually no particles are expected to achieve radiative saturation~($\gx < \gradt$), and X-point acceleration, while unaffected by radiative cooling on the short \ts[] on which it occurs~[$t_{\rm X}(\gx) < \tcoolt(\gx)$], does produce particles of sufficiently high energies~($> \gcool$) to be strongly cooled.
  
Next we arrive at scale hierarchies where at least a few particles exhibit saturated cooling. One such domain is~$\gmax > \gx > \gradt > \gcool > \langle \gamma \rangle$. Here, some particles are promptly accelerated near X-points to the upper-limit energy~$\gradt < \gx$, but the bulk particles, with~$\langle \gamma \rangle < \gcool$, barely radiate even on global~$L/c$ \ts[s.]This is perhaps the most extreme example of how vastly different the cooling rates can be for different reconnection-energized particles. However, unless~$\gx$ can substantially exceed its nominal~$4 \sigc$ value, this regime may not be realized in astrophysical contexts. This is because, if~$\gx = 4 \sigc$, then~$\gx / \langle \gamma \rangle \simeq 16$, implying that~$\gradt / \gcool < 16$, and, through equation~(\ref{eq:gcool}), that~$\gmax / \gradt < 16$, whereas~$\gmax$,~$\gradt$, and~$\gcool$ are each usually separated by several decades in astrophysical systems (see section~\ref{sec:astroestimates}).

Moving to the last two possible orderings, we have~$\gmax > \gx > \gradt > \langle \gamma \rangle > \gcool$, in which the high-energy particles accelerated near X-points attain radiative saturation~($\gradt < \gx$) and the bulk particles are strongly cooled~($\gcool < \langle \gamma \rangle$). Finally, there is an \textit{extremely radiative} regime,~$\gmax > \gx > \langle \gamma \rangle > \gradt > \gcool$. Here, IC losses firmly cap the acceleration of nearly all particles -- not even the formal mean energy,~$\langle \gamma \rangle$, available per particle can be attained -- and should have dramatic effects on the large-scale reconnection dynamics \citep{u16}. Table~\ref{table:tregimes} summarizes the radiative reconnection regimes discussed in this section.

To complete our tour of the Thomson IC reconnection landscape, we review some of the previously identified physical effects that occur in these regimes. Several systematic PIC studies of Thomson radiative reconnection have been conducted in recent years, including those by \citet{wpu19}, \citet{sb20}, and \citet{mwu20}. There have also been radiative PIC studies of reconnection with strong synchrotron cooling \citep[e.g.][]{cwu13, cwu14b, cwu14a, ynz16} and QED effects \citep[like pair-production;][]{sgu19, hps19}. In some cases, the qualitative features of reconnection regimes mediated by different radiative processes are similar, but, in this section, which is intended chiefly as a jumping-off point for our more general discussion of Klein-Nishina physics to come, we focus only on those effects studied within the context of Thomson IC losses.

The three radiative PIC studies conducted by \citet{wpu19}, \citet{sb20}, and \citet{mwu20} explored a number of regimes outlined in this section. \citet{wpu19} studied the effect of Compton losses on large-scale reconnection dynamics and on \nonthermal[]particle acceleration, exploring all the way from the non-radiative regime to that of fully saturated high-energy cooling and strong bulk cooling~($\gmax > \gx > \gradt > \langle \gamma \rangle > \gcool$). They found that radiation steepens the high-energy part of the \nonthermal[]tail of reconnection-accelerated particles but that the overall reconnection rate is virtually unaffected. \citet{mwu20} focused on the angular distributions of high-energy particles, showing that particles approaching radiative saturation~($\gradt \lesssim \gx$) exhibit energy-dependent collimation in momentum space, forming narrow beams at the highest energies. Strong radiative losses thus appear to be an essential ingredient in mediating this \quoted{kinetic beaming}effect, which was first discovered by \citet{cwu12}. Numerically exploring the scenario first outlined by \citet{b17}, \citet{sb20} focused primarily on the regime~$\gmax > \gx > \gradt > \langle \gamma \rangle > \gcool$ where the bulk particles are strongly cooled and the highest-energy particles saturate at~$\gamma = \gradt$. They found that, here, plasmoids are generally filled with cold plasma that has already released much of its energy through Compton losses. The plasma kinetic energy inside plasmoids is then dominated by bulk motion. This motion is also subject to radiative drag and, hence, is slower than in the non-radiative case. \citet{sb20} also confirmed that the highest-energy particles are accelerated near reconnection X-points and top out at Lorentz factors close to~$\gradt$.  

\section{Overview of Klein-Nishina radiative reconnection}
\label{sec:knradreconn}
We now \generaliz[e]our discussion to Klein-Nishina IC losses, focusing first on single-particle cooling and then on collective effects.

\subsection{Single-Particle Klein-Nishina IC Cooling}
We begin with some guiding intuition. One can readily infer that the quadratic Thomson scaling of the scattered photon energy,~$\epsilon_{\rm f} \sim \gamma^2 \epsilon$, must break down at some point: the particle cannot emit a photon of greater energy than its own~$\gamma \me c^2$ (ignoring the small initial energy~$\epsilon$). A new physical regime must take over when~$\gamma^2 \epsilon$ becomes of order~$\gamma \me c^2$. At that point, the particle can no longer radiate continuously; it will lose an order-unity fraction of its energy in a single scattering event. Moreover, for even higher Lorentz factors, the photon energy can scale, at most, linearly with~$\gamma \me c^2$. The following analysis shows how these basic observations are borne out quantitatively. 

At high energies, when the Thomson limit begins to break down, the seed photon energy~$\epsilon$ becomes a dynamically important variable, influencing not just the spectrum of Comptonized photons, but also the power radiated by a particle~$P_{\rm IC}(\gamma)$. To simplify our treatment in the presence of this complication, we specialize to a monochromatic distribution of background radiation
\begin{align}
    u_{\rm ph}(\epsilon) = \uph \delta(\epsilon - \eph) \, .
    \label{eq:umono}
\end{align}

To quantify the IC cooling domain, it is useful to define a critical Lorentz factor
\begin{align}
    \gkn \equiv \frac{\me c^2}{4 \eph}
    \label{eq:gkndef}
\end{align}
and a Klein-Nishina parameter
\begin{align}
    q \equiv \frac{4 \gamma \eph}{\me c^2} = \frac{\gamma}{\gkn} \, .
    \label{eq:qdef}
\end{align}
Scattering particles suffer little recoil from individual photons when~$q \ll 1$~($\gamma \ll \gkn$):  IC radiation proceeds in the Thomson regime. The opposite, deep Klein-Nishina limit is when~$q \gg 1$~($\gamma \gg \gkn$). The crossover point~$q = 1$~($\gamma = \gkn$) corresponds to setting the maximum Thomson emission energy,~$4 \gamma^2 \eph$, equal to the Comptonizing particle energy,~$\gamma \me c^2$.

The Lorentz factor~$\gkn$, like~$\gradt$ for~$\uph$ in the Thomson limit, is the fundamentally new energy scale introduced by Klein-Nishina physics. It serves as a proxy for the underlying physical parameter~$\eph$. By ordering~$\gkn$ with respect to our other fundamental energy scales discussed in the preceding section,~$\gradt$,~$\sigc$, and~$\gmax$, we can determine what new radiative regimes of reconnection are accessible once Klein-Nishina effects have been added to our physical framework.

\begin{figure}
    \centering
    \includegraphics[width=\columnwidth]{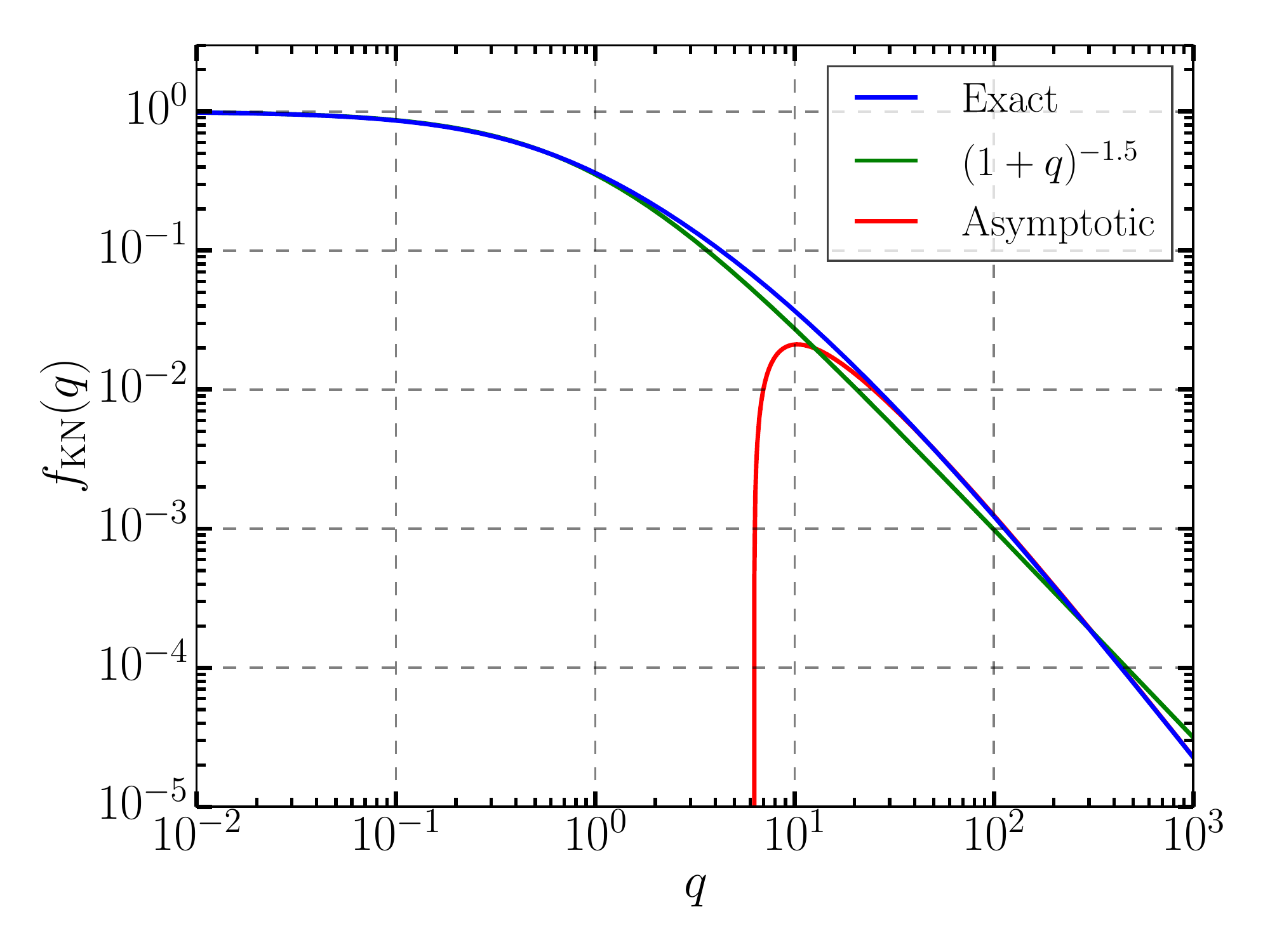}
    \caption{The function~$f_{\rm KN}(\knp)$ as defined in equation~(\ref{eq:fkn}) in blue, its asymptotic form~(\ref{eq:fknasym}) in red, and its approximation~(\ref{eq:fknapprox}) in green.}
    \label{fig:fkn}
\end{figure}

Before embarking on that task, however, we focus purely on the radiative physics (ignoring collective plasma effects), to build our intuition for how individual particles experience IC losses. In the presence of the seed photon distribution~(\ref{eq:umono}), the IC power radiated by a single particle becomes
\begin{align}
    P_{\rm IC}(\gamma) &= P_{\rm T}(\gamma) \, f_{\rm KN} ( \gamma / \gkn ) \, ,
    \label{eq:pic}
\end{align}
This is the same as the Thomson expression~(\ref{eq:pthom}) but modified by the dimensionless function of~$\knp$ (cf. \citealt{j68}; \citealt{nyc18}; Appendix~\ref{sec:knfuncs})
\begin{align}
    f_{\rm KN}(\knp) &= \frac{9}{\knp^3} \left[ \left( \frac{\knp}{2} + 6 + \frac{6}{\knp} \right) \ln \left( 1 + \knp \right) \right. \notag \\
    &- \left. \frac{1}{\left(1 + \knp\right)^2} \left( \frac{11}{12}\knp^3 + 6 \knp^2 + 9 \knp + 4 \right) - 2 + 2 \Li_{2}(-\knp) \right] \, ,
    \label{eq:fkn}
\end{align}
where~$\Li_2$ is the dilogarithm. Figure~\ref{fig:fkn} displays~$f_{\rm KN}(\knp)$ together with its asymptotic large-argument limit
\begin{align}
    f_{\rm KN}(\knp \gg 1) \simeq (9/2 \knp^2) \left[ \ln(q) - 11/6 \right]
    \label{eq:fknasym}
\end{align}
and its approximate form \citep[e.g.][]{msc05}
\begin{align}
    f_{\rm KN}(\knp) \simeq \frac{1}{(1+\knp)^{1.5}} \, ,
    \label{eq:fknapprox}
\end{align}
which is roughly correct up to~$\knp \sim 10^4$ (at~$\knp \simeq 10^4$, the error reaches a factor of~$3$ and begins increasing rapidly). The function~$f_{\rm KN}(\knp)$ tends to unity as~$\knp$ becomes small, as required in the Thomson-limit~$P_{\rm IC}(\gamma) \to P_{\rm T}(\gamma)$. For large arguments,~$f_{\rm KN}(\knp)$ falls off quadratically with a logarithmic correction: the scattering \crosssection[]is suppressed in the deep Klein-Nishina limit.

Let us now examine the \quoted{discreteness}of radiative losses as~$\knp$ increases through unity. To do so, we note that, similar to the radiated power~$P_{\rm IC}(\gamma)$, the rate~$R_{\rm IC}(\gamma)$ at which a single electron or positron scatters photons distributed according to~(\ref{eq:umono}) can be written as the corresponding Thomson rate,~$R_{\rm T}(\gamma) = R_{\rm T} = c \sigma_{\rm T} \uph / \eph$, times a dimensionless function:
\begin{align}
    R_{\rm IC} (\gamma) = R_{\rm T} \, g_{\rm KN}(\gamma / \gkn) \, ,
    \label{eq:ric}
\end{align}
where~$g_{\rm KN}(\knp)$ reads (Appendix~\ref{sec:knfuncs})
\begin{align}
    g_{\rm KN}(\knp) &= \frac{3}{2 \knp^2} \left[ \left( \knp + 9 + \frac{8}{\knp} \right) \ln( 1 + \knp ) \right. \notag \\
    &\left. - \frac{1}{1 + \knp} \left( \frac{\knp^2}{2} + 9 \knp + 8 \right) + 4 \Li_2 (-\knp) \right] 
    \label{eq:gkn}
\end{align}
and has asymptotic form
\begin{align}
    g_{\rm KN}(\knp \gg 1) \simeq \frac{3}{4 \knp} \left[ 2 \ln(\knp) - 1 \right] \, .
    \label{eq:gknasym}
\end{align}

Using~(\ref{eq:ric}), one can write down the average photon energy~$\langle \epsilon \rangle (\gamma)$ emitted by a particle with Lorentz factor~$\gamma$:
\begin{align}
    \langle \epsilon \rangle(\gamma) \equiv \frac{P_{\rm IC}(\gamma)}{R_{\rm IC}(\gamma)} = \left[ \frac{\gamma}{3 \gkn} \frac{f_{\rm KN}(\gamma / \gkn)}{g_{\rm KN}(\gamma / \gkn)} \right] \gamma \me c^2 \, .
    \label{eq:epsavg}
\end{align}
The quantity in square brackets here is the \quoted[~$\langle \epsilon \rangle(\gamma) / \gamma \me c^2$,]{inelasticity}the typical fraction of a particle's energy lost in a single scattering encounter \citep{msc05}.
Using the small-argument limits~$f_{\rm KN}(\knp \ll 1) \to 1$ and~$g_{\rm KN}(\knp \ll 1) \to 1$, one may read off the well-known Thomson inelasticity
\begin{align}
    \lim_{\gamma \ll \gkn} \frac{\langle \epsilon \rangle(\gamma)}{\gamma \me c^2} = \frac{\gamma}{3 \gkn} \ll 1 \, ,
    \label{eq:thominelasticity}
\end{align}
or, more commonly,
\begin{align}
    \lim_{\gamma \ll \gkn} \langle \epsilon \rangle(\gamma) = \frac{\gamma^2 \me c^2}{3 \gkn} = \frac{4}{3}\gamma^2 \eph \, .
    \label{eq:thomavgen}
\end{align}
Similarly, plugging in the asymptotic forms for~$f_{\rm KN}$ and~$g_{\rm KN}$ verifies that the inelasticity approaches unity as~$q$ is taken to infinity:
\begin{align}
    \lim_{\gamma \to \infty} \frac{\langle \epsilon \rangle(\gamma)}{\gamma \me c^2} = \lim_{\gamma \to \infty} \frac{\ln(\gamma / \gkn) - 11/6}{\ln(\gamma / \gkn) - 1/2} = 1 \, .
    \label{eq:inelasticitylim}
\end{align}
However, the limiting value is approached quite slowly, for the ratio in~(\ref{eq:inelasticitylim}) is only~$\simeq 1$ when~$\ln(\gamma / \gkn) \gg 11/6$. In fact,~$\langle \epsilon \rangle(\gamma)$ does not surpass~$3 \gamma \me c^2 / 4$ until~$\gamma \gtrsim 300 \gkn$. Nevertheless, the inelasticity does obtain a value \textit{of order} unity for much more modest~$\knp$. For example,~$\langle \epsilon \rangle(\gamma) \geq \gamma \me c^2 / 4$ when~$\gamma \gtrsim 2 \gkn$.

\begin{figure}
    \centering
    \includegraphics[width=\columnwidth]{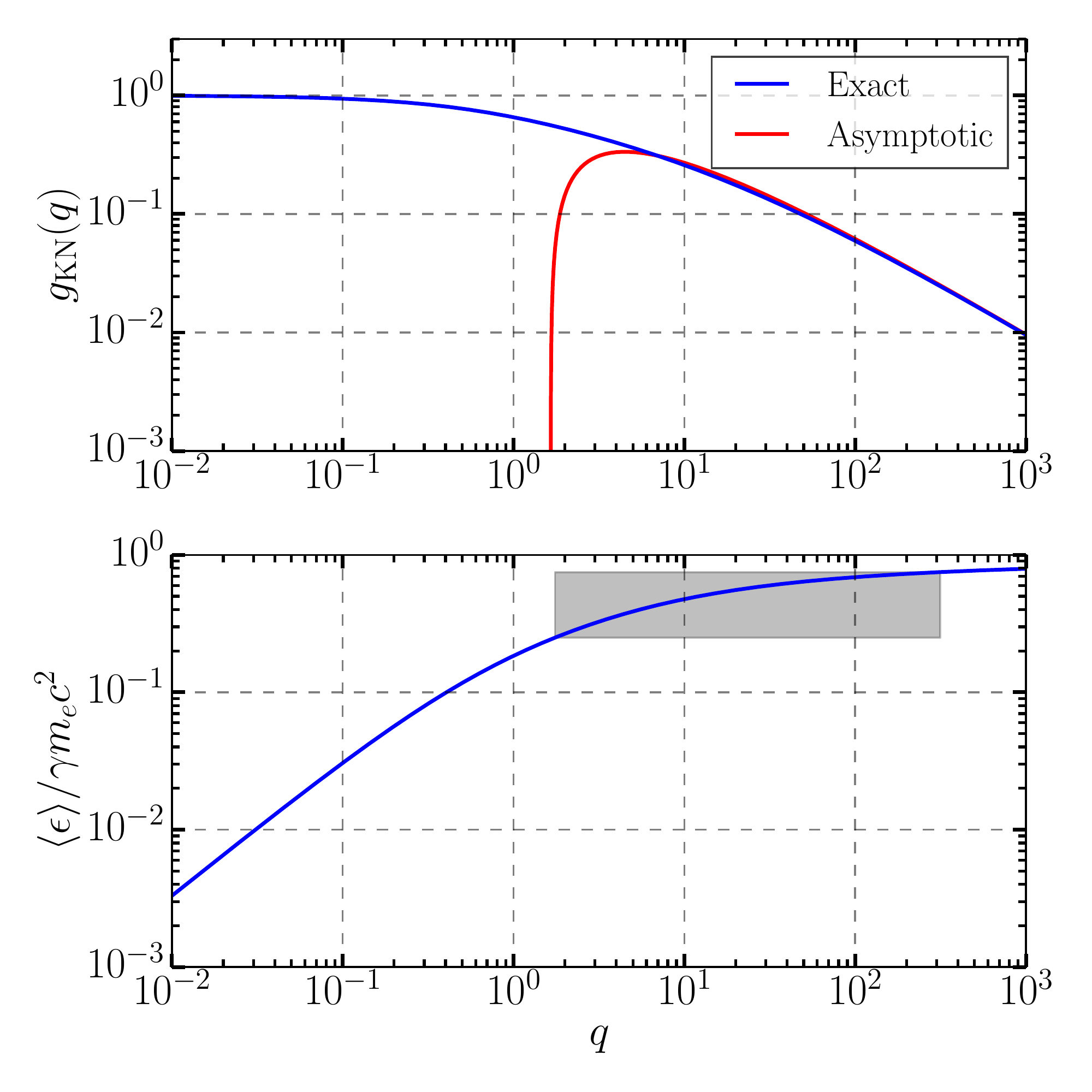}
    \caption{Top panel: the function~$g_{\rm KN}(\knp)$ [equation~(\ref{eq:gkn})] in blue and its asymptotic form~(\ref{eq:gknasym}) in red. Bottom panel: the inelasticity~$\langle \epsilon \rangle(\gamma) / \gamma \me c^2 = \knp f_{\rm KN}(\knp) / 3 g_{\rm KN}(\knp)$ [equation~(\ref{eq:epsavg})]. The shaded region shows the appreciable range~$\knp \in [1.75, 315]$ over which~$\langle \epsilon \rangle$ is between~$\gamma \me c^2 / 4$ and~$3 \gamma \me c^2 / 4$ (i.e. where it is close to~$\gamma \me c^2 / 2$). This indicates that, although~$\langle \epsilon \rangle(\gamma)$ eventually approaches its asymptotic limit of~$\gamma \me c^2$, it does so rather slowly.}
    \label{fig:gkn}
\end{figure}

Both of these effects -- the inelasticity slowly approaching, but rapidly rising to the vicinity of, unity for~$\knp \gtrsim 1$ -- as well as the function~$g_{\rm KN}(\knp)$ and its asymptotic form~(\ref{eq:gknasym}), are plotted in Fig.~\ref{fig:gkn}. In the figure, one sees that~$\langle \epsilon \rangle(\gamma) \simeq \gamma \me c^2 / 2$ for a wide range of~$\knp$. Therefore, when we later need~$\langle \epsilon \rangle$ for estimates, we adopt~$\langle \epsilon \rangle(\gamma) \simeq \gamma \me c^2 / 2$ rather than~$\langle \epsilon \rangle \simeq \gamma \me c^2$ whenever~$\knp \gtrsim 1$. The latter becomes a more accurate approximation than the former for~$\knp \gtrsim 300$, but it is unclear that the astrophysical systems we attempt to model contain particles at such high energies. (However, even if~$\knp \gg 300$ in some systems, the distinction here between factors of order unity is well within the uncertainty of all of the estimates in this paper.) Furthermore, our model (section~\ref{sec:knradreconn_mod}) is mostly concerned with~$\knp < 280$ [equation~(\ref{eq:gct2})]. 

We have now verified our qualitative expectations for the deep Klein-Nishina regime: radiative losses become discrete when~$\knp \gtrsim 1$, with particles losing an appreciable fraction of their energy to single photons. Moreover, the scattered photon energy scales approximately linearly with the pre-collision energy of the particle (rather than quadratically, as in the Thomson limit).

Next, because we are concerned primarily with IC radiation in this paper, we consider the circumstances in which synchrotron losses may be neglected. For an isotropic particle pitch-angle distribution, the average synchrotron power radiated per particle is
\begin{align}
    P_{\rm s} = \frac{4}{3} \sigma_{\rm T} c \beta^2 \gamma^2 U_{\rm B} \, . 
    \label{eq:ps}
\end{align}
Note that this is the same as the Thomson IC power~(\ref{eq:pthom}) but with the ambient radiation energy density~$\uph$ replaced by the magnetic field energy density~$U_{\rm B} \sim B_0^2 / 8 \pi$ (we approximate the magnetic field strength throughout the reconnection system by its upstream value~$B_0$). Equation~(\ref{eq:ps}) gives a total (IC + synchrotron) radiated power per particle
\begin{align}
    P_{\rm tot}(\gamma) &= P_{\rm IC}(\gamma) + P_{\rm s}(\gamma) \notag \\
        &= P_{\rm IC}(\gamma) \left( 1 + \frac{U_{\rm B}}{\uph f_{\rm KN}(\gamma / \gkn)} \right) \, .
    \label{eq:ptot}
\end{align}
Clearly, IC losses dominate if
\begin{align}
    f_{\rm KN}(q) > \frac{U_{\rm B}}{\uph} \, .
    \label{eq:iccriterion}
\end{align}
Because~$f_{\rm KN}(q) \leq 1$, this criterion can only be met for systems whose ambient radiation energy density exceeds the magnetic field energy density. And, importantly, because Klein-Nishina effects begin to suppress IC cooling for~$\gamma > \gkn$, even when~$\uph \gg U_{\rm B}$, there is always a high-energy Lorentz factor~$\gamma_{\rm s}$ above which synchrotron losses dominate. Using the approximate form~$f_{\rm KN}(q) \simeq (1+q)^{-1.5}$ [equation~(\ref{eq:fknapprox})], one has \citep[cf][]{msc05}
\begin{align}
    \gamma_{\rm s} \simeq \gkn \left[ \left(\frac{\uph}{U_{\rm B}} \right)^{2/3} - 1 \right] \, .
    \label{eq:gsynapprox}
\end{align}

Thus, neglecting synchrotron losses is justified when~$\gamma_{\rm s}$ exceeds the highest Lorentz factors reached in the system. We assume that this is indeed the case for the remainder of this work. We return to discuss the effects of finite~$\gamma_{\rm s}$ as an effective limitation of our analysis in section~\ref{sec:applications}.

\subsection{Klein-Nishina IC effects on collective plasma behavi\spellor}
\label{sec:collickn}
We now examine how the collective reconnection dynamics are influenced by Klein-Nishina radiation-reaction. We focus especially on differences from the case of purely Thomson radiative cooling.

Previously (in the Thomson regime), the highest Lorentz factor to which a particle could be accelerated was~$\gradt$. However, from equation~(\ref{eq:pic}), radiative losses are suppressed once~$\knp$ exceeds unity. This enables acceleration beyond~$\gradt$, and our definition of the radiative \cutoff Lorentz factor can be \generaliz[ed]to include this effect. By equating the force from the reconnection electric field~$0.1 e B_0$ to the (Klein-Nishina) Compton radiation reaction force~$P_{\rm IC}(\gamma) / c$, one may define a \generaliz[ed]\cutoff[~$\gradk$]through
\begin{align}
    1 \equiv \frac{\gradk^2}{\gradt^2} f_{\rm KN}\left( \frac{\gradk}{\gkn} \right) = \frac{\gradk^2}{\gradt^2} f_{\rm KN} \left( \frac{\gradk}{\gradt} \frac{\gradt}{ \gkn} \right) \, .
    \label{eq:gradk}
\end{align}
The second equality explicitly shows that the ratio~$\gradk / \gradt$ is determined solely by the ratio~$\gradt / \gkn$. Thus,~$\gradk$ is a derived scale; it is fixed by the other radiative Lorentz factors~$\gkn$ and~$\gradt$ (or, equivalently, through the physical parameters~$\eph$ and~$\uph$).

\begin{figure}
    \centering
    \includegraphics[width=\columnwidth]{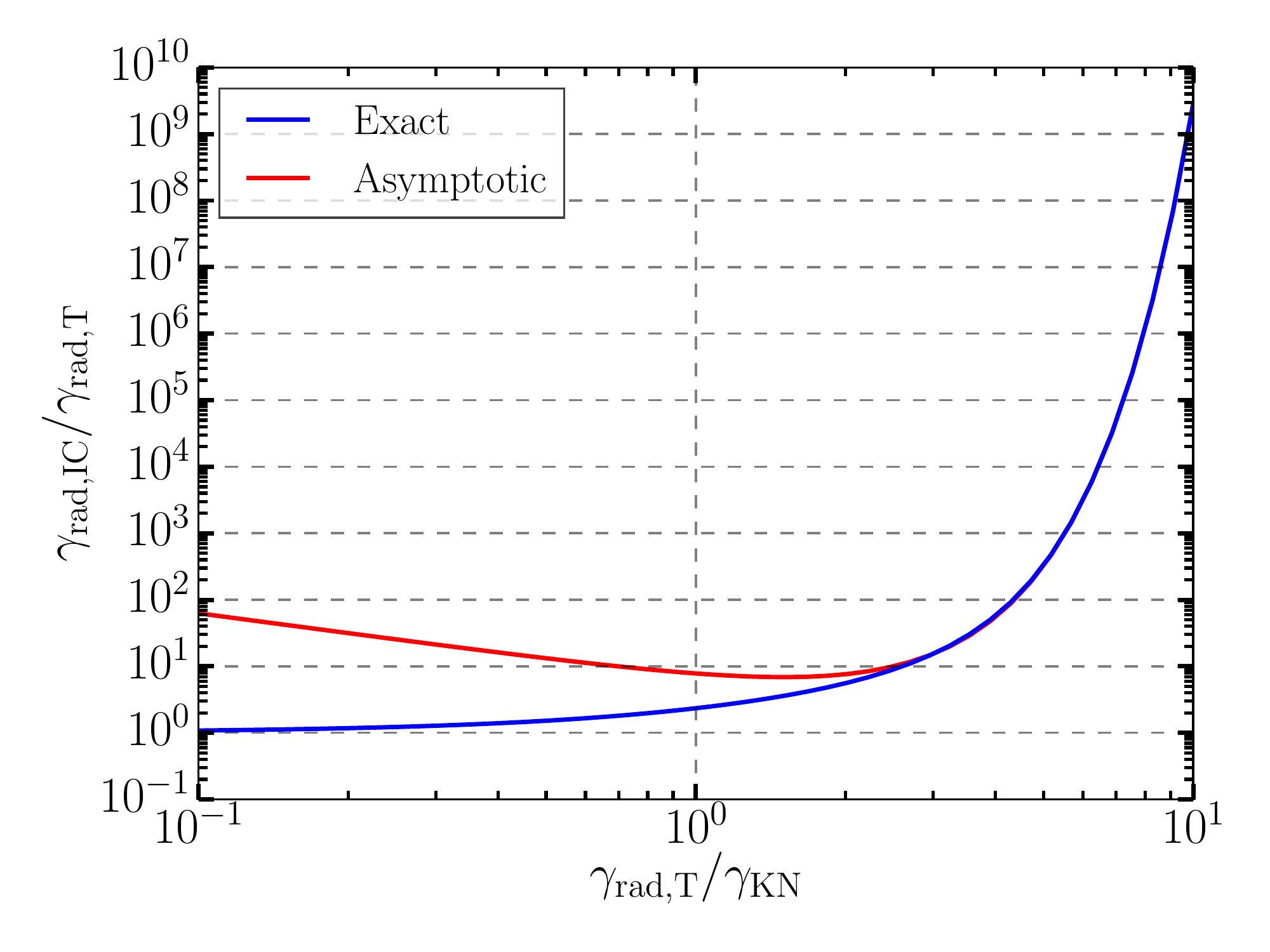}
    \caption{The Klein-Nishina \cutoff Lorentz factor~$\gradk$ [equation~(\ref{eq:gradk})] and its asymptotic form~(\ref{eq:gradkasym}), both \normaliz[ed]by the Thomson equivalent~$\gradt$. The asymptotic form is already accurate when~$\gradt \simeq 2.5 \gkn$.}
    \label{fig:gradk}
\end{figure}

Equation~(\ref{eq:gradk}) can always be satisfied for a finite~$\gradk$, and a numerical solution to the equation is displayed in Fig.~\ref{fig:gradk}. When~$\gradt \ll \gkn$, the equation is satisfied by~$\gradk \simeq \gradt$ because, in that case,~$f_{\rm KN}(\gradt / \gkn \ll 1) \simeq 1$. However, when~$\gradt$ becomes greater than~$\gkn$, the \cutoff[~$\gradk$]becomes a rapidly increasing function of~$\gradt / \gkn$. In fact, in the limit~$\gradt \gg \gkn$, for which the large-argument approximation to~$f_{\rm KN}$ [equation~(\ref{eq:fknasym})] applies,~$\gradk$ grows super-exponentially:
\begin{align}
    \gradk \simeq \gkn \exp \left[ \frac{2}{9} \left( \frac{\gradt}{\gkn} \right)^2 + \frac{11}{6} \right] \, .
    \label{eq:gradkasym}
\end{align}
As shown in Fig.~\ref{fig:gradk}, this limiting form gives a good approximation to~$\gradk$, even when~$\gradt$ only slightly exceeds~$\gkn$. 
We note that the rapid transition between the~$\gkn \gg \gradt$ and~$\gkn \ll \gradt$ limits is smoothed out in the presence of an extended (e.g.\ power-law) distribution of ambient radiation. Then,~$\gradk$ does not grow super-exponentially until~$\me c^2 / 4 \epsilon_{\rm low} \ll \gradt$, where~$\epsilon_{\rm low}$ is the energy of the softest ambient photons.

Hence, when~$\gkn$ becomes less than~$\gradt$, Klein-Nishina physics inhibits radiative cooling from competing with rapid acceleration near reconnection X-points. It is then highly likely that the \cutoff[]energy for X-point acceleration is set intrinsically rather than by radiative cooling (one expects~$\gx \ll \gradk$).

\begin{figure}
    \centering
    \includegraphics[width=\columnwidth]{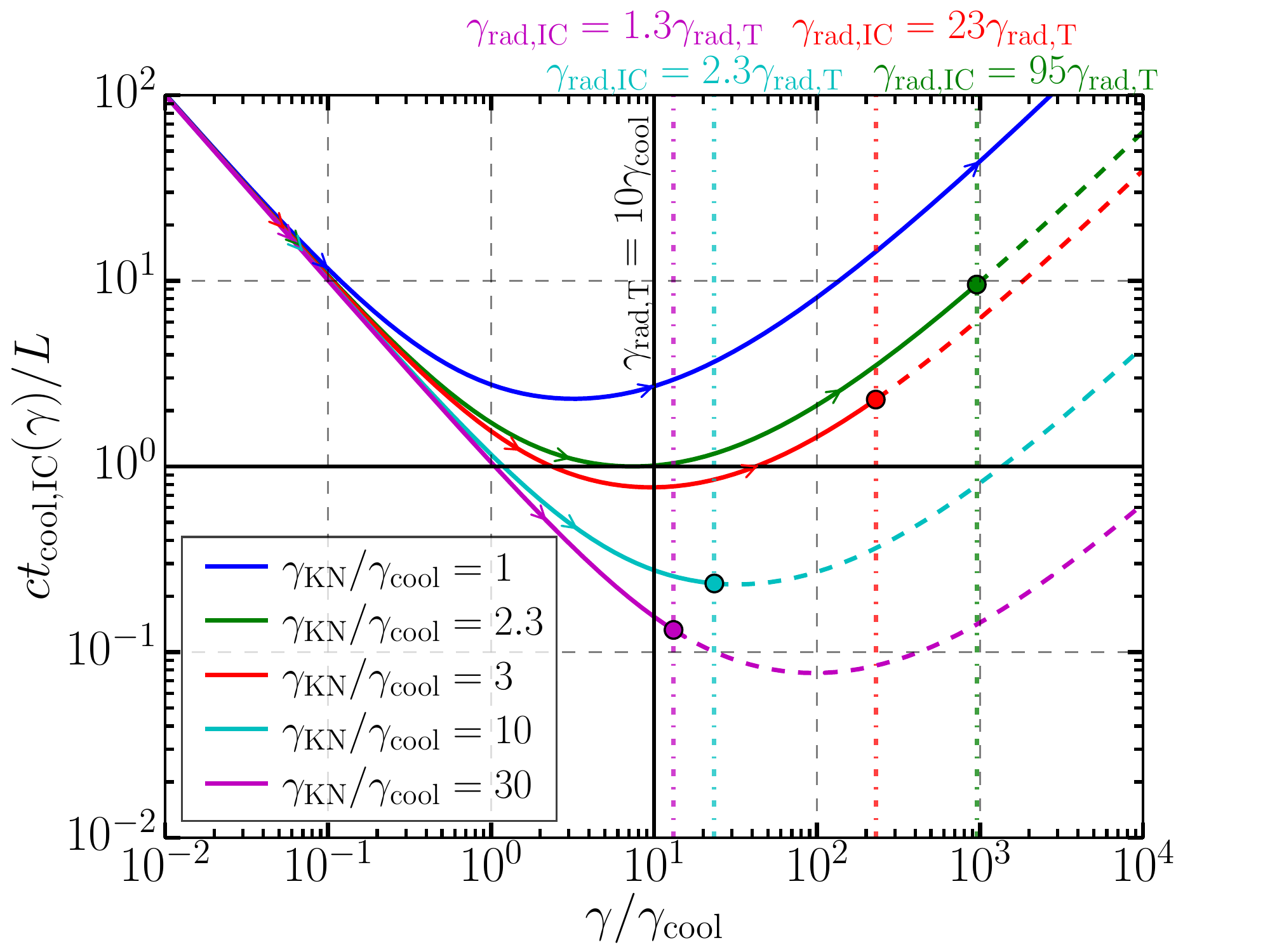}
    \caption{The IC cooling time~$c \tcoolk(\gamma)/L$ [equation~(\ref{eq:tcoolic})]. In the plot,~$\gcool$ is fixed,~$\gmax = 10 \gradt = 100 \gcool$, and~$\gkn$ varies between curves. Arrows indicate the progression of particle cooling times during acceleration near reconnection X-points, which terminates at~$\gamma = \gradk$ (denoted by filled circles). For reference, each curve continues past~$\gradk$ as a dashed line. The cooling time is non-monotonic. Depending on~$\gkn$, some (or even all) of the particles may attain high enough energies not to radiate [possessing cooling times~$\tcoolk(\gamma) > L/c$]. If~$\gkn < 2.3 \gcool$, then all particles are non-radiative. If~$\gkn > 2.3 \gcool$, then only the low-energy~($\gamma < \gcooli{1} \sim \gcool$) and high-energy~($\gamma > \gcooli{2}$) particles are non-radiative. For particles to exceed~$\gcooli{2}$, the condition~$\gradk \geq \gcooli{2}$ is necessary, but not sufficient (see text).}
    \label{fig:tcoolic}
\end{figure}

We now frame the super-exponential divergence in~$\gradk$ from a different perspective: that of competing acceleration and radiative cooling \ts[s.]We then predict whether divergences occur in the \cutoff[]energies of secondary acceleration channels by similarly comparing their \ts[s]against the cooling time. Including Klein-Nishina effects, the IC cooling \ts[]is\footnote{As shown later [equation~(\ref{eq:taugg})], the factor~$\gcool / \gkn$ in~(\ref{eq:tcoolic}) equals~$3/5\tau_{\gamma\gamma}$ where~$\tau_{\gamma\gamma}$ is the characteristic pair-production optical depth of the system.}
\begin{align}
    \tcoolk(\gamma) &\equiv \frac{\gamma \me c^2}{P_{\rm IC}(\gamma)} = \frac{\gcool}{\gamma f_{\rm KN}(\gamma / \gkn)} \frac{L}{c} \notag \\
    &= \frac{\gcool}{\gkn} \left[ (\gamma / \gkn) f_{\rm KN}(\gamma  / \gkn) \right]^{-1} \frac{L}{c} \, ,
    \label{eq:tcoolic}
\end{align}
where~$\gcool$ [equation~(\ref{eq:gcool})] is the Lorentz factor of a particle with Thomson cooling time~$\tcoolt(\gamma) = \gamma \me c^2 / P_{\rm T}(\gamma)$ equal to~$L/c$.
A plot of~$\tcoolk(\gamma)$ is presented in Fig.~\ref{fig:tcoolic}.

The relationship~(\ref{eq:tcoolic}) encodes a wealth of information. In the Thomson regime~$\gamma \ll \gkn$,~$f_{\rm KN} \to 1$, and~$\tcoolk \to \tcoolt = \gcool L / \gamma c$, which decreases inversely with~$\gamma$. However, for~$\gamma \gg \gkn$,~$f_{\rm KN}$ assumes its asymptotic form~(\ref{eq:fknasym}), inducing the scaling
\begin{align}
    \tcoolk(\gamma \gg \gkn) \simeq \frac{2 L \gcool}{9 c \gkn} \frac{\gamma / \gkn}{\ln(\gamma / \gkn) - 11 / 6} \, .
    \label{eq:tcoolkasym}
\end{align}
Thus, in the deep Klein-Nishina limit,~$\tcoolk(\gamma)$ \textit{increases} linearly in~$\gamma$ with a logarithmic correction. This is the source of the super-exponential divergence of~$\gradk$ with~$\gradt$ in equation~(\ref{eq:gradkasym}). Equation~(\ref{eq:tx}) shows that the \ts[,~$t_{\rm X}$,]on which X-point acceleration occurs is also linear in~$\gamma$. Upon equating~$\tcoolk$ and~$t_{\rm X}$, which defines the radiative \cutoff[~$\gradk$,~$\gamma$]cancels. Thus,~$t_{\rm X}$ can only surpass~$\tcoolk$ if~$\ln(\gamma / \gkn)$ grows large enough, inducing the super-exponential scaling in equation~(\ref{eq:gradkasym}).

Importantly, a similar situation does not arise for secondary acceleration channels, which are slower, possessing \ts[s]super-linear in~$\gamma$ \citep[e.g.][]{mwu20}. As an example, consider the secondary process described in section~\ref{sec:thomradreconn}, where particles inside contracting plasmoids are gradually energized. The Lorentz factors of such particles grow as~$\gamma(t) \propto \sqrt{t}$ \citep{ps18, hps20}, yielding the acceleration time~$t_{\rm sec}(t) = \gamma(t) / \dot{\gamma}(t) \propto \gamma^2(t)$. Thus, it is much easier for this secondary mechanism -- and any other for which~$t_{\rm sec} \propto \gamma^\zeta$ with~$\zeta > 1$ -- to radiatively saturate, as we now show.

For the sake of generality, suppose that~$t_{\rm sec} = C \gamma^\zeta$ with~$C$ a constant independent of~$\eph$ and~$\uph$ (and, hence, of~$\gcool$,~$\gradt$,~$\gkn$, and~$\gradk$; cf.\ the argument in \citealt{mwu20}). Then, when IC cooling proceeds deep into the Klein-Nishina regime, the equality~$\tcoolk = t_{\rm sec}$ reduces to~$\gamma_{\rm sec}^{\zeta - 1} \propto \gcool / \gkn^2$. The key difference from the direct X-point acceleration channel is that here we can ignore the~$\ln(\gamma)$ correction -- its dependence on~$\gamma$ is much weaker than~$\gamma^{\zeta - 1}$. As a result, the \cutoff[~$\gamma_{\rm sec}$]scales merely polynomially in~$\gcool$ (i.e., in~$\gradt$) and~$\gkn$:~$\gamma_{\rm sec} \propto (\gcool / \gkn^2)^{1/(\zeta-1)}$. Plugging in~$\zeta = 2$ for the adiabatic plasmoid compression process gives~$\gamma_{\rm sec} \propto \gcool / \gkn^2$ (different from~$\gamma_{\rm sec}$ reported in section~\ref{sec:thomradreconn} because we are now considering deep Klein-Nishina cooling). Thus, Klein-Nishina physics may effectively remove the high-energy radiation-reaction cap on impulsive X-point acceleration, but \textit{not} on other processes, potentially increasing the relative importance of the primary direct energization channel.

As discussed above,~$\tcoolk(\gamma)$ is non-monotonic, decreasing with~$\gamma$ when~$\gamma \ll \gkn$ and increasing when~$\gamma \gg \gkn$. It reaches the minimum
\begin{align}
    \min_{\gamma} \left( \frac{c \tcoolk(\gamma)}{L} \right) \simeq 2.32 \frac{\gcool}{\gkn}
    \label{eq:tcoolicmin}
\end{align}
at a critical \textit{fastest-cooling Lorentz factor}
\begin{align}
    \argmin[\gamma] \left( \frac{c \tcoolk(\gamma)}{L} \right) \simeq 3.20 \gkn \, ,
    \label{eq:tcoolicamin}
\end{align}
[i.e.~$\min_{\gamma}(\tcoolk(\gamma)) = \tcoolk(3.20 \gkn)$].
The minimum cooling time~(\ref{eq:tcoolicmin}) implies that, when~$\gkn < 2.32 \gcool$, \textit{all} of the particles in the system radiate weakly (they have cooling times exceeding~$L/c$). Even if~$\gkn$ falls above this threshold and, hence,~$\min (\tcoolk) < L /c$, some high-energy particles may radiate weakly. Namely, if a particle surpasses the fastest-cooling Lorentz factor~$3.20 \gkn$ by a sufficient amount, it reaches a high-energy domain with~$\tcoolk(\gamma) > L /c$. This effect does not occur in the Thomson regime.

These remarks are illustrated in Fig.~\ref{fig:tcoolic}. The figure displays~$c \tcoolk(\gamma) / L$ for fixed~$\gcool = \gradt / 10 = \gmax / 100$ and several~$\gkn$. On each curve for which~$\gkn > 2.32 \gcool$, the line~$\tcoolk = L / c$ is crossed twice, once at a low Lorentz factor~$\gcooli{1}$ and once at a high Lorentz factor~$\gcooli{2}$. We \analyz[e~$\gcooli{1}$]and~$\gcooli{2}$ shortly, but we point out some basic features of Fig.~\ref{fig:tcoolic} beforehand. First, when~$\gradk \simeq \gradt \leq \gkn$ (implying~$\gkn > 2.32 \gcool$ because~$\gkn > \gradt = 10 \gcool > 2.32 \gcool$), particles may access only the Thomson portion of a cooling curve where~$\tcoolk \propto \gamma^{-1}$. The~$\gkn = 30 \gcool = 3 \gradt$ case illustrates this. Next, in the opposite limit, when~$\gkn$ becomes smaller than~$\gradt$, the radiative \cutoff[~$\gradk$]begins to grow rapidly, opening up the portion of a curve that bends upward. Eventually, at the critical Lorentz factor~$\gcooli{2}$, the cooling time~$\tcoolk$ once again equals~$L/c$. Thus, if~$\gradk > \gcooli{2}$, particles accelerated near X-points could break into the high-energy weakly radiative regime. However, in reality, whether particles will actually cross this boundary does not depend solely on whether~$\gradk$ surpasses~$\gcooli{2}$. That is just a necessary condition. In addition, the intrinsic X-point acceleration Lorentz factor~$\gx$ must exceed~$\gcooli{2}$, or -- if it does not -- secondary acceleration channels must be able to energize particles against radiative cooling past~$\gcooli{2}$.

\begin{figure}
    \centering
    \includegraphics[width=\columnwidth]{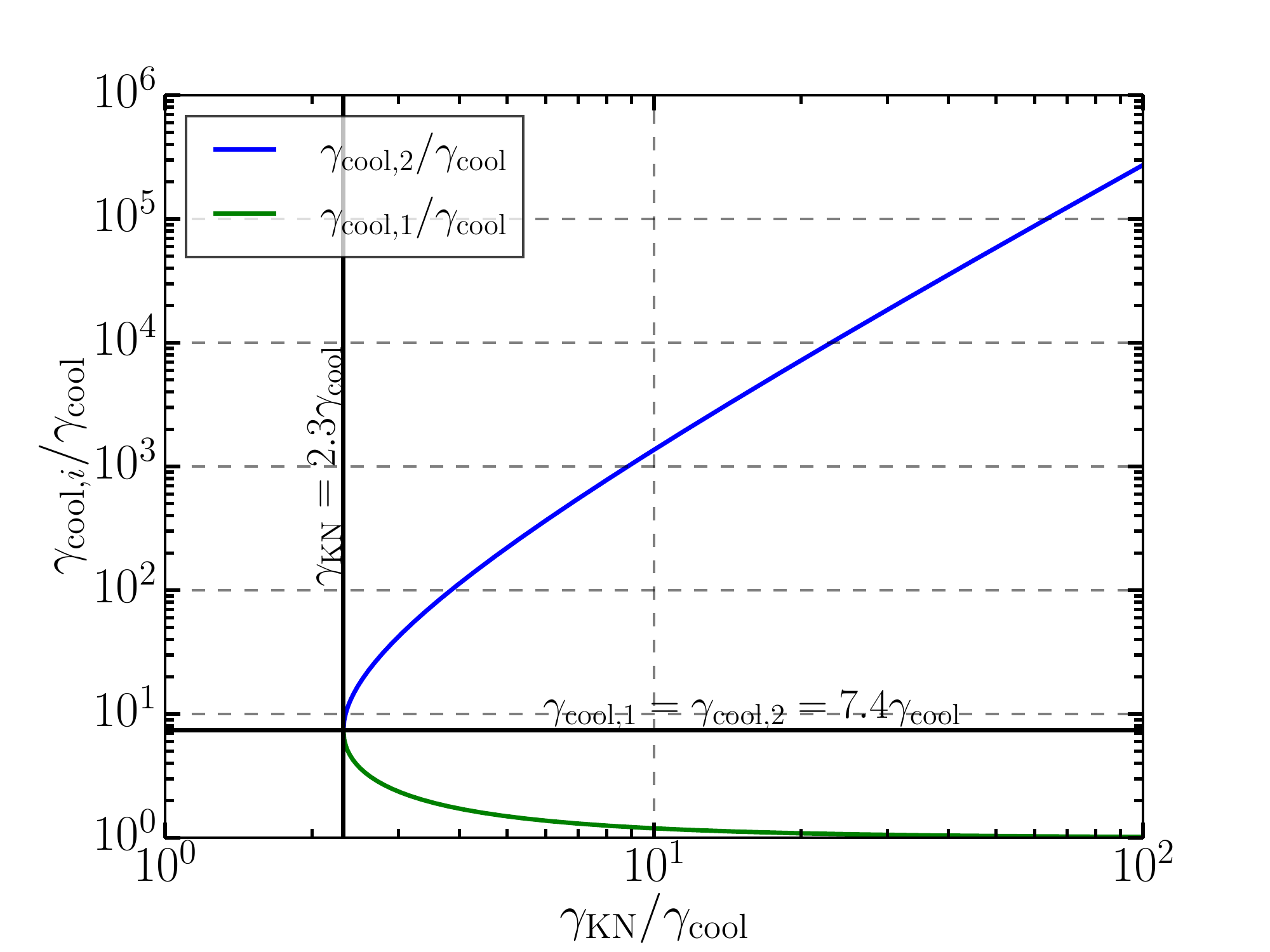}
    \caption{The Lorentz factors~$\gcooli{1}$ and~$\gcooli{2}$ where~$\tcoolk(\gamma)$ dips below and above one dynamical time~$L/c$, respectively. Since~$\gcooli{1}$ occurs almost entirely in the Thomson regime, it is almost always quite close to~$\gcool$, differing from~$\gcool$ by at most the factor~$7.4 = 3.20 \times 2.32$ [equations~(\ref{eq:tcoolicmin}) and~(\ref{eq:tcoolicamin})] when~$\gkn$ approaches the non-radiative value~$2.32 \gcool$. In contrast~$\gcooli{2}$ grows approximately quadratically in~$\gkn / \gcool$ (with a logarithmic correction).}
    \label{fig:gcooli}
\end{figure}

Both~$\gcooli{1}$ and~$\gcooli{2}$ are illustrated in Fig.~\ref{fig:gcooli} as functions of~$\gkn / \gcool$ (they only depend on~$\gradt$ and~$\gmax$ through~$\gcool$). In general,~$\gcooli{1}$ is close to~$\gcool$, because, as illustrated in Fig.~\ref{fig:tcoolic}, particles are almost completely in the Thomson limit when~$\tcoolk$ crosses~$L/c$ from above. In contrast,~$\gcooli{2}$, where~$\tcoolk$ crosses~$L/c$ in the opposite direction, depends rather strongly on~$\gkn$. Because~$\gcooli{2}$ occurs fairly deep into the Klein-Nishina regime,  one may employ expression~(\ref{eq:tcoolkasym}) to see that~$\gcooli{2}$ satisfies
\begin{align}
    \frac{9 \gkn}{2 \gcool} \simeq \frac{\gcooli{2} / \gkn}{\ln(\gcooli{2} / \gkn) - 11 / 6} \, ,
    \label{eq:gcool2asym}
\end{align}
implying that~$\gcooli{2} / \gcool \propto (\gkn / \gcool)^2 \times \mathcal{O}(\ln(\gcooli{2} / \gkn))$.

Thus,~$\gcooli{2}$ is approximately quadratic in~$\gkn$. This differs from~$\gradk$, which grows as~$\gkn$ is \textit{reduced}. Let us imagine that one starts with~$\gkn$ high enough that~$\gcooli{2} \gg \gmax \gg \gradk$ (for example, as on the~$\gkn = 30 \gcool$ curve in Fig.~\ref{fig:tcoolic}). Then, dialing down~$\gkn$, eventually~$\gcooli{2}$ will cross~$\gradk$ and~$\gmax$ from above. It turns out that, by definition, these crossings occur simultaneously. For, if~$\gcooli{2} = \gmax$, then~$t_{\rm X}(\gmax) \equiv L/c$ [equation~(\ref{eq:gmaxalt})] and~$\tcoolk(\gcooli{2}) \equiv L/c$ [equation~(\ref{eq:tcoolic})]. Consequently,~$\tcoolk(\gcooli{2}) = t_{\rm X}(\gcooli{2})$, which implies~$\gmax = \gcooli{2} = \gradk$. By similar reasoning, one can show that, if~$\gcooli{2} = \gradk$, then~$\gcooli{2} = \gmax$. Continuing to reduce~$\gkn$ beyond this \quoted{triple point}yields the scale ordering~$\gradk \gg \gmax \gg \gcooli{2}$ (typified in the~$\gkn = 3 \gcool$ curve of Fig.~\ref{fig:tcoolic}, although the separation of scales is rather small in that example). The first of these inequalities,~$\gradk \gg \gmax$, means that no particles achieve radiative saturation; all have Lorentz factors~$\gamma < \gradk$ (see section~\ref{sec:thomradreconn} discussion). The second inequality~$\gcooli{2} \ll \gmax$, potentially allows various acceleration channels (e.g.\ X-point acceleration if~$\gx > \gcooli{2}$) to break a population of high-energy particles through the weakly radiative~$\gamma = \gcooli{2}$ barrier.

To \summariz[e]up to this point, we have encountered several stark departures from the Thomson picture of radiative cooling induced by finite~$\gkn$. Not only do particles with Lorentz factors~$\gamma > \gkn$ radiate their energy in discrete chunks, but their radiative cooling times can actually be quite long. When~$\gkn \lesssim \gradt$, the effective \cutoff[]Lorentz factor~$\gradk$ begins to grow rapidly, and comes, with just a small change in~$\gkn$, to exceed~$\gmax$, the analog of the Hillas Lorentz factor for relativistic magnetic reconnection. When~$\gkn$ is decreased even more, eventually it falls below~$\gcool$, and no particles in the system radiate efficiently. Thus, even in a nominally strongly radiative Thomson scale ordering, by making~$\gkn$ small enough, a non-radiative regime can be reached. At intermediate~$\gkn$, a variety of intriguing and exotic physical effects can occur, which we elucidate later in this study. But first we cover one additional piece of physics that is entirely new to the Klein-Nishina realm: pair production.

\subsubsection{Pair-production in Klein-Nishina reconnection}
\label{sec:knradreconnpp}
A Comptonized photon of energy~$\epsilon$ may collide with a background photon (energy~$\eph$) to produce an electron-positron pair if the threshold criterion,~$\epsilon \eph \geq (\me c^2)^2$, is met. In this work, we assume that~$\gkn = \me c^2 / 4 \eph \gg 1$~($\eph \ll 100 \, \rm keV$), implying that~$\epsilon \gg \me c^2 \gg \eph$ is required to reach pair-production threshold. Such a high-energy photon can only be emitted in the Klein-Nishina IC regime. (The first author would here like to acknowledge Beno\^{i}t Cerutti, who originally pointed this out to him.) If one assumes that the IC scattering occurs in the Thomson limit, then a contradiction arises because~$1 \leq \epsilon \eph / (\me c^2)^2 \sim (\gamma \eph / \me c^2)^2 \sim \knp^2$. In contrast, assuming a Klein-Nishina scaling~$\epsilon \sim \gamma \me c^2 / 2$ yields the self-consistent result,~$\knp \equiv 4 \gamma \eph / \me c^2 \sim 8 \epsilon \eph / (\me c^2)^2 \geq 8$. We adopt~$\gth \equiv 8 \gkn$ (used mainly in section~\ref{sec:knradreconn_mod}) as the characteristic minimum particle Lorentz factor to emit above-threshold photons.

Now, the pair-production cross section~$\sigma_{\gamma\gamma}$ is zero precisely at threshold,~$\epsilon \eph = (\me c^2)^2$, but, for the (isotropic, monochromatic) background distribution~(\ref{eq:umono}), it soon peaks at~$\sigma_{\gamma\gamma} \simeq \sigma_{\rm T} / 5$ when~$\epsilon \eph \simeq 3.6 (\me c^2)^2$ \citep{gs67}. For such photons~$\knp \sim 3.6 \times 8 \simeq 30$. Hence, although the energy scales at which Klein-Nishina IC cooling and pair production occur are both set, fundamentally, by~$\gkn$, they are offset from one another by a factor of about~$10-30$. The former kicks in when~$\knp \sim 1$ and the latter when~$\knp \sim 8-30$. In this sense, the energy scale~$\gkn$ \quoted[,]{splits}similarly to~$\sigc$ (section~\ref{sec:thomradreconn}), into two that are offset by a fixed ratio.

However, it is not clear that astrophysical reconnection accelerates particles to energies that are high enough to stray from the Thomson limit but not to emit pair-producing photons. If, contrary to our simplified monochromatic assumption, the seed photons have any spread in energy, then photons Comptonized from the high-energy end of the background will more easily pair-produce with the lower-energy component. Even for a thermal radiation bath, the two frequencies where the Planck spectrum attains half its maximum value are offset from each other by about a factor of~$5$, reducing the effective splitting of~$\gkn$ from a factor of~$10-30$ to~$2-6$. The \cutoff[]in the reconnection-energized particle distribution would then have to fall precisely in a narrow range for Klein-Nishina effects to kick in but for pair-production to remain impossible. And, even in this case, only the very highest-energy sliver of particles would experience Klein-Nishina IC losses; most of the particles would still be cooled in the Thomson regime. Thus, from here on, we assume that Klein-Nishina IC scattering coincides with the emission of above-threshold photons in reconnection.

However, just because a high-energy photon is above threshold does not mean that it gets absorbed inside the reconnection system. One must also consider the optical depth,~$\tau_{\gamma\gamma} = \uph \sigma_{\gamma\gamma} L / \eph$, to pair-production. For simplicity, we evaluate~$\tau_{\gamma\gamma}$ at the peak cross section~$\sigma_{\gamma\gamma} \simeq \sigma_{\rm T} / 5$, which is attained when~$\epsilon \eph \simeq 3.6 \me c^2$. Thus
\begin{align}
    \tau_{\gamma\gamma} \equiv \frac{\uph \sigma_{\rm T} L}{5 \eph} = \frac{3}{5} \frac{\gmax \gkn}{\gradt^2} = \frac{3}{5} \frac{\gkn}{\gcool} \, .
    \label{eq:taugg}
\end{align}
We have already encountered~$\tau_{\gamma\gamma}$. It is the (inverse of the) prefactor in the expression for the cooling time~$\tcoolk(\gamma)$ in equation~(\ref{eq:tcoolic}). Thus, the condition~$\gkn > 2.32 \gcool$ [equation~(\ref{eq:tcoolicmin})], which ensures that at least some particles cool in times shorter than~$L/c$, is the same as the optically thick condition~$\tau_{\gamma\gamma} > (3/5) \times 2.32 \simeq 1$.

This means that there is an appreciable range of parameters where one expects both dynamically-important Klein-Nishina radiative cooling and pair-production. Both mechanisms may actively feed back on the reconnection process when~$\gcool \ll \gkn \ll \gradt$. The first relationship,~$\gcool \ll \gkn$, is necessary both for~$\tau_{\gamma\gamma} \gg 1$ and for~$\min[\tcoolk(\gamma)] \ll L / c$. The second criterion~$\gkn \ll \gradt$ is required for at least some particles to enter the regime where Klein-Nishina effects begin to impact their radiative cooling, also enabling them to emit photons above pair threshold.

\begin{figure}
    \centering
    \includegraphics[width=\columnwidth]{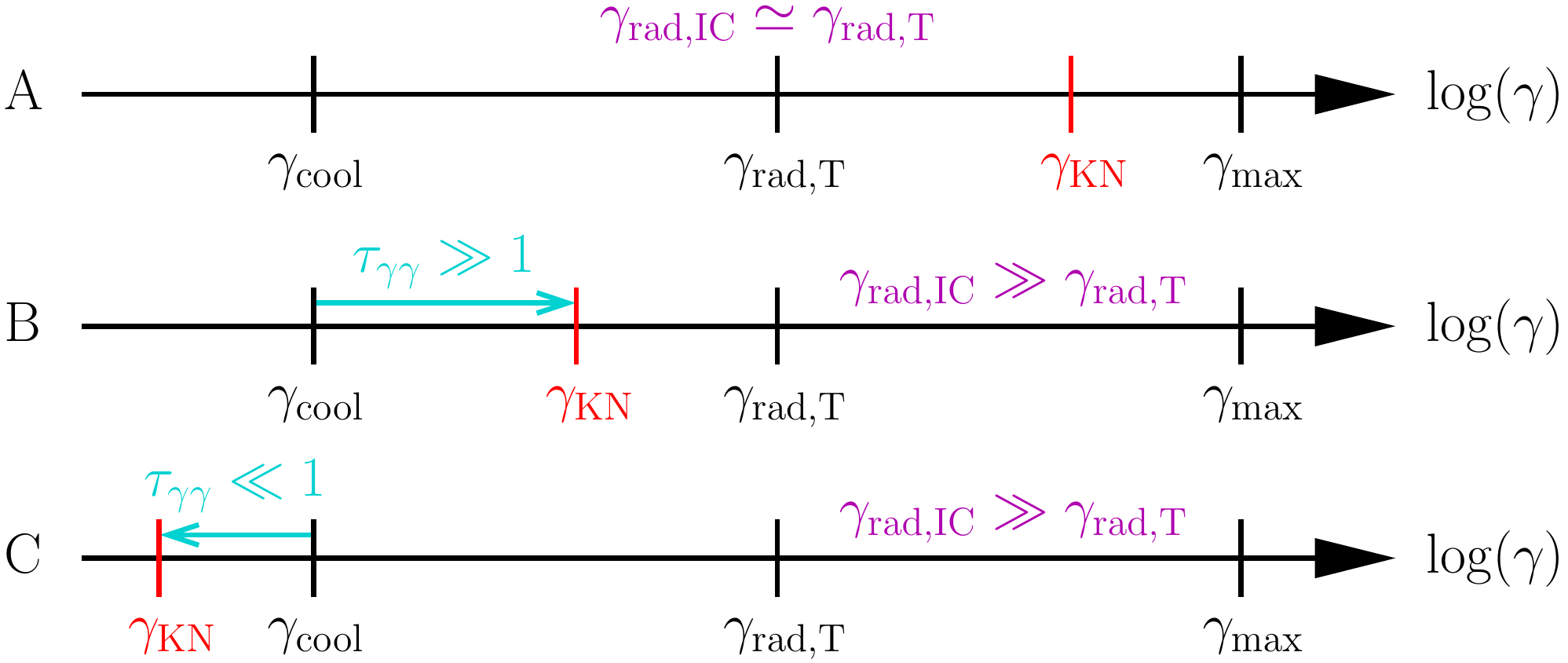}
    \caption{Basic orderings of~$\gkn$ with respect to~$\gcool$,~$\gradt$, and~$\gmax$. Note that~$\gradt \equiv (\gcool \gmax)^{1/2}$ [equation~(\ref{eq:gcool})]. The case~$\gmax < \gradt < \gcool$, which is non-radiative irrespective of~$\gkn$, is not considered. The pair-production optical depth~$\tau_{\gamma\gamma} \sim \gkn / \gcool$ is roughly the separation between~$\gkn$ and~$\gcool$. The \cutoff[~$\gradk$]is virtually identical to~$\gradt$ when~$\gradt \ll \gkn$; otherwise, it is much larger than~$\gradt$. Case~A: A Thomson radiative reconnection ordering (section~\ref{sec:thomradreconn});~$\gradt \ll \gkn$ prevents particles from accessing the Klein-Nishina regime and from emitting above-threshold photons. Case~B: A scale hierarchy where Klein-Nishina IC losses (because~$\gkn \ll \gradt$) and pair production (because~$\gcool \ll \gkn$ implies~$\tau_{\gamma\gamma} \gg 1$) are both likely. Case~C: All orderings with~$\gkn \ll \gcool$ are non/weakly-radiative because~$\tcoolk > L/c$. Even the few photons radiated above pair threshold are not absorbed because~$\tau_{\gamma\gamma} \ll 1$.}
    \label{fig:knppnumlines}
\end{figure}

We are thus equipped with a simple rule for deciding when Klein-Nishina and pair-production physics become important in reconnection. We just assemble all of our energy scales:~$\gmax$,~$\sigc$ (i.e.~$\gx$ and~$\langle \gamma \rangle$),~$\gradt$, and~$\gcool$, arrange these into a familiar Thomson hierarchy (as in section~\ref{sec:thomradreconn}), and insert~$\gkn$ into a relevant location. If~$\gkn$ is larger than~$\gradt$, Klein-Nishina effects are absent because~$\gradt$ imposes a hard upper bound on particle acceleration, and, consequently, no particles ever reach~$\gkn$. If, on the other hand,~$\gkn < \gcool$, then Klein-Nishina effects suppress cooling so much that the whole system becomes non-radiative. Only if~$\gcool \ll \gkn \ll \gradt$ are Klein-Nishina IC cooling and pair-production both important. And, in that case, it is also necessary to consider how~$\gradt$,~$\gkn$, and~$\gcool$ are ordered with respect to the other scales in the problem. These remarks are illustrated in Fig.~\ref{fig:knppnumlines} and elaborated in the next subsection.

\subsubsection{Regimes of Klein-Nishina radiative reconnection}
\label{sec:knregimes}
We now systematically explore, as we did for Thomson IC cooling in section~\ref{sec:thomregimes}, how to classify regimes of Klein-Nishina radiative reconnection. As an example, consider the Thomson ordering~$\gmax \geq \gradt \geq \gx \gg \langle \gamma \rangle \gg \gcool$~($4$th row in Table~\ref{table:tregimes}). If we insert~$\gkn$ between~$\gradt$ and~$\gx$, then Klein-Nishina effects do not affect primary X-point acceleration. They only come into play if secondary energization channels can push particles up to~$\gamma \sim \gkn$. If~$\gkn$ is instead placed between~$\gx$ and~$\langle \gamma \rangle$, Klein-Nishina radiative cooling definitely impacts high-energy particles accelerated near X-points, and these particles are also likely to emit pair-producing photons. Klein-Nishina and pair-production physics become even more important if~$\gkn$ is made smaller than~$\langle \gamma \rangle \sim \sigc / 4$. Then, the bulk of the accelerated particles -- not just the high-energy tail -- emit in the Klein-Nishina regime and, likely, many pairs are produced.

\begin{figure}
    \centering
    \includegraphics[width=\columnwidth]{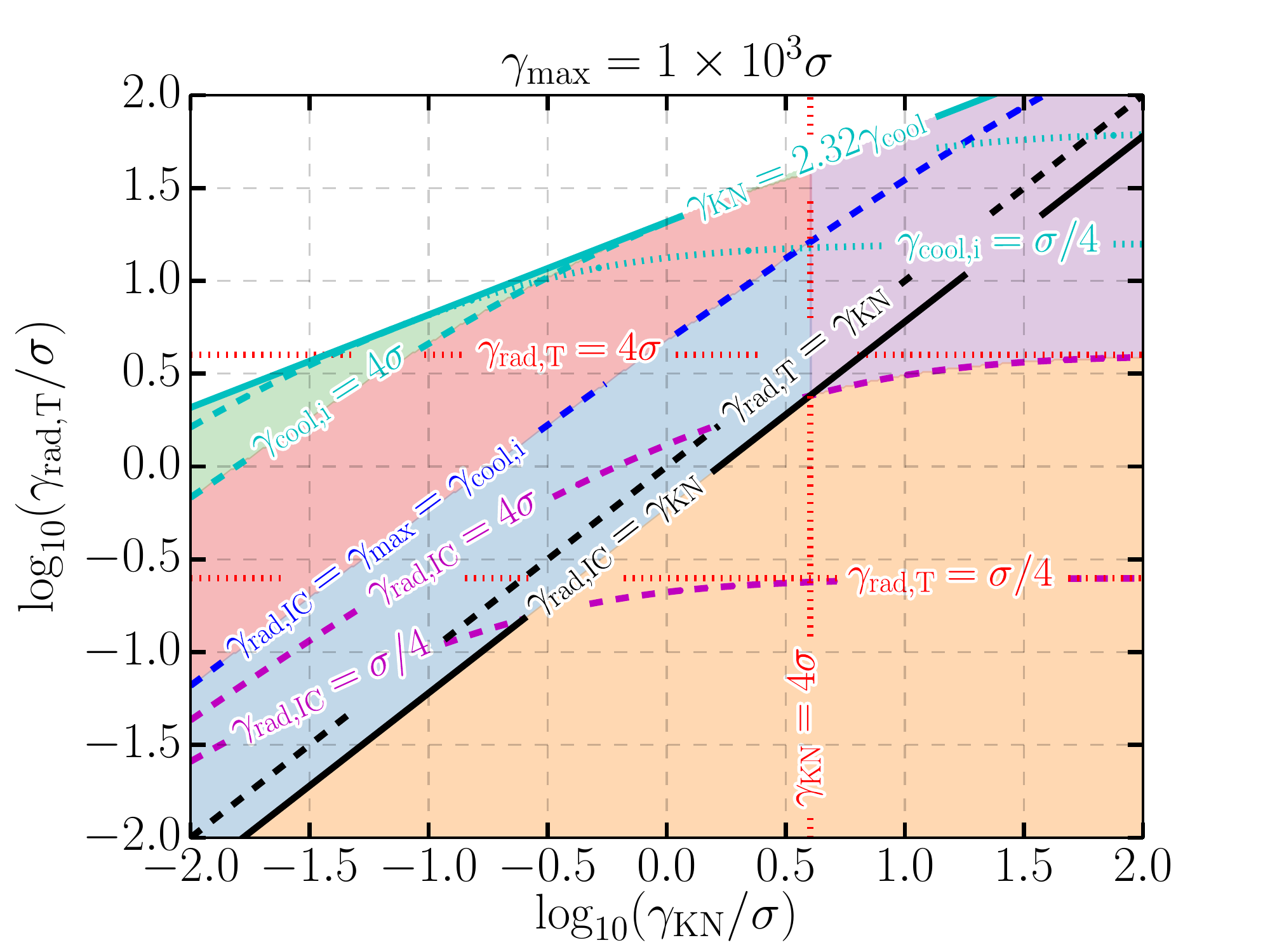}
    \caption{Radiative phase diagram for Klein-Nishina relativistic magnetic reconnection. The diagram is inherently 3-dimensional -- it depends on~$\gradt / \sigc$,~$\gkn / \sigc$, and~$\gmax / \sigc$ -- but we display a~2D slice where~$\gmax = 10^3 \sigc$. In the plot~$\sigc$ is abbreviated as~$\sigma$. Contours show important equalities between energy scales. The~$\gcooli{i} = \rm const.$ contours transition, when they strike the curve~$\gkn = 2.32 \gcool$, from a dashed portion along which~$\gcooli{2}$ is fixed, to a dotted portion where~$\gcooli{1}$ is fixed to the same value. Regions with a common col\spellor[]exhibit similar qualitative behavi\spellor[.]The col\spellor[s]are inspired by assuming that~$\gx = 4 \sigc$ and, fictitiously, that X-point acceleration is the only operative energization mechanism. Thus, the highest Lorentz factor attained is~$\min(\gradk, 4 \sigc)$. In the purple region,~$4 \sigc < \gkn, \gradk$. Hence, radiative cooling does not furnish an upper bound on particle acceleration, and no particles suffer Klein-Nishina losses. In the orange region,~$\gradk < \gkn, 4 \sigc$. Here, Thomson IC losses firmly cap the maximum particle energy to~$\gradk \simeq \gradt < \gkn$, and, consequently, the highest energy particles are radiatively saturated. In the blue region,~$\gkn < \gx, \gradk$ and~$\gradk < \gmax < \gcooli{2}$. Because~$\gkn < \gx$ some particles reach~$\knp \geq 1$. In addition, in the subset of the region where~$\gradk < \gx = 4 \sigc$, the highest energy particles are radiatively saturated. The red region is defined by~$\gradk > \gmax > \gcooli{2} > 4 \sigc$. Thus, radiative losses do not bound particle acceleration, and, as a result, there is likely a broad spectrum of particles above the threshold energy~$\gamma \simeq 8 \gkn$ to emit pair-producing radiation. In the green~$\gcooli{2} < 4 \sigc$ region, Klein-Nishina effects suppress cooling so that some high-energy particles are weakly radiative, having~$\gamma > \gcooli{2}$ and, hence,~$\tcoolk > L/c$. In the white~$\gkn < 2.32 \gcool$ region, all particles have~$\tcoolk > L/c$.}
    \label{fig:radphasespace}
\end{figure}

Because~$\gkn$ is not the only new scale, but also introduces a few derived scales (e.g.~$\gcooli{2}$ and~$\gradk$), exhaustively discussing all possible regimes like in section~\ref{sec:thomregimes} is prohibitively tedious. Even in the preceding paragraph, we did not consider subtleties such as whether~$\gx > \gcooli{2}$, in which case some high-energy particles radiate inefficiently. In lieu of an exhaustive discussion, we supply Fig.~\ref{fig:radphasespace}, a \quoted[,]{phase diagram}in the~$\gradt/\sigc$--$\gkn / \sigc$ plane, of the complex radiative parameter space for Klein-Nishina reconnection. The parameter space is, in reality,~$3$-dimensional, depending also on~$\gmax / \sigc$. To display it in~2D, we set~$\gmax = 10^3 \sigc$ in Fig.~\ref{fig:radphasespace}.

In the figure, contours highlight important values of energy scales and often distinguish different-col\spellor[ed]regimes of interest. We caution that the col\spellor scheme is somewhat arbitrary. Almost every sliver of parameter space enclosed within a set of contours is its own physical regime, and only a subset of the relevant contours are shown. Without \analyz[ing]every possible contour-enclosed region, the best we can do is group regions based on similar expected qualitative behavi\spellor[,]a heuristic that guides the col\spellor[-coding]in Fig.~\ref{fig:radphasespace}. However, this exercise is ultimately subjective. A given grouping is useful for conceptualizing some physical similarities, but may need to be reevaluated if the physics of main interest changes.

We now describe the (col\spellor[-coded)]grouping of regions adopted in Fig.~\ref{fig:radphasespace}. We begin with the fundamentally new domains corresponding to case~B in Fig.~\ref{fig:knppnumlines},~$\gcool < \gkn < \gradt$, where Klein-Nishina effects feature prominently. One of these is the blue area, in which~$\gkn < \gradk$ and~$\gradk < \gmax < \gcooli{2}$. In this area, the overall radiative \cutoff[]Lorentz factor~$\gradk$ is finite (less than~$\gmax$) while~$\gcooli{2}$ is not~($\gcooli{2} > \gmax$). This means that, despite Klein-Nishina suppression of the IC cross section, radiative losses may still regulate the highest achievable energies. To illustrate this, the contour~$\gradk = 4 \sigc = \gx$ is drawn (in this discussion, we assume that~$\gx = 4 \sigc$ for definiteness), below which IC radiation limits direct acceleration near reconnection X-points. Straying from the blue region across the line~$\gmax = \gradk = \gcooli{2}$ lands one in the red area. Here, the ordering of~$\gcooli{2}$ and~$\gradk$ about~$\gmax$ is flipped:~$\gcooli{2} < \gmax < \gradk$. Thus the radiative \cutoff[]energy~$\gradk$ is no longer finite, but the energy~$\gcooli{2}$, beyond which particles are weakly radiative (cooling on \ts[s]longer than~$L/c$), is now accessible. Here, it may be possible for high-energy particles to surpass~$\gcooli{2}$ and enter into a weakly radiative regime. As one moves to the northwest through the red region,~$\gcooli{2}$ becomes lower. Eventually, when the contour~$\gcooli{2} = 4 \sigc$ is crossed,~$\gcooli{2}$ falls below~$\gx \, (= 4 \sigc)$, guaranteeing that some particles venture into the high-energy Klein-Nishina weakly radiative limit. Continuing even farther upward in the diagram, one eventually crosses the~$\gkn = 2.32 \gcool$ line, where the whole system becomes virtually non-radiative (case~C in Fig.~\ref{fig:knppnumlines}).

Finally, let us discuss the {Fig.-\ref{fig:radphasespace}} regions corresponding to case~A of Fig.~\ref{fig:knppnumlines}~($\gcool < \gradt < \gkn, \gmax$). The first of these is the orange area, where~$\gradk \simeq \gradt < \gkn, \gx$. Here, IC radiation limits X-point energization to below~$\gkn$. In the final remaining region, the purple area, we have~$4 \sigc = \gx < \gradk, \gkn$. In this regime, radiative losses do not inhibit X-point acceleration, and X-point acceleration also cannot promote particles to high enough energies to stray outside the Thomson IC cooling limit.

Having overviewed the rich radiative parameter space available to IC-cooled relativistic reconnection, we now specialize to one as-yet relatively unexplored regime where Klein-Nishina physics profoundly impacts the overall dynamics. Here, pair production and Klein-Nishina radiative cooling can conspire together to form an important self-regulation mechanism. We devote the following section to a theoretical exploration of this pair-regulated Klein-Nishina radiative reconnection. We discuss applications to reconnection-driven emission from ADCe and FSRQ jets in section~\ref{sec:applications}.

\section{A model of pair-regulated Klein-Nishina reconnection}
\label{sec:knradreconn_mod}
This section explores technical aspects of reconnection with Klein-Nishina radiative cooling and pair production. The general picture is that pairs are primarily born into the upstream region, where they load the plasma energetically (i.e.\ the pairs are hot) but not from a number density standpoint (i.e.\ the pairs are tenuous). Before diving in, we state the following basic assumptions to clarify the relevant region of radiative phase space (in the sense of Fig.~\ref{fig:radphasespace}):
\begin{enumerate}
     \item Radiation takes place in the Klein-Nishina regime, where~$\gkn < \gradk$ and~$\gkn < \gx$. \label{en:knregime}
     \item The reconnection region is radiatively efficient, with all particles accelerated above~$\gcooli{1} \simeq \gcool$ cooling in less than a dynamical time~$L/c$ and most particles reaching these energies:~$\langle \gamma \rangle \gg \gcool$. \label{en:efficientradiation}
     \item The pair-production mean free paths~$\lmfp$ of all gamma-rays above pair threshold are
         \begin{enumerate}[topsep=0pt]
             \item independent of photon energy, and \label{en:lmfpenindependent}
             \item between the full thickness~$\thickness$ of \textit{\radzone[s]}-- the parts of the reconnection layer where above-threshold photons are produced -- and the layer's full length~$L$, i.e.~$\thickness \ll \lmfp \ll L$. \label{en:lmfpintermediate}
         \end{enumerate}
\end{enumerate}
As a reminder, we refer to as the \quoted{layer}the region of the system threaded with reconnected magnetic flux.
 
\begin{figure}
    \centering
    \includegraphics[width=\columnwidth, trim=0 0 105 0, clip]{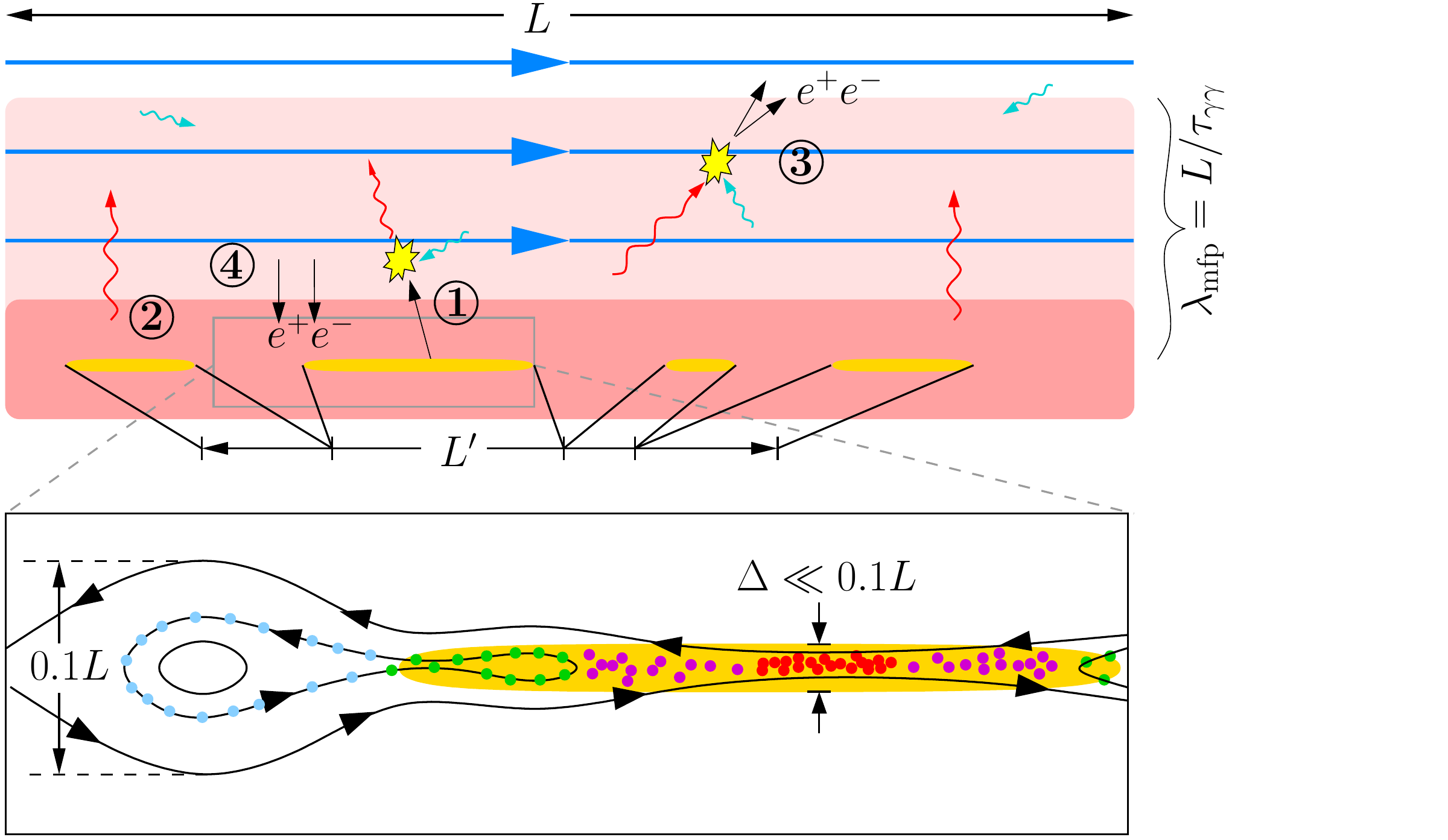}
    \caption{A more detailed view of the Klein-Nishina radiative reconnection system. The layer (region threaded with reconnected flux) contains subregions -- \textit{\radzone[s]}-- where particles are radiating photons above pair-production threshold (gold). These occupy a combined fraction~$L'/L<1$ of the layer's full length. They are also thin, having full transverse width,~$\thickness \ll 0.1 L$, much smaller than the expected size of the largest plasmoids. A detailed view of a hypothetical \radzone[]is displayed on the lower half of the plot. The thinness of the region ultimately stems from the kinetic-scale current sheets where particles (red dots) are accelerated near reconnection X-points. Following their impulsive acceleration in these locations, the particles (magenta dots) are magnetized by the reconnected magnetic field, which carries them away from the vicinity of an X-point. We concentrate on particles (green dots) confined to one reconnected field line. Before these particles are able to fill an entire ring in a large plasmoid where they eventually end up, they cool down below the minimum energy~$\gth$ (light blue dots) to radiate photons above pair-production threshold. The transition point where particles cool below~$\gth$ -- where the col\spellor[ing]changes from green to light blue -- determines both the length of the \radzone[](distance from the X-point) and its width, i.e\ the spread in red/magenta/green particles (those with~$\gamma > \gth$) about the midplane.}
    \label{fig:detaileddiagram}
\end{figure}

Assumption~\ref{en:knregime} places us in the Klein-Nishina -- i.e. blue or red -- region of the radiative phase diagram (Fig.~\ref{fig:radphasespace}). Assumption~\ref{en:efficientradiation} excludes the white and green regions, implying that all of the accelerated particles -- from the average energy~$\langle \gamma \rangle$ to the \cutoff[]energy -- are between~$\gcooli{1}$ and~$\gcooli{2}$, and hence are strongly cooled. Statement~\ref{en:lmfpenindependent} is not strictly true, but the pair-production cross section~$\sigma_{\gamma\gamma}(\epsilon)$ varies relatively weakly with energy beyond its peak~$\sigma_{\gamma\gamma}(\epsilon) \simeq \sigma_{\rm T} / 5$ when~$\epsilon \simeq 3.6 (\me c^2)^2 / \eph$. For example,~$\sigma_{\gamma\gamma}(100 \me^2 c^4 / \eph) \simeq 0.08 \sigma_{\rm T}$. Finally, the inequality~$\lmfp \ll L$ in assumption~\ref{en:lmfpintermediate} means that almost all above-threshold photons produced in the system are also absorbed in the system, and~$\thickness \ll \lmfp$ further means that absorption predominantly occurs in the inflow (upstream) plasma. Note that we distinguish between the effective full thickness,~$\thickness$, of the reconnection layer itself, which could be taken as the width of the largest plasmoids~$\simeq 0.1 L$ \citep{uls10}, and the thickness of the \radzone[,]which (as discussed below in section~\ref{sec:upstreamendens}) could be much thinner, even approaching the thicknesses of interplasmoid current layers. We illustrate the difference between~$\thickness$ and~$0.1L$ in Fig.~\ref{fig:detaileddiagram}.

In addition to all of these assumptions, we ignore effects due to synchrotron radiation. These enter at energy scales~$\gamma > \gamma_{\rm s}$ [see equation~(\ref{eq:gsynapprox}) and its surrounding discussion]. We estimate~$\gamma_{\rm s}$ for certain astrophysical systems, and comment on the consequent limitations on the applicability of our model, in section~\ref{sec:applications}.

To investigate the basic features of reconnection in this radiative regime, we begin (section~\ref{sec:upstreamendens}) with some relatively simple energy-budget arguments. Based on energy considerations alone, we show that a self-regulated steady state or limit cycle should emerge -- irrespective, even, of whether a pair cascade develops in the upstream region. We then decorate this basic picture by \analyz[ing]the number of produced pairs. This shows (section~\ref{sec:nocascades}) that an exponential pair cascade, with each generation containing a constant factor~$> 1$ more particles than the previous one, is not expected except for (almost unrealistically) efficient particle acceleration in the reconnection layer. We further apply detailed information on the distribution of newborn pairs, showing that (section~\ref{sec:upstreammult}), for~$\sigc \ll \gkn$, these should be fewer than those originally present in the upstream region.

\subsection{The large energy density of newborn upstream pairs}
\label{sec:upstreamendens}
Because we assume a radiatively efficient reconnection layer~\ref{en:efficientradiation}, a sizeable fraction \citep[e.g. one half;][]{wpu19} of the inflowing Poynting flux is promptly emitted. A fraction~$\mathcal{F}$ of the radiated energy lies above pair threshold with the ambient photon bath. This fraction penetrates a distance~$\lmfp$ [assumption~\ref{en:lmfpintermediate}] into the upstream plasma on both sides of the layer.\footnote{For simplicity, we ignore kinetic beaming \citep{ucb11, cwu12, mwu20}, which produces potentially important anisotropy in the distributions of high-energy particles and their emitted photons. We comment on expected consequences of this beaming in Appendix~\ref{sec:fnoesc} but ultimately defer its full treatment to a future simulation study.} There, it is recaptured as newborn hot pairs and, ultimately, readvected into the layer. If the energy density of fresh pairs is high enough, the overall enthalpy density~$w = w_0 + w_{\gamma\gamma}$ of inflowing material substantially increases. This reduces the effective hot magnetization
\begin{align}
    \sighgen \equiv \frac{B_0^2}{4 \pi w} 
    \label{eq:sigheffdef}
\end{align}
below~$\sigh = B_0^2 / 4 \pi w_0$ [equation (\ref{eq:sighdef})], which \characteriz[es]the \textit{far upstream} region (beyond~$\lmfp$ from the layer). In our convention, subscript~\quoted{0}denotes far upstream quantities and subscript~\quoted{$\gamma\gamma$}quantities sourced by pair creation within~$\lmfp$ of the layer. Corresponding naked symbols (e.g.~$\sighgen$ or~$w$) are decided by a combination of pair-creation-sourced and far upstream values.

A reduced effective~$\sighgen$ may strongly suppress the efficiency of \nonthermal[]particle acceleration (NTPA) in the layer \citep[e.g.][]{ss14, sgp16, gld14, gld15, wuc16, wu17, wub18, bso18}. This enables a negative feedback loop, in which a layer fed initially by highly magnetized~($\sigh \gg 1$) plasma efficiently accelerates particles to gamma-ray emitting energies. The gamma-rays, in turn, produce pairs in the inflow region, reducing its effective magnetization and, hence, suppressing subsequent NTPA (cf.\ \citealt{hps19}; see Fig.~\ref{fig:knppdiagram}). In this section, we calculate the fixed point~$\sighgen$ for this feedback loop. Additionally, we determine the conditions governing whether the system asymptotically approaches its fixed point in a late-time steady state. We further show that, if the fixed point is not reached, the system exhibits undamped, large-amplitude cycles of copious pair creation followed by shutdown of NTPA.

The Poynting flux delivered to the reconnection layer (per unit length in the out-of-plane direction) is
\begin{align}
    P_{\rm Poynt} \sim 2 L \beta_{\rm rec} c \frac{B_0^2}{4 \pi} \, .
    \label{eq:ppoynt}
\end{align}
The leading factor of~$2$ results from Poynting flux entering the reconnection region from two directions. If half of this power is given to particles that quickly [within~$L/c$; assumption~\ref{en:efficientradiation}] radiate it away through the IC process, the volume-averaged IC emissivity~$j_{\rm IC}$ (power radiated per unit volume) in the reconnection layer satisfies
\begin{align}
    j_{\rm IC} \thickness L' \sim \frac{1}{2} P_{\rm Poynt} \sim L \beta_{\rm rec} c \frac{B_0^2}{4 \pi} \, ,
    \label{eq:jic}
\end{align}
where~$L'$ is the combined length of all \radzone[s]in the reconnection layer. One can ignore all plasmoid/current-sheet substructure, taking the entire layer to be one large \radzone[,]by setting~$L' = L$. However, given our assumption~\ref{en:efficientradiation} of a radiatively efficient reconnection system,~$L'$ may actually be shorter than~$L$. This is because particles may cool to below the minimum energy,~$\gth \equiv 8 \gkn$ (section~\ref{sec:knradreconnpp}), to emit pair-producing photons before trave\ling[]far from their primary X-point acceleration sites. Moreover, as particles travel away from an X-point, they also spread out about the reconnection midplane. Thus, a cooling limit on the combined length of \radzone[s](such that~$L' < L$) also limits their effective thickness,~$\thickness$, potentially keeping them much thinner than the characteristic large-plasmoid width (e.g.~$\thickness < 0.1 L$; Fig.~\ref{fig:detaileddiagram}).

To determine the total enthalpy density~$w$ and, from it, the effective magnetization~$\sighgen$ [equation~(\ref{eq:sigheffdef})], we need to know the fraction~$\mathcal{F}$ of power radiated away from the reconnection layer above pair-production threshold (and, hence, captured in the upstream region as electron-positron pairs). Using~$\gth$ along with the distribution function of radiating layer particles~$\dif N / \dif \gamma$, ~$\mathcal{F}$ reads 
\begin{align}
    \mathcal{F} = \frac{\int_{\gth}^\infty \dif \gamma \, \dif N / \dif \gamma \, P_{\rm IC}(\gamma)}{\int_1^\infty \dif \gamma \, \dif N / \dif \gamma \, P_{\rm IC}(\gamma)} \, .
    \label{eq:scriptf}
\end{align}
To evaluate~$\mathcal{F}$, we insert a power-law reconnection-energized pair-plasma distribution:
\begin{align}
    \frac{\dif N}{\dif \gamma} = A \begin{cases}
        \gamma^{-p} &\gamma_1 \leq \gamma \leq \gf \\
        0 &\mathrm{otherwise}
    \end{cases} \, ,
    \label{eq:dndgam}
\end{align}
where~$A$ is a \normaliz[ation]factor and~$\gamma_1 \ll \gkn$ is assumed. If~$p < 3$, the~$\gamma^2$-dependence of~$P_{\rm IC}(\gamma)$ when~$\gamma \ll \gkn$ suppresses the dependence of~$\mathcal{F}$ on the onset energy~$\gamma_1 \ll \gkn$ of the power law, and~$\gamma_1$ can thus be taken to unity. If, instead,~$p \geq 3$, the onset energy,~$\gamma_1$, can also be ignored -- the same dependence,~$P_{\rm IC}(\gamma \ll \gkn) \propto \gamma^2$, pushes~$\mathcal{F}$ to zero independently of~$\gamma_1$. Thus, our assumption~$\gamma_1 \ll \gkn$ is equivalent to setting~$\gamma_1 = 1$.

\begin{figure*}
    \centering
    \begin{subfigure}{0.49\textwidth}
        \centering
        \includegraphics[width=\linewidth]{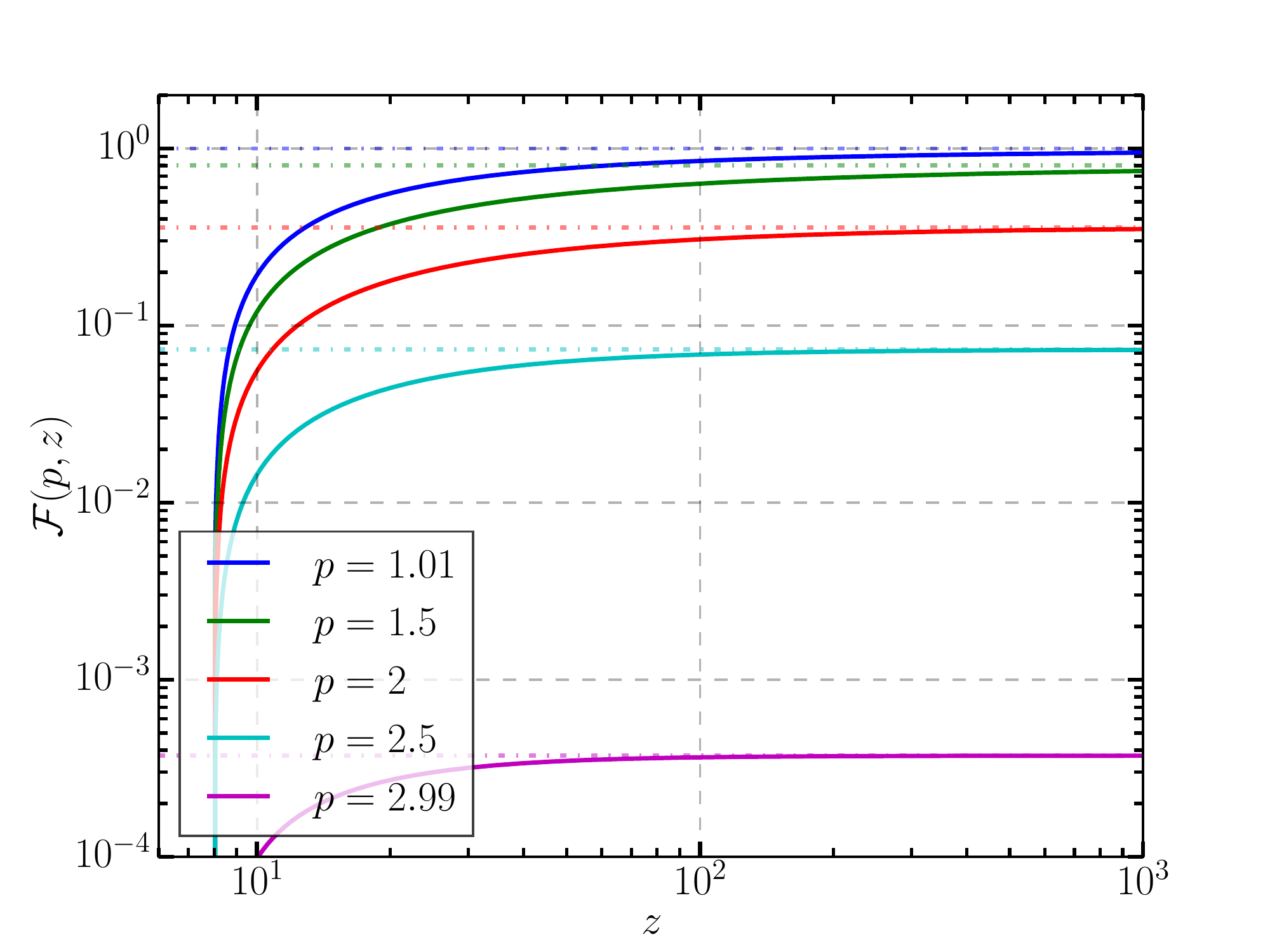}
    \end{subfigure}
    \begin{subfigure}{0.49\textwidth}
        \centering
        \includegraphics[width=\linewidth]{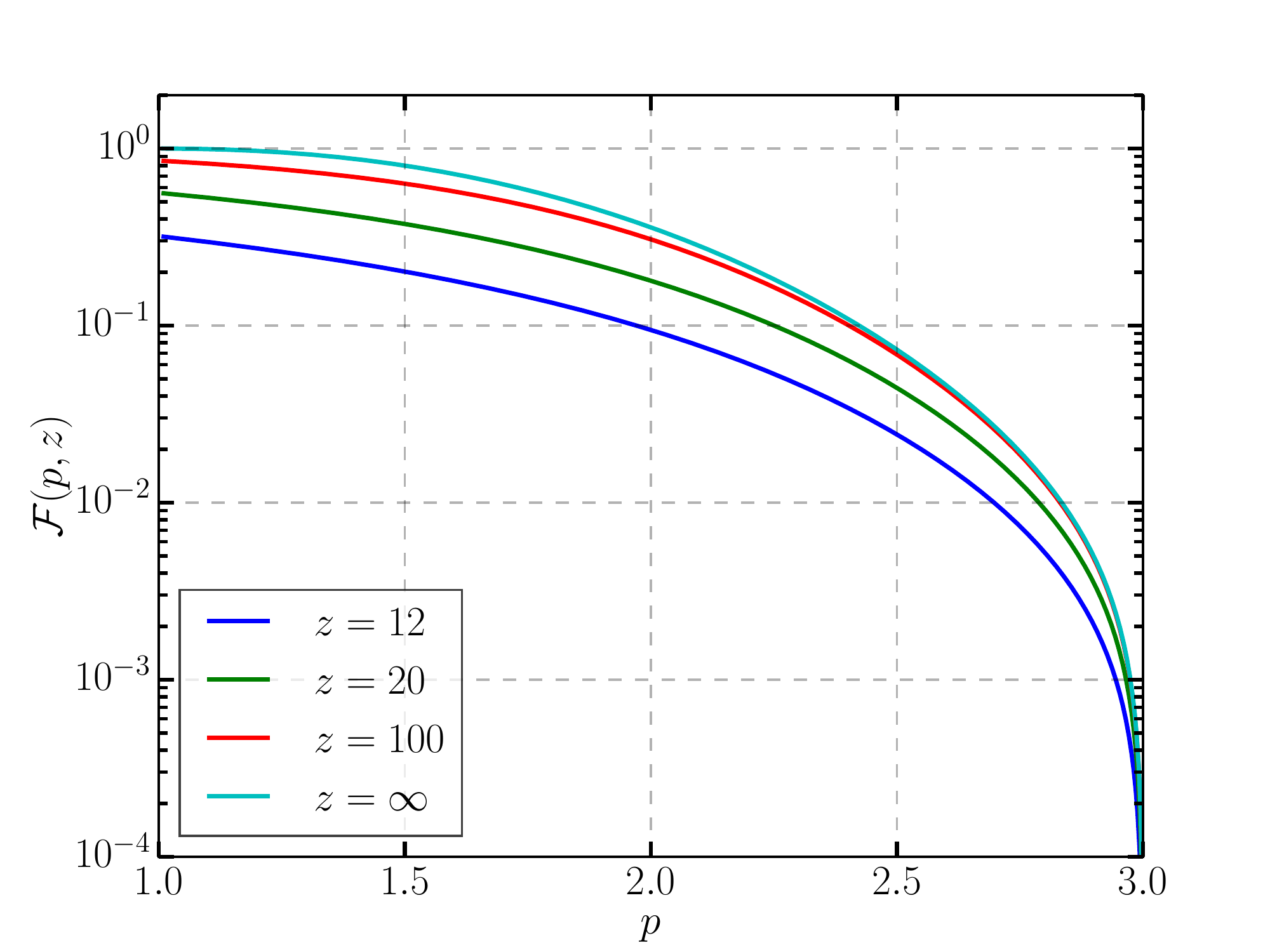}
    \end{subfigure}
    \caption{Left: A plot of~$\mathcal{F}(p, z)$ as a function of~$z \equiv \gf / \gkn$ for several values of~$p$. Dot-dashed lines indicate~$\lim_{z \to \infty}\mathcal{F}(p,z)$. Right: A plot of~$\mathcal{F}(p,z)$ as a function of~$p$ for several values of~$z$. The fraction~$\mathcal{F}$ is strongly dependent on~$p$ but not on~$z$ (at least after the threshold value~$z = 8$ is crossed).}
    \label{fig:fplaw}
\end{figure*}

Substituting, now,~(\ref{eq:pic}) and~(\ref{eq:dndgam}) into~(\ref{eq:scriptf}), as well as putting~$\gth = 8 \gkn$ and~$\gamma_1 / \gkn = 0$, gives
\begin{align}
    \mathcal{F}(p, z) = \frac{\int_{8}^{z} \dif x \, x^{-p + 2} f_{\rm KN}(x)}{\int_0^{z} \dif x \, x^{-p+2} f_{\rm KN}(x)} \, ,
    \label{eq:fplaw}
\end{align}
where~$z \equiv \gf / \gkn$.
Fig.~\ref{fig:fplaw} displays~$\mathcal{F}(p,z)$ computed according to~(\ref{eq:fplaw}). The graphs confirm the above argument that~$\lim_{p \to 3}\mathcal{F}(p,z) = 0$ for all~$z$. Furthermore, because~$f_{\rm KN}(x) \sim \ln(x) / x^2$ as~$x \to \infty$, when~$p \leq 1$, the integrals in~(\ref{eq:fplaw}) diverge with~$z$, but in such a way that~$\mathcal{F} = 1$. This signals that virtually all radiation from the layer is emitted above pair threshold.
Fig.~\ref{fig:fplaw} also shows that, modulo a strong~$z$-dependence near pair threshold~$z = 8$,~$\mathcal{F}(p,z)$ becomes nearly~$z$-independent once~$z \gtrsim 12$. Essentially,~$\mathcal{F}(p, z \gtrsim 12) \simeq \mathcal{F}(p, \infty)$. 

Next, we explicitly connect the fraction~$\mathcal{F}$ to the effective hot magnetization~$\sighgen$. The power~$\mathcal{F} j_{\rm IC} \thickness L'$ shining out of the reconnection layer's \radzone[s]penetrates a distance~$\lmfp$ back into the upstream area before being deposited as hot pairs. Assuming this deposition is approximately uniform in space up to a distance~$\lmfp$ above and below the reconnection layer, hot pairs add to the upstream plasma energy density at a rate~$\dif u_{\gamma\gamma} / \dif t$ satisfying
\begin{align}
    \frac{\dif u_{\gamma\gamma}}{\dif t} (2 \lmfp) L &\sim \mathcal{F} j_{\rm IC} \thickness L' \notag \\
    \Rightarrow \frac{\dif u_{\gamma\gamma}}{\dif t} &\sim \frac{\mathcal{F} j_{\rm IC}}{2} \frac{\thickness}{\lmfp} \frac{L'}{L} \sim \mathcal{F} \frac{\beta_{\rm rec} c}{\lmfp} \frac{B_0^2}{8 \pi} \, .
    \label{eq:duggdt}
\end{align}
In the first line, we assume that, as the radiation propagates away from the layer, it also fills in the gaps between radiation zones so that the upstream region receives pairs approximately uniformly across its length~$L$. The second line in~(\ref{eq:duggdt}) is obtained from the first by plugging in equation~(\ref{eq:jic}). The factor of~$2$ accounts for radiated energy being absorbed both below and above the layer. 

Consider a plasma parcel with initial energy density~$u_0$ that starts far upstream,~$\abs{y} \gg \lmfp$, of the layer. The parcel is advected inward at transverse velocity $v_{y} = -\sign(y) \beta_{\rm rec} c \simeq -\sign(y) 0.1 c$, and, upon reaching the pair-creation zone,~$\,\abs{y} \sim \lmfp$, begins accruing additional energy at the rate~$\dif u_{\gamma\gamma} / \dif t$. The extra energy acquired in transit from~$\, \abs{y} \sim \lmfp$ to the layer~($\,\abs{y} \sim \thickness$) is simply
\begin{align}
    u_{\gamma\gamma}^+ \sim \frac{1}{\beta_{\rm rec} c} (\lmfp - \thickness) \frac{\dif u_{\gamma\gamma}}{\dif t} \simeq \frac{\lmfp}{\beta_{\rm rec} c} \frac{\dif u_{\gamma\gamma}}{\dif t} \sim \mathcal{F} \frac{B_0^2}{8 \pi} \, .
    \label{eq:uggp}
\end{align}
The accumulated internal energy density~$u_{\gamma\gamma}^+$ is less than~$B_0^2/8\pi$ because~$\mathcal{F} \leq 1$, but it can still far exceed~$u_0$ given sufficient magnetization~$\sigh \gg 1$. The superscript~\quoted[]{$+$}denotes that this is only energy \textit{added} to the plasma; we have not yet considered that some energy may be lost en route to the layer -- either through radiation or because particles physically escape the system.

Importantly, the \quoted[~$\tread \equiv \lmfp/\beta_{\rm rec}c$]{readvection time}cancels in~(\ref{eq:uggp}). Thus, whether the pair-creation zone is truly confined to transverse distances~$\, \abs{y} \sim \lmfp$ or occupies a much larger region (for example, for an~$N$-generation pair cascade, one expects~$\abs{y} \sim \sqrt{N} \lmfp \gg \lmfp$ -- a possibility that we entertain in section~\ref{sec:nocascades}),~$u_{\gamma\gamma}^+$ remains approximately the same. For reference, the readvection time is related to the global dynamical time~$L/c$ through
\begin{align}
    \tread \equiv \frac{\lmfp}{\beta_{\rm rec} c} \simeq \frac{10}{\tau_{\gamma\gamma}} \frac{L}{c} \, ,
    \label{eq:readvectdivlc}
\end{align}
where we used~$\lmfp = L / \tau_{\gamma\gamma}$ and~$\beta_{\rm rec} \simeq 0.1$. Note that the prefix \quoted{re}in \quoted{readvection}applies only to the energy, which is captured \textit{again} by the reconnection layer. The pairs that carry this energy, by contrast, are advected into the layer for their first time.

We now estimate~$u_{\gamma\gamma}$, the energy density retained by the fresh plasma swept into the reconnection layer. This yields the enthalpy density~$w_{\gamma\gamma}$ and, through equation~(\ref{eq:sigheffdef}), the effective hot magnetization~$\sighgen$. Now,~$u_{\gamma\gamma}$ is less than the deposited energy density~$u_{\gamma\gamma}^+$ because, while trave\ling[]to the layer, newborn pairs may both radiatively cool and escape the system. To account for this, we define the \textit{energy recapture efficiency},~$\xi \equiv u_{\gamma\gamma} / u_{\gamma\gamma}^+ \leq 1$, and write 
\begin{align}
    \xi = \fnc f_{\rm noesc} \, .
    \label{eq:xifacs}
\end{align}
Here,~$\fnc$ and~$f_{\rm noesc}$ are, respectively, the fraction of the accumulated energy that is not radiated away~($\fnc$) and that is not lost through escaping particles~($f_{\rm noesc}$). 

We calculate the cooling factor~$\fnc$ in detail in Appendix~\ref{sec:feedbackdetails}. There, we identify a physically allowed range~$\fnc \in [3/400, 1]$ and show how, within this interval,~$\fnc$ depends on the other parameters in the problem (on the effective magnetization~$\sighgen$ and on the \cutoff[~$z$).] While that calculation allows us to compute~$\sighgen$ self-consistently (since, in reality,~$\fnc$ depends on~$\sighgen$), it is mathematically complicated. Furthermore, we find that the main qualitative features of self-regulated Klein-Nishina reconnection are captured by treating~$\fnc$ as an independent parameter and scanning it across the allowed interval~$[3/400, 1]$. That is the approach we adopt in this section.

In addition to this simplified prescription for~$\fnc$, we set the escape factor~$f_{\rm noesc}$ to unity, effectively putting~$\xi = \fnc$. This is what one expects if the time~$L/c$ for a relativistic particle to stream out of the system is longer than the readvection time~(\ref{eq:readvectdivlc}), which is true for~$\tau_{\gamma\gamma} \gtrsim 10$ (and hence for a broad range of radiative parameters). We comment more thoroughly on the many additional kinetic effects that may influence~$f_{\rm noesc}$ in Appendix~\ref{sec:fnoesc}. However, because most of these effects tend to push~$f_{\rm noesc}$ toward unity, we simply leave~$f_{\rm noesc} \simeq 1$ from here onward.

Using~$\xi$, the energy density of fresh pairs entering the reconnection layer is
\begin{align}
    u_{\gamma\gamma} = \xi u_{\gamma\gamma}^+ \sim \xi \mathcal{F} \frac{B_0^2}{8 \pi} \, .
    \label{eq:ugg}
\end{align}
If these pairs are relativistically hot, then~$p_{\gamma\gamma} = u_{\gamma\gamma} / 3$ and~$w_{\gamma\gamma} = p_{\gamma\gamma} + u_{\gamma\gamma} \simeq (4/3)u_{\gamma\gamma}$; otherwise~$w_{\gamma\gamma} = u_{\gamma\gamma}$. We take~$w_{\gamma\gamma} = (4/3) u_{\gamma\gamma}$ -- still a good approximation in the non-relativistic limit. 

The effective inflowing plasma magnetization~$\sighgen$ is then
\begin{align}
    \sighgen \equiv \frac{B_0^2}{4 \pi (w_0 + w_{\gamma\gamma})} = \frac{B_0^2 / 4 \pi w_0}{1 + w_{\gamma\gamma} / w_0} \sim \frac{\sigh}{1 + 2 \xi \mathcal{F} \sigh / 3} \, .
    \label{eq:sigheff}
\end{align}
Equation~(\ref{eq:sigheff}) encodes two main possible fixed points for~$\sighgen$. The first is when~$\xi \mathcal{F} \sigh \ll 1$. Then, pair-production is too inefficient to load the upstream plasma substantially and the solution to~(\ref{eq:sigheff}) is simply~$\sighgen \simeq \sigh$. The other regime is when~$\xi \mathcal{F} \sigh \gg 1$. In this situation, hot pairs suppress~$\sighgen$ to a universal value
\begin{align}
    \sighgen \sim \frac{3}{2 \xi \mathcal{F}} \, ,
    \label{eq:sigheffuniv}
\end{align}
which is entirely independent of~$\sigh$. Not only is~(\ref{eq:sigheffuniv}) universal, but, in principle, it can be solved to yield self-consistent values of~$\sighgen$ and~$p$. This is because the effective magnetization governs the efficiency of NTPA \citep[cf.][]{wuc16, wu17, wub18, bso18} and ultimately specifies the power-law index~$p$. One only needs to know the reconnection NTPA \quoted[,~$p(\sighgen)$.]{equation of state}

Let us assume that a suitable~$p(\sighgen)$ can be borrowed from non-radiative reconnection studies. We take
\begin{align}
    p(\sighgen) = 1 + 2 / \sqrt{\sighgen} \, ,
    \label{eq:eos}
\end{align}
which can be obtained from fitting the data in fig.~$3$ of \citet{wuc16} to the general form~$p = C_0 + C_1 / \sqrt{\sighgen}$ used by \citet{wu17} \citep[see also][]{wub18, bso18}. We acknowledge that the distribution~$\dif N / \dif \gamma$ in equation~(\ref{eq:dndgam}) is the instantaneous distribution of radiating particles in the reconnection layer, which -- in our radiative context -- may differ from the injected (non-radiative) power-law distribution \characteriz[ed]by~$p(\sighgen)$. Later on, we account for approximate radiative modifications to the distribution of emitting particles. For mathematical transparency, however, in this first calculation, we plug~(\ref{eq:eos}) directly into our expression for~$\mathcal{F}$.

To simplify further, we take~$z = \gf / \gkn \to \infty$ when evaluating~$\mathcal{F}$ even though calculating~$\sighgen$ and~$p$ runs the same for any~$z$. As previously remarked, the fraction~$\mathcal{F}(p,z)$ is relatively~$z$-independent as long as~$z \gtrsim 12$, so taking~$z \to \infty$ gives a solution representing a wide range of likely values (i.e.\ almost all values beyond those very close to the threshold~$z = 8$ for pair production to turn on).

\begin{figure}
    \centering
    \includegraphics[width=\columnwidth]{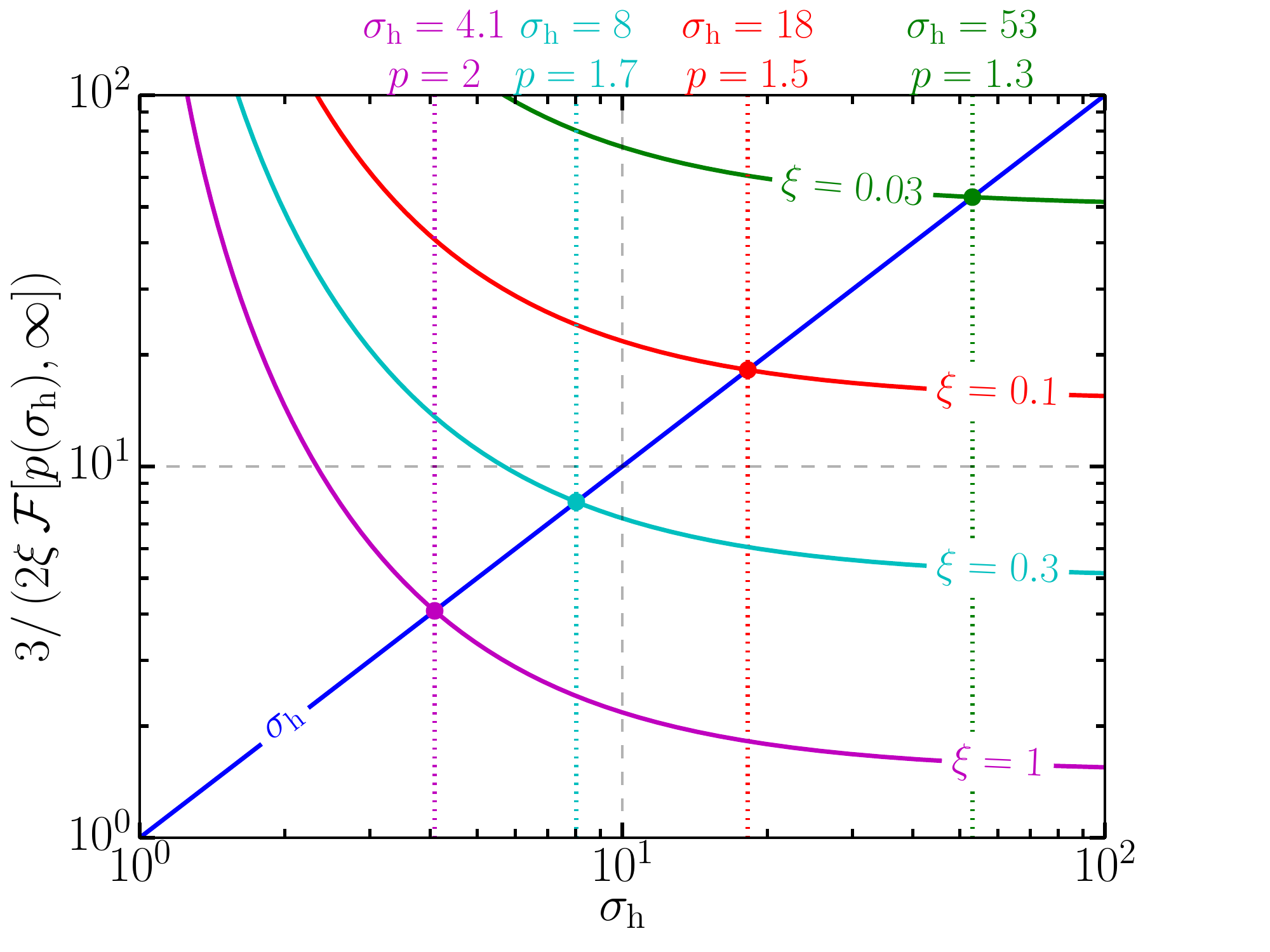}
    \caption{Solutions to equations~(\ref{eq:sigheffuniv}) and~(\ref{eq:eos}) for several~$\xi$ values. Lower~$\xi$ causes the pairs born into the upstream region to cool more. This somewhat inhibits the feedback mechanism by reducing the overall enthalpy density~$w = w_0 + w_{\gamma\gamma} \simeq w_{\gamma\gamma}$ of plasma arriving at the layer.}
    \label{fig:sighunivnorad}
\end{figure}

We now solve~(\ref{eq:sigheffuniv}) and~(\ref{eq:eos}) for a variety of~$\xi$ values and graphically present the solutions in Fig.~\ref{fig:sighunivnorad}. A lower~$\xi$ (lower~$f_{\rm nocool}$) increases radiative cooling of newborn pairs as they travel toward the reconnection layer. This diminishes their enthalpy density,~$w_{\gamma\gamma}$ (which, nevertheless, still dominates over the initial plasma because~$w = w_0 + w_{\gamma\gamma} \simeq w_{\gamma\gamma}$), enhancing the effective magnetization~$\sighgen$, and, through~$p(\sighgen)$, hardening the resulting distribution of reconnection-energized particles.

\begin{figure}
    \centering
    \includegraphics[width=\columnwidth]{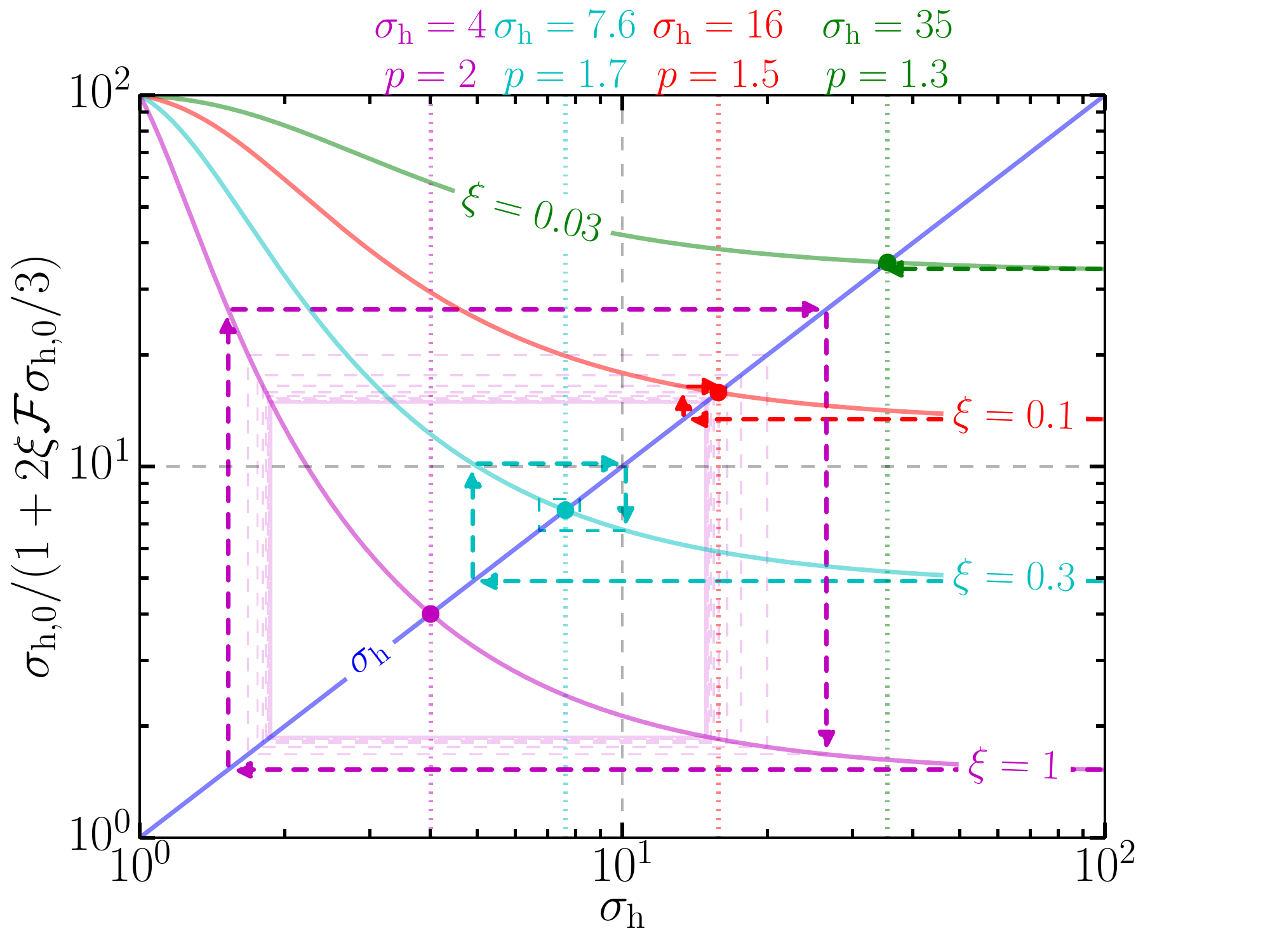}
    \caption{Solutions to the more general equation~(\ref{eq:sigheff}) and equation~(\ref{eq:eos}) using~$\sigh = 100$. Unlike equation~(\ref{eq:sigheffuniv}), equation~(\ref{eq:sigheff}) does not assume \textit{a priori} that~$\sighgen \ll \sigh$. The figure also depicts the stability of each solution. Systems starting with magnetization~$\sigh$ transition to a new magnetization~$\sighn{1}=h(\sigh)$ after approximately one readvection time~$\lmfp / \beta_{\rm rec} c$. This is illustrated for each~$\xi$ by a horizontal line running left from~$h(\sigh)$ to the corresponding new value of~$\sighn{1}$ on the blue diagonal [which represents the curve~$x = h(x)$]. Then, after another readvection time, the second modified magnetization~$\sighn{2} = h(\sighn{1})$ is reached. This is represented by both an upward-running dashed line from~$\sighn{1}$ to~$h(\sighn{1})$ and a horizontal line, from left to right, terminating on the corresponding value~$\sighn{2}$ on the diagonal. Further transitions are denoted by one vertical and one horizontal dashed line: either down-then-left or up-then-right. For low~$\xi$, the system approaches the fixed point~$\sighgen$ after just a few transitions. For high~$\xi$, pair feedback is so efficient that the system gets stuck in a two-state swing cycle.}
    \label{fig:sighunivnoradwithevolution}
\end{figure}

In Fig.~\ref{fig:sighunivnorad}, we solve equations~(\ref{eq:sigheffuniv}) and~(\ref{eq:eos}) rather than the more general form~(\ref{eq:sigheff}). This presupposes that the solution~$\sighgen$ is much smaller than the original (far upstream) hot magnetization~$\sigh$. To illustrate the effect of a finite~$\sigh$, we also display solutions to equation~(\ref{eq:sigheff}) for~$\sigh = 100$ in Fig.~\ref{fig:sighunivnoradwithevolution}. As expected, a finite~$\sigh$ has relatively little impact on the value of~$\sighgen$ when~$\sighgen \ll \sigh$ -- the universal regime in which the solution is insensitive to the far upstream magnetization. However, as the resulting solution~$\sighgen$ gets closer to~$\sigh$, the approximate solution obtained from~(\ref{eq:sigheffuniv}) becomes less accurate. This occurs roughly when~$\xi \sim 1 / \sigh$ (compare, for example, the solutions obtained for~$\xi = 0.03 = 3 / \sigh$ in Figs.~\ref{fig:sighunivnorad} and~\ref{fig:sighunivnoradwithevolution}). In addition, Fig.~\ref{fig:sighunivnoradwithevolution} illustrates the stability of the fixed point~$\sighgen$, which is the topic of the next section.

\subsubsection{Stability of the pair-loaded~$\sighgen$}
Now that we know how to calculate the fixed point~$\sighgen$, we can also begin to ask whether a system that starts from the initial magnetization~$\sigh$ actually approaches~$\sighgen$ at some late time. We call the fixed point \quoted{stable}if the plasma feeding the reconnection layer approaches a quasi-steady magnetization~$\sighgen$, and \quoted{unstable}otherwise. Where necessary, we further distinguish \quoted[,]{global stability}which refers to the notion of stability just described, from \quoted[,]{local stability}which is simpler, and only determines whether a system that starts at a magnetization some infinitesimal distance~$\delta \sighgen$ away from~$\sighgen$ approaches~$\sighgen$.

To discuss stability quantitatively, we abbreviate the right-hand-side of~(\ref{eq:sigheff}) as
\begin{align}
    h(x) \equiv \frac{\sigh}{1 + 2 \xi \, \mathcal{F}[p(x), \infty] \, \sigh / 3} \, .
    \label{eq:hdef}
\end{align}
A system that begins with initial magnetization~$\sigh$ (i.e. before any pairs have been produced) accelerates a distribution of particles in the reconnection layer with a power-law index~$p(\sigh)$ given by equation~(\ref{eq:eos}). These particles then radiate photons, some of which are above pair threshold with the background radiation and consequently -- after a time~$\sim \lmfp / c$ -- pair-produce somewhere in the zone~$\abs{y} \lesssim \lmfp$. The newborn pairs are then advected toward the layer, reaching it after a readvection time~$\tread = \lmfp / \beta_{\rm rec} c$. At that point the layer witnesses a new effective magnetization~$\sighn{1}$ that is determined by~$\sigh$ via~$\sighn{1} = h(\sigh)$. Assuming that the time for the layer to respond to a new magnetization is smaller than~$\lmfp / \beta_{\rm rec} c$, the lag time between when the layer starts processing a~$\sigh$-plasma and when it starts to see a~$\sighn{1}$-plasma is~$\lmfp / c + \lmfp / \beta_{\rm rec} c \simeq \lmfp / \beta_{\rm rec} c$. Thus, the readvection time \characteriz[es]the transition period from the initial magnetization~$\sigh$ to the first modified magnetization~$\sighn{1}$.

By similar reasoning, after about~$n \lmfp / \beta_{\rm rec} c$, the layer witnesses effective magnetization
\begin{align}
    \sighn{n} = h(\sighn{n-1}) \, .
    \label{eq:sighn}
\end{align}
Thus, the system approaches the fixed point~$\sighgen$ as an asymptotic steady state if~$\lim_{n \to \infty} \sighn{n} = \sighgen$. Of course, if the convergence is slow, then the system may not actually reach the fixed point~$\sighgen$ (even though it is stable) before reconnection finishes. All that is certain in that case is that oscillatory swings about~$\sighgen$ are damped: with each successive readvection time, the system reaches a magnetization that is somewhat closer to the asymptotic steady state. However, if the long-time limit of the iterated map~$\sighn{n} = h(\sighn{n-1})$ does not approach the fixed point~$\sighgen$, then some other behavi\spellor occurs. That late-time outcome is large-amplitude, undamped oscillations about~$\sighgen$ between two states, one with a low magnetization~$\sighn{<} < \sighgen$ and one with a high value~$\sighn{>} > \sighgen$. The high-$\sighn{>}$ state drives efficient NTPA in the reconnection layer and, consequently, pair production in the upstream region. When the created hot pairs reach the layer, they initiate the low-$\sighn{<}$ state, where NTPA is shut down, pair creation ceases, and, eventually (after another~$\lmfp / \beta_{\rm rec} c$), the high-$\sighn{>}$ state is restored. The cycle repeats from there.

Both behavi\spellor[s]-- an asymptotic steady state and a late-time two-state cycle -- are illustrated in Fig.~\ref{fig:sighunivnoradwithevolution}. For low energy recapture efficiency~$\xi$, the system tends toward a steady state; when~$\xi$ is increased beyond a critical threshold, the system bifurcates, asymptotically fav\spellor[ing]large amplitude swings. This can be understood as follows. For efficient feedback, so much of the energy radiated from the layer is caught in, and then recaptured from, the upstream region that the system overshoots the fixed point~$\sighgen$ by a wide margin. This chokes subsequent pair-producing emission from the layer effectively enough that the system violently ricochets back to a high magnetization, getting  caught between two extreme magnetizations~$\sighn{<}$ and~$\sighn{>}$ as described above.

We now show how this physical picture is encoded in our mathematical machinery. As is well known from the theory of iterated maps, the fixed point~$\sighgen = h(\sighgen)$ is locally stable if~$\, \abs{h'(\sighgen)} \leq 1$ (meaning, as discussed before, that magnetizations~$\sighgen + \delta \sighgen$ move toward~$\sighgen$ with each successive iteration). Higher~$\xi$ lowers the curve~$h(x)$, which increases the magnitude of the slope~$h'(\sighgen)$ (i.e.\ makes it more negative) at the location of the fixed point (see Fig.~\ref{fig:sighunivnoradwithevolution}), pushing the system toward the instability threshold~$\, \abs{h'(\sighgen)} = 1$ for swing cycles. Even though this is only a local stability condition, we argue in Appendix~\ref{sec:hstability} that, for the restricted class of monotonically decreasing~$h(x)$ functions, local instability is sufficient for global instability:~$\sighgen$ is never approached by systems that start at~$\sigh$ when~$\, \abs{h'(\sighgen)} > 1$. In that case, the late-time dynamics consist of a two-state swing cycle. In contrast, if~$\, \abs{h'(\sighgen)} < 1$, the system may or may not converge toward~$\sighgen$ depending on the details of the map function~$h(x)$. However (as also argued in Appendix~\ref{sec:hstability}), at late times, all monotonically decreasing~$h(x)$ functions result in either a steady state or a swing cycle. For such functions, \textit{no other late-time dynamics are allowed.}

\begin{figure}
    \centering
    \includegraphics[width=\columnwidth]{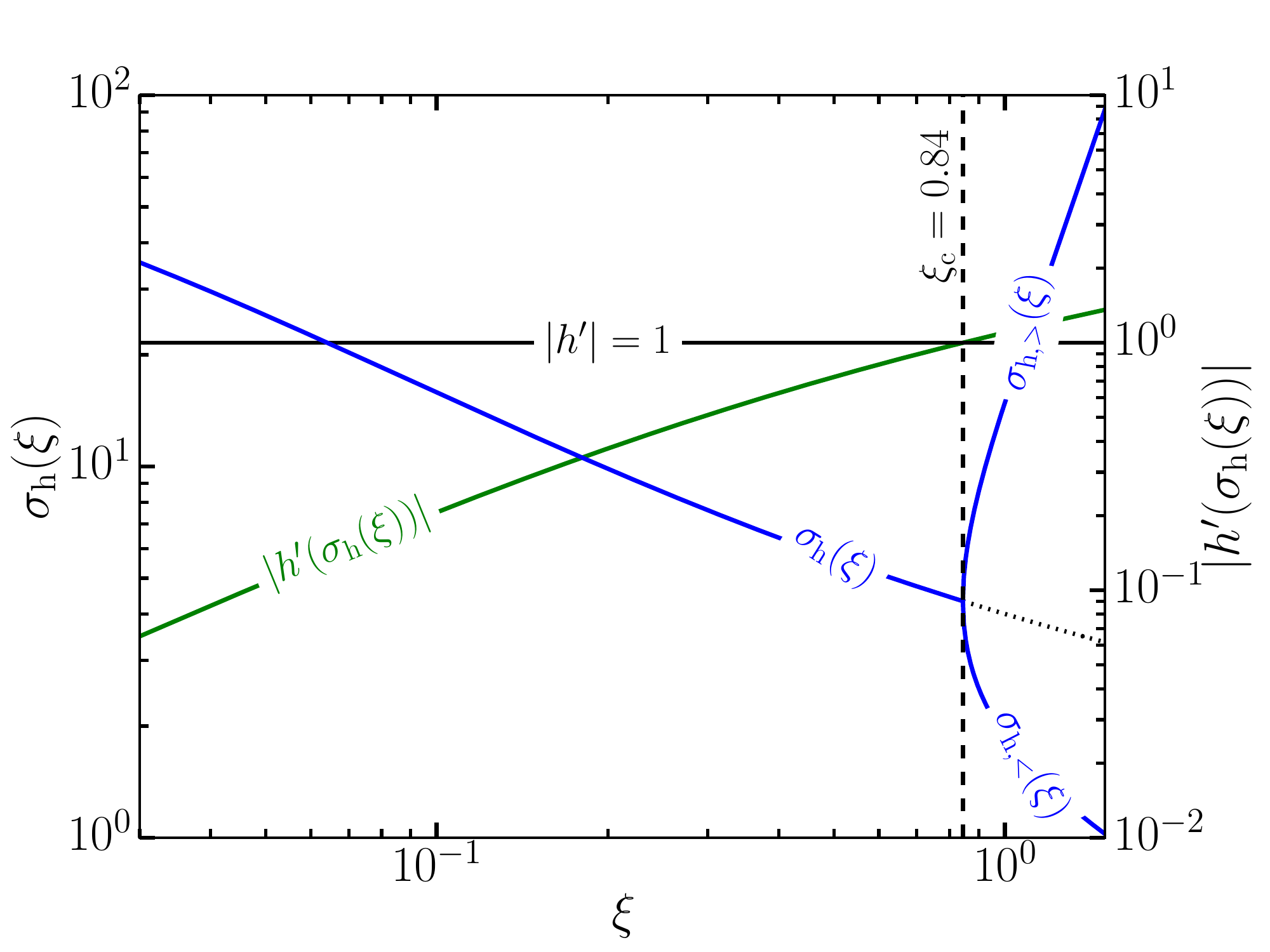}
    \caption{The fixed point~$\sighgen$ of the reconnection layer [i.e.\ of the iterated map~$\sighn{n} = h(\sighn{n-1})$ with~$h(x)$ as defined in equation~(\ref{eq:hdef})] plotted as a function of~$\xi$. Also displayed are the magnitude of the slope of~$h(x)$ evaluated at the fixed point and the low~$\sighn{<}$ and high~$\sighn{>}$ magnetizations associated with two-state cycles. At the critical value~$\xi = \xi_{\rm c} \simeq 0.84$, the system undergoes a bifurcation: the fixed point~$\sighgen$ goes unstable, as signaled by~$\, \abs{h'(\sighgen(\xi))}$ crossing above~$1$. The fixed point still exists when~$\xi > \xi_{\rm c}$ (and is indicated by a dotted line in that case), but does not control the late-time dynamics. Instead, the system comes to hop between two magnetizations~$\sighn{<} < \sighgen$ and~$\sighn{>} > \sighgen$.}
    \label{fig:sighvsxi}
\end{figure}

To illustrate these remarks, we supply Fig.~\ref{fig:sighvsxi}. That figure displays the fixed point~$\sighgen$ as a function of~$\xi$. The system bifurcates at a critical value of~$\xi = \xi_{\rm c} \simeq 0.84$: the fixed point~$\sighgen$ becomes unstable {[$\, \abs{h'(\sighgen)}$ goes above~$1$]} and a new attractor for the late-time dynamics appears -- a two-state cycle characterized by a low magnetization~$\sighn{<} < \sighgen$ and a high magnetization~$\sighn{>} > \sighgen$. Importantly, the condition~$\, \abs{h'} > 1$ implies that the system traps into a limit cycle at late times. While, in general, it is \textit{not} the case that~$\, \abs{h'} < 1$ implies convergence to the fixed point~$\sighgen$, for this particular system that happens to be true. (We discuss a case where this does not hold in Appendix~\ref{sec:feedbackdetails}.)

\subsubsection{Calculating~$\sighgen$ with more realistic radiative feedback}
Having established the mathematical techniques for computing the self-regulated magnetization~$\sighgen$ and determining its stability, we now repeat our calculation with more physical realism. In particular, we modify our assumed layer particle distribution~$\dif N / \dif \gamma$ from a single power law, as in~(\ref{eq:dndgam}), to a broken power law
\begin{align}
    \frac{\dif N}{\dif \gamma} = A \begin{cases}
        (\gamma/\gkn)^{-\pt} &\gamma \leq \gkn \\
        (\gamma/\gkn)^{-\pk} &\gamma > \gkn
    \end{cases} \, .
    \label{eq:dndgambroken}
\end{align}
This allows us to account for the different expected modifications to the power-law scaling of the distribution function in both the Thomson~($\gamma \leq \gkn$) and deep Klein-Nishina~($\gamma > \gkn$) regimes.

Given a steady injected particle distribution with power-law index~$p$, the realized distribution, if cooled in the Thomson limit, is steepened to index~$\pt = p + 1$. In contrast, if one adopts the approximate form for~$f_{\rm KN}(q)$ from equation~(\ref{eq:fknapprox}), then, for~$\gamma > \gkn$, the actual distribution of radiating particles, in fact, hardens, attaining a power-law scaling~$\pk \simeq p - 0.5$ \citep{msc05}. We therefore take~$\pt = p(\sighgen) + 1$ and~$\pk = p(\sighgen) - 0.5$ where~$p(\sighgen) = 1 + 2 / \sqrt{\sighgen}$ as in equation~(\ref{eq:eos}). Admittedly, this is still very crude. It ignores the finite \ts[]of particle energization in the reconnection layer and the finite time it takes for the full energy range of the distribution function to respond to radiative losses. It ignores, moreover, the bursty nature of magnetic reconnection at the highest energies \citep[e.g.][]{wpu19, mwu20}. However, these relationships for~$\pt$ and~$\pk$ are at least a first step toward capturing some of the qualitative effects that radiative cooling may have on the pair feedback mechanism.

\begin{figure*}
    \centering
    \begin{subfigure}{0.49\textwidth}
        \centering
        \includegraphics[width=\linewidth]{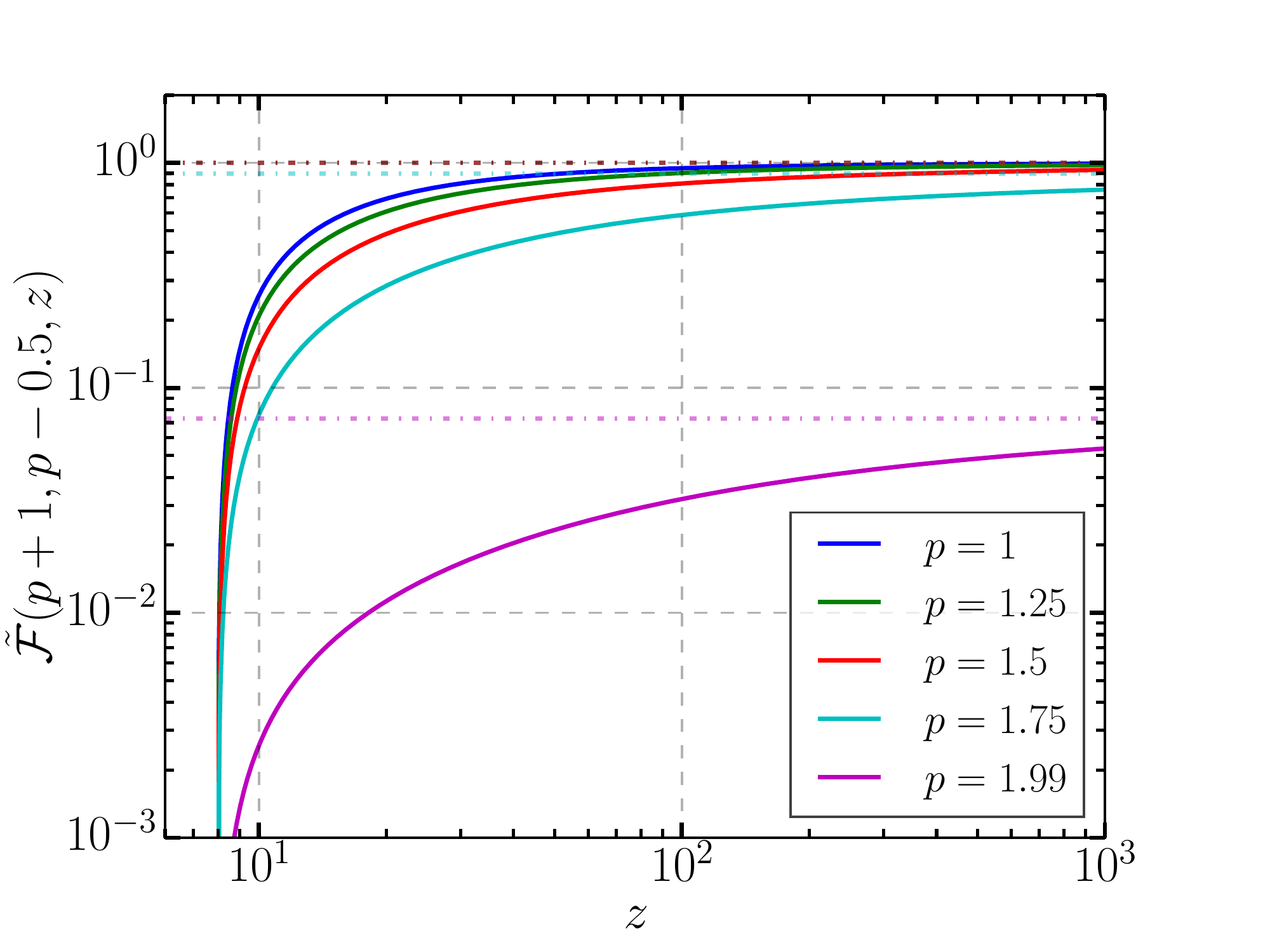}
    \end{subfigure}
    \begin{subfigure}{0.49\textwidth}
        \centering
        \includegraphics[width=\linewidth]{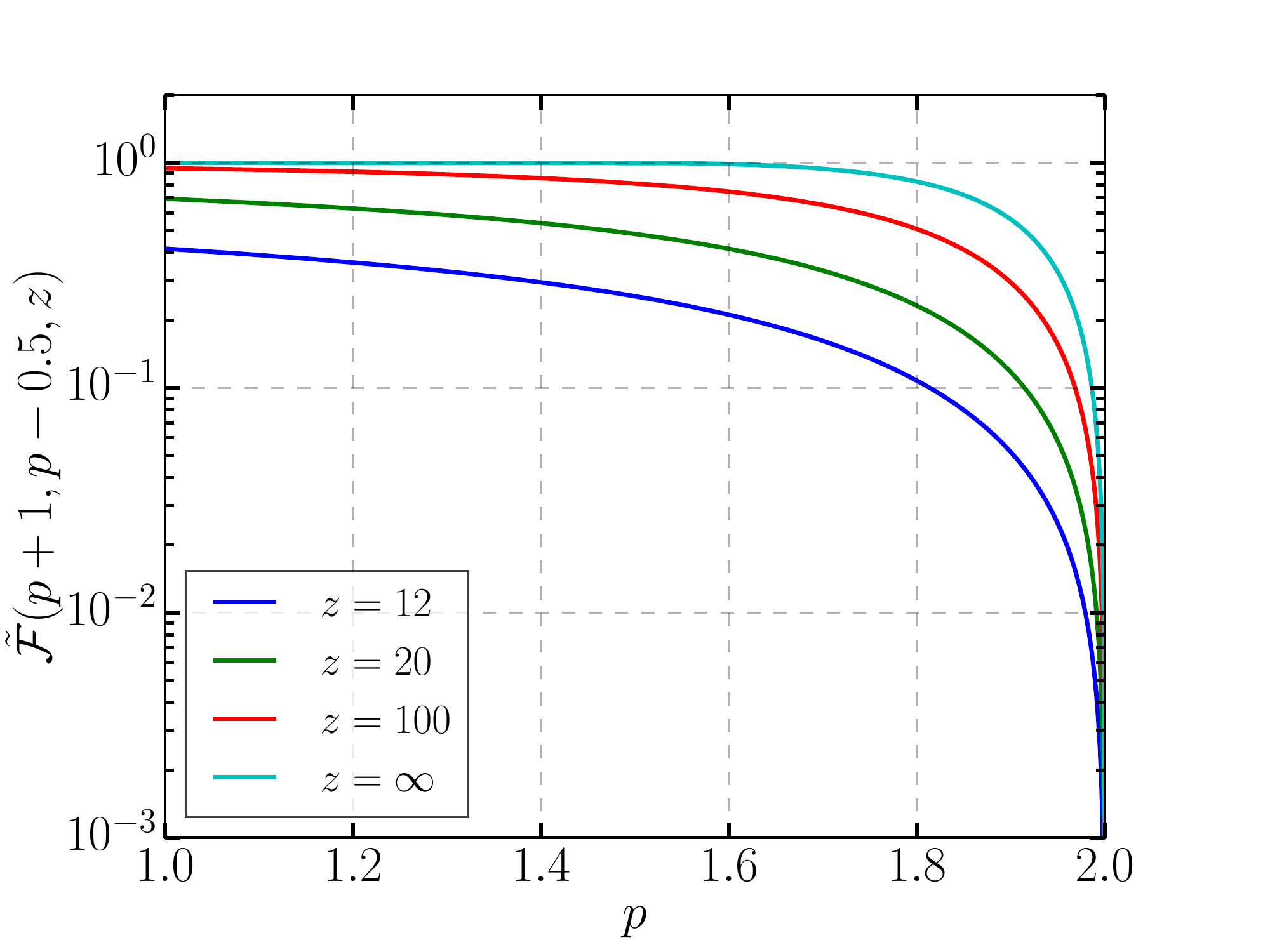}
    \end{subfigure}
    \caption{Left: A plot of~$\tilde{\mathcal{F}}(p + 1, p - 0.5, z)$ as a function of~$z \equiv \gf / \gkn$ for several~$p$. Dot-dashed lines indicate~$\lim_{z \to \infty}\tilde{\mathcal{F}}(p + 1, p - 0.5, z)$. Right: A plot of $\tilde{\mathcal{F}}(p,z)$ as a function of~$p$ for several~$z$. Due to the opposite influences of radiative cooling in the Thomson and deep Klein-Nishina regimes, where the particle energy distribution tends to be softened and hardened, respectively, the range in~$p$ over which~$\tilde{\mathcal{F}}$ significantly varies compresses relative to that of~$\mathcal{F}$~($1.5 < p < 2$ for the former instead of~$1.5 < p < 3$ for the latter).}
    \label{fig:fplaw_broken}
\end{figure*}

Using~(\ref{eq:dndgambroken}), the power fraction becomes [cf.\ equation~(\ref{eq:fplaw})]
\begin{align}
    \tilde{\mathcal{F}}(\pt, \pk, z) \equiv \frac{\int_8^z \dif x \, x^{-\pk + 2} f_{\rm KN}(x)}{\int_0^1 \dif x \, x^{-\pt + 2} f_{\rm KN}(x) + \int_1^z \dif x \, x^{-\pk + 2} f_{\rm KN}(x)} \, .
    \label{eq:fplaw_broken}
\end{align}
Note that~$\tilde{\mathcal{F}}(\pt, \pk, z \leq 8) = 0$, and so it is not necessary to consider the case~$z < 1$ in the denominator. The fraction~$\tilde{\mathcal{F}}$, with~$\pt = p + 1$ and~$\pk = p - 0.5$ is displayed as a function of~$z$ and~$p$ in Fig.~\ref{fig:fplaw_broken}. The opposite impacts of radiative losses in the Thomson and deep Klein-Nishina limits -- where, respectively, the particle energy distribution tends to steepen and become shallower -- have a pronounced effect on the~$p$-dependence of~$\tilde{\mathcal{F}}$. Due to the steepening of the distribution at lower energies, the denominator diverges when~$p > 2$~($\pt > 3$). Meanwhile, due to the hardening in the Klein-Nishina regime, the fraction~$\tilde{\mathcal{F}}$ goes to~$1$ as~$z \to \infty$ when~$p < 1.5$~($\pk < 1$). This compresses the effective range in~$p$ over which~$\tilde{\mathcal{F}}$ varies. Whereas our original power fraction~$\mathcal{F}(p, \infty)$ in equation~(\ref{eq:fplaw}) starts to decline from~$1$ only once~$p \gtrsim 1.5$, not reaching zero until~$p = 3$, our new fraction~$\tilde{\mathcal{F}}(p + 1, p - 0.5, \infty)$ departs from~$1$ when~$p > 1.5$ but hits zero already by the time~$p = 2$.

\begin{figure*}
    \centering
    \begin{subfigure}{0.49\textwidth}
        \centering
        \includegraphics[width=\linewidth]{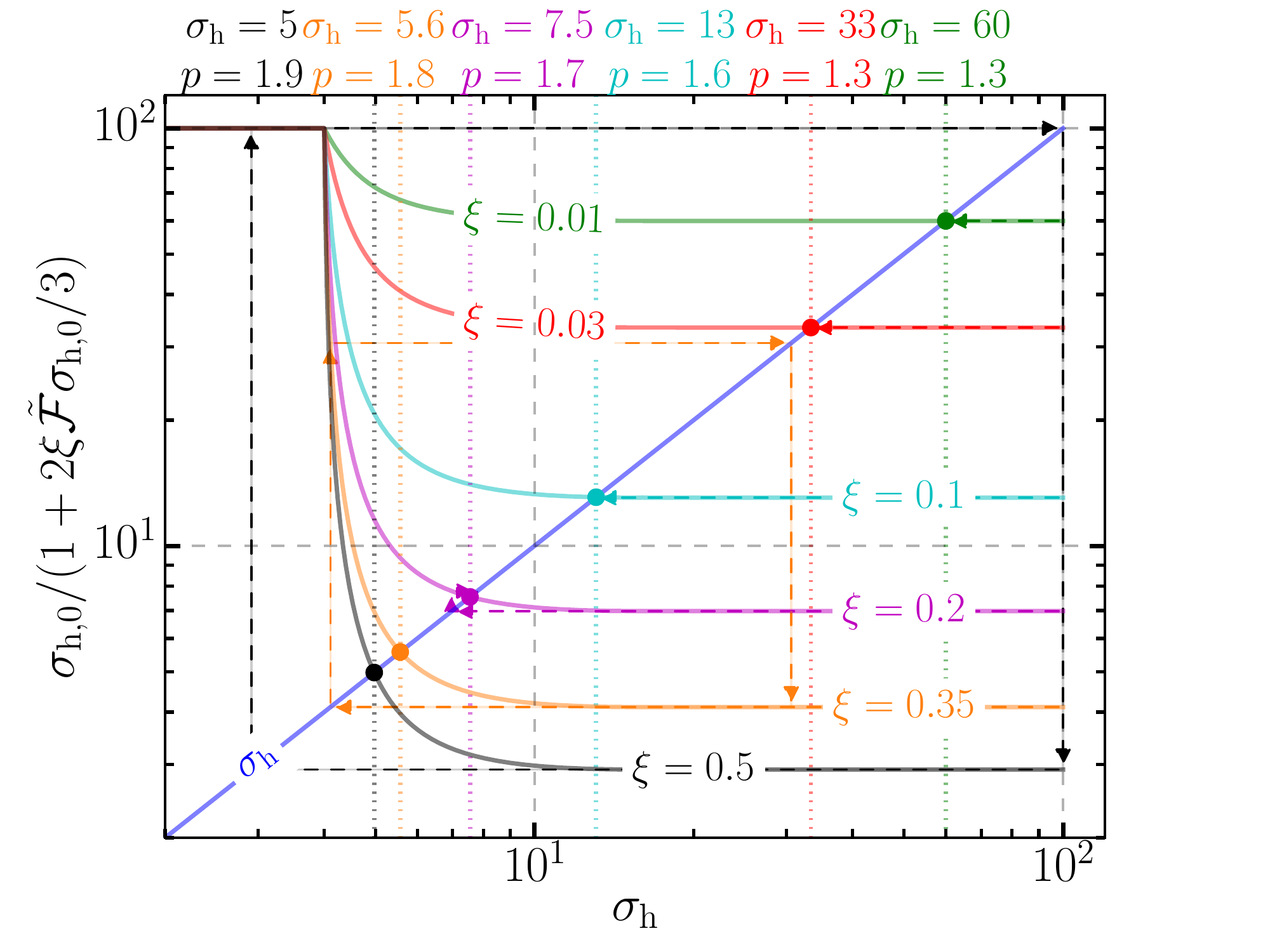}
    \end{subfigure}
    \begin{subfigure}{0.49\textwidth}
        \centering
        \includegraphics[width=\linewidth]{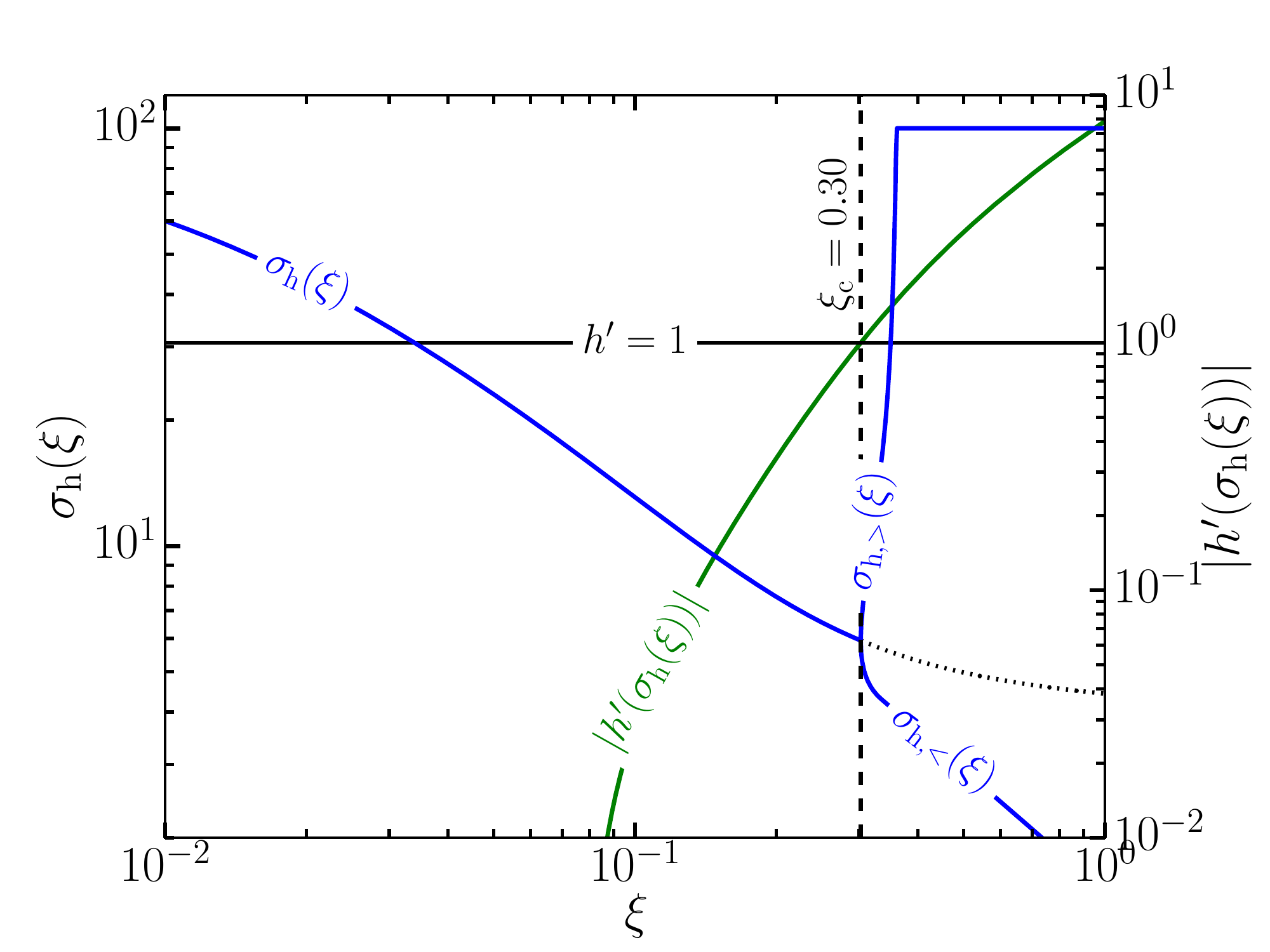}
    \end{subfigure}
    \caption{Left: The analogue of Fig.~\ref{fig:sighunivnoradwithevolution} for~$\tilde{\mathcal{F}}$ [as defined in equation~(\ref{eq:fplaw_broken})] instead of~$\mathcal{F}$ [as defined in equation~(\ref{eq:fplaw})]. Right: The analogue of Fig.~\ref{fig:sighvsxi} for~$\tilde{\mathcal{F}}$. The tendency for radiative losses to steepen the particle distribution in the Thomson regime and, in contrast, to harden it in the deep Klein-Nishina limit creates a sharp and dramatic bifurcation. The critical value of the energy recapture efficiency~$\xi_{\rm c} \simeq 0.30$ is smaller than~$\xi_{\rm c} = 0.84$ obtained in Fig.~\ref{fig:sighvsxi}, indicating that radiation back reaction makes the system more \quoted[:]{touchy}more susceptible to late-time limit cycles. Once~$\xi$ exceeds this threshold, the system rapidly transitions to a two-state cycle featuring very large-amplitude swings between the initial magnetization~$\sighn{>} = \sigh$ and a small magnetization~$\sighn{<} \sim 1$.}
    \label{fig:sighsolnrad}
\end{figure*}

This dramatically impacts the fixed points~$\sighgen(\xi)$ and their stabilities. In analogy to equation~(\ref{eq:hdef}), we define
\begin{align}
    \tilde{h}(x) = \frac{\sigh}{1 + 2 \xi \, \tilde{\mathcal{F}}[p(x) + 1, p(x) - 0.5, \infty] \sigh / 3} \, .
    \label{eq:htildedef}
\end{align}
Fig.~\ref{fig:sighsolnrad} displays a few~$\tilde{h}(x)$ curves, each one with different energy recapture efficiency~$\xi$. As promised,~$\tilde{h}(x)$ is flat when~$p(x) = 1 + 2 / \sqrt{x} > 1.5$ (i.e. when~$x > 16$) because it is here that~$\tilde{\mathcal{F}} = 1$. At the same time,~$\tilde{h}(x < 4) = \sigh$ since~$p(x < 4) > 2$, and, hence,~$\tilde{\mathcal{F}} = 0$. The rapid transition in~$\tilde{h}(x)$ from nearly vertical to nearly flat between~$x = 4$ and~$x = 16$ induces a sharp transition in the stability of the fixed point~$\sighgen$. The fixed point becomes unstable at~$\xi = \xi_{\rm c} \simeq 0.30$, and further increasing~$\xi$ beyond this point yields much more dramatic swing cycles than did incremental changes in~$\xi$ beyond the threshold~$\xi_{\rm c} \simeq 0.84$ of~$h$ in equation~(\ref{eq:hdef}). By the time~$\xi \gtrsim 0.5$, the reconnection layer \textit{starts} in a swing cycle, hopping between its \textit{initial} magnetization~$\sigh = \sighn{>}$ and a much lower magnetization~$\sighn{<} < 4$. In the low-magnetization state, pair-production completely ceases~($\tilde{\mathcal{F}} = 0$) and, after one readvection time, the plasma flowing into the layer contains no newborn pair component and once again possesses the initial magnetization~$\sigh$.

Although we are still a long way from a detailed model, in this section, we have uncovered what appears to be a robust mechanism for pair feedback on Klein-Nishina relativistic reconnection. An initially highly-magnetized system~$\sigh \gg 1$ can efficiently accelerate gamma-ray-radiating leptons in the reconnection layer. These gamma-rays collide with ambient background photons to produce a hot newborn pair component in the upstream plasma. Subsequently, the new pairs are advected into the layer where they suppress NTPA and, hence, the production of additional pairs. This mechanism operates even when radiative cooling of the layer particles is taken into account. If the energy recapture efficiency~$\xi$ approaches order unity, then the system undergoes undamped, large-amplitude oscillations between a high magnetization (even as high as the initial value~$\sigh$) -- sourcing copious upstream pairs -- and a low magnetization where pair-production is completely shut down. For~$\xi \ll 1$ however, the system approaches a steady state -- \characteriz[ed] by a balance between upstream pair-loading and particle energization in the layer -- after just one or a few readvection times~$\tread = \lmfp / \beta_{\rm rec} c$. 

\subsection{Upstream pair cascades are not generally expected}
\label{sec:nocascades}
So far, we have described a fixed point solution in which the reconnection layer regulates itself, maintaining a universal effective hot magnetization~$\sighgen$, or, potentially, exhibiting large-amplitude two-state limit cycles. The arguments that allowed us to \characteriz[e]this behaviour were based solely upon tracing the flow of energy through the system. We now undertake an analogous program, tracing the plasma particles instead of the energy.

Here, we must deal with an additional complication that was not present in the preceding energy-based arguments. Namely, because of pair production, particle number is not conserved, and one cannot simply equate the number of pairs born into the inflow plasma with (twice) the number of above-threshold photons radiated away from the layer. In principle, additional pair creation in the upstream region can occur: newborn pairs can radiate additional photons that are themselves above pair-production threshold and capable of producing secondary pair generations. We address this issue in this section. As an immediate byproduct of the analysis, we show that a pair cascade is not generally expected -- the number of particles only grows exponentially in each subsequent pair generation under certain optimal conditions.

An important quantity here is the distribution of pairs injected in the inflow plasma as the result of high-energy photons, originally emitted from the reconnection layer, getting absorbed in the upstream region. We denote this as
\begin{align}
    \Qgg[1](\gamma) = \Qnorm[1] \begin{cases}
        \gamma^{-\pgg[1]} &\gon[1] \leq \gamma < \gfn[1] \\
        0 &\mathrm{otherwise}
    \end{cases} \, .
    \label{eq:qggdef}
\end{align}
Here, the superscript~\quoted{$(1)$}(subscript~\quoted{$1$}on~$\pgg[1]$ and~$\Qnorm[1]$) indicates the first generation. In this section, we calculate
\begin{align}
    \Qgg[n](\gamma) = \Qnorm[n] \begin{cases}
        \gamma^{-\pgg[n]} & \gon[n] \leq \gamma < \gfn[n] \\
        0 & \mathrm{otherwise}
    \end{cases}
    \label{eq:qggndef}
\end{align}
the injected distribution of~$n$th-generation pairs, in terms of~$\Qgg[1](\gamma)$. The dimensions of~$\Qgg[n](\gamma)$ are that of a rate: particles per unit time per unit energy. We have not \normaliz[ed]by spatial volume. Thus,~$\Qgg[n](\gamma)$ is averaged over the spatial region where~$n$th-generation pair-production is active. For now, we assume that~$\Qgg[n](\gamma)$ is a power law and check this \textit{a posteriori}. 

\subsubsection{Newborn pair generations: basic observations}
Before calculating the~$\Qgg[n](\gamma)$ distributions in detail, we offer a few guiding remarks. These primarily pertain to the range of energies present in each generation -- i.e.\ to~$\gon[n]$ and~$\gfn[n]$ -- and to the number of generations that the upstream region can support.

As discussed in section~\ref{sec:knradreconn}, a particle with Lorentz factor~$\gamma \geq \gth = 8 \gkn$ emits photons of characteristic energy~$E_{\gamma} \sim \gamma \me c^2 / 2$. Upon absorption by the soft radiation background, these photons each create two particles with approximate Lorentz factors~$E_{\gamma} / 2 \me c^2 \sim \gamma / 4$. Thus, the power law~(\ref{eq:qggdef}) begins and ends at energies about a factor of~$4$ less than the corresponding energies of layer particles:~$\gon[1] \sim \gth / 4 = 2 \gkn$ and~$\gfn[1] \sim \gf / 4$,~[$\gf$ is the \cutoff[]layer particle energy; see equation~(\ref{eq:dndgam})].

In general, the low and high injected energies in each successive pair generation follow by identical reasoning to those of the first generation. One has~$\gon[n] \sim \gon[1] \sim \gth / 4 = 2 \gkn$ and~$\gfn[n] \sim \gf / 4^n$. The last generation,~$\ngen$, possesses \cutoff[]at or below the threshold to emit pair-producing radiation:~$\gth = 8 \gkn \geq \gfn[\ngen] = \gf / 4^\ngen$. This caps the number of generations to~$\ngen = \ceil{\log_4(\gf / 8 \gkn)}$, where~$\ceil{x}$ rounds~$x$ up to the nearest integer. Note that, if~$\log_4(\gf / 8 \gkn)$ is not an integer, then a partial additional generation is produced from the subset of pairs in the preceding generation with Lorentz factors above~$\gth$. For example, if~$\log_4(\gf / 8 \gkn)$ is~$2.4$ -- and, hence,~$\ngen = 3$ -- then two full generations are created in addition to a partial third generation. 

One may also place a potentially firmer limit on the number of generations. Note that newborn upstream particles with~$\tcoolk(\gamma) > \tread$ do not radiate before entering the layer, and, thus, do not yield additional upstream pairs. This condition is marginally satisfied if~$\gamma = \gct$ such that
\begin{align}
    \tcoolk(\gct) \equiv \frac{10}{\tau_{\gamma\gamma}} \frac{L}{c} \simeq \frac{\lmfp}{\beta_{\rm rec} c}\, ,
    \label{eq:gctdef}
\end{align}
where, for convenience, we use the approximate form~$\tread \simeq 10 L / \tau_{\gamma\gamma} c$ [equation~(\ref{eq:readvectdivlc})]. Note that the cooling time~$\tcoolk$ is the product of the readvection time with a dimensionless function of~$\knp = \gamma / \gkn$: namely,~$\tcoolk(\gamma) = (3/50)(10 L/ \tau_{\gamma\gamma} c) [\knp f_{\rm KN}(\knp)]^{-1}$, which follows from equations~(\ref{eq:tcoolic}) and~(\ref{eq:readvectdivlc}). This implies that the solution to~(\ref{eq:gctdef}) depends only on~$\gkn$ -- it is independent of all other system parameters (e.g.~$\gmax$,~$\gradt$, and~$\sigc$). Moreover, because~$\tcoolk(\gamma)$ is non-monotonic [see equation~(\ref{eq:tcoolic}), Fig.~\ref{fig:tcoolic}, and the surrounding discussion], there are actually two solutions to~(\ref{eq:gctdef}). These can be found numerically and are
\begin{align}
    \gcti{1} \simeq 0.066 \gkn \simeq (3/50) \gkn \, ,
    \label{eq:gct1}
\end{align}
which is in the Thomson limit, and
\begin{align}
    \gcti{2} \simeq 280 \gkn \, ,
    \label{eq:gct2}
\end{align}
which is in the deep Klein-Nishina regime.  

All pairs born into the upstream region with~$\gamma > \gcti{2}$ have~$\tcoolk(\gamma) > \tread$ and thus do not radiate before being swept into the reconnection layer. However, pairs born with~$\gamma < \gcti{2}$ (which also have~$\gamma > \gcti{1}$ because~$\gamma \geq \gon[n] \sim 2 \gkn > \gcti{1}$) have~$\tcoolk(\gamma) < \tread$. They maintain this condition as they radiatively cool all the way until they reach~$\gamma = \gcti{1}$, at which point~$\tcoolk(\gamma) = \tread$. Thus, pairs born between the energies~$\gcti{1}$ and~$\gcti{2}$ may cool all the way down to~$\gcti{1}$ before entering the layer. (They may not cool quite this far if not born a full mean-free-path~$\lmfp$ upstream of the layer.)

Because first-generation pairs whose parent layer particles had Lorentz factors~$\geq 4 \gcti{2}$ do not spawn additional pairs, the number of generations is limited to~$\ngen = \ceil{\log_4(4 \gcti{2} / 8 \gkn)} \simeq \ceil{\log_4(282 / 2)} = 4$. Combining this with the previous limit yields
\begin{align}
    \ngen = \min\left[ \ceil*{\log_4\left( \frac{\gf}{8 \gkn} \right)}\, , 4\right] \, ,
    \label{eq:ngen}
\end{align}
which slightly modifies the \cutoff[]energies from~$\gfn[n] \sim \gf / 4^n$ to
\begin{align}
    \gfn[n] \sim \begin{cases}
        \gf / 4 & n = 1 \\
        \min(\gf, 4\gcti{2}) / 4^n & n > 1
    \end{cases} \, .
    \label{eq:gfntrue}
\end{align}
While the first generation's \cutoff[]is determined entirely in terms of the layer \cutoff[~$\gf$,]only particles in the first generation with~$\gamma < \gcti{2}$ can give rise to further generations before entering the layer.

The maximum number of generations~$\ngen$ places an important constraint on the self-regulated reconnection system. The mean excursion of~$n$th-generation photons (those that create~$n$th-generation pairs) from the reconnection layer is~$\, \abs{\bar{y}^{(n)}} \sim \sqrt{n} \lmfp$, which scales with~$n$ as a~1D random walk. Thus, photons would begin to escape the system if~$\ngen \geq (L / \lmfp)^2 \gg 1$. But, because~$\ngen$ cannot exceed~$4$, this is not expected. 

\subsubsection{Newborn pair generations: detailed calculation}
We now calculate, in detail, the~$\Qgg[n](\gamma)$ distributions in terms of~$\Qgg[1](\gamma)$. To do this, we set up a system of coupled differential equations for~$\Ngg[n](\gamma)$, the (volume-integrated) distributions of pairs born into the inflow plasma, and for~$\Nph[n](\epsilon)$, the (volume-integrated) distributions of photons residing in the upstream region. To formulate these equations, we temporarily add time-dependence to the distributions though we ultimately specialize to the steady state. The time-derivatives of the~$\Ngg[n]$'s can be expressed as
\begin{align}
    \frac{\partial}{\partial t} \Ngg[n](\gamma,t) &= (\mathrm{injection}) - (\mathrm{cooling}) \notag \\
    &= \Qgg[n](\gamma, t) - \frac{\partial}{\partial \gamma} \left( \dot{\gamma} \Ngg[n](\gamma, t) \right) \, .
    \label{eq:fppairskn}
\end{align}
Here, we have fictitiously assumed that the discrete Klein-Nishina-limit cooling of particles can be represented continuously. However, this approximation has proven to be quite accurate for non-mono-energetic pair distributions~$\Ngg[n]$ \citep{z89, msc05}. The signed cooling rate is given by~$-\dot{\gamma} = \abs{\dot{\gamma}} = \gamma / \tcoolk(\gamma)$.

Simultaneously, the time-derivatives of the~$\Nph[n]$'s are
\begin{align}
    \frac{\partial}{\partial t} \Nph[n](\epsilon, t) &= (\mathrm{emission}) - (\mathrm{annihilation}) \notag \\
    &= 2 \frac{\Ngg[n-1](2 \epsilon, t)}{\tcoolk(2 \epsilon)} - \frac{\Nph[n](\epsilon, t)}{\lmfp / c} \, , 
    \label{eq:fpphotonskn}
\end{align}
where, in this section only, we write photon energies~$\epsilon$ in units of~$\me c^2$. The leading factor~$2$ on the first term comes from the Jacobian~$\dif \gamma / \dif \epsilon = 2$ corresponding to~$\gamma = 2 \epsilon$.
Equation~(\ref{eq:fpphotonskn}) is written explicitly in the Klein-Nishina regime, where each upstream pair cools in time~$\tcoolk(\gamma)$, and, when it does, emits a photon of energy~$\gamma \me c^2 / 2$. Then, each photon travels a distance~$\lmfp$ in time~$\lmfp / c$ before annihilating against the background to produce a new pair. Because~$2$ particles are injected into the~$n$th generation upon each annihilation of an~$n$th-generation photon, one has
\begin{align}
    \Qgg[n](\gamma, t) = 2 \times 2 \frac{\Nph[n](2 \gamma, t)}{\lmfp / c} \, .
    \label{eq:qggformula}
\end{align}
Here again, one factor of~$2$ comes from the Jacobian from particle to photon energies.

Equations~{(\ref{eq:fppairskn})-(\ref{eq:qggformula})} achieve a steady state if the readvection time, over which the~$n=1$ injection term is constant, is longer than the cooling times of newborn pairs and the photon streaming time~$\lmfp / c$. The first condition is satisfied because all pairs (at least for~$n > 1$) are born with Lorentz factors~$\gamma < \gcti{2}$, while the latter is satisfied because the~$n$th-generation readvection time is~$\sqrt{n} \lmfp / \beta_{\rm rec} c \sim 10 \sqrt{n} \lmfp / c \gg \lmfp / c$. (The particles in the~$n=1$ generation with~$\gamma > \gcti{2}$ do not produce additional pairs and so can be excluded from these equations.) Let us therefore specialize to the steady state of equations~(\ref{eq:fppairskn}) and~(\ref{eq:fpphotonskn}). This gives, upon plugging~(\ref{eq:qggformula}) into~(\ref{eq:fpphotonskn}),
\begin{align}
    \Qgg[n](\gamma) = 8 \frac{\Ngg[n-1](4 \gamma)}{\tcoolk(4 \gamma)} \, ,
    \label{eq:qggsteady}
\end{align}
which, in turn, allows one to write
\begin{align}
    \frac{\partial}{\partial \gamma} \left( \dot{\gamma} \Ngg[n](\gamma) \right) = -\frac{1}{8} \frac{\partial}{\partial \gamma} \left( \gamma \Qgg[n+1](\gamma/4) \right) \, ,
    \label{eq:pgamrewrite}
\end{align}
where we have used~$-\dot{\gamma} = \gamma / \tcoolk(\gamma)$.

Then, inserting~(\ref{eq:pgamrewrite}) into the steady state of~(\ref{eq:fppairskn}) gives a recursive formula for the~$\Qgg[n](\gamma)$'s:
\begin{align}
    -\frac{1}{8} \frac{\partial}{\partial \gamma} \left( \gamma \Qgg[n+1](\gamma/4) \right) = \Qgg[n](\gamma) \, .
    \label{eq:qggrecursediff}
\end{align}
Let us integrate this equation from~$\gamma \geq \gth$ to~$\gamma = \infty$. The upper bound gives zero on the left-hand-side of~(\ref{eq:qggrecursediff}) since~$\Qgg[n](\infty) = 0$. The lower bound does \textit{not} give zero on the left because~$\gamma / 4$ exceeds the lowest energy~$\gon[n] = 2 \gkn$ in the~$(n+1)$st injected distribution. Hence,
\begin{align}
    \Qgg[n+1](\gamma) = \frac{2}{\gamma} \int_{4 \gamma}^\infty \Qgg[n](\gamma') \dif \gamma' \, .
    \label{eq:qggrecurseint}
\end{align}

Let us now use~(\ref{eq:qggrecurseint}) to determine~$\Qgg[n+1](\gamma)$ explicitly, assuming~$\Qgg[n](\gamma)$ is given by a power law as in equation~(\ref{eq:qggndef}). Restricting to~$2 \gkn = \gon[n+1] \leq \gamma < \gfn[n+1] = \gfn[n] / 4$, we have
\begin{align}
    \Qgg[n+1](\gamma) &= \frac{2 \Qnorm[n]}{\gamma} \int_{4 \gamma}^{\gfn[n]} (\gamma')^{-\pgg[n]} \dif \gamma' \notag \\
    &= \frac{2 \Qnorm[n]}{\gamma (\pgg[n] - 1)} (4 \gamma)^{-\pgg[n] + 1} \left[ 1 - \left( \frac{4 \gamma}{\gfn[n]} \right)^{\pgg[n] - 1} \right] \, .
    \label{eq:qggrecurse}
\end{align}
For~$\gamma$ not between~$2 \gkn$ and~$\gfn[n] / 4$, the distribution~$\Qgg[n](\gamma)$ is zero. Strictly speaking, for~$n = 1$,~$\gfn[1] = \gf / 4$ needs to be modified to~$\min(\gfn[1], \gcti{2}) / 4$, since only first-generation particles with energies less than~$\gcti{2}$ can spawn additional pairs.

Noting that the term in square brackets in~(\ref{eq:qggrecurse}) is roughly unity except when~$\gamma \sim \gfn[n] / 4$, we see that~$\Qgg[n+1](\gamma)$ is just a power law with a \cutoff[]at~$\gfn[n+1] = \gfn[n] / 4$, in agreement with equation~(\ref{eq:gfntrue}). This verifies the assumed power-law form~(\ref{eq:qggndef}) provided the initial injected distribution~$\Qgg[1](\gamma)$ is also a power-law. In that case,
\begin{align}
    \pgg[n+1] = \pgg[n] = \pgg[1] \equiv \pgg
    \label{eq:pggrecurse}
\end{align}
and
\begin{align}
    \Qnorm[n+1] = \mathcal{A}(\pgg) \Qnorm[n] = \left[ \mathcal{A}(\pgg) \right]^{n} \Qnorm[1] \, ,
    \label{eq:qnormrecurse}
\end{align}
where we have defined
\begin{align}
    \mathcal{A}(\pgg) \equiv \frac{8}{4^{\pgg} (\pgg - 1)} \, .
    \label{eq:scripta}
\end{align}

These recurrence formulae imply that the total pair injection rate in the upstream region is
\begin{align}
    \sum_{n = 1}^\ngen \int \Qgg[n](\gamma) \dif \gamma &= \Qnorm[1] \sum_{n = 1}^\ngen \left[ \mathcal{A}(\pgg) \right]^{n - 1} \frac{\gamma_{\rm min}^{-\pgg + 1}}{\pgg - 1} \left[ 1 - \left( \frac{\gamma_{\rm min}}{\gfn[n]} \right)^{\pgg - 1} \right] \notag \\
    &\simeq \frac{\Qnorm[1] \gamma_{\rm min}^{-\pgg + 1}}{\pgg - 1} \sum_{n = 0}^{\ngen - 1} \left[ \mathcal{A}(\pgg) \right]^{n} \notag \\
    &\simeq \left[ \int \Qgg[1](\gamma) \dif \gamma \right] \sum_{n = 0}^{\ngen - 1} \left[ \mathcal{A}(\pgg) \right]^n \notag \\
    &= \frac{1 - \left[ \mathcal{A}(\pgg) \right]^{\ngen}}{1 - \mathcal{A}(\pgg)} \int \Qgg[1](\gamma) \dif \gamma \, ,
    \label{eq:totalinject}
\end{align}
where, for convenience, we define~$\gamma_{\rm min} \equiv \gon[n] = 2 \gkn$.
In the second and third lines, we assume that~$\pgg > 1$ and, hence, that the term~$(\gamma_{\rm min} / \gfn[n])^{\pgg - 1}$ can be neglected. The final line is the crux of this section. It determines under what conditions a true pair cascade develops -- when the total number of injected pairs in the upstream region is exponential in the number of generations~$\ngen$. We have,
\begin{align}
    \mathcal{A}(\pgg) > 1 &\Longleftrightarrow \mathrm{pair\, cascade} \notag \\
    \mathcal{A}(\pgg) < 1 &\Longleftrightarrow \mathrm{no\, pair\, cascade} \, .
    \label{eq:ccriterion}
\end{align}
The multiplication factor~$\mathcal{A}(\pgg)$ is less than unity for~$\pgg > 1.73$. Whether a pair cascade develops comes down to the expected value of~$\pgg$ in the first-generation injected distribution.

We expect that~$\Qgg[1](\gamma)$ inherits its power-law scaling from the distribution of photons emitted into the upstream region from the layer. For a power-law distribution~(\ref{eq:dndgambroken}) of radiating layer particles, the emitted photon spectrum is, approximately, a power law with index~$\pgg = \pk + 1$ (plus a logarithmic correction; see \citealt{bg70} and \citealt{aa81}). Since we expect the scaling of the layer particle distribution in the Klein-Nishina regime to be~$\pk = p - 0.5$, which corresponds to an intrinsic particle acceleration index~$p$ hardened due to Klein-Nishina IC losses, we predict~$\min(\pgg) = \min(p + 0.5)$. Now, according to~(\ref{eq:eos}),~$\min(p) = 1$, and thus the distribution of first-generation injected pairs should be no harder than~$\min(\pgg) = 1.5$. This gives a maximum multiplication factor of~$\max(\mathcal{A}(\pgg)) = \mathcal{A}(\min (\pgg)) = \mathcal{A}(1.5) = 2$, for which the number of injected pairs doubles in each successive generation.

A shallow enough scaling~$\pgg$ to bring~$\mathcal{A}(\pgg)$ above unity is only achieved for~$p < 1.7 - 0.5 = 1.2$. Using equation~(\ref{eq:eos}), this implies a magnetization~$\sighgen > 80$. Also, the theory we have presented thus far ignores the possibility of a guide field, which tends to suppress reconnection-powered NTPA \citep[e.g.][]{wu17}. Thus, a pair cascade is possible, but requires quite optimal combinations of parameters~(high~$\sighgen$ and small guide field, for example). 

\subsection{The small number density of upstream pairs}
\label{sec:upstreammult}
In this section, we employ our tally of the upstream pair creation rate -- given by the~$\Qgg[n](\gamma)$ distributions -- to answer another important question associated with Klein-Nishina radiative reconnection. Namely, we calculate the pair-production \textit{multiplicity}
\begin{align}
    \eta \equiv \frac{n_{\gamma\gamma}}{n_0} \, ,
    \label{eq:multdef}
\end{align}
the ratio of the number density~$n_{\gamma\gamma}$ of newborn pairs entering the reconnection layer to that~$n_0$ of pairs in the far upstream region.

Let us define the volumetric (per unit time per unit volume) rate of pair production~$\dif \newbornn[n] / \dif t$ into the~$n$th generation. Assuming that~$n$th-generation pairs are deposited uniformly up to a transverse distance~$\avgy[n]$ (e.g.~$\avgy[n] \sim \sqrt{n} \lmfp$) away from the layer across the full width~$L$ of the system, we have
\begin{align}
    \frac{\dif \newbornn[n]}{\dif t} &\sim \frac{1}{2 \avgy[n] L} \int \dif \gamma \, \Qgg[n](\gamma) \, .
    \label{eq:dnggdt}
\end{align}
Noting that a parcel of plasma travels from~$\, \abs{y} \sim \avgy[n]$ to the layer at~$\, \abs{y} \sim \thickness \ll \avgy[n]$ over time~$\avgy[n] / \beta_{\rm rec} c$, during which it accrues~$n$th-generation pairs at the rate~$\dif \newbornn[n] / \dif t$, we have
\begin{align}
    \newbornsym \sim f_{\rm noesc} \sum_{n = 1}^{N} \frac{\avgy[n]}{\beta_{\rm rec} c} \frac{\dif \newbornn[n]}{\dif t} \sim \frac{1}{2 \beta_{\rm rec} c L} \int \dif \gamma \sum_{n = 1}^{N} \Qgg[n](\gamma) \, .
    \label{eq:nggdef}
\end{align}
In the second step, we assume~$f_{\rm noesc} = 1$. Similar to the cancellation of the readvection time in~(\ref{eq:uggp}),~$\avgy[n]$ cancels in~(\ref{eq:nggdef}).

To evaluate~(\ref{eq:nggdef}), we \normaliz[e]the~$\Qgg[n](\gamma)$'s by balancing the pair creation rate in the \textit{first} generation with twice the number of above-threshold photons emitted from the reconnection layer:
\begin{align}
    \int \dif \gamma \, \Qgg[1](\gamma) = 2 \thickness L'\int_{\gth} \dif \gamma \, R_{\rm IC}(\gamma) \dif N / \dif \gamma \, ,
    \label{eq:pairsfromlayer}
\end{align}
where~$\dif N / \dif \gamma$ is the distribution of radiating particles as in equations~(\ref{eq:dndgam}) and~(\ref{eq:dndgambroken}). In turn, one may \normaliz[e]the distribution~$\dif N / \dif \gamma$ [to find~$A$ in equation~(\ref{eq:dndgam}) or~(\ref{eq:dndgambroken})] by setting the power radiated from the layer to~$P_{\rm Poynt} / 2$ [as in equations~(\ref{eq:ppoynt}) and~(\ref{eq:jic})]:
\begin{align}
    \frac{1}{2} P_{\rm Poynt} &= L \beta_{\rm rec} c \frac{B_0^2}{4 \pi} \sim j_{\rm IC} \thickness L' \notag \\
    &= \thickness L' \int \dif \gamma P_{\rm IC}(\gamma) \dif N / \dif \gamma \, .
    \label{eq:dndgamnorm}
\end{align}

In keeping with the earlier parts of this study, we consider two cases: one in which the radiation reaction force on layer particles is ignored and one in which it is approximately included. In the former case, we assume a single power-law form for~$\dif N / \dif \gamma$, as in equation~(\ref{eq:dndgam}), with index~$p(\sighgen) = 1 + 2 / \sqrt{\sighgen}$ [equation~(\ref{eq:eos})]. In the latter, we adopt a broken power law, as in equation~(\ref{eq:dndgambroken}), with separate indices~$\pt = p(\sighgen) + 1$ and~$\pk = p(\sighgen) - 0.5$ in the Thomson~($\gamma < \gkn$) and Klein-Nishina~($\gamma > \gkn$) regimes, respectively. To save space, we simultaneously conduct both analyses by plugging in equation~(\ref{eq:dndgambroken}) for~$\dif N / \dif \gamma$ and leaving~$\pt$ and~$\pk$ unspecified until the end of the calculation.

Proceeding in this manner, we find that
\begin{align}
    n_{\gamma\gamma} &\sim n_0 \frac{\sigc}{\gkn} \tilde{\mathcal{M}}(\pt, \pk, z) \, ,
    \label{eq:ngg}
\end{align}
where~$z = \gf / \gkn$. The right-hand-side does not depend on~$n_0$: the~$n_0$ in the numerator cancels against that in the definition of~$\sigc$. The \textit{multiplicity function}~$\tilde{\mathcal{M}}(\pt, \pk, z)$ is derived, as discussed, by evaluating~(\ref{eq:nggdef}) whilst enforcing~(\ref{eq:pairsfromlayer}) and~(\ref{eq:dndgamnorm}). It reads
\begin{align}
    \tilde{\mathcal{M}}(\pt, \pk, z) &\equiv \frac{3 \int_8^z \dif x \, g_{\rm KN}(x) x^{-\pk}}{\int_0^1 \dif x \, f_{\rm KN}(x) x^{-\pt + 2} + \int_1^z \dif x \, f_{\rm KN}(x) x^{-\pk + 2}} \notag \\
    &\times \frac{\int \dif \gamma \sum_{n = 1}^{N} \Qgg[n](\gamma)}{\int \dif \gamma \, \Qgg[1](\gamma)} \, .
    \label{eq:scriptm}
\end{align}
The factor on the second line encodes the possibility of a pair cascade and is roughly equal to~$[1 - \mathcal{A}(\pgg)^{N}] / [1 - \mathcal{A}(\pgg)]$, as in equation~(\ref{eq:totalinject}). However, here we evaluate this factor by maintaining finite~$z$-dependent \cutoff[s]in each term, writing
\begin{align}
    \int \dif \gamma \, \Qgg[n](\gamma) \simeq \frac{[\mathcal{A}(\pgg)]^{n-1} \Qnorm[1] \gamma_{\rm min}^{-\pgg + 1}}{\pgg - 1} \left[ 1 - \left( \frac{\gamma_{\rm min}}{\gfn[n]} \right)^{\pgg - 1} \right] \, .
    \label{eq:qinjexplicit2}
\end{align}
We do not simplify the right-hand-side to~$\mathcal{A}(\pgg)^{n-1} \Qnorm[1] \gamma_{\rm min}^{-\pgg + 1} / (\pgg - 1)$, as done in section~\ref{sec:nocascades} [equation~(\ref{eq:totalinject})]. This ensures that contributions from each generation turn on gradually (as they do in reality), thereby keeping the multiplicity function continuous.

To denote the case where we ignore radiative feedback on the layer particles, in which we put~$\pt = \pk = p$, we write
\begin{align}
    \mathcal{M}(p, z) \equiv \tilde{\mathcal{M}}(p, p, z) \, .
    \label{eq:scriptmt}
\end{align}

\begin{figure*}
    \centering
    \begin{subfigure}{0.49\textwidth}
        \centering
        \includegraphics[width=\linewidth]{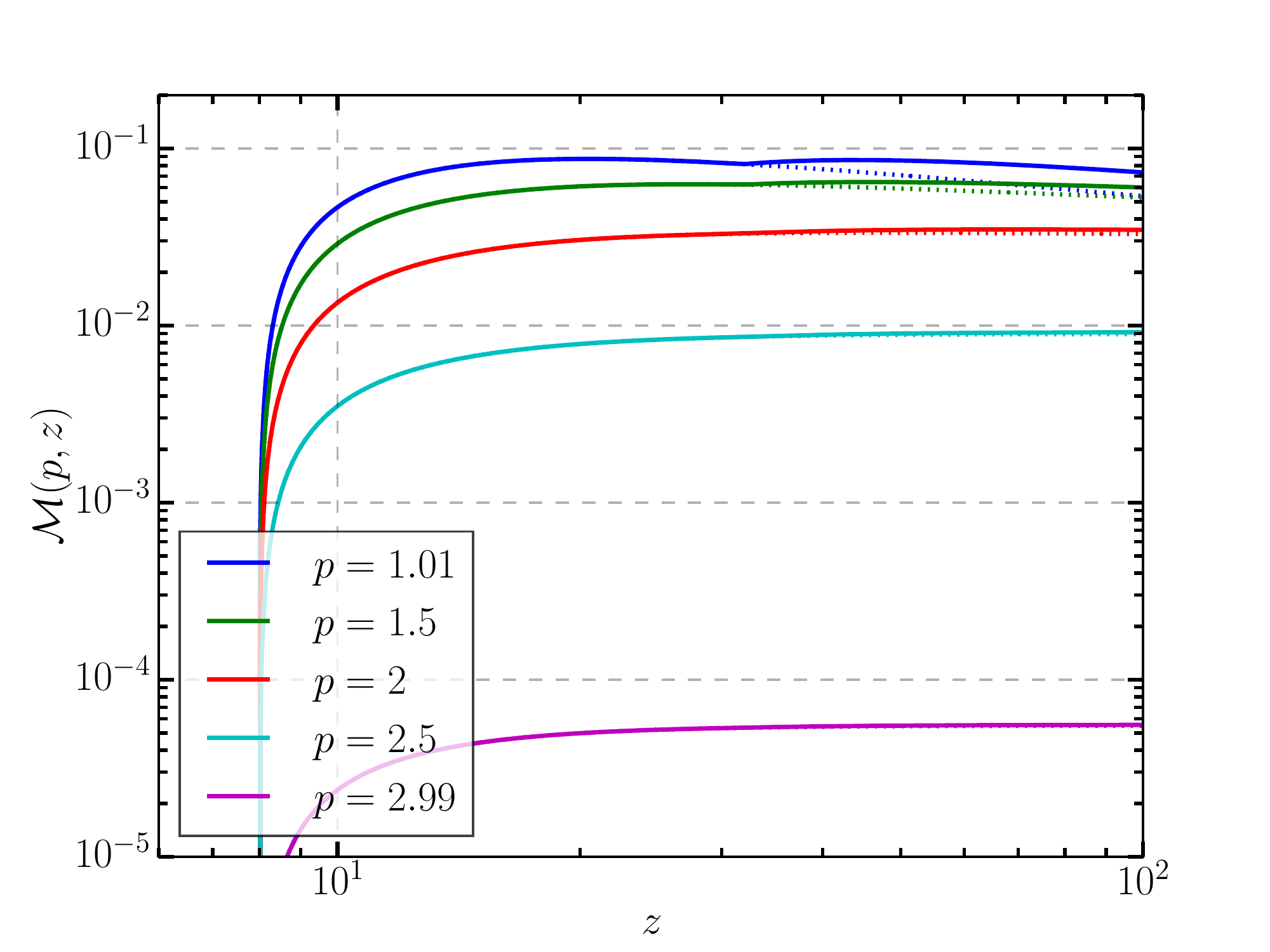}
    \end{subfigure}
    \begin{subfigure}{0.49\textwidth}
        \centering
        \includegraphics[width=\linewidth]{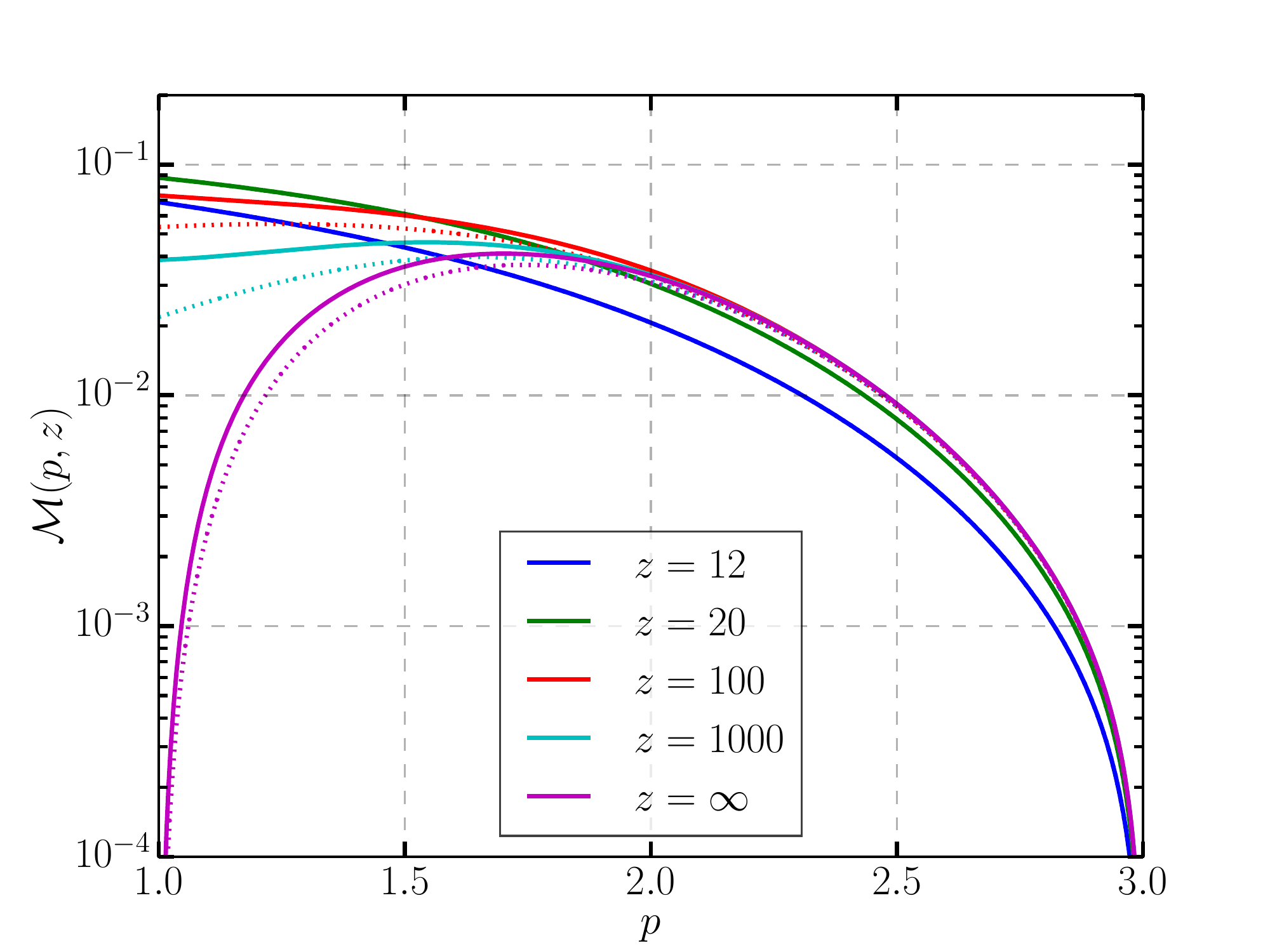}
    \end{subfigure} \\
    \begin{subfigure}{0.49\textwidth}
        \centering
        \includegraphics[width=\linewidth]{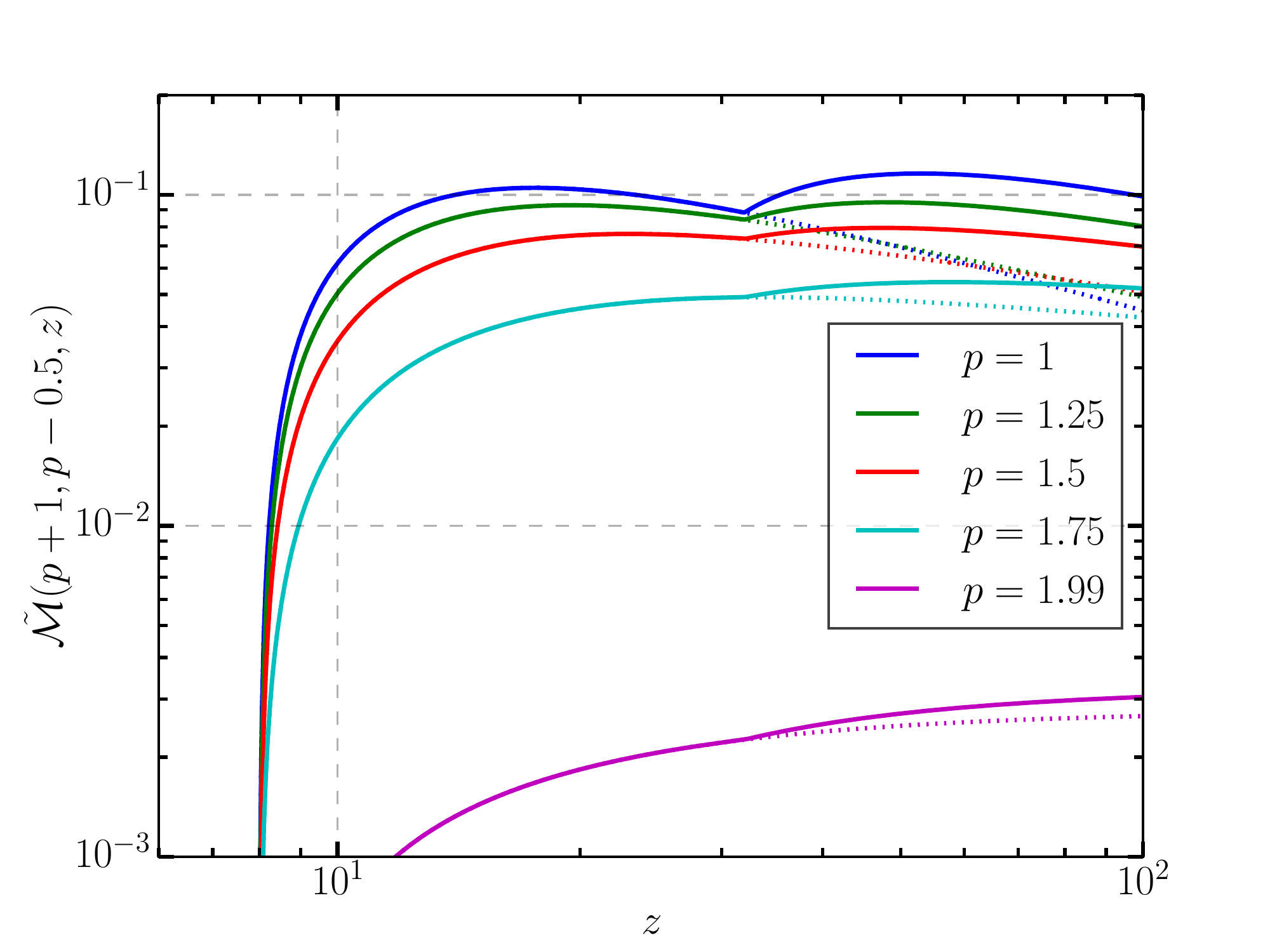}
    \end{subfigure}
    \begin{subfigure}{0.49\textwidth}
        \centering
        \includegraphics[width=\linewidth]{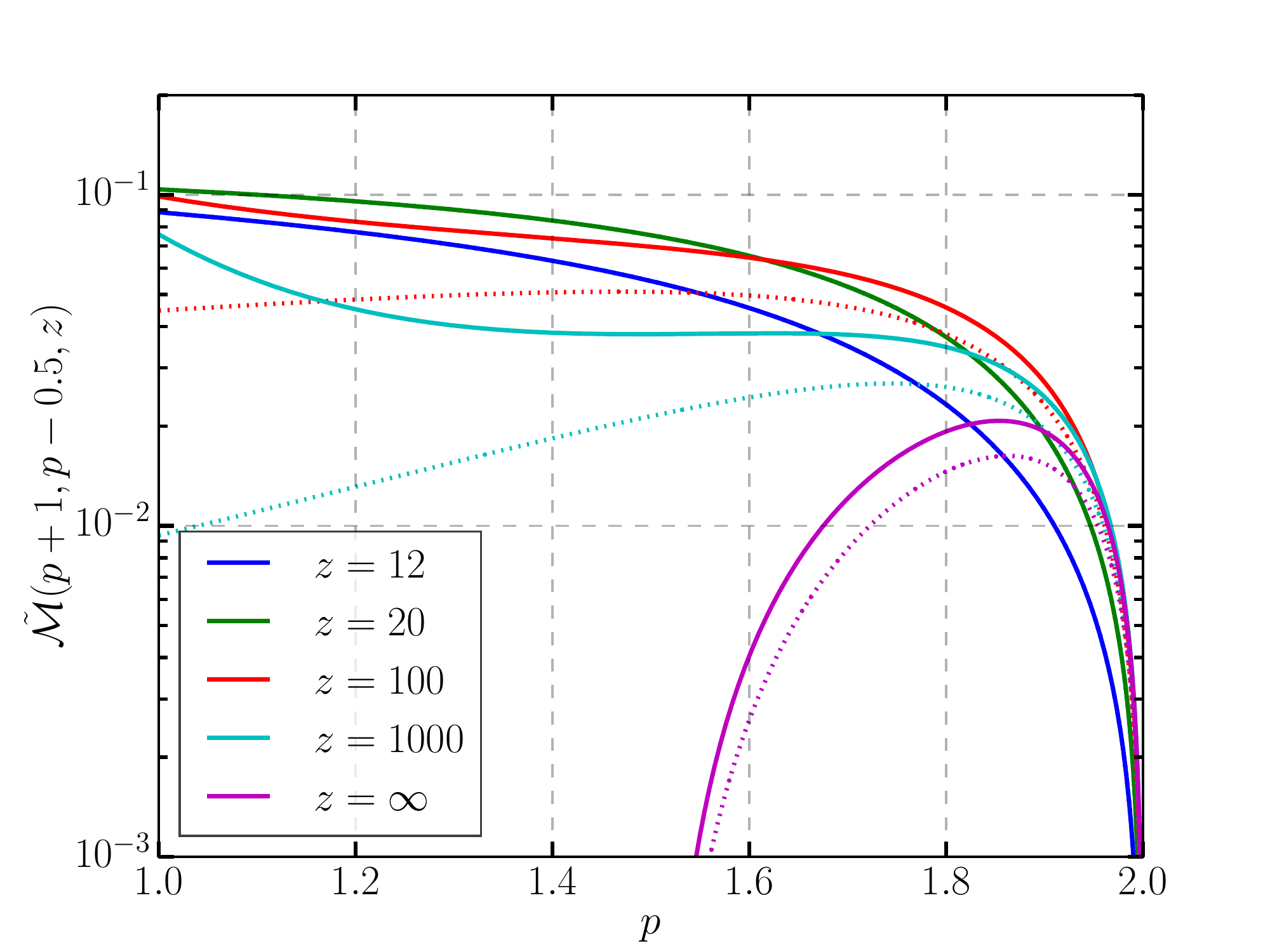}
    \end{subfigure}
    \caption{The analogue of Figs.~\ref{fig:fplaw} and~\ref{fig:fplaw_broken}, which respectively display~$\mathcal{F}$ and~$\tilde{\mathcal{F}}$, for~$\mathcal{M}$ and~$\tilde{\mathcal{M}}$. In all plots, dotted lines show the value of the multiplicity function divided by the pair cascade factor on the second line of equation~(\ref{eq:scriptm}), demonstrating what the multiplicity function would be if only accounting for pairs injected into the first upstream generation (i.e.\ ignoring a possible pair cascade). Top left:~$\mathcal{M}(p, z) \equiv \tilde{\mathcal{M}}(p, p, z)$ displayed as a function of~$z$ for several~$p$. Top right:~$\mathcal{M}(p, z)$ displayed as a function of~$p$ for several~$z$. Bottom left:~$\tilde{\mathcal{M}}(p+1,p-0.5,z)$ displayed as a function of~$z$ for several~$p$. Bottom right:~$\tilde{\mathcal{M}}(p+1,p-0.5,z)$ displayed as a function of~$p$ for several~$z$. While the dependence of~$\mathcal{M}$ and~$\tilde{\mathcal{M}}$ on~$p$ and~$z$ is more complicated (in particular, non-monotonic) than that of~$\mathcal{F}$ and~$\tilde{\mathcal{F}}$, the former are always small -- no larger than~$0.1$ -- and therefore the multiplicity~$\eta$ is small whenever~$\gkn > \sigc / 10$. Pair cascades significantly influence the multiplicity for large~$z$ and small~$\pk$, and, hence, have a more pronounced impact on~$\tilde{\mathcal{M}}$, where~$\pk = p - 0.5$, than on~$\mathcal{M}$, where~$\pk = p$.}
    \label{fig:scriptm}
\end{figure*}

We display the functions~$\mathcal{M}(p,z)$ and~$\tilde{\mathcal{M}}(p + 1, p - 0.5, z)$ in Fig.~\ref{fig:scriptm}. Similar to the case of~$\mathcal{F}$ and~$\tilde{\mathcal{F}}$, the separate impacts of radiative cooling in the Thomson and deep Klein-Nishina limits contract the range in~$p$ across which the multiplicity function varies. Just as for the power fraction~$\mathcal{F}$, the steepening of the layer particle distribution~$\pt = p + 1$ in the Thomson regime completely shuts down pair production whenever~$\pt > 3$ (i.e.~$p > 2$). However, unlike the power fraction~$\mathcal{F}$, the multiplicity function exhibits a much more complicated non-monotonic dependence on the underlying parameters~$p$ and~$z$. In addition to this, while the power fraction attains order unity for a wide range of~$p$ and~$z$, the multiplicity function never does -- it is never larger than~$\simeq 0.1$ for any parameter combination, and is very often much smaller than this.

This last fact means that, even when a pair cascade truly does develop in the upstream region -- and the total newborn pair count is exponential in the number of generations -- the overall multiplicity may still be \textit{small}. In fact, because the multiplicity function obtains a global maximum of order~$10^{-1}$, we can map out precisely the parameters for which~$\eta$ is guaranteed to be small:~$\gkn > \sigc / 10$. As we show in section~\ref{sec:applications}, this condition is roughly satisfied for reconnection in both FSRQ jets and black hole accretion disc coronae.
 
This concludes our detailed discussion of pair-regulated Klein-Nishina magnetic reconnection. Next, we examine the potential astrophysical ramifications of reconnection in this regime.

\section{Astrophysical implications}
\label{sec:applications}
In this section, we discuss observational aspects of pair-regulated Klein-Nishina magnetic reconnection. Our approach has two stages. First, we elaborate observable consequences of our model that are generic, not requiring an explicit global astrophysical context. This grounds our subsequent discussion where we estimate parameters for concrete astrophysical systems -- FSRQs and black hole ADCe -- and comment on observations that the model may help to explain.

\subsection{General observable features}
Here, we discuss the appearance of Klein-Nishina radiative reconnection as viewed through a telescope. We assume that the reconnection system is not spatially resolved.

\subsubsection{Observed radiation comes mostly from the layer}
\label{sec:layerdominance}
In radiative reconnection without pair-production feedback, the radiative output is dominated by high-energy particles in the layer (i.e.\ the downstream region permeated by reconnected magnetic flux). However, with pair feedback, the upstream and downstream plasmas are radiatively coupled: a substantial portion of the energy emitted from the layer may be intercepted upstream and reprocessed into high-energy newborn pairs. While en route to the layer, these pairs emit potentially observable light (if~$f_{\rm nocool} < 1$) that could, in principle, outshine the escaping (below-threshold) radiation from the layer.

Therefore, we wish to determine whether the light that an observer sees comes predominantly from the upstream or downstream (layer) plasmas. We denote the respective \textit{observable} luminosities (i.e.\ only of below-threshold and, hence, escaping radiation) of these two regions per unit length in the out-of-plane direction by~$L_{\rm upstream}$ and~$L_{\rm layer}$. The first luminosity is the fraction of the layer's above-threshold radiated power that is captured by the upstream region and reemitted below threshold before flowing back into the layer:
\begin{align}
    L_{\rm upstream} &\sim (1 - f_{\rm nocool}) (2 L \lmfp) \frac{\dif u_{\gamma\gamma}}{\dif t} \notag \\
    &\sim (1 - f_{\rm nocool}) \mathcal{F} \beta_{\rm rec} c L \frac{B_0^2}{4 \pi}
    \label{eq:lupstream}
\end{align}
[cf.\ equation~(\ref{eq:duggdt})].
Meanwhile, the fraction of the layer's power emitted below pair threshold is [cf.\ equation~(\ref{eq:jic})]
\begin{align}
    L_{\rm layer} \sim (1 - \mathcal{F}) \frac{1}{2} P_{\rm Poynt} \sim (1 - \mathcal{F}) \beta_{\rm rec} c L \frac{B_0^2}{4 \pi} \, .
    \label{eq:llayer}
\end{align}
Note that one may substitute~$\tilde{\mathcal{F}}$ for~$\mathcal{F}$ in these expressions provided the same symbol is used in both. The luminosity ratio is
\begin{align}
    \frac{L_{\rm upstream}}{L_{\rm layer}} \sim (1 - f_{\rm nocool}) \frac{\mathcal{F}}{1 - \mathcal{F}} \, .
    \label{eq:lratinphase}
\end{align}
For the upstream region to outshine the layer, both significant above-threshold radiation~($\mathcal{F} > 1/2$) and relatively low energy recapture efficiency~($f_{\rm nocool} \ll 1$) are required.

However, it is also possible that the system enters a two-state cycle, swinging between a low and a high effective upstream magnetization (see section~\ref{sec:upstreamendens}). Then, the upstream domain may appear brightest in one magnetization state while the downstream region shines the most in the other state. In this case, the relevant luminosities to compare are probably the brightest luminosities achieved by each region (even if in opposite states). This is certainly appropriate if the readvection time~$\tread$ is not resolved by the observations. However, even if the readvection time is resolved, different locations along the layer's surface (e.g. different areas in the horizontal direction of Fig.~\ref{fig:detaileddiagram}) may be decorrelated from each other. One radiating zone may be in a high-magnetization phase while a neighb\spellor[ing]zone is in the corresponding low-magnetization state. Then, the observed luminosities from each zone are averaged and (assuming, for simplicity, a 50 per cent duty cycle) dominated by the brighter state of the more luminous region (upstream or downstream) in that zone.

Thus, the luminosity ratio in equation~(\ref{eq:lratinphase}) may not accurately describe a reconnection layer prone to swing cycles. In fact, according to equation~(\ref{eq:llayer}), the layer is observationally brightest in its \textit{low} magnetization state -- this is when~$\mathcal{F}$ is smallest and, hence, when most of the incident Poynting flux is radiated at low enough energies (below pair threshold) to escape the system. In contrast, the upstream region becomes brightest when it receives an abundant supply of above-threshold photons -- when the layer is in its \textit{high} magnetization phase. These are processed into an energetic fresh pair plasma component that subsequently radiates below threshold [unless~$\fnc(\sighn{>}) \simeq 1$]. Thus we expect that, in a two-state swing cycle, the appropriate luminosity ratio is
\begin{align}
    \frac{L_{\rm upstream}(\sighn{>})}{L_{\rm layer}(\sighn{<})} &\sim [1 - \fnc(\sighn{>})] \frac{\mathcal{F}(\sighn{>})}{1 - \mathcal{F}(\sighn{<})} \notag \\
    &\sim [1 - \fnc(\sighn{>})] \leq 1 \, ,
    \label{eq:lratoutphase}
\end{align}
where, in the last step, we loosely approximated~$\mathcal{F}(\sighn{>}) \sim (1 - \mathcal{F}(\sighn{<})) \sim 1$, which is often roughly correct in swing cycles (see section~\ref{sec:knradreconn_mod}). In the above, we also consider~$\fnc$ to depend on the magnetization. The detailed dependence (discussed in Appendix~\ref{sec:feedbackdetails}) is not important here beyond that, generally,~$\fnc(\sighn{<}) \ll 1$ and~$\fnc(\sighn{>}) \sim 1$. Thus, in a swing cycle, the layer probably dominates the observed radiation.

An exceptional case occurs if the swing cycle high-state has energy recapture efficiency~$\xi(\sighn{>}) \simeq \fnc(\sighn{>})$ close enough to unity to render the upstream region brightest in the low-magnetization state \{i.e.\ if~$[1 - \fnc(\sighn{>})] \mathcal{F}(\sighn{>}) < [1 - \fnc(\sighn{<})] \mathcal{F}(\sighn{<})$\}. Then the relevant luminosity ratio, instead of equation~(\ref{eq:lratoutphase}), is~$L_{\rm upstream}(\sighn{<})/L_{\rm layer}(\sighn{<})$, which evaluates to~$[1 - \fnc(\sighn{<})] \mathcal{F}(\sighn{<}) / [1 - \mathcal{F}(\sighn{<})] \sim \mathcal{F}(\sighn{<}) \ll 1$, and is quite dominated by the layer.

Thus, the layer generally produces most of the observable radiation from pair-regulated Klein-Nishina reconnection. The upstream luminosity only dominates in a steady state if~$\mathcal{F} > 1/2$ and~$f_{\rm nocool} \ll 1$, and it may at most be comparable to the layer luminosity in an asymptotic swing cycle. This seems to be corroborated by observations because the expected spectrum produced from the upstream region has spectral index~$\alpha = 1/2$ (calculated in Appendices~\ref{sec:feedbackdetails} and~\ref{sec:fpexactsoln}), but the objects we discuss below have steeper scalings (as the layer might produce in a low magnetization state).

\subsubsection{Klein-Nishina physics may promote rapid variability through kinetic beaming}
\label{sec:kbprospects}
In our earlier work \citep{mwu20}, we performed a detailed study of the interplay between radiative physics (Thomson IC radiation reaction) in reconnection and the kinetic beaming phenomenon first discovered by \citet{cwu12}. As detailed in both references, in collisionless relativistic reconnection, the electromagnetic fields near reconnection X-points tend to simultaneously accelerate \textit{and} collimate particles. The higher energy particles are focused more tightly than the lower energy particles, making this beaming inherently kinetic. A collimated beam of high-energy particles may then sweep across an observer's line of sight, and the synchrotron or IC emission of the bunch -- also emitted as a beam -- may then create a dramatic blip in the measured lightcurve: a \quoted[.]{lighthouse effect} 

In \citet{mwu20}, we showed that efficient radiative losses play a critical role in enabling the kinetic beaming mechanism to impact observations. Without strong radiative cooling, collimated beams of particles isotropize before dumping their reconnection-acquired energy into energetic photons, and thus most of their radiation is emitted quasi-isotropically. Only efficiently cooled beamed particle bunches can radiate their energy before dispersing, leading to sweeping beams of light that may manifest as rapid flares.

We here recall two specific criteria from our earlier work that may be necessary for, or at least promote, observable signatures of kinetic beaming:
\begin{enumerate}
    \item Particles may need to have cooling times within a certain multiple of their gyroperiods in the reconnecting magnetic field~$B_0$. (If this multiple is~$1$, then this is the \textit{saturated cooling} condition from section~\ref{sec:thomregimes} -- particles are at their radiation-limited energy~$\gradk$.) The number of gyroperiods before a collimated bunch of particles isotropizes may be large but still finite. In fact, one main result from \citet{mwu20} is that kinetic beaming only manifests in the emission from particles with energies within about a decade of the radiative saturation Lorentz factor~$\gradt$, corresponding to~$\tcoolt \lesssim 100$ gyroperiods. \label{en:coolingrat} 
    \item Direct acceleration (by the reconnection electric field) near reconnection X-points must deliver particles to higher energies than secondary acceleration channels (in the language of sections~\ref{sec:thomradreconn} and~\ref{sec:knradreconn},~$\gamma_{\rm sec} \ll \gradk, \gx$ -- secondary channels must radiatively stall before both the X-point radiative and intrinsic acceleration limits). Otherwise, secondary -- and presumably more isotropic -- energization processes may wash out signatures of kinetic beaming in the reconnection-energized distribution of particles. \label{en:secondarysuppression} 
\end{enumerate}

Our present work provides at least~$3$ reasons why pair-regulated Klein-Nishina reconnection could meet these criteria, perhaps leading to even more pronounced kinetic beaming than when particles are subject to purely Thomson radiative cooling.

(1) The first reason pertains to point~\ref{en:coolingrat} above. Namely, Klein-Nishina effects render a particle's \textit{cooling ratio} -- the ratio of its cooling length to its Larmor radius -- very insensitive, when~$\gamma \gg \gkn$, to its Lorentz factor. To wit,
\begin{align}
    \lim_{\gamma \gg \gkn} &\frac{c \tcoolk(\gamma)}{2 \pi \gamma \rho_0} = \lim_{\gamma \gg \gkn} \left( \frac{\gradt}{\gamma} \right)^2 \frac{1}{f_{\rm KN}(\gamma / \gkn)} \notag \\
    &= \frac{10}{2 \pi} \frac{2}{9} \frac{( \gradt/\gkn )^2}{\ln \left( \gamma/\gkn \right) - 11/6} \simeq 0.4 \frac{( \gradt/\gkn )^2}{\ln \left( \gamma/\gkn \right) - 11/6} \, ,
    \label{eq:coolingrat}
\end{align}
where we used approximation~(\ref{eq:fknasym}). Even when Lorentz factors as high as~$\gradk$, for which~$c \tcoolk / 2 \pi \gamma \rho_0 = 10 / 2 \pi \sim 1$, are not accessible, particles may still reach a high-energy~($\gamma \gg \gkn$) regime where their cooling times come within some moderately large, beaming-favoring multiple of their gyroperiods, and where their cooling ratios become essentially~$\gamma$-independent. In such a scenario, a broad range~(e.g.\ from~$\gkn$ to~$\gx$) of energetic particles may radiate efficiently enough to be kinetically beamed. This range could possibly be broader than in the Thomson regime, for which~$\tcoolt(\gamma)$ declines as~$\gamma^{-1}$, making it increasingly difficult to accelerate particles up to higher and higher energies. However, one should invoke this argument with some caution because the smallest accessible cooling ratio in the Klein-Nishina regime may be quite large, scaling, according to~(\ref{eq:coolingrat}), as~$(\gradt / \gkn)^{2}$.

(2) A second way Klein-Nishina effects may promote kinetic beaming pertains to item~\ref{en:secondarysuppression} above. As argued in section~\ref{sec:collickn}, Klein-Nishina radiative cooling selectively suppresses secondary, slower acceleration channels relative to rapid, impulsive acceleration near reconnection X-points. In the Klein-Nishina regime, the IC cooling time~$\tcoolk(\gamma)$ grows as~$\gamma/\ln(\gamma)$~[equation~(\ref{eq:tcoolkasym})], which effectively removes the radiative cap on X-point acceleration [because~$t_{\rm X}$ is also proportional to~$\gamma$; see equation~(\ref{eq:tx})]. In contrast, secondary energization mechanisms generally operate on \ts[s]that grow more quickly with~$\gamma$ (e.g.\ as~$\gamma^2$), and so maintain a finite \cutoff[,]even for deeply Klein-Nishina radiative cooling.

(3) Finally, and also relevant to point~\ref{en:secondarysuppression} above, the energy distribution of produced pairs is very broad and~\nonthermal[](section~\ref{sec:nocascades}). These particles serve precisely as the pre-accelerated upstream population that, as argued by \citet{mwu20}, may help overcome the conventional~$\gx \simeq 4 \sigcgen$ limit. Upon entering the reconnection layer, newborn particles possess Larmor radii~($\geq \gcti{1} \rho_0$) that may exceed those,~$\sim \sigcgen \rho_0$, of typical accelerated particles -- most of which come from the much colder and more numerous particle population (section~\ref{sec:upstreammult}) that was already present in the far upstream region. If so, then newborn pairs sample larger field structures than the elementary current layers and plasmoids, with size scale~$\sigcgen \rho_0$, at the bottom of the plasmoid hierarchy \citep{wuc16, u20}. Unlike the vast majority of initially present cold upstream particles, these pairs may surf across many elementary layers -- that together comprise a much larger acceleration region -- becoming energized well beyond~$4 \sigcgen$ before finally becoming magnetized. Thus,~$\gx$ is effectively raised. This is important because higher~$\gx$ enables direct acceleration to energize and collimate higher-energy particles, potentially increasing the energy range of kinetically beamed particles and photons. 

Detailed predictions for kinetic beaming are outside the scope of this paper; our analysis only traces the largest-scale bulk flow of energy and particles through the reconnection system. However, these simple observations provide a target for future dedicated simulations, which may study, in detail, the effects of Klein-Nishina and pair-production physics on kinetic beaming.

\subsubsection{Caution against detailed spectral predictions}
While we believe that the basic qualitative features of our model presented in section~\ref{sec:knradreconn_mod} are fairly robust, the model is not presently quantitatively accurate enough to warrant making specific spectral predictions for various astrophysical sources. The specific power-law scalings of the particle and photon spectra depend on a number of uncertain details.

For one thing, the NTPA \quoted[~$p(\sighgen) = 1 + 2 / \sqrt{\sighgen}$]{equation of state}from equation~(\ref{eq:eos}) is quite crude and only intended for illustration. The steady-state power-law scalings~$\pt = p + 1$ and (especially)~$\pk = p - 0.5$ are also just rough estimates. For example, the bursty nature of reconnection is already known to modify~$\pt$ when Thomson IC cooling is quite strong \citep{wpu19}.

As a related issue, the power-law index of the photon spectrum radiated by particles above~$\gkn$ (in the~$\pk$ portion of the particle distribution) is only very loosely given by~$\pgg = \pk + 1 = p + 0.5$. In fact, \citet{msc05} estimate a harder power-law~$\pgg = p$ and predict little change between the power-law scaling of the emitted photon spectrum from the Thomson to Klein-Nishina regimes. Thus, given the level of quantitative uncertainty in our model, one should avoid inferring numerical values of plasma parameters (like~$\sighgen$ and~$z = \gf / \gkn$) from observed photon spectra, especially at the highest energies (above~$\gkn \me c^2$). It is also probably not warranted to test the model by searching for a prominent Klein-Nishina spectral break. 

Having discussed now some generic observational features of our model as well as some of its quantitative limitations (with respect to interpreting power-law scalings of photon spectra), we now turn to some concrete classes of astrophysical systems.

\subsection{Consequences for specific astrophysical systems}
\label{sec:astroestimates}
In this section, we explicitly estimate reconnection energy scales~($\gmax$,~$\sigc$,~$\gradt$,~$\gkn$, etc.) for two types of astrophysical systems -- flat-spectrum radio quasar (FSRQ) jets and black hole accretion disc coronae (ADCe). We also comment on the prospects for our model either to explain certain observational phenomena or to help constrain astrophysical details that are difficult to pin down from observations alone.

\subsubsection{The radiative environments of FSRQs}
\label{sec:fsrqenvs}
FSRQs form a subclass of blazars, which are active galactic nuclei that launch counterpropagating relativistic jets, one of which is directed toward the observer. Blazar spectra generally feature a characteristic double-humped structure \citep{fmc98, g11}. The low-energy component is usually understood as synchrotron radiation by electrons and positrons in the jet. The high-energy peak is also commonly associated with jet leptons but via a different process: IC scattering of soft seed photons \citep{bfr08, brs13, ms16}. Leptonic models differ on whether the particles producing both spectral components are cospatial and on the source of Compton seed photons. In BL Lacs, lower-luminosity blazars with low- and high-energy humps peaked at relatively high frequencies \citep[UV/X-ray and hundreds of GeV gamma-rays, respectively;][]{g11, ms16}, the seed photons are generally thought to be furnished by synchrotron emission within the jet itself \citep[synchrotron self-Compton models; e.g.][]{mgc92, bm96}. For the higher-luminosity sources with lower-energy spectral peaks \citep[often falling in the IR at low energies and a few hundred MeV in the gamma-rays;][]{g11, ms16}, the FSRQs, typical models envision Compton seed photons impinging from the circumnuclear environment \citep{bs87, mk89, sbr94}.

We focus on FSRQs. In addition to their higher luminosities and lower-energy spectra (relative to BL Lacs), FSRQs possess strong emission lines attributed to an accretion disc. The disc illuminates the jet directly and also shines onto clouds of circumnuclear material -- the broad emission line region (also \quoted[;]{broad-line region}BLR) and dusty torus (also \quoted[;]{hot dust region}HDR). These structures reprocess the accretion disc light and redirect some of its energy back onto the jet, thereby providing strong sources of external radiation situated relatively far from the central engine \citep[e.g.][]{nbs14, ms16}.

The smaller of the two circumnuclear regions is the BLR, which is made of gas that is partially ionized by the accretion disc radiation and hence illuminates the jet with UV line emission \citep[most prominently Ly~$\alpha$;][]{tg08}. Thus, the characteristic BLR photon energy is
\begin{align}
    \epsilon_{\rm BLR} \sim 10 \, \rm eV \, .
    \label{eq:eblr}
\end{align}
The BLR extends out to a radius
\begin{align}
    r_{\rm BLR} \sim 0.1 L^{1/2}_{\mathrm{d,}46} \, \mathrm{pc} \,
    \label{eq:rblr}
\end{align}
where~$L_{\mathrm{d,}46}$ is the luminosity of the accretion disc,~$L_{\rm d}$, in units of~$10^{46} \, \rm erg \, s^{-1}$ \citep{tg08, ssm09, nbc12}.
At jet propagation distances~$r < r_{\rm BLR}$ from the central engine, the jet traverses roughly isotropic ambient radiation, sourced by the BLR, of galaxy-frame energy density 
\begin{align}
    U_{\rm BLR} \sim \frac{L_{\rm BLR}}{4 \pi r_{\rm BLR}^2 c} \sim 6 \times 10^{-3} \, \rm erg \, cm^{-3}
    \label{eq:ublr}
\end{align}
\citep[][]{tg08, ssm09, nbc12}. Here we assume that the broad-line region intercepts and reprocesses a certain fraction~\citep[$2$ per cent;][]{tbg11} of the accretion disc light, and, therefore, that~$L_{\rm BLR} \propto L_{\rm d}$. Thus,~$U_{\rm BLR}$ is insensitive to~$L_{\rm d}$.

Farther removed from the nucleus than the BLR is the hot dust region, which, radiatively heated by the accretion disc, shines a quasi-thermal spectrum of temperature~$T_{\rm HDR} \simeq 1200 \, \rm K$ onto the jet. Thus \citep[see][]{nsi08a, nsi08b, nbc12},
\begin{align}
    \epsilon_{\rm HDR} \sim 3 k_{\rm B} T_{\rm HDR} = 0.3 \, \rm eV \, .
    \label{eq:ehdr}
\end{align}
The HDR extends out to a distance from the central engine of
\begin{align}
    r_{\rm HDR} \sim 4 \, L^{1/2}_{\mathrm{d,}46} T^{-2.6}_{\mathrm{HDR,}3} \, \rm pc \, ,
    \label{eq:rhdr}
\end{align}
where~$T_{\mathrm{HDR,}3} \equiv T_{\rm HDR} / 1000 \, \mathrm{K} = 1.2$ \citep{nsi08a, nsi08b, ssm09}.
Thus, when~$r < r_{\rm HDR}$, the HDR radiation energy density traversed by the jet is roughly isotropic and approximately
\begin{align}
    U_{\rm HDR} \sim \frac{L_{\rm HDR}}{4 \pi r_{\rm HDR}^2 c} \sim 9 \times 10^{-5} \, \rm erg \, cm^{-3} \, .
    \label{eq:uhdr}
\end{align}
Here, we again take a fixed fraction \citep[in this case~$10$ per cent;][]{mmj11} of the disc radiation to be reprocessed by the circumnuclear structure. Thus~$U_{\rm HDR}$, like~$U_{\rm BLR}$, lacks~$L_{\rm d}$-dependence.

We posit that at least some quiescent and flaring blazar gamma-ray observations are powered by magnetic reconnection \citep[][]{gub09, ngb11, g13, spg15, pgs16, wub18, nyc18, cps19, on20, sns21}, either induced by macroscopic field polarity reversals \citep[e.g.][]{gu19, srn21} or taking place at the small-scale terminus of a turbulent cascade \citep[e.g.][]{zup13, zuw20, cs18, cs19, lb20, bl20, nb20, sns21}. 

The blazar emission zone -- the distance~$r$ where most of the emission is produced -- is an important but difficult-to-constrain quantity in blazar research: by nature of being collimated along the observing line of sight, blazar jets appear point-like on the sky, and so it is not possible to deduce~$r$ directly from observations. We therefore allow~$r$ to vary over an appreciable range. If~$r < r_{\rm BLR}$, the BLR dominates the ambient radiation bathing the jet and the emission-powering reconnection zone; if~$r_{\rm BLR} < r < r_{\rm HDR}$, the HDR dominates.\footnote{In reality, the BLR intensity falls off smoothly with~$r$ and thus dominates to distances slightly exceeding~$r_{\rm BLR}$ (but still much less than $r_{\rm HDR}$; e.g. \citealt{nbs14}).} We consider only blazar zones far enough from the central engine \citep[$r \gtrsim 0.01 \, \rm pc$;][]{ds02, ssm09, nbs14} that the direct accretion disc light, which illuminates the jet from behind, is redshifted in the jet rest frame to a lower energy density than the BLR and HDR radiation fields. (Unlike the accretion disc light, the BLR and HDR photons impinge quasi-isotropically on the jet in the galaxy frame and are thus blueshifted when boosted to the jet frame). We do not consider~$r \gg r_{\rm HDR}$. 

The BLR and HDR choke gamma-rays above the energies
\begin{align}
    \epsilon_{\rm c,BLR} = \frac{(\me c^2)^2}{\epsilon_{\rm BLR}} \sim 30 \, \rm GeV 
    \label{eq:ecblr}
\end{align}
and
\begin{align}
    \epsilon_{\rm c, HDR} = \frac{(\me c^2)^2}{\epsilon_{\rm HDR}} \sim 0.9 \, \rm TeV \, ,
    \label{eq:echdr}
\end{align}
respectively, since the pair-production optical depths suffered by a photon traversing these regions are
\begin{align}
    \tau_{\rm BLR} = \frac{U_{\rm BLR} \sigma_{\rm T} r_{\rm BLR}}{5 \epsilon_{\rm BLR}} \sim 15
    \label{eq:taublr}
\end{align}
and
\begin{align}
    \tau_{\rm HDR} = \frac{U_{\rm HDR} \sigma_{\rm T} r_{\rm HDR}}{5 \epsilon_{\rm HDR}} \sim 300 \, .
    \label{eq:tauhdr}
\end{align}
Note that these optical depths are~$\textit{not}$ equal to those,~$\tau_{\gamma\gamma,\rm BLR}$ and~$\tau_{\gamma\gamma, \rm HDR}$, of the corresponding reconnection sites where the emission is sourced. Those optical depths are evaluated later using the smaller size of the emission region and with the seed photon number density~$U_i / \epsilon_i$~($i = \mathrm{BLR}\, \mathrm{or}\, \mathrm{HDR}$) boosted to the jet frame. Since quiescent FSRQs are generally not observed at very high energies~($\gtrsim 0.1 \, \rm TeV$), the most relevant model for them may be irradiation by the BLR. However, FSRQ flares are sometimes observed at up to several hundred GeV \citep[][]{magic08, magic11, hess13, sbb15, veritas15, magic15} -- still below~$\epsilon_{\rm c,HDR}$. In these cases, emission between~$r_{\rm BLR}$ and~$r_{\rm HDR}$, where the external photons come from the HDR, is most likely.

\subsubsection{FSRQ jet parameters}
In order to estimate reconnection parameters, we must now make a number of further assumptions about the nature of the jet: including its speed, shape, and magnetic field strength. 

We assume that reconnection occurs in the rest-frame of the jet and that the jet travels with relativistic bulk Lorentz factor~$\Gamma_{\rm j} \gg 1$ with respect to the host galaxy. We denote jet-frame quantities with primes. However, we exclude individual particle Lorentz factors (including cold magnetizations~$\sigc$) from this convention, writing them exclusively in the jet frame and without primes. We adopt a relatively high jet Lorentz factor~$\Gamma_{\rm j} = 40$. This is the same as in our earlier model \citep{mwu20} of the rapidly variable very high-energy~($\gtrsim 0.1 \rm \, TeV$) flare from PKS 1222+21 \citep{magic11}. A large~$\Gamma_{\rm j}$ may be reconciled with more typical Lorentz factors (e.g.~$\Gamma_{\rm j} \simeq 10$) by invoking a structured jet \citep[e.g.][]{gtc05, bfr08, srb16, tg16, t17, srn21} in which an inner spine region, moving quickly~($\Gamma_{\rm j} \simeq 40$) and carrying the reconnection current sheet, is surrounded by a slower-moving sheath. This explicit structure is not necessary to our ensuing discussion, however.\footnote{Unlike in our earlier model \citep{mwu20}, we do not entertain the sheath as a potential source of Compton seed photons here.}

We imagine that the jet is conical with opening angle~$\theta_{\rm j} \sim 1 / 5 \Gamma_{\rm j}$ \citep{pkl09} and model the comoving magnetic field strength by assuming
\begin{align}
    B_0'(r) \sim 0.1 \left( \frac{r}{1 \, \rm pc} \right)^{-1} \, \rm G
    \label{eq:br}
\end{align}
\citep{nbc12, mwu20}, corresponding to a total jet Poynting flux luminosity of~$(\theta_{\rm j} r)^2 (\Gamma_{\rm j}^2 B_0'^2) c \sim 1 \times 10^{43} \rm erg\, s^{-1}$ that is conserved in~$r$ and independent of~$\Gamma_{\rm j}$. If the Poynting flux is instead dissipated (as indeed a reconnection scenario suggests), then the scaling of~$B_0'$ with~$r$ should be steeper, but perhaps not by much. For example, the fiducial striped jet model of \citet{gu19}, which includes magnetic dissipation through reconnection, gives a power law close to~$B_0' \propto r^{-5/4}$.

We note that the typical isotropic FSRQ luminosities~$L_{\rm iso} \sim 10^{48} \, \rm erg \, s^{-1}$ require minimum intrinsic jet power, presumably carried by magnetic fields in a reconnection scenario, of~$L_{\rm iso} / \Gamma_{\rm j}^2 \sim 6 \times 10^{44} \rm erg \, s^{-1}$. We have checked that increasing our fiducial magnetic field strength [e.g.~$B_0'(1 \, \rm pc) \sim 1 \, \rm G$] to supply such jet power does not affect our conclusions in this section, but we adopt~(\ref{eq:br}) to maintain continuity with our previous work \citep{mwu20}.

\subsubsection{Reconnection parameters in FSRQs}
\label{sec:fsrqest}
We now explicitly estimate parameters for reconnection in an FSRQ jet illuminated by either the broad-line region or the dusty torus.

First, the maximum (system-size-limited) Lorentz factor is
\begin{align}
    \gmax \equiv \frac{0.1 L}{\rho_0} &\sim \frac{0.1 e B_0'(r) r \theta_{\rm j}}{\me c^2} \notag \\
    &\sim 9 \times 10^{10} \left( \frac{\Gamma_{\rm j}}{40} \right)^{-1} \, .
    \label{eq:gmaxfsrq}
\end{align}
Note that this is independent of~$r$ because we take the reconnection layer length~$L$ to be comparable to the transverse width~$r \theta_{\rm j}$ of the jet, and thus the dependence of~$L$ on~$r$ cancels against that of the magnetic field strength in equation~(\ref{eq:br}).

\begin{figure*}
    \centering
    \begin{subfigure}{0.49\textwidth}
        \centering
        \includegraphics[width=\linewidth]{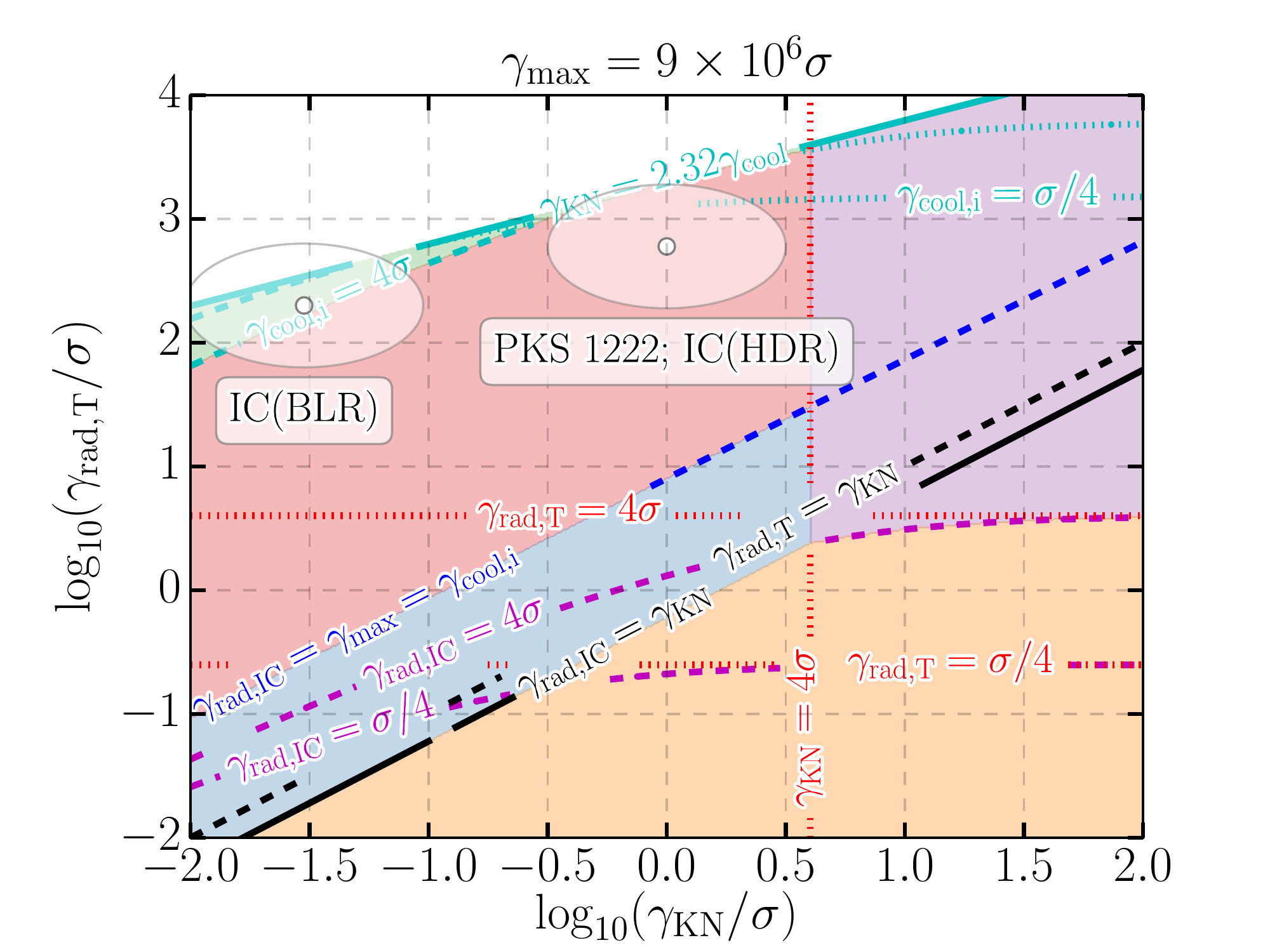}
    \end{subfigure}
    \begin{subfigure}{0.49\textwidth}
        \centering
        \includegraphics[width=\linewidth]{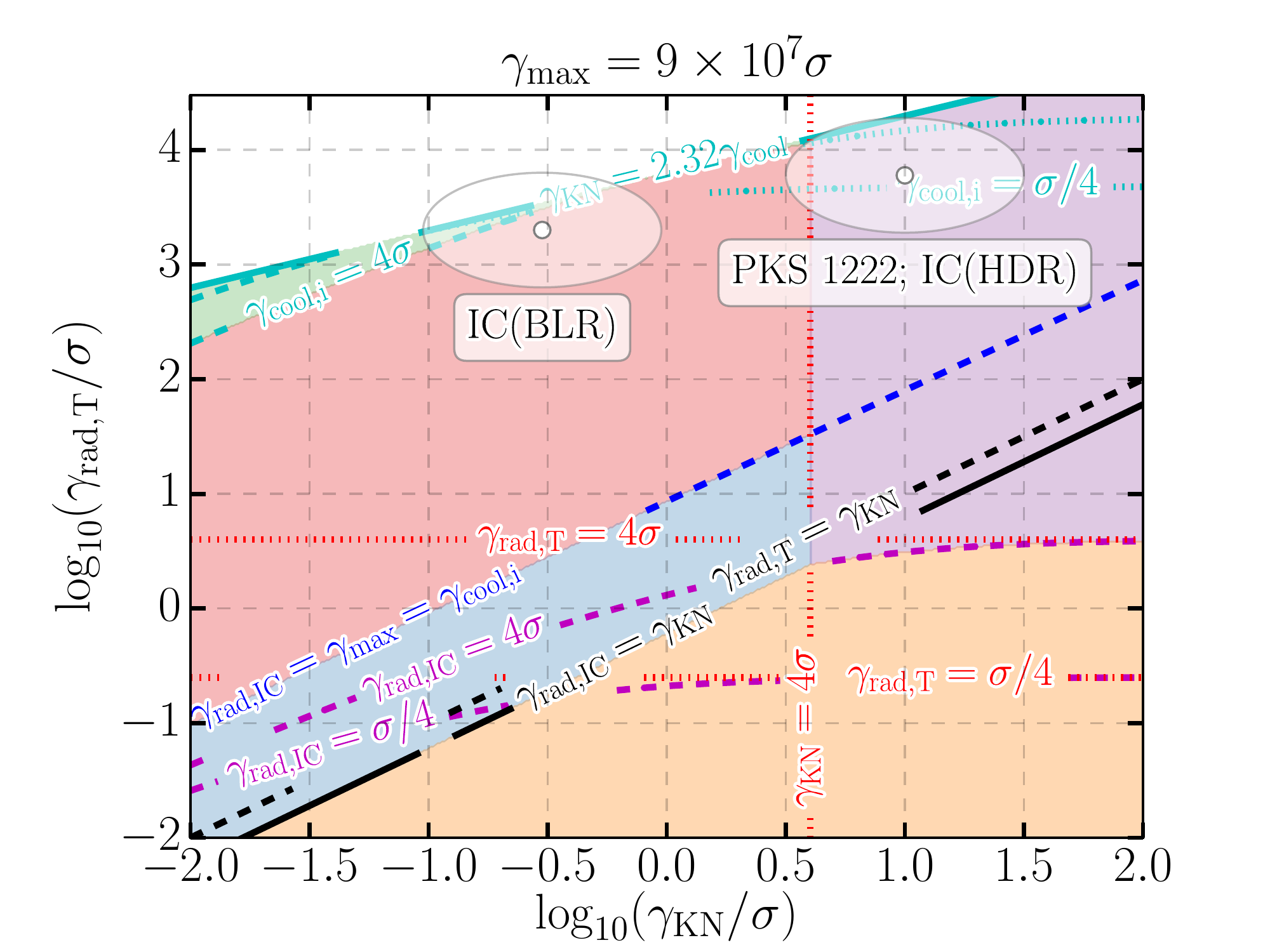}
    \end{subfigure}
    \caption{Radiative reconnection phase diagrams for the case~$\gmax = 9 \times 10^{10}$ [equation~(\ref{eq:gmaxfsrq})] and two different values of~$\sigc$ (abbreviated as~$\sigma$ in the plot):~$\sigc = 10^4$ (left) and~$\sigc = 10^3$ (right). Region col\spellor[ings]have the same meaning as in Fig.~\ref{fig:radphasespace}. In each diagram, white points indicate the parameters of equations~{(\ref{eq:blrradparamsfirst})-(\ref{eq:blrradparamslast})} and~{(\ref{eq:hdrradparamsfirst})-(\ref{eq:hdrradparamslast})}, where gamma-ray emission comes primarily through Comptonization of BLR photons [IC(BLR)] and HDR photons [IC(HDR)], respectively. Larger shaded white ellipsoids indicate one order-of-magnitude uncertainties in each direction. The IC(HDR) parameters are particularly relevant to the rapid very high-energy flare from PKS 1222+21 \citep{magic11} discussed in our earlier work \citep{mwu20}. Because the three parameters~$\gradt$,~$\gkn$, and~$\gmax$ are all proportional to~$\Gamma_{\rm j}^{-1}$, changing~$\Gamma_{\rm j}$ is equivalent to changing~$\sigc$ by the same factor.}
    \label{fig:pks1222pspace}
\end{figure*}

Next we discuss the jet magnetization. In lieu of estimating a fiducial jet electron number density (which is needed to estimate~$\sigc$), we take inspiration from previous studies in which the jet is moderately magnetized at the parsec scale \citep[e.g.][]{g13, gu19}. If the jet is mass-dominated by protons, this still allows individual electrons/positrons to attain high Lorentz factors. Assuming an electron-proton jet -- and that our main results from earlier sections carry over to electron-proton plasmas containing ultrarelativistic initial magnetic energy per electron -- we adopt a cold \textit{ion} magnetization (in the jet frame) of~$5$, which corresponds to a fiducial jet-frame cold \textit{electron} magnetization~$\sigc \sim 10^4$. To illustrate our high degree of uncertainty in this quantity, in Fig.~\ref{fig:pks1222pspace}, we also present the case~$\sigc \sim 10^3$, which corresponds to a cold ion magnetization of~$0.5$ (or of~$5$ but with~$10$ positrons per proton in the presence of mixed composition). In addition, we crudely assume that the cold magnetization does not change in~$r$, which is consistent with Poynting flux conservation if one also has particle flux conservation~[$\pi (r\theta_{\rm j})^2 n_0(r) = \, \rm constant$].

Now we estimate the key radiative parameters. For reconnection illuminated by the broad-line region~($r < r_{\rm BLR}$), we have, plugging the jet-frame seed photon energy~$\epsilon_{\rm BLR}' \sim \Gamma_{\rm j} \epsilon_{\rm BLR}$ and energy density~$U_{\rm BLR}' \sim \Gamma_{\rm j}^2 U_{\rm BLR}$ [with~$\epsilon_{\rm BLR}$ and~$U_{\rm BLR}$ given in equations~(\ref{eq:eblr}) and~(\ref{eq:ublr})] into definitions~(\ref{eq:gradt}),~(\ref{eq:gcool}),~(\ref{eq:gkndef}) and~(\ref{eq:taugg}):
\begin{align}
    \gradt^{(\rm BLR)} &\sim 2 \times 10^6 \left( \frac{r}{0.1 \, \rm pc} \right)^{-1/2} \left( \frac{\Gamma_{\rm j}}{40} \right)^{-1} \, , \label{eq:blrradparamsfirst} \\
    \gamma_{\rm cool, BLR} &\sim 60 \left( \frac{r}{0.1 \, \rm pc} \right)^{-1} \left( \frac{\Gamma_{\rm j}}{40} \right)^{-1} \, , \\
    \gamma_{\rm KN, BLR} &\sim 300 \left( \frac{\Gamma_{\rm j}}{40} \right)^{-1} \quad \mathrm{and} \\
    \tau_{\gamma\gamma, \rm BLR} &\sim 3 \left( \frac{r}{0.1 \, \rm pc} \right) \, .
    \label{eq:blrradparamslast}
\end{align}
The observed photon energies emitted by particles with these radiative Lorentz factors are
\begin{align}
    \epsilon_{\rm obs}^{(\rm BLR)}\left(\gradt^{(\rm BLR)}\right) &\sim \frac{1}{2} \Gamma_{\rm j} \gradt^{(\rm BLR)} \me c^2 \sim 20 \left( \frac{r}{0.1 \, \rm pc} \right)^{-1} \, \rm TeV \, ,\label{eq:blrphtensfirst} \\
    \epsilon_{\rm obs}^{(\rm BLR)}\left(\gamma_{\rm cool,BLR} \right) &\sim \frac{4}{3} \Gamma_{\rm j}^2 \gamma_{\rm cool,BLR}^2 \epsilon_{\rm BLR} \notag \\
    &\sim 80 \left( \frac{r}{0.1 \, \rm pc} \right)^{-2} \, \rm MeV \quad \mathrm{and} \\
    \epsilon_{\rm obs}^{(\rm BLR)}\left(\gamma_{\rm KN,BLR} \right) &\sim \frac{4}{3} \Gamma_{\rm j}^2 \gamma_{\rm KN,BLR}^2 \epsilon_{\rm BLR} \sim 2 \, \rm GeV \, .
    \label{eq:blrphtenslast}
\end{align}
These photon energies are independent of the jet Lorentz factor~$\Gamma_{\rm j}$. One of them,~$\epsilon_{\rm obs}^{(\rm BLR)}(\gamma_{\rm KN,BLR}) \sim 2 \, \rm GeV$, depends on no unknowns and implies that virtually all the~$> 1 \, \rm GeV$ emission, routinely observed by \textit{Fermi} \citep{g11}, is emitted in the Klein-Nishina regime (if Comptonized from BLR photons). Note that, in evaluating the photon energies emitted by~$\gamma = \gkn$ particles, one may use either the Thomson scaling~$4 \Gamma_{\rm j}^2 \gkn^2 \eph / 3$ (as we have done) or the Klein-Nishina scaling~$\Gamma_{\rm j} \gkn \me c^2 / 2$; this choice modifies the estimate by less than a factor of~$2$.

If, instead,~$r_{\rm BLR} < r < r_{\rm HDR}$, the radiation from the dusty torus is strongest, yielding, from the HDR estimates~(\ref{eq:ehdr}) and~(\ref{eq:uhdr}),
\begin{align}
    \gradt^{(\rm HDR)} &\sim 6 \times 10^6 \left( \frac{r}{1 \, \rm pc} \right)^{-1/2} \left( \frac{\Gamma_{\rm j}}{40} \right)^{-1} \, , \label{eq:hdrradparamsfirst} \\
    \gamma_{\rm cool, HDR} &\sim 400 \left( \frac{r}{1 \, \rm pc} \right)^{-1} \left( \frac{\Gamma_{\rm j}}{40} \right)^{-1} \, , \\
    \gamma_{\rm KN, HDR} &\sim 1 \times 10^4 \left( \frac{\Gamma_{\rm j}}{40} \right)^{-1} \quad \mathrm{and} \\
    \tau_{\gamma\gamma, \rm HDR} &\sim 20 \left( \frac{r}{1 \, \rm pc} \right) \, .
    \label{eq:hdrradparamslast}
\end{align}
The corresponding observed photon energies are
\begin{align}
    \epsilon_{\rm obs}^{(\rm HDR)}\left(\gradt^{(\rm HDR)}\right) &\sim \frac{1}{2} \Gamma_{\rm j} \gradt^{(\rm HDR)} \me c^2 \sim 60 \left( \frac{r}{1 \, \rm pc} \right)^{-1} \, \rm TeV \, ,\label{eq:hdrphtensfirst} \\
    \epsilon_{\rm obs}^{(\rm HDR)}\left(\gamma_{\rm cool,HDR} \right) &\sim \frac{4}{3} \Gamma_{\rm j}^2 \gamma_{\rm cool,HDR}^2 \epsilon_{\rm HDR} \notag \\
    &\sim 100 \left( \frac{r}{1 \, \rm pc} \right)^{-2} \, \rm MeV \quad \mathrm{and} \\
    \epsilon_{\rm obs}^{(\rm HDR)}\left(\gamma_{\rm KN,HDR} \right) &\sim \frac{4}{3} \Gamma_{\rm j}^2 \gamma_{\rm KN,HDR}^2 \epsilon_{\rm HDR} \sim 70 \, \rm GeV \, .
    \label{eq:hdrphtenslast}
\end{align}
The photon energy~$\epsilon_{\rm obs}^{(\rm HDR)}(\gamma_{\rm KN,HDR}) \sim 70 \, \rm GeV$ [like the corresponding energy~$\epsilon_{\rm obs}^{(\rm BLR)}(\gamma_{\rm KN,BLR})$ for Comptonization within the BLR] does not depend on any unknowns and implies that IC(HDR) TeV emission from FSRQs is produced almost entirely in the Klein-Nishina regime. Irrespective of the dominant radiation field (BLR or HDR), the overall IC radiative \cutoff[]energy,~$\gradk$, is effectively infinite, scaling according to equation~(\ref{eq:gradkasym}) as~$\gradk \propto \gkn \exp[ (2/9) (\gradt / \gkn)^2 ]$.

Fig.~\ref{fig:pks1222pspace} displays the BLR~[(\ref{eq:blrradparamsfirst})-(\ref{eq:blrradparamslast})] and HDR~[(\ref{eq:hdrradparamsfirst})-(\ref{eq:hdrradparamslast})] energy scales by plotting them as points in a radiative reconnection \quoted{phase diagram}in the style of Fig.~\ref{fig:radphasespace}. This clearly illustrates the main result of this section: \textit{reconnection in FSRQ jets proceeds in the regime governed by Klein-Nishina and pair production physics}. This does not depend on whether the dominant seed photons stem from the BLR or the HDR. Broadly speaking (we discuss caveats and technical points below), reconnection proceeds in the red region. Here, X-point acceleration is not inhibited by radiative losses~($\gradk \gg \gmax \gg \gx$), but subsequently most, if not all, particles cool strongly -- on \ts[s]shorter than~$L/c$ -- rendering reconnection efficiently radiative. Furthermore, because~$\gx > \gkn$, many particles become impulsively energized up to energies where they radiate in the Klein-Nishina limit, producing above-threshold radiation that may activate pair feedback (section~\ref{sec:knradreconn_mod}).

Before moving to additional observational implications, we discuss a few secondary technical details to flesh out this basic picture. First, though one might worry that the IC(HDR) model, depending on~$\sigc$, crosses into the purple region of Fig.~\ref{fig:pks1222pspace} where the nominal intrinsic X-point acceleration \cutoff[]energy,~$\gx \simeq 4 \sigc$, falls below~$\gkn$, this is not a huge concern. The observed photon energies~$\epsilon_{\rm obs}^{(\rm BLR)}(\gamma_{\rm KN, BLR}) \sim 2 \, \rm GeV$ and~$\epsilon_{\rm obs}^{(\rm HDR)}(\gamma_{\rm KN, HDR}) \sim 70 \, \rm GeV$ are routinely observed in quiescent FSRQ spectra \citep[for the former;][]{g11, ms16} and in TeV outbursts \citep[for the latter;][]{magic08, magic11, hess13, sbb15, veritas15, magic15}. Thus, in a reconnection scenario, observations, in fact, suggest that~$\gx$, whether through high~$\sigc$ or through circumventing the conventional~$4 \sigc$-limit (as discussed in section~\ref{sec:kbprospects}), is high enough to accelerate particles beyond~$\gkn$, pushing reconnection into the Klein-Nishina radiative regime.

Next, one should note that the pair-production optical depth of the BLR-illuminated reconnection layer is not very large:~$\tau_{\gamma\gamma,\rm BLR} \sim 3$. According to equation~(\ref{eq:tcoolicmin}), this means that even the most strongly radiative particles possess cooling times~$\tcoolk(\gamma) \sim L / c \tau_{\gamma\gamma}$ only marginally faster than a dynamical time. This somewhat limits the overall radiative efficiency of reconnection, as signa\led[]by one of the IC(BLR) points in Fig.~\ref{fig:pks1222pspace} bordering on the green region where Klein-Nishina suppression of the IC \crosssection[]begins to shut down radiative cooling for the highest energy~($\gamma > \gcooli{2}$) particles. Plus, since~$\tau_{\gamma\gamma} < 10$, some newborn pairs may escape the system before being swept into the layer~($f_{\rm noesc}$ drops below unity; see section~\ref{sec:upstreamendens}), which diminishes the likelihood of swing cycles for the IC(BLR) scenario. 

Regarding the multiplicity of produced pairs (section~\ref{sec:upstreammult}), the estimates of this section yield~$\gamma_{\rm KN, BLR} / \sigc \geq 0.03$ and~$\gamma_{\rm KN, HDR} / \sigc \geq 1$ for~$\sigc \leq 10^4$ and~$\Gamma_{\rm j} \leq 40$. Thus, the requirement~$\gkn > \sigc / 10$ to ensure a pair multiplicity (the ratio of number densities of newborn pairs to originally present pairs),~$\eta$, less than unity is roughly satisfied. The BLR may marginally support~$\eta \gtrsim 1$ because~$\max (\eta) \simeq \sigc / 10 \gkn$ (section~\ref{sec:upstreammult}). However this is doubtful for several reasons:~(i)~$\gamma_{\rm KN, BLR} / \sigc = 0.03$ is a lower bound (increasing for smaller~$\sigc$ or~$\Gamma_{\rm j}$);~(ii)~$\eta$ may never attain its global maximum; and~(iii) the result~$\max (\eta) \simeq \sigc / 10 \gkn$ ignores particle escape, which may be an issue for reconnection inside the BLR, as mentioned in the previous paragraph. Thus, our estimates support the picture of tenuous newborn pairs that control the energy density of the inflow plasma but not the inflowing particle count.

\subsubsection{Relevance to FSRQ observations}
\label{sec:fsrqobs}
Next, we comment on a few observational issues that our model of pair-regulated Klein-Nishina reconnection may help to address. First, and as discussed in section~\ref{sec:kbprospects}, Klein-Nishina reconnection may enhance the range of photon energies that are kinetically beamed. This provides an attractive explanation for rapid very high-energy FSRQ flares \citep[of the kind first observed by][]{magic11}. A kinetic beaming scenario has already been advocated on energy-budget grounds by \citet{nbc12}, and has recently been explored by us in the context of Thomson radiative reconnection \citep{mwu20}. However, our earlier work requires a dense population of seed photons -- much denser than those supplied by the BLR and HDR -- in order to facilitate the high degree of radiative efficiency needed by kinetic beaming. Therefore, \citet{mwu20} disfav\spellor[]a single-zone external IC(HDR) scenario and suggest a two-zone (e.g.\ spine-sheath) configuration where the seed photons are supplied by the jet itself. However, if Klein-Nishina effects ease the radiative efficiency requirement, an IC(HDR) model may still be compatible with kinetic beaming.  

Next, although we would like to refrain from trying to make detailed spectral interpretations, we note that the spectra of flaring FSRQs in the very high-energy~(VHE; $\gtrsim 0.1 \, \rm TeV$) band are generally quite steep, with intrinsic (deabsorbed) photon number index~$\pgg[\rm VHE] \gtrsim 2.5$ \citep{magic08, magic11, magic15}. If one assumes that there is no significant spectral break between the Thomson and deep Klein-Nishina regimes \citep{msc05}, then~$\pgg[\rm VHE] \simeq (\pt + 1) / 2 \simeq (p + 2)/2$. This suggests a steep injected index~$p \gtrsim 3$, which is realized in our reconnection \quoted[~$p(\sighgen) = 1 + 2 / \sqrt{\sighgen}$,]{equation of state}equation~(\ref{eq:eos}), if~$\sigh \lesssim 1$. This may occur when the ions are moderately magnetized and dominate the initial hot magnetization \citep[e.g.][]{wub18}. In this situation, pair feedback, because it cannot reduce~$\sighgen$ to below order-unity [equation~(\ref{eq:sigheffuniv})], would probably not strongly modify the spectrum.

However, if~$\sigh \gg 1$, pair feedback may produce the observed steep spectral indices during dramatic swing cycles (e.g.\ Fig.~\ref{fig:sighsolnrad}), wherein the observable emission is dominated by the layer (section~\ref{sec:layerdominance}) in the low-magnetization~$\sighn{<} \sim 1$ part of the cycle. The requirement~$\sigh \gg 1$ could be realized if either: (i) the initial hot magnetization governing the distribution of accelerated electrons/positrons decouples from the (order-unity) ion magnetization, or (ii) a highly magnetized pair-dominated jet region furnishes the upstream material for reconnection.

In addition, to support swing cycles, the layer generally requires high energy recapture efficiency~$\xi \simeq \fnc \sim 1$ (e.g.\ Fig.~\ref{fig:sighsolnrad}). In Appendix~\ref{sec:feedbackdetails}, we show that order-unity energy retention factors~$\fnc$ require a large \cutoff[~$z = \gf / \gkn$](e.g.~$z \gtrsim 1000$ in Fig.~\ref{fig:scx_stability}). Such high \cutoff[s]might not be realized because the observed Compton dominance in FSRQs is often~$\uph / U_{B} \sim 100$, suggesting that the particle energy,~$\gamma_{\rm s}$, beyond which synchrotron losses outcompete IC losses and may hence limit further particle acceleration, is about~$\gamma_{\rm s} \sim 30 \gkn$ [equation~(\ref{eq:gsynapprox})].\footnote{Generally,~$\uph / U_{B}$ is equated to the observed ratio of IC-to-synchrotron luminosities. This is sound in one-zone emission scenarios where Klein-Nishina effects do not suppress IC emission near the IC spectral peak. The latter requirement is generally satisfied because Compton FSRQ peaks typically fall at lower energies than~$\epsilon_{\rm obs}^{(\rm BLR)}(\gamma_{\rm KN,BLR})$ and~$\epsilon_{\rm obs}^{(\rm HDR)} (\gamma_{\rm KN,HDR})$.}

However, there are a few effects that may serve to promote swing cycles in spite of nonzero synchrotron losses. First, the most energetic particles responsible for emitting pair-producing photons are likely accelerated near reconnection X-points deep inside the current layer. There, the magnetic field is weaker \citep{ucb11, cub12}, which reduces synchrotron radiation relative to IC losses, potentially allowing the latter to remain dominant to higher energies than~$\gamma_{\rm s}$. A second possibility is that swing cycles actually set in at lower values of~$z$ than we predict. This is because we conservatively estimate a steep distribution of pair-producing photons penetrating the upstream plasma~$\pgg = p + 0.5$ in section~\ref{sec:nocascades} \citep[rather than, e.g.,~$\pgg = p$ as in][]{msc05}, and therefore may artificially underestimate~$\fnc$. Finally, and related to the previous point, we find (not presented here) that the threshold on~$z$ for swing cycles to kick in is quite sensitive to the precise dependence of~$\fnc$ on~$z$. These considerations demonstrate the need for radiative kinetic simulations to examine in detail whether swing cycles are possible, when they occur, and how they impact the spectrum of observed radiation.

As a completely separate prospect, our model may connect to the matter-antimatter balance in FSRQ jets, which is difficult to constrain from observations alone \citep{ms16}. This ratio may be initially imprinted at the base of the jet -- for example, by magnetospheric spark-gap discharges \citep[e.g.][]{bz77, bip92, fkm18, cyy18, ccp20} or by interaction between the nascent jet and the accretion flow \citep[e.g.][]{rbp20, wrg21} -- but it may also be modified \textit{in situ} as the jet propagates. Although we expect relatively few newborn pairs compared to the initial number of upstream particles in each reconnection episode (section~\ref{sec:upstreammult}), continuous, repeated reconnection occurring as the jet ploughs through ambient radiation fields could still lead to secular growth in the number of pairs present in the plasma. This may be relevant to FSRQs in their quiescent states, where their spectral \cutoff[s]are broadly consistent with gamma-ray absorption in the broad-line region. Furthermore, because~$\tau_{\rm BLR} > 1$, even if the reconnection region itself is optically thin~($\tau_{\gamma\gamma,\rm BLR} < 1$), emitted gamma-rays may still produce pairs in the jet before exiting the BLR (though they do not strongly impact the reconnection dynamics).

In summary, reconnection in FSRQ jets illuminated by either the broad-line region or the dusty torus is expected to occur in the radiative regime governed by pair-production and Klein-Nishina physics. Here, kinetic beaming and pair-feedback-initiated swing cycles may explain the \ts[s]of rapid TeV flares and typical FSRQ VHE spectral indices, respectively. However, confirming whether these mechanisms are active in Klein-Nishina radiative reconnection depends on several unknowns and remains a key open question for future simulations to address. In addition, \textit{in situ} pair production driven by reconnection could contribute to the pair content of quiescent FSRQ jets. 

\subsubsection{Reconnection parameters in black hole ADCe}
\label{sec:adceparams}
Magnetic reconnection may also power emission from a highly magnetized corona sandwiching a black hole accretion disc \citep{lp77, grv79, d98, ug08, gu08, u16, b17, wpu19, sb20}. Here, we focus on the high/soft states of black hole X-ray binaries, showing that coronal reconnection in this context likely proceeds in a highly radiative Klein-Nishina regime. We term as high/soft any state where the quasi-thermal~($\sim 1 \, \rm keV$) spectral component -- which is attributed to an optically thick, geometrically thin accretion disc \citep{ss73} -- strongly dominates over the \nonthermal[]component, which extends to much higher, hard X- and gamma-ray energies \citep[e.g.][]{rm06}. Because reconnection intrinsically gives rise to a \nonthermal[]distribution of particle energies, a reconnection scenario does not rely on repeated scatterings by an isothermal plasma of moderate optical depth \citep[e.g.][]{sle76, el76, rl79} to explain the observed \nonthermal[]spectrum.

The bright quasi-thermal disc emission in the high/soft state provides an intense soft photon bath to the coronal reconnection region, and we assume that Comptonization of disc photons is responsible for the observed \nonthermal[]spectrum. We now estimate the energy scales~$\gamma_{\rm KN,\corlab}$ and~$\gamma_{\rm rad,T}^{(\rm \corlab)}$ associated with the disc seed photons, as well as the energy scales~$\sigma_{\rm c,0}$ and~$\gamma_{\rm max}$. Determining these then allows us to accomplish our main goal for this section: \characteriz[ing]the radiative regime of reconnection.

Beginning with~$\gamma_{\rm KN,\corlab}$, the characteristic seed photon energy emitted from the disc is 
\begin{align}
    \epsilon_{\rm \corlab} \sim 1 \, \rm keV \, ,
    \label{eq:edisc}
\end{align}
as both theorized \citep{ss73} and observed \citep[e.g.\ in Cyg~X-1;][]{mzb02, rm06}.
Through equation~(\ref{eq:gkn}), the Klein-Nishina Lorentz factor is then
\begin{align}
    \gamma_{\rm KN,\corlab} \equiv \frac{\me c^2}{4 \epsilon_{\rm \corlab}}\sim 100 \, .
    \label{eq:gkncor}
\end{align}
The corresponding energy to which photons encountering~$\gamma_{\rm KN,\corlab}$-particles are upscattered is
\begin{align}
    \epsilon_{\rm obs}^{(\rm \corlab)}\left( \gamma_{\rm KN,\corlab} \right) \sim 20 \, \rm MeV \, .
    \label{eq:egkncor}
\end{align}
Emission at these energies has been observed by the \textit{Fermi} Large Area Telescope (LAT) from Cyg X-1 \citep{zmc17}.

We now move on to estimate~$\gamma_{\rm rad,T}^{(\rm \corlab)}$. This requires [equation~(\ref{eq:gradt})] the energy density of the emission intercepted by the corona at a distance~$r$ from the black hole, which is roughly
\begin{align}
    U_{\rm \corlab} \sim \frac{L_{\rm disc}}{4 \pi r^2 c} \, .
    \label{eq:uphcor}
\end{align}
Here,~$L_{\rm disc}$ is the luminosity of the disc (emitted mostly near its inner edge) and~$r$ is assumed to exceed the innermost disc orbit. 

In the high/soft state~$L_{\rm disc}$ can be quite high -- up to several per cent of the Eddington limit,~$L_{\rm Edd} = 4 \pi G M_{\rm BH} m_p c / \sigma_{\rm T} = 4 \pi r_g m_p c^3 / \sigma_{\rm T} \simeq 1 \times 10^{39} (M_{\rm BH} / 10 M_{\odot}) \, \rm erg \, s^{-1}$ \citep[e.g.\ 4 per cent for Cyg X-1;][]{mzb02}. Furthermore,~$r$ is at least several gravitational radii,~$r_g \equiv G M_{\rm BH} / c^2 \simeq 1 \times 10^{6} (M_{\rm BH} / 10 M_{\odot}) \, \rm cm$. Thus, an upper bound and fiducial scale for the energy density of ambient radiation shining onto the reconnection layer is \citep[cf.][]{b17}
\begin{align}
    U_{\star} \equiv \frac{L_{\rm Edd}}{4 \pi r_g^2 c} = \frac{m_p c^2}{r_g \sigma_{\rm T}} \simeq 2 \times 10^{15} \left( \frac{M_{\rm BH}}{10 M_{\odot}} \right)^{-1} \, \mathrm{erg}\, \mathrm{cm}^{-3} \, .
    \label{eq:udmax}
\end{align}
Here, the fiducial black hole mass is~$10 M_\odot$. In the following, we estimate~$r \sim r_g$ and~$U_{\rm \corlab} \sim 0.04 U_\star$, corresponding to~$L_{\rm disc} \simeq 0.04 L_{\rm Edd}$ observed from Cyg~X-1 \citep{mzb02}. The parameter~$U_{\star}$ also provides a fiducial scale for the magnetic field energy density in the disc required to transport angular momentum outward \citep{ss73, bbr84, gu08, b17}. 

Beyond the fiducial scale~$U_\star$, there are a number of uncertain geometric factors that control how the radiation and magnetic field energy densities decay as one moves from the vicinity of the black hole in the disc to the reconnection region in the corona. This leads to uncertainty in~$\gamma_{\rm rad,T}^{(\rm \corlab)}$, which depends on both~$U_{\rm \corlab}$ and the coronal magnetic field energy density,~$U_{B,\rm cor}$. We \parameteriz[e]these uncertainties by exhibiting, in the expressions to follow,~$U_{\rm \corlab}$ as a multiple of~$U_\star$ and~$U_{B,\rm cor}$ as~$U_{\rm \corlab} / C_{\rm d}$, where~$C_{\rm d} \equiv U_{\rm \corlab} / U_{B, \rm cor}$ is the nominal Compton dominance. Without knowing the geometric factors governing how the seed photon and magnetic energy densities decay from the disc to the coronal reconnection region, we assume for simplicity that they fall off in the same way relative to~$U_{\star}$, adopting a fiducial value~$C_{\rm d} = 1$. We discuss consequences on Klein-Nishina radiative reconnection if~$C_{\rm d}$ is truly~$\sim 1$ at the end of this section.

Following these conventions, one may write the fiducial upstream magnetic field strength as \citep[e.g.][]{gu08, u16, b17}
\begin{align}
    B_0 &\equiv \sqrt{8 \pi U_{B,\rm cor}} = \sqrt{8 \pi U_{\rm \corlab}/ C_{\rm d}} \notag \\
    &\sim 4 \times 10^7 \left( \frac{U_{\rm \corlab}}{0.04 U_\star(M_{\rm BH})} \right)^{1/2} \left( \frac{M_{\rm BH}}{10 M_\odot} \right)^{-1/2} \left( \frac{C_{\rm d}}{1} \right)^{-1/2} \, \rm G \,
    \label{eq:b0adc}
\end{align}
and then use equation~(\ref{eq:gradt}) to write
\begin{align}
    \gradt^{\rm (\corlab)} &\equiv \sqrt{\frac{0.3 e B_{0,\rm cor}}{4 \sigma_{\rm T} U_{\rm \corlab}}}= \left( \frac{0.3 e}{4 \sigma_{\rm T}} \sqrt{ \frac{8 \pi}{C_{\rm d} U_{\rm \corlab} } } \right)^{1/2} \notag \\
    &\sim 6 \times 10^{3} \left( \frac{U_{\rm \corlab}}{0.04 U_\star (M_{\rm BH})} \right)^{-1/4} \left( \frac{C_{\rm d}}{1} \right)^{-1/4} \left( \frac{M_{\rm BH}}{10 M_\odot} \right)^{1/4} \, .
    \label{eq:gradtcor}
\end{align}
Though the complicated geometry precludes determining~$U_{\rm \corlab}$ and~$U_{B, \rm cor}$ precisely, the actual value of~$\gradt^{(\rm \corlab)}$ depends only weakly on these quantities. 

From equations~(\ref{eq:gkncor}) and~(\ref{eq:gradtcor}), one sees that~$\gradt^{(\rm \corlab)}$ is much higher than~$\gamma_{\rm KN,\corlab}$. This opens up the possibility that radiative reconnection operates in the Klein-Nishina regime, since particles are not radiatively inhibited from becoming accelerated beyond~$\gamma_{\rm KN,\corlab}$. Additionally, the radiative particle acceleration limit is not equal to~$\gradt^{(\rm \corlab)}$, but is strongly modified by Klein-Nishina effects to~$\gradk^{(\rm \corlab)} \gg \gradt^{(\rm \corlab)}$ [equation~(\ref{eq:gradkasym})]. For reference, particles at~$\gamma_{\rm rad,T}^{(\rm \corlab)}$ Comptonize disc photons to energy
\begin{align}
    \epsilon_{\rm obs}^{(\rm \corlab)}&\left(\gradt^{(\rm \corlab)}\right) \sim \frac{1}{2} \gradt^{(\rm \corlab)} \me c^2 \notag \\
    &\sim 2 \left( \frac{U_{\rm \corlab}}{0.04 U_\star (M_{\rm BH})} \right)^{-1/4} \left( \frac{C_{\rm d}}{1} \right)^{-1/4} \left( \frac{M_{\rm BH}}{10 M_\odot} \right)^{1/4}  \, \rm GeV \, .
    \label{eq:egratcor}
\end{align}

Next, to fully specify the radiative reconnection regime, we estimate the remaining parameters~$\gamma_{\rm max}$ and~$\sigma_{\rm c,0}$. We assume that the characteristic current sheet length is
\begin{align}
    r_g \simeq 1 \times 10^6 \left( \frac{M_{\rm BH}}{10 M_\odot} \right) \, \rm cm \, .
    \label{eq:lcor}
\end{align}
This gives an extremely large system-size-limited Lorentz factor
\begin{align}
    \gamma_{\rm max} &\equiv \frac{0.1 r_g e B_0}{\me c^2} = \frac{0.1 r_g e}{\me c^2} \sqrt{\frac{8 \pi U_{\rm \corlab}}{C_{\rm d}}} \notag \\
    &\simeq 3 \times 10^{9} \left( \frac{U_{\rm \corlab}}{0.04 U_\star (M_{\rm BH})} \right)^{1/2} \left( \frac{C_{\rm d}}{1} \right)^{-1/2} \left( \frac{M_{\rm BH}}{10 M_\odot} \right)^{1/2} \, .
    \label{eq:gmaxadc}
\end{align}
With~$\gamma_{\rm KN,\corlab}$,~$\gamma_{\rm rad,T}^{(\rm \corlab)}$, and~$\gamma_{\rm max}$ estimated,~$\gamma_{\rm cool,\corlab}$ [equation~(\ref{eq:gcool})] and~$\tau_{\gamma\gamma, \rm \corlab}$ [equation~(\ref{eq:taugg})] follow. Respectively, they are
\begin{align}
    \gamma_{\rm cool,\corlab} &\equiv \frac{3 \me c^2}{4 U_{\rm \corlab} \sigma_{\rm T} r_g} \sim \frac{3}{4} \frac{U_\star}{U_{\rm \corlab}} \frac{\me}{m_{\rm p}} \notag \\
    &\simeq 0.01 \left( \frac{U_{\rm \corlab}}{0.04 U_\star(M_{\rm BH})} \right)^{-1} 
    \label{eq:gcoolcor}
\end{align}
and
\begin{align}
    \tau_{\gamma\gamma, \rm \corlab} \equiv \frac{3 \gamma_{\rm KN,\corlab}}{5 \gamma_{\rm cool,\corlab}}\sim 8 \times 10^{3} \left( \frac{U_{\rm \corlab}}{0.04 U_\star(M_{\rm BH})} \right) \, .
    \label{eq:tauggcor}
\end{align}
Neither~$\gamma_{\rm cool,\corlab}$ nor~$\tau_{\gamma\gamma,\rm\corlab}$ depend on any unknowns besides the ratio~$U_{\rm \corlab} / U_\star$. Unlike in FSRQ jets, the formal~$\gcool$ is now small. Thus, it is not a physical Lorentz factor, but just signals that all particles cool to non-relativistic energies faster than~$L/c$. In other words, the system is highly compact:~$\ell_{\rm cor} \sim 1/\gamma_{\rm cool,cor} \gg 1$. For particles that have cooled to non-relativistic energies, IC scattering reduces to Thomson scattering and the observed photon energy is just~$\epsilon_{\rm obs}^{(\rm cor)}(1) = \epsilon_{\rm cor}$.

We move now to our final estimate: the cold electron magnetization. As in the case of FSRQ jets, we are fairly uncertain of this parameter because we do not know the background coronal electron density~$n_0$. To proceed, let us recast~$n_0$ in terms of the Thomson optical depth~$\tau_{\rm T} = n_0 \sigma_{\rm T} r_g$ along the reconnection layer [equation~(\ref{eq:lcor})]. This coincides with the optical depth of the corona itself assuming the coronal scale height is~$\sim r_g$. The cold \textit{electron} magnetization can then be compactly written in terms of~$\tau_{\rm T}$ and the \textit{magnetic compactness},
\begin{align}
    \ell_{\rm B} &\equiv \frac{U_{\rm B,cor} \sigma_{\rm T} r_g}{\me c^2} = \frac{U_{\rm \corlab} \sigma_{\rm T} r_g}{C_{\rm d} \me c^2} = \frac{U_{\rm \corlab} m_{\rm p}}{C_{\rm d} U_{\star} \me} \notag \\
    &\sim 70 \left( \frac{U_{\rm \corlab}}{U_{\star}(M_{\rm BH})} \right) \left( \frac{C_{\rm d}}{1} \right)^{-1} \, ,
    \label{eq:bcompact}
\end{align}
as
\begin{align}
    \sigc \equiv \frac{2 U_{B, \rm cor}}{n_0 \me c^2} \equiv 2 \frac{\ell_{\rm B}}{\tau_{\rm T}} \, .
    \label{eq:sigcadc}
\end{align}
Thus, our uncertainty of~$n_0$ is shifted onto~$\tau_{\rm T}$. We consider a range of typical values~$10^{-2} \leq \tau_{\rm T} \leq 1$ commonly inferred for coronae of accreting black holes, including in the high/soft state \citep[e.g.][]{el76, pkr97, gzp99, mzb02, gu08, b17}. Then, rewriting the right-hand-side of~(\ref{eq:sigcadc}) as
\begin{align}
    \sigc \equiv 2 \frac{\ell_{\rm B}}{\tau_{\rm T}} = \frac{2 U_{\rm \corlab} m_{\rm p}}{C_{\rm d} U_{\star} \tau_{\rm T} \me} 
    \label{eq:sigcadcalt}
\end{align}
and plugging in~$10^{-2} \leq \tau_{\rm T} \leq 1$ gives
\begin{align}
    1 \times 10^2 \leq \sigc \left( \frac{C_{\rm d}}{1} \right) \left( \frac{U_{\rm \corlab}}{0.04 U_\star(M_{\rm BH})} \right)^{-1} \lesssim  1 \times 10^4 \, .
    \label{eq:sigcval}
\end{align}
The corresponding range in the number density is~$n_0 = \tau_{\rm T} / \sigma_{\rm T} r_g \in [10^{16} \, \mathrm{cm}^{-3}, 10^{18} \, \mathrm{cm}^{-3}]$ \citep[cf.][]{gu08}.

\begin{figure*}
    \centering
    \begin{subfigure}{0.49\textwidth}
        \centering
        \includegraphics[width=\linewidth]{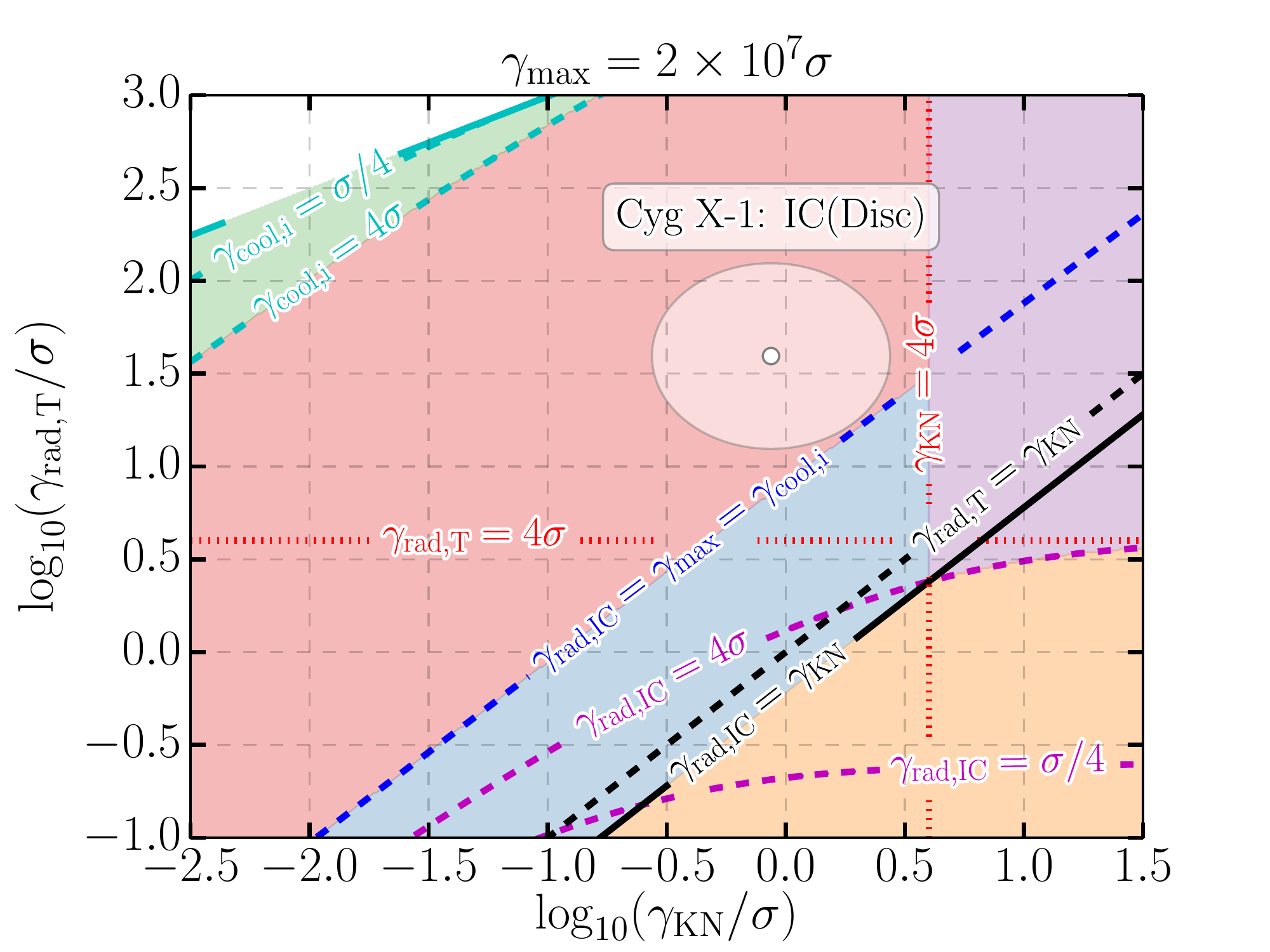}
    \end{subfigure}
    \begin{subfigure}{0.49\textwidth}
        \centering
        \includegraphics[width=\linewidth]{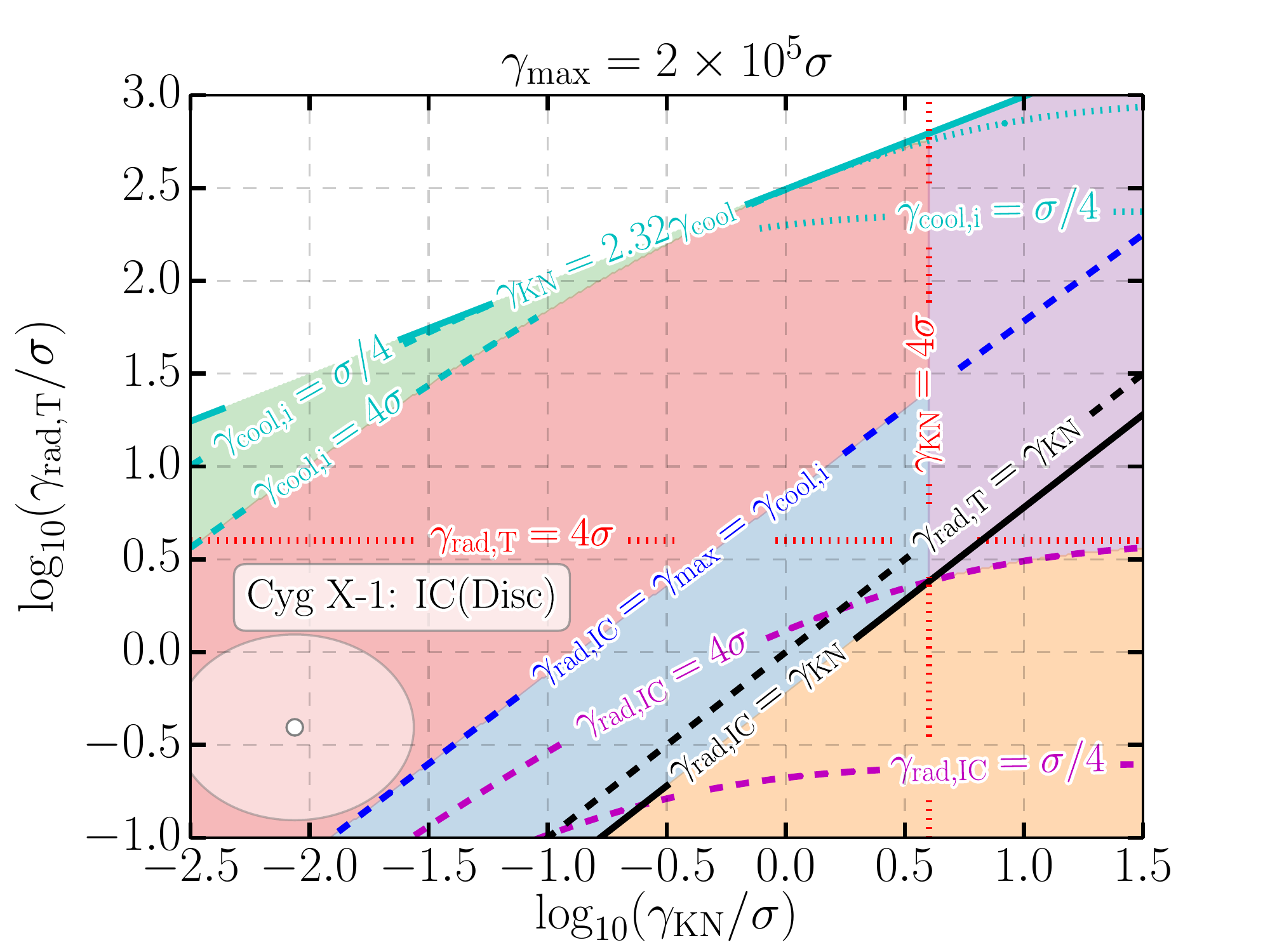}
    \end{subfigure}
\caption{Radiative reconnection phase diagrams for parameters relevant to accreting black holes in their high/soft states (particularly Cyg X-1). Color-coding is the same as in Figs.~\ref{fig:radphasespace} and~\ref{fig:pks1222pspace}. As in those figures,~$\sigc$ is abbreviated as~$\sigma$. White points indicate the parameters~$\gradt^{(\rm \corlab)}$ and~$\gamma_{\rm KN,\corlab}$ estimated in this section; they are surrounded by ellipses that show an arbitrarily chosen uncertainty of one decade in each direction. We display two diagrams corresponding to a range~(\ref{eq:sigcval}) of the cold magnetization~$\sigc$ (based on typical Thomson optical depths~$\tau_{\rm T}$). These are~$\sigc = 10^2$~($\tau_{\rm T} = 1$; left) and~$\sigc = 10^{4}$~($\tau_{\rm T} = 10^{-2}$; right). Throughout the~$\sigc$ range, reconnection operates in a domain strongly impacted by Klein-Nishina effects and pair production.}
    \label{fig:cygx1pspace}
\end{figure*}

We display the estimates in this section in a radiative reconnection phase diagram, as done for FSRQs earlier, in Fig.~\ref{fig:cygx1pspace}. 
This illustrates the main point of all these estimates: \textit{coronal reconnection in the high/soft states of accreting black holes is likely strongly impacted by Klein-Nishina and pair-production physics}. For~$\tau_{\rm T} \in [10^{-2}, 1]$, reconnection occurs in the red region of the phase diagram, where~$\gradk > \gmax > \gcooli{2} > 4 \sigc > \gkn > \gcooli{1}$. Thus, radiative losses do not hinder acceleration near reconnection X-points, but virtually all particles are still cooled strongly (on \ts[s]much shorter than~$L/c$). In addition, a broad distribution of particles develops at energies~$\gg \gkn$. These particles emit above-threshold photons that then produce pairs which may feed back on reconnection. 

We now mention a few subtler details associated with this picture. First, we note that our model of pair-regulated Klein-Nishina reconnection in black hole ADCe is complimentary to the reconnection scenario laid out in the context of the low/hard states of black hole X-ray binaries by \citet{b17}. Pair production also features in that earlier work, but, there, it occurs when two layer-Comptonized photons, both carrying energy~$< \me^2 c^4 / \eph$, collide. In contrast, in this study, a pair is produced when a single Comptonized photon with even higher energy,~$> \me^2 c^4 / \eph$, is absorbed by the seed radiation field. 

Next, we mention the implications of our estimates for the pair multiplicity,~$\eta$. Equations~(\ref{eq:gkncor}) and~(\ref{eq:sigcval}) show that the requirement~$\gamma_{\rm KN, cor} / \sigc \geq 0.1$ for~$\eta$ to be less than unity (section~\ref{sec:upstreammult}) is not always satisfied. Thus, the pairs sourced from gamma-ray absorption by the disc seed photons may sometimes reach a number density comparable to the background upstream plasma. However, a stronger nominal Compton dominance~$C_{\rm d} \equiv U_{\rm \corlab} / U_{B,\rm cor}$ drives down~$\sigc$ [equation~(\ref{eq:sigcval})] and pushes reconnection into the regime where the produced pairs are tenuous.

Lastly, we discuss the implications of the assumption~$C_{\rm d} \sim 1$. As mentioned earlier,~$U_{\rm \corlab}$ and~$U_{B,\rm cor}$ are both set, modulo complicated geometric factors, by the scale~$U_\star$ (or, in our specific estimates, by~$0.04 U_\star$). Thus, it is possible that~$C_{\rm d}$ may be close to unity. However, our model of pair-regulated Klein-Nishina reconnection (section~\ref{sec:knradreconn_mod}) ignores synchrotron losses, and so, in the presence of such modest~$C_{\rm d}$, synchrotron cooling may need to be suppressed in order for the picture of Compton cooling and pair regulation to remain valid. There are several ways for this to occur. Near reconnection X-points, the magnetic field is weaker, and so synchrotron cooling is reduced there (\citealt{u11, cub12}; see discussion of FSRQs). In addition, as discussed in Appendix~\ref{sec:fnoesc}, the angular distribution of particles radiating above-threshold photons from the layer may cause the birth velocities of newborn upstream pairs to be nearly parallel to the reconnection midplane. Because the upstream magnetic field is also parallel to this plane, the pitch angles of such newborn pairs may be small, allowing their cooling to be IC-dominated. Finally, synchrotron self-absorption may inhibit synchrotron cooling among the lower-energy particles \citep[e.g.\ with~$\gamma \lesssim 10$;][]{b17}.

However, if~$C_{\rm d} > 1$, then our estimates numerically change, but the ordering of scales (region of the radiative phase diagram occupied by reconnection) remains essentially the same. For reference,~$\gkn / \sigc \propto C_{\rm d}$;~$\gradt / \sigc \propto C_{\rm d}^{3/4}$; and~$\gmax / \sigc \propto C_{\rm d}^{1/2}$. If, for illustration, one scales up~$C_{\rm d}$ to~$10$, the ordering~$\gradk > \gmax > \gcooli{2} > \gradt > 4 \sigc > \gkn$ is preserved except at the {high-$\sigc$} {(low-$\tau_{\rm T}$)} end of the interval~(\ref{eq:sigcval}). There,~$\gkn$ switches places with~$4 \sigc$, which brings reconnection into the magenta region (e.g. Fig.~\ref{fig:cygx1pspace}) of the phase diagrams where the nominal impulsive X-point acceleration limit,~$4 \sigc$, falls below the minimum energy,~$\gkn$, for Klein-Nishina effects kick in. However, like for FSRQs (section~\ref{sec:fsrqest}), we may infer from observations [equation~(\ref{eq:egkncor})] that particles are accelerated at least up to~$\gkn$, suggesting that, if~$4 \sigc < \gkn$, the nominal~$4 \sigc$ X-point acceleration limit is circumvented and that reconnection may still be strongly influenced by Klein-Nishina radiative physics.

\subsubsection{Relevance to observations of black hole X-ray binaries}
\label{sec:adceobs}
We now describe the implications of our reconnection model for high/soft states of black hole ADCe in accretion-powered X-ray binaries. We are aware of two such systems for which gamma-ray observations have been reported: Cyg X-1 \citep[detected by the \textit{Fermi} LAT,][]{zmc17} and Cyg X-3 (detected by \textit{AGILE}, \citealt{agile09}, and by the \textit{Fermi} LAT, \citealt{fermi09, zmd18}; we adopt the hypothesis that the Cyg X-3 compact object is a black hole).

For Cyg X-1, gamma-ray observations place the high/soft-state photon energy \cutoff[]at~$40-80 \, \rm MeV$ \citep{zmc17}. This is fairly consistent with absorption of gamma-rays propagating through the quasi-thermal radiation field of the disc. Naively, one expects absorption to kick in at around~$\sim (\me c^2)^2 / \epsilon_{\rm \corlab} \sim 300 \, \rm MeV$. However, the quasi-thermal spectrum is not monochromatic, but falls off smoothly at energies beyond its~$1$-keV peak. In fact, the disc dominates the observed spectrum at up to~$3$ or~$4$ keV \citep[e.g.][]{mzb02}, which brings the absorption \cutoff[]into the observed range of~$40-80 \, \rm MeV$. Note that the pair-production optical depth~$\tau_{\gamma\gamma,\rm \corlab}$ [equation~(\ref{eq:tauggcor})] decreases from~$\sim 10^4$ at photon energies beyond the spectral peak, but is still much larger than unity at~$3-4$ keV. Alternatively, the photon energy \cutoff[]could correspond to the \cutoff[]in the distribution of radiating particles.

Unlike Cyg X-1, Cyg X-3 exhibits gamma-ray emission far above its expected absorption \cutoff[.]However, these high-energy gamma-rays are thought to come from Comptonization of the companion star's radiation by jet electrons at much larger distances \citep{dch10, zmd18}. Thus, gamma-ray observations of Cyg X-3 do not provide a strong constraint on the \cutoff[]in the coronal emission, which may be similar to Cyg X-1 and also consistent with pair absorption.

As for their hard X-ray spectra, Cyg~X-1 and Cyg~X-3 both display high/soft states with power-law hard X-ray photon number indices~$\Gamma_{\rm X,HS} \simeq 2.5$ (\citealt{mzb02, zg04, szm08}; here \quoted{HS}stands for \quoted[).]{high/soft}In addition, as detailed by \citet{rm06}, many black hole X-ray binaries are observed with photon indices clustered around this value in their high/soft states (and also in \quoted{steep power-law}states where the observed thermal and \nonthermal[]luminosities are comparable). As discussed for FSRQ jets (section~\ref{sec:fsrqobs}), the relatively steep index~$\pgg[\rm X,HS] \simeq 2.5$ (corresponding to a reconnection-powered injected electron energy distribution power-law index~$p \gtrsim 3$; section~\ref{sec:fsrqobs}) could be associated with swing cycles in pair-regulated Klein-Nishina reconnection. 

Like FSRQs, accreting black holes also exhibit rapid flares. These are observed in the X-rays and have been witnessed on \ts[s]as short as milliseconds \citep{gz03} -- close to the \lc[]time of an orbit at~$r = 10r_g$ from a black hole with mass~$M_{\rm BH} = 10 M_\odot$. Kinetic beaming in the context of Klein-Nishina reconnection may help explain such rapid variability by delivering observed \ts[s]as short as~$0.1L/c$ \citep{cwu13, cwu14a, cwu14b}. However, this has not been explored in detail as in the case of rapid TeV FSRQ flares and requires further attention. 

In summary, pair-regulated Klein-Nishina reconnection in the accretion disc coronae of black hole X-ray binaries may explain the observed steep spectra from these objects, including the gamma-ray \cutoff[](observed in Cyg X-1, \citealt{zmc17}, though this may also result from the \cutoff[]in the distribution of radiating particles). Kinetic beaming facilitated by Klein-Nishina reconnection may additionally explain short flaring \ts[s,]but this requires much more thorough investigation.

\section{Conclusions}
\label{sec:conclusions}
Collisionless astrophysical reconnection is often radiative, with radiative processes not only producing the observed light, but also coupling to the reconnection dynamics. Many interesting systems realize such a regime where an intense but soft radiation field bathes the reconnection region, and reconnection-energized particles Comptonize the ambient photons to observed X- and gamma-ray energies.

Sometimes, IC losses proceed purely in the Thomson limit and the Comptonized photons freely stream out of the system -- a regime we discuss in section~\ref{sec:thomradreconn}. Even then, however, radiation is far from passive. The emitting particles may still lose a significant amount of their energy on sub-dynamical \ts[s,]and this has the potential to modify both reconnection itself and the qualitative features of the received light \citep[e.g.][]{u16, b17, wpu19, sb20, mwu20}. 

However, there are some astrophysical systems -- such as FSRQ blazar jets and the coronae of accreting black holes -- where the physical picture is even richer. In these cases, Comptonization of ambient photons can enter the Klein-Nishina regime, and this both qualitatively changes the physics of radiative cooling and opens up an entirely new channel for radiative feedback on reconnection: pair production. By incorporating Klein-Nishina and pair-production physics into the basic conceptual framework of radiative reconnection (section~\ref{sec:knradreconn}), we find a fundamentally new, self-regulated, Klein-Nishina reconnection regime (section~\ref{sec:knradreconn_mod}) that stems from a negative feedback loop (Fig.~\ref{fig:knppdiagram}). In this loop, gamma-ray photons Comptonized in the reconnection layer propagate into the upstream plasma where they produce hot pairs by colliding with soft (unscattered) ambient photons, thus lowering the effective upstream magnetization. This inhibits further particle acceleration in the reconnection layer, closing the feedback loop. 

The pair-regulation mechanism gives rise to an effective fixed-point upstream magnetization~$\sighgen$ that is \textit{universal} -- independent of the initial value~$\sigh$. However, the fixed point is not necessarily reached by the system. If pair feedback is efficient enough, the system overshoots its natural solution,~$\sighgen$, by a wide margin and enters a limit (or \quoted[)]{swing}cycle, constantly oscillating between a high- and a low-magnetization state.

In section~\ref{sec:knradreconn_mod}, we also \analyz[e]the number of particles produced in the upstream region. A pair cascade, in which the population in each subsequent newborn generation grows exponentially, is not generally expected (except for unrealistically optimal reconnection-driven NTPA). Even in the presence of a cascade, there is a wide parameter range where the total created pair count is dwarfed by that of the pairs originally present. Thus, remarkably, though the newborn pairs are so hot that they can dominate the upstream energy budget, they are often also so tenuous that they contribute negligibly to the upstream lepton density. This feature, in particular, distinguishes Klein-Nishina radiative reconnection from other pair-regulated regimes \citep{l96, hps19}.

Finally, in section~\ref{sec:applications}, we discuss the observable aspects of pair-regulated Klein-Nishina reconnection. We expect that emission from the layer (i.e.\ the region threaded with reconnected flux) dominates that from the newborn component of the upstream plasma (section~\ref{sec:layerdominance}). We also (section~\ref{sec:kbprospects}) identify several reasons to expect that Klein-Nishina physics may promote kinetic beaming, which is an important mechanism for facilitating rapid flaring variability in reconnection \citep{cwu12, cwu13, cwu14b, cwu14a, nbc12, mwu20}. Furthermore, we explicitly estimate the parameters (i.e.\ the energy scales introduced in sections~\ref{sec:thomradreconn} and~\ref{sec:knradreconn}) governing the radiative regime of reconnection in FSRQ blazar jets (sections~{\ref{sec:fsrqenvs}-\ref{sec:fsrqobs}}) and ADCe of black hole X-ray binaries (sections~{\ref{sec:adceparams}-\ref{sec:adceobs}}). Reconnection in both types of objects is very likely to be strongly impacted by Klein-Nishina physics, and may even enter the pair-regulated regime modeled in this paper. In fact, the generally steep \nonthermal[]spectra produced by FSRQ jets and black hole ADCe appear consistent with strong pair loading in our model, and the observed instances of rapid variability could be caused by the kinetic beaming mechanism. Pair-regulated reconnection may further provide a source for \textit{in situ} pair production in FSRQ jets, where the lepton content is difficult to constrain observationally.

However, many of our observational remarks, as well as the basic physical features of our model, require additional testing through first-principles radiative PIC simulations, which can address some of the following key physical and observational questions:
\begin{itemize}
    \item How robust is the pair feedback mechanism? In particular,
        \begin{itemize}
            \item Is the universal magnetization solution ever realized?
            \item Can an order-unity energy recapture efficiency~$\xi$ be achieved?
            \item Is the feedback strong enough to induce late-time limit cycles that jump between high- and low-magnetization states?
        \end{itemize}
    \item How many pairs are produced? Are they indeed few compared to the number of background particles, and does their number density scale with the reconnection parameters as predicted by our model?
    \item Are the emitted spectra produced by reconnection in the steady state (or in a limit cycle) consistent with our predictions? Are they consistent with observations?
    \item Can kinetic beaming operate in Klein-Nishina reconnection as we argue? If so, is beaming more or less prominent than in Thomson radiative reconnection \citep[as diagnosed by][]{mwu20}?
    \item To what extent does the physical picture change in 3D, in the presence of ions, with a finite guide field, and (especially) with finite synchrotron losses?
\end{itemize}

In addition to addressing these immediate follow-up questions, it would be interesting to \generaliz[e]our theory in several additional ways. For example, one might consider a broader (e.g.~power-law) ambient photon spectrum (here we only treat monochromatic ambient seed photons). Beyond shedding light on a broader class of astrophysical systems, such an exercise steps toward a theory of synchrotron self-Compton reconnection, where the dominant population of seed photons for Comptonization is the synchrotron photons produced inside the reconnection system. This, in turn, would be very important for understanding reconnection-powered emission from BL Lac blazars \citep{ms16}. As a completely separate extension of this work, kinetic simulations mimicking global jet-like geometries \citep[e.g.][]{azf18, srn21} and also incorporating Klein-Nishina and pair-production physics would determine whether and how this type of reconnection occurs in an explicitly global context. The same could be said for accretion-like numerical setups \citep[e.g.][]{ccd20}. 

Overall, this study poses a rich set of important questions for future inquiry. In the short term, direct numerical tests of our model will facilitate more straightforward translation to observations. More remotely, global simulations may reveal explicitly how reconnection as modeled here might occur in nature. Finally, the novel radiative physics treated in this work \citep[but see also][]{l96, b17, wpu19, sgu19, hps19, sb20} may help to pave the way to theories of increasingly realistic reconnection regimes with applications to the high-energy universe.

\section*{Acknowledgements}
The authors express their gratitude to Beno\^{i}t Cerutti for fruitful discussions. This work is supported by NASA and the NSF, grant numbers NASA ATP NNX16AB28G, NASA ATP NNX17AK57G, NASA ATP 80NSSC20K0545, NSF AST 1411879, and NSF AST 1903335.

\section*{Data Availability}
No new data were generated or analysed in support of this research.

\bibliographystyle{mnras}
\bibliography{ref}

\appendix
\section{The Klein-Nishina scattering power and rate}
\label{sec:knfuncs}
Here, we sketch the derivations of the functions~$f_{\rm KN}(q)$ and~$g_{\rm KN}(q)$ as defined in equations~(\ref{eq:fkn}) and~(\ref{eq:gkn}). \citet{j68} and \citet{bg70} report the Klein-Nishina scattering kernel
\begin{align}
    \frac{\dif N}{\dif t \dif r} &\dif r = c \sigma_{\rm T} \frac{\uph}{\eph} \notag \\
    &\times \frac{3}{(1 + \knp r)^2} \left[ 2 r \ln r + (1 + 2 r)(1 - r) + \frac{1}{2} \frac{(\knp r)^2}{1 + \knp r}(1 - r) \right] \dif r \notag \\
    &\equiv c \sigma_{\rm T} \frac{\uph}{\eph} K(r, \knp) \dif r \, .
    \label{eq:scatkern}
\end{align}
This is the number of photons scattered per unit time by a particle with Klein-Nishina parameter~$\knp = \gamma / \gkn = 4 \gamma \eph / \me c^2$ to final photon energies between~$r$ and~$r + \dif r$. The parameter~$r$ is not actually equal to the scattered photon energy~$\epsilon_1$ but is defined in terms of it through
\begin{align}
    r \equiv \frac{\epsilon_1 / \gamma \me c^2}{\knp \left( 1 - \epsilon_1 / \gamma \me c^2 \right)} \, .
    \label{eq:rdef}
\end{align}
Equivalently,
\begin{align}
    \epsilon_1 = \gamma \me c^2 \frac{\knp r}{1 + \knp r} \, .
    \label{eq:scatenvsr}
\end{align}
It is convenient to take integrals over~$r$ as a proxy for~$\epsilon_1$ because, as~$\epsilon_1$ spans its kinematically allowed range~$[\eph, \gamma \me c^2 \knp / (1 + \knp)]$,~$r$ runs approximately from~$0$ to~$1$ \citep{j68, bg70}.

For completeness, we mention the validity conditions for the kernel~(\ref{eq:scatkern}):~$\gamma \me c^2 / \eph = 4 \gkn \gamma \gg 1$ and~$\gamma \gg 1$ \citep{j68}. The first condition is always satisfied for~$\gkn \gg 1$, which is true of every astrophysical system to which we apply our results (and also of many others). The relativistic requirement~$\gamma \gg 1$ is then taken care of because the general Klein-Nishina expressions are only needed for~$\gamma \gg \gkn \gg 1$. At non-relativistic energies~$\gamma \simeq 1$, the Thomson prefactors in equations~(\ref{eq:pic}) and~(\ref{eq:ric}) are all that remain~[$f_{\rm KN}(\knp \ll 1) \to 1$ and~$g_{\rm KN}(\knp \ll 1) \to 1$], and these are non-relativistically correct. Thus, for~$\gkn \gg 1$, equations~(\ref{eq:pic}) and~(\ref{eq:ric}) are correct even in the non-relativistic case. 

To obtain the rates at which a scattering particle loses energy [equation~(\ref{eq:pic})] and encounters soft seed photons [equation~(\ref{eq:ric})] requires integrating over~$\epsilon_1$ -- or, equivalently, over~$r$ -- while keeping~$\gamma$,~$\eph$, and, hence,~$\knp = \gamma / \gkn$ fixed. One obtains
\begin{align}
    \begin{bmatrix}
        R_{\rm IC}(\gamma) \\
        P_{\rm IC}(\gamma)
    \end{bmatrix}
    &= c \sigma_{\rm T} \frac{\uph}{\eph} \int_0^1 \dif r \, K(r, \knp)
    \begin{bmatrix}
        1 \\
        \epsilon_1(r)
    \end{bmatrix} \notag \\
    &=
    \begin{bmatrix}
        c \sigma_{\rm T} \uph / \eph \\
        (4/3) c \sigma_{\rm T} \gamma^2 \uph
    \end{bmatrix}
    \int_0^1 \dif r \, K(r, \knp)
    \begin{bmatrix}
        1 \\
        3 r / (1 + \knp r)
    \end{bmatrix} \, .
    \label{eq:knfuncresult}
\end{align}
Then, using~$\beta^2 = 1 - 1/\gamma^2 \simeq 1$ to write~$P_{\rm T}(\gamma) \simeq (4/3) c \sigma_{\rm T} \gamma^2 \uph$, one can read off~$g_{\rm KN}(\knp)$ and~$f_{\rm KN}(\knp)$ from equation~(\ref{eq:knfuncresult}):
\begin{align}
    g_{\rm KN}(\knp) = \int_0^1 \dif r \, K(r, \knp)
    \label{eq:gknint}
\end{align}
and
\begin{align}
    f_{\rm KN}(\knp) = 3 \int_0^1 \dif r \, K(r, \knp) \frac{r}{1 + \knp r} \, .
    \label{eq:fknint}
\end{align}
The integrals can be evaluated to yield equations~(\ref{eq:fkn}) and~(\ref{eq:gkn}). Should the reader wish to verify by explicit computation, we find that the identity
\begin{align}
    \Li_2 \left( \frac{1}{1+\knp} \right) = \Li_2 (-\knp) + \frac{\pi^2}{6} + \frac{1}{2} \log^2(\knp) - \frac{1}{2} \log^2 \left( \frac{\knp}{1 + \knp} \right)
    \label{eq:dilogid}
\end{align}
is useful.

\section{The energy retention factor~$\fnc$}
\label{sec:feedbackdetails}
Recall that in section~\ref{sec:upstreamendens}, we left~$\xi$ as a free parameter. This stemmed chiefly from our uncertainty regarding the fraction of energy~$\fnc$ retained by the fresh pair plasma as it travels toward the reconnection layer. In order to demonstrate the effect of this parameter without knowing it precisely, we entertained a class of models where~$\xi$ (specifically $f_{\rm nocool}$) was constant: independent of NTPA in the layer [i.e.\ of~$p(\sighgen)$ and~$z$]. However, in principle, we can explicitly compute the distributions~$\Ngg[n](\gamma)$ of newborn pairs (by straightforwardly extending the analysis of section~\ref{sec:nocascades}), and, from them, calculate the energy carried by fresh plasma reaching the layer. This can then be compared to the energy \textit{deposited} into the upstream region to yield~$\fnc$.

In this section, we adopt this strategy to calculate~$\xi = \fnc$. We find that, like~$\mathcal{F}$,~$\xi$ is a function of~$p(\sighgen)$ and~$z$:~$\xi = \Xi[p(\sighgen), z]$. Furthermore, the basic intuition gleaned in section~\ref{sec:upstreamendens} remains intact: extremely efficient energy delivery to the layer,~$\xi \sim 1$, may still push the system into a 2-state swing cycle, and radiative feedback on NTPA still makes the system more prone to these oscillations. Through the present analysis, we merely gain a more precise notion of the values of physical parameters -- particularly the \cutoff[~$z$]in the distribution of layer particles -- required to initiate swing cycles versus those that cause rapid progression toward the fixed point~$\sighgen$.

Our first step is to write down a quantitative expression for~$\fnc$. In order to do this, we define several auxiliary quantities: the~$n$th-generation steady-state particle count
\begin{align}
    \Ngen[n] \equiv \int \Ngg[n](\gamma) \dif \gamma \, ,
    \label{eq:Ngendef}
\end{align}
which we distinguish from~$\Ngg[n](\gamma)$ by omitting the functional argument; the~$n$th-generation particle injection rate
\begin{align}
    \Qinj[n] \equiv \int \Qgg[n](\gamma) \dif \gamma \, ,
    \label{eq:qinjdef}
\end{align}
also distinguished from~$\Qgg[n](\gamma)$ by argument omission;
the average particle energy \textit{injected} into the~$n$th generation
\begin{align}
    \avgenninj[n] \equiv \frac{\int \gamma \Qgg[n](\gamma) \dif \gamma}{\Qinj[n]} \, ;
    \label{eq:avgenninjdef}
\end{align}
and the average particle energy \textit{carried} by the~$n$th generation in its steady state
\begin{align}
    \avgenn[n] \equiv \frac{\int \gamma \Ngg[n](\gamma) \dif \gamma}{\Ngen[n]} \, . 
    \label{eq:avgenndef}
\end{align}
In terms of these quantities, one has
\begin{align}
    \fnc &\equiv \frac{\mathrm{newborn\, pair\, energy\, carried\, into\, layer}}{\mathrm{energy\, injected\, upstream}} \notag \\
    &= \frac{\sum_{n = 1}^{\ngen} \Ngen[n] \avgenn[n]}{\sum_{n = 1}^{\ngen} \Ngen[n] \avgenninj[n]} \notag \\
    &= \frac{\sum_{n = 1}^{\ngen} \Ngen[n] \avgenninj[n] \fncn[n]}{\sum_{n = 1}^{\ngen} \Ngen[n] \avgenninj[n]}  \, ,
    \label{eq:fncdef}
\end{align}
where
\begin{align}
    \fncn[n] \equiv \frac{\Ngen[n] \avgenn[n]}{\Ngen[n] \avgenninj[n]} = \frac{\avgenn[n]}{\avgenninj[n]}
    \label{eq:fncndef}
\end{align}
is the average fractional energy retained by the~$n$th generation.

Before we calculate~$\fnc$ in detail using equation~(\ref{eq:fncdef}), let us briefly argue why, as claimed in section~\ref{sec:upstreamendens},~$\fnc$ should be physically confined to the range~$[3/400, 1]$. The upper bound is trivial:~$\fnc$ cannot exceed one by definition. Let us now see how the lower bound arises.

In the case where~$\fnc$ is as small as possible, all particles born into the upstream region are quite energetic, with~$\gkn \ll \avgenninj[n] \leq \gcti{2}$, but they cool quickly -- until their cooling times match their readvection time~$\tread$, and hence until their Lorentz factors equal~$\gcti{1}$ [equation~(\ref{eq:gct1})]. However, when these particles' energies exceed~$\gth$, none of their emitted photons escape the system. Rather, the power~$\Qinj[1] \avgenninj[1] \me c^2$ injected into the upstream region losslessly converts to secondary pairs, and therefore matches the steady-state energy flux~$\injsym' \gth \me c^2$ into the particle energy bin~$\gamma < \gth < \gamma + \dif \gamma$. Here,~$\injsym'$ is the (generationally summed) particle flux into the same bin. Only once particles cool past~$\gth$ does their emission leak out of the system. Thus, if the injected particles have high initial energies but all cool down to~$\gcti{1}$,~$\fnc$ reaches the (lowest possible) value
\begin{align}
    \fncmin &\equiv 1 - \frac{\mathrm{power\, radiated\, by\, upstream\, particles}}{\mathrm{power\, injected\, upstream}} \notag \\
        &= 1 - \frac{\injsym' (\gth - \gcti{1})}{\Qinj[1] \avgenninj[n]} \notag \\
        &= 1 - \frac{\gth - \gcti{1}}{\gth} \notag \\
        &= \frac{\gcti{1}}{\gth} = \frac{1}{8} \frac{3}{50} = \frac{3}{400} \, .
    \label{eq:fncmin}
\end{align}

The reason that (as we show quantitatively in this section)~$\fncmin = 3/400$ is an overly pessimistic estimate for~$\fnc$ is that, in reality, upstream particles are constantly being replenished at high energies even as they rapidly cool toward~$\gcti{1}$. This results in a pronounced high-energy tail of hot upstream particles entering the reconnection layer, greatly enhancing~$\fnc$, even, in some cases, to order unity.

\subsection{The steady-state pair distributions~$\Ngg[n](\gamma)$}
Evaluating equation~(\ref{eq:fncdef}) for~$\fnc$ explicitly requires the steady-state pair distributions~$\Ngg[n](\gamma)$. We now retrieve these distributions by extending the analysis of section~\ref{sec:nocascades}. Rearranging equation~(\ref{eq:qggsteady}), one may write
\begin{align}
    \Ngg[n](\gamma) = \frac{1}{8} \tcoolk(\gamma) \Qgg[n+1](\gamma/4) \, .
    \label{eq:nggsteady}
\end{align}
Plugging in~$\tcoolk(\gamma) \propto [ \gamma f_{\rm KN}(\gamma / \gkn) ]^{-1}$ and using the approximate expression~$f_{\rm KN}(\knp) \simeq (1 + \knp)^{-1.5}$ from equation~(\ref{eq:fknapprox}), one sees that~$\Ngg[n](\gamma) \propto \gamma^{-\pgg + 0.5}$ provided~$\Qgg[n] \propto \gamma^{-\pgg}$ [i.e.\ as in~(\ref{eq:qggndef})] and~$\gamma \gg \gkn$. The latter condition is satisfied because the deep Klein-Nishina regime is assumed in writing equations~(\ref{eq:fppairskn}) and~(\ref{eq:fpphotonskn}), from which~(\ref{eq:nggsteady}) follows.

However, this is not the full story for~$\Ngg[n](\gamma)$. Even though~$\Qgg[n](\gamma)$ is zero when~$\gamma < \gamma_{\rm min} \equiv 2 \gkn$, which comes from the fact that no pairs are injected at energies below~$\gamma_{\rm min}$,~$\Ngg[n](\gamma)$ is \textit{not} zero at these low energies. Instead, particles in every generation cool continuously -- in the Thomson regime -- once they reach energies~$\sim \gamma_{\rm KN} = \gamma_{\rm min} / 2$, populating a low-energy component of each distribution~$\Ngg[n](\gamma)$.

To model this situation, we assume that, in each generation, a constant flux of particles leaks from above to below~$\gkn$ in energy space [through the advective term in equation~(\ref{eq:fppairskn}), which remains nonzero in the Thomson regime even though~$\Qgg[n](\gamma)$ vanishes]. Once there, the rate of change of each particle's energy is approximately Thomson:~$-\dot{\gamma} \simeq -\dot{\gamma}_{\rm T} \equiv \gamma / \tcoolt(\gamma)$. To determine~$\Ngg[n](\gamma)$ in the Thomson limit~$\gamma < \gkn$ then requires solving a simplified version of equation~(\ref{eq:fppairskn}),
\begin{align}
    \frac{\partial}{\partial t} \Ngg[n](\gamma, t) + \frac{\partial}{\partial \gamma} \left( \dot{\gamma}_{\rm T} \Ngg[n](\gamma, t) \right) = \Qinj[n] \delta(\gamma - \gkn) \, ,
    \label{eq:fppairst}
\end{align}
where the source term~$\Qinj[n] \delta(\gamma - \gkn) = \delta(\gamma - \gkn) \int \Qgg[n](\gamma) \dif \gamma$ gives the flux into~$\gkn$ of particles from higher energies.\footnote{Strictly speaking, the first generation of pairs~$\Qgg[1](\gamma)$ may have particles at energies between~$\gcti{2}$ and~$\gfn[1] = \gf / 4$ if~$\gf > 4 \gcti{2}$. In that case, taking~$\Qinj[n] = \int \Qgg[n](\gamma) \dif \gamma$, and not cutting off the integral at~$\gcti{2}$, overestimates the number of particles in the low-energy Thomson regime, artificially reducing our estimate of~$\avgenn[n]$ and, hence, of~$\fncn[n]$.} 
The solution to~(\ref{eq:fppairst}) can be obtained exactly, and the steps are detailed in Appendix~\ref{sec:fpexactsoln}. The result is a low-energy distribution of particles arriving at the layer in each generation of
\begin{align}
    \Ngg[n](\gamma \lesssim \gkn) = \Qinj[n] \gcool \frac{L}{c} \begin{cases}
        \gamma^{-2} & \gamma \geq \gcti{1} \\
        0 & \gamma < \gcti{1} 
    \end{cases} \, .
    \label{eq:nggsteadyt}
\end{align}
Equivalently, one can use~$\gcool = (10 / \tau_{\gamma\gamma}) \gcti{1} = (10 / \tau_{\gamma\gamma}) (3/50) \gkn$ to express this as
\begin{align}
    \Ngg[n](\gamma \lesssim \gkn) = \Qinj[n] \frac{3}{50} \frac{10 L}{\tau_{\gamma\gamma} c} \begin{cases}
        \gkn / \gamma^{2} & \gamma \geq \gcti{1} \\
        0 & \gamma < \gcti{1} 
    \end{cases} \, ,
    \label{eq:nggsteadyt2}
\end{align}
which will be a slightly more useful form later.
For reference, the average energy of this distribution is
\begin{align}
    \frac{\int_1^{\gamma_{\rm min}} \gamma \Ngg[n](\gamma) \dif \gamma}{\int_1^{\gamma_{\rm min}} \Ngg[n](\gamma) \dif \gamma} &= \gcti{1} \ln \left( \frac{2 \gkn}{\gcti{1}} \right) \notag \\
    &\simeq 3.5 \gcti{1} \, ,
    \label{eq:nggavgglo}
\end{align}
where the integrals are taken through~$\gamma_{\rm min} = 2 \gkn$ because, as detailed below, this is where we match the Thomson solution~(\ref{eq:nggsteadyt2}) to the one we will obtain in the deep Klein-Nishina regime.

Next, we determine the high-energy~($\gamma > \gkn$) part of the steady-state distributions~$\Ngg[n](\gamma)$. We note that~$\Qinj[n]$ can be written [cf.\ equation~(\ref{eq:totalinject})] as
\begin{align}
    \Qinj[n] \simeq \Qnorm[n] \frac{\gamma_{\rm min}^{-\pgg + 1}}{\pgg - 1} \left[ 1 - \left( \frac{\gamma_{\rm min}}{\gfn[n]} \right)^{\pgg - 1} \right] \simeq \Qnorm[n] \frac{\gamma_{\rm min}^{-\pgg + 1}}{\pgg - 1} \, .
    \label{eq:qinjexplicit}
\end{align}
Plugging this into the right-hand-side of~(\ref{eq:nggsteady}) gives
\begin{align}
    \Ngg[n](&\gamma \gtrsim \gkn) = \frac{1}{8} \left[ \tcoolk(\gamma) \right] \left[ \Qgg[n+1](\gamma/4) \right] \notag \\
    &= \frac{1}{8} \left[ \frac{3}{50} \frac{10 L}{\tau_{\gamma\gamma} c} \frac{1}{q f_{\rm KN}(q)} \right] \left[ \Qnorm[n+1] \left( \frac{\gamma}{4} \right)^{-\pgg} \right] \notag \\
    &= \frac{1}{8} \left[ \frac{3}{50} \frac{10 L}{\tau_{\gamma\gamma} c} \frac{1}{q f_{\rm KN}(q)} \right] \left[ \mathcal{A}(\pgg) \Qnorm[n] 4^{\pgg} \gamma^{-\pgg} \right] \notag \\
    &\simeq \frac{1}{8} \left[ \frac{3}{50} \frac{10 L}{\tau_{\gamma\gamma} c} \frac{1}{q f_{\rm KN}(q)} \right] \left[ \mathcal{A}(\pgg) \Qinj[n] 4^{\pgg} (\pgg - 1) \gamma_{\rm min}^{-1} \left( \frac{\gamma}{\gamma_{\rm min}} \right)^{-\pgg} \right] \notag \\
    &= \frac{1}{2 \gkn} \Qinj[n] \left[ \frac{3}{50} \frac{10 L}{\tau_{\gamma\gamma} c} \frac{1}{q f_{\rm KN}(q)} \right] \left( \frac{\gamma}{\gamma_{\rm min}} \right)^{-\pgg} \, .
    \label{eq:nggsteadyexplicit}
\end{align}
The third line follows from~(\ref{eq:qnormrecurse}), the fourth line from~(\ref{eq:qinjexplicit}), and the last line from~(\ref{eq:scripta}) along with~$\gamma_{\rm min} = 2 \gkn$. Rigorously matching the low-energy Thomson solution~(\ref{eq:nggsteadyt2}) to the high-energy Klein-Nishina result~(\ref{eq:nggsteadyexplicit}) requires a detailed analysis of the radiative physics near~$\gamma \sim \gkn$. We expect this, at most, to modify~(\ref{eq:nggsteadyt2}) and~(\ref{eq:nggsteadyexplicit}) by order-unity factors near~$\gamma \sim \gkn$, since the assumptions from which these solutions are derived are fairly robust in their respective limits (far from~$\gkn$). Moreover, the solutions are already of similar scale at a natural matching point,~$\gamma = \gamma_{\rm min} = 2 \gkn$, with equation~(\ref{eq:nggsteadyexplicit}) a factor of about~$1/f_{\rm KN}(2) \sim 5$ larger than~(\ref{eq:nggsteadyt2}) when both are evaluated at~$\gamma_{\rm min}$. This is independent of~$\pgg$. Thus, as a rough estimate, we take the overall distribution of~$n$th-generation pairs (including both~$\gamma > \gkn$ and~$\gamma < \gkn$) to be
\begin{align}
    \Ngg[n](\gamma) &\simeq \Qinj[n] \frac{3}{50} \frac{10 L}{\tau_{\gamma\gamma} c} \notag \\
    &\times \begin{cases}
        \gamma_{\rm min} / 2 \gamma^2 & \gcti{1} \leq \gamma < \gamma_{\rm min} \\
        [ \gamma_{\rm min} q f_{\rm KN}(q)]^{-1} ( \gamma / \gamma_{\rm min} )^{-\pgg} & \gamma_{\rm min} \leq \gamma < \gfn[n] \\
        0 & \mathrm{otherwise}
    \end{cases} \, .
    \label{eq:nggsteadyfull}
\end{align}
If one leverages the fact that~$q > 2$ when~$\gamma > \gamma_{\rm min}$ to approximate~$q f_{\rm KN}(q) \simeq q^{-0.5}$, equation~(\ref{eq:nggsteadyfull}) becomes
\begin{align}
    \Ngg[n](\gamma) &\simeq \Qinj[n] \frac{3}{50} \frac{10 L}{\tau_{\gamma\gamma} c} \notag \\
    &\times \begin{cases}
        \gamma_{\rm min} / 2 \gamma^2 & \gcti{1} \leq \gamma < \gamma_{\rm min} \\
        (\sqrt{2} / \gamma_{\rm min}) ( \gamma / \gamma_{\rm min} )^{-\pgg + 0.5} & \gamma_{\rm min} \leq \gamma < \gfn[n] \\
        0 & \mathrm{otherwise}
    \end{cases} \, .
    \label{eq:nggsteadyapprox}
\end{align}

\subsection{Evaluating~$\fnc$}
Armed with the pair distributions~$\Ngg[n](\gamma)$, and knowing, from section~\ref{sec:nocascades}, the injected distributions~$\Qgg[n](\gamma)$, we can now explicitly evaluate equation~(\ref{eq:fncdef}) to obtain~$\fnc$.
Assembling all of the ingredients -- equations~(\ref{eq:qggndef}),~{(\ref{eq:Ngendef})-(\ref{eq:avgenndef})},~(\ref{eq:qinjexplicit}), and~(\ref{eq:nggsteadyapprox}) -- we have
\begin{align}
    \frac{\avgenninj[n]}{\gth} &= \frac{1}{4} \frac{\pgg - 1}{\pgg - 2} \left[ \frac{1 - \left( z / 4^{n+1/2} \right)^{2 - \pgg}}{1 - \left( z / 4^{n+1/2} \right)^{1 - \pgg}} \right] \, ,
    \label{eq:avgenninjeval} \\
    \Ngen[n] &= \Qinj[n] \frac{10 L}{\tau_{\gamma\gamma} c} \left\{ 1 + \sqrt{2} \frac{3 / 50}{\pgg - 1.5} \left[ 1 - \left( \frac{z}{4^{n+1/2}} \right)^{1.5 - \pgg} \right] \right\} \, ,
    \label{eq:Ngeneval}
\end{align}
and
\begin{align}
    \frac{\avgenn[n]}{\gth} &= \frac{3}{400} \left\{ 3.5 + \frac{ 2 \sqrt{2}}{2.5 - \pgg} \left[ \left( \frac{z}{4^{n + 1/2}} \right)^{2.5 - \pgg} - 1 \right] \right\} \notag \\
    &\times \left\{ 1 + \sqrt{2} \frac{3/50}{\pgg - 1.5} \left[ 1 - \left( \frac{z}{4^{n+1/2}} \right)^{1.5 - \pgg} \right] \right\}^{-1} \, .
    \label{eq:avgenneval}
\end{align}
Some special numbers that appear in these expressions are:~$3.5 \simeq \ln(2 \times 3 / 50) = \ln(\gamma_{\rm min} / \gcti{1})$ in the numerator of~(\ref{eq:avgenneval}) [cf.\ equation~(\ref{eq:nggavgglo})];~$1/4 \equiv \gamma_{\rm min} / \gth$, the prefactor of~(\ref{eq:avgenninjeval}); and~$3/400 \equiv \gcti{1} / \gth \equiv \fncmin$, the prefactor of~(\ref{eq:avgenneval}). In writing these expressions, we assume that the layer particle distribution cuts off at~$\gf \leq 4 \gcti{2}$ and, therefore, that we are justified to replace~$\gfn[n] / \gamma_{\rm min}$ with~$\gf / 2 \gkn 4^{n} = z / 4^{n+1/2}$, as we have done.

Some important features and limits of these formulae are as follows. First, noting that~$\pgg \simeq \pk + 1 \simeq p + 0.5$, we see that~$\pgg \in [1.5, 3.5]$. Thus, the second term in the braces of equation~(\ref{eq:Ngeneval}) is almost always small, and, generally,~$\Ngen[n] \sim \Qinj[n] 10 L / \tau_{\gamma\gamma c}$: the number of particles in the steady state is dominated by those in the Thomson regime. Next, when~$\pgg$ is on the softer side~($\pgg \gtrsim 2.5$), all expressions become virtually~$z$-independent (unless~$z \sim 4^{n+1/2}$). This is because the distributions of pairs are so steep that all quantities are dominated by the low energies [near~$\gamma_{\rm min}$ for the~$\Qgg[n](\gamma)$'s and near~$\gcti{1}$ for the~$\Ngg[n](\gamma)$'s]. In this limit, the weights~$\avgenninj[n] \Ngen[n] / \sum_k \avgenninj[k] \Ngen[k]$ used to average the~$\fncn[n]$'s in~(\ref{eq:fncdef}) exhibit a very simple~$n$-dependence:~$\avgenninj[n] \Ngen[n] \propto \Qinj[n] \propto \mathcal{A}(\pgg)^n \ll 1$. Thus,~$\fnc$ is dominated by the first generation, tending to
\begin{align}
    \lim_{\pgg \to 3.5} \fnc &= \lim_{\pgg \to 3.5} \fncn[1] = \lim_{\pgg \to 3.5} \frac{\avgenn[n]}{\avgenninj[n]} \notag \\
    &= \frac{3}{400} \frac{3.5 + 2 \sqrt{2} }{(3.5-1)/4(3.5-2)} \simeq 0.1 \, .
    \label{eq:fnclim}
\end{align}
This is much higher than our pessimistic estimate~$\fncmin = 3/400$, and owes to the extended nature of the power-law distribution of particles. Though dominated by the low energies, this distribution still carries an average particle energy that is a factor of several higher than~$\gcti{1}$.

Finally, we note that~$\avgenn[n]$ becomes highly~$z$-dependent when~$\pgg < 2.5$, and so does~$\avgenninj[n]$ when~$\pgg < 2$. As the result of this, the highest values of~$z$ that we study -- all the way up to~$z = 4 \gcti{2} / \gkn \simeq 1100$ -- produce~$\fnc$ of order unity (between~$0.4$ and~$0.5$). Thus, in our quantitative framework,~$\fnc$ is rather weakly dependent on the physical parameters~$z$ and~$p$, only varying between about~$0.1$ and~$0.5$ over the broad parameter space,~$(p, z) \in [1, 3] \times [8, 1100]$. However, one should bear in mind that we have made many simplifying assumptions in our model, and so a wider range of~$\fnc$ may be possible in reality. Thus we would like to stress the overall qualitative insight -- that~$\fnc$ is never really too small (always at least several per cent) and may potentially reach order unity -- more than our exact quantitative values.
So, even though we now repeat our analysis from section~\ref{sec:upstreamendens} using our functional form for~$\fnc$ (via those for~$\avgenn[n]$,~$\avgenninj[n]$ and~$\Ngen[n]$), we aim to stress general features, showing how the (now somewhat more self-consistent) dynamics predicted by our model depend in a very qualitative sense on~$z$ and~$p$.

\subsection{Reconnection dynamics with self-consistent~$\fnc$}
\begin{figure*}
    \centering
    \begin{subfigure}{0.49\textwidth}
        \centering
        \includegraphics[width=\linewidth]{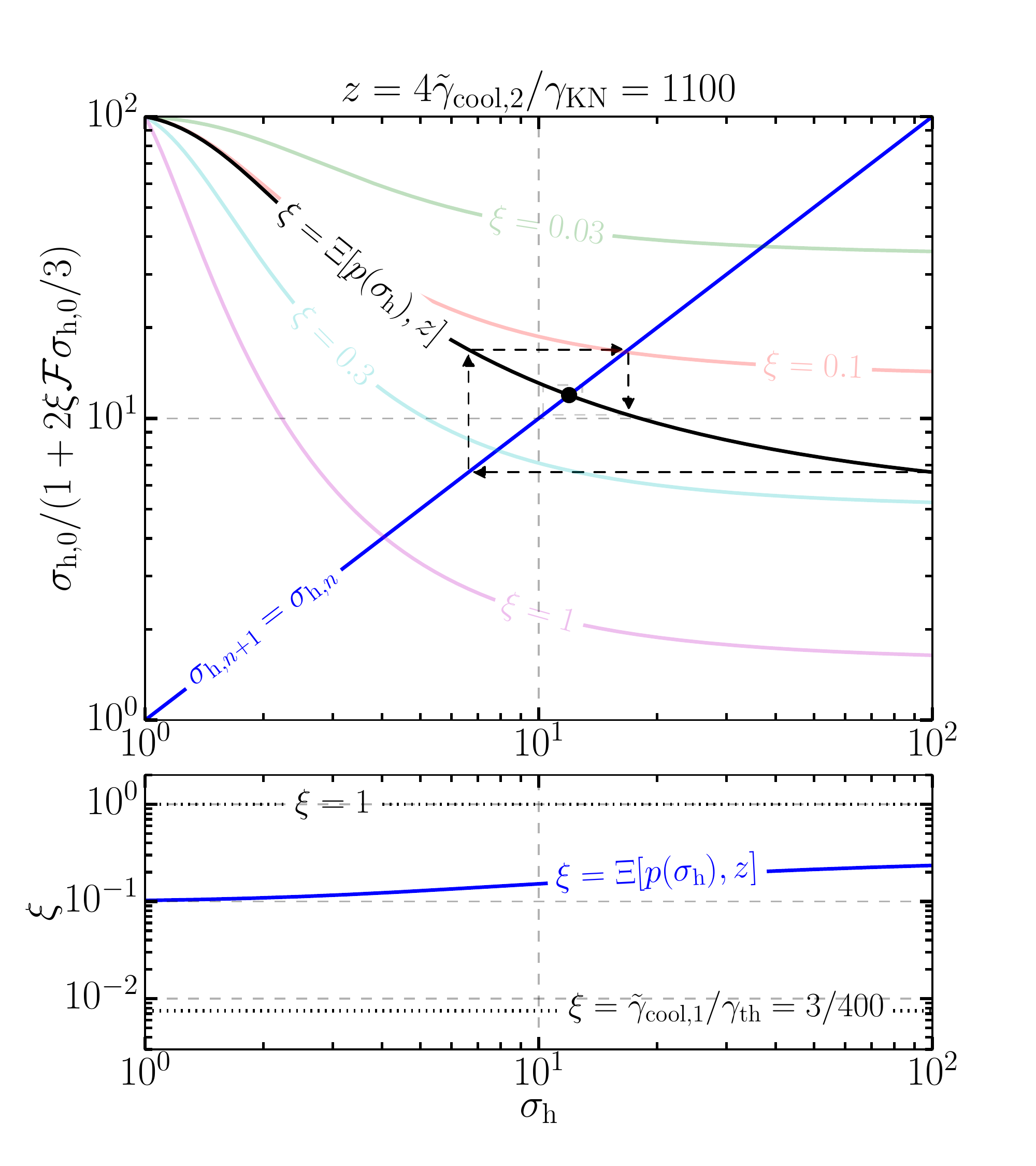}
    \end{subfigure}
    \begin{subfigure}{0.49\textwidth}
        \centering
        \includegraphics[width=\linewidth]{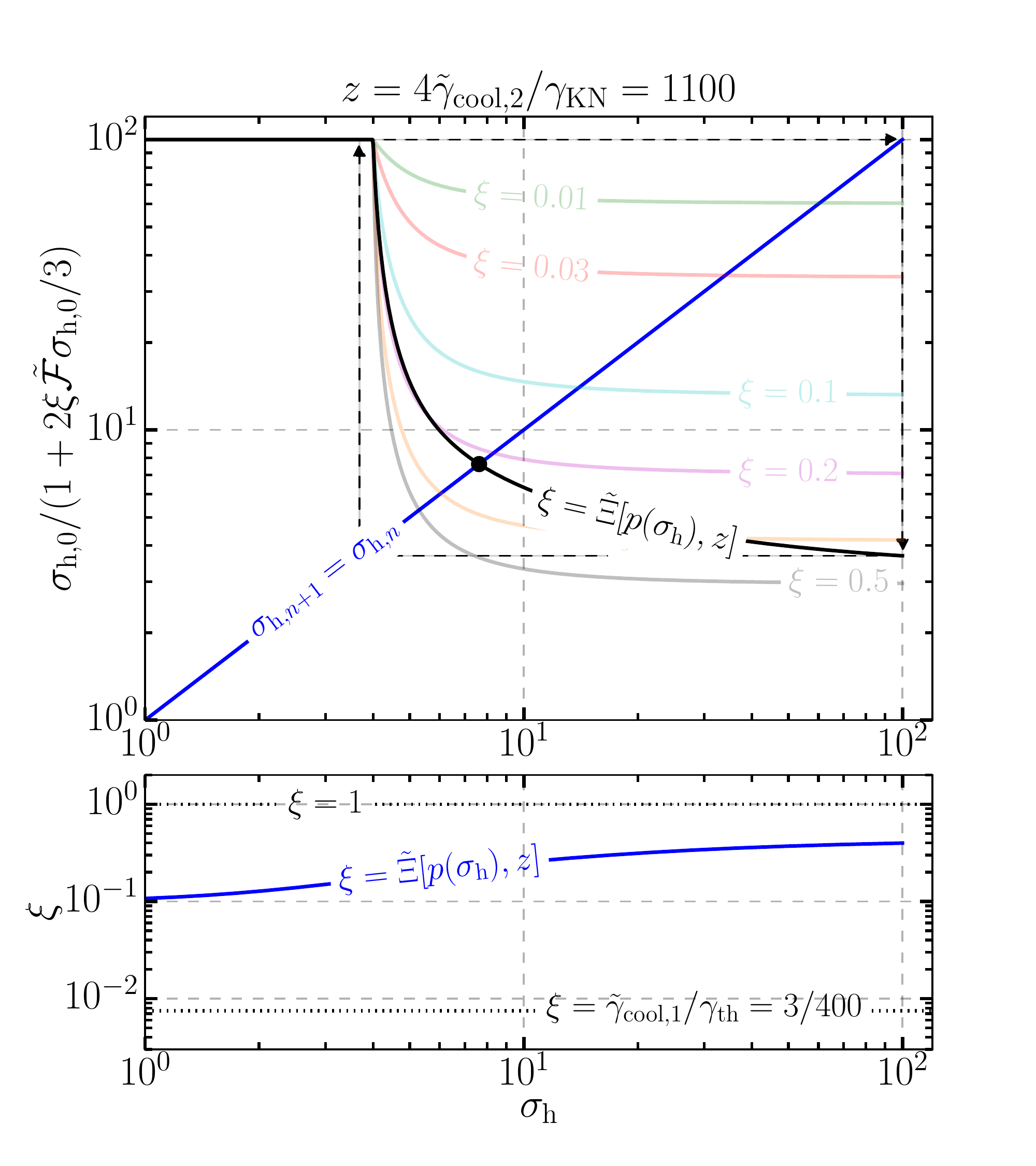}
    \end{subfigure}
    \caption{Top subpanels: On the left (right), the function~$H(\sighgen)$~[$\tilde{H}(\sighgen)$], which includes a calculation of~$\xi$ including (excluding) the effect of radiation-reaction on the distribution of radiating particles in the reconnection layer, for the case~$z = \gf / \gkn = 1100$. This function is overlaid on several {constant-$\xi$} contours of~$h(\sighgen)$~[$\tilde{h}(\sighgen)$], which are taken from Fig.~\ref{fig:sighunivnoradwithevolution} (Fig.~\ref{fig:sighsolnrad}, left panel). When a contour is crossed, it means that the value of~$\xi = \Xi$~($\xi = \tilde{\Xi}$) equals the corresponding contour value. Bottom subpanels: The function~$\Xi$~($\tilde{\Xi}$) plotted explicitly, alongside the minimum and maximum allowed values,~$3/400 \sim 0.01$ and~$1$, respectively. The energy recapture efficiency~$\xi$ is always much higher than the minimum, regardless of the assumptions made about radiation reaction, and, if radiation reaction is accounted for, may even approach order unity.}
    \label{fig:scx_constoverlay}
\end{figure*}

\begin{figure*}
    \centering
    \begin{subfigure}{0.49\textwidth}
        \centering
        \includegraphics[width=\linewidth]{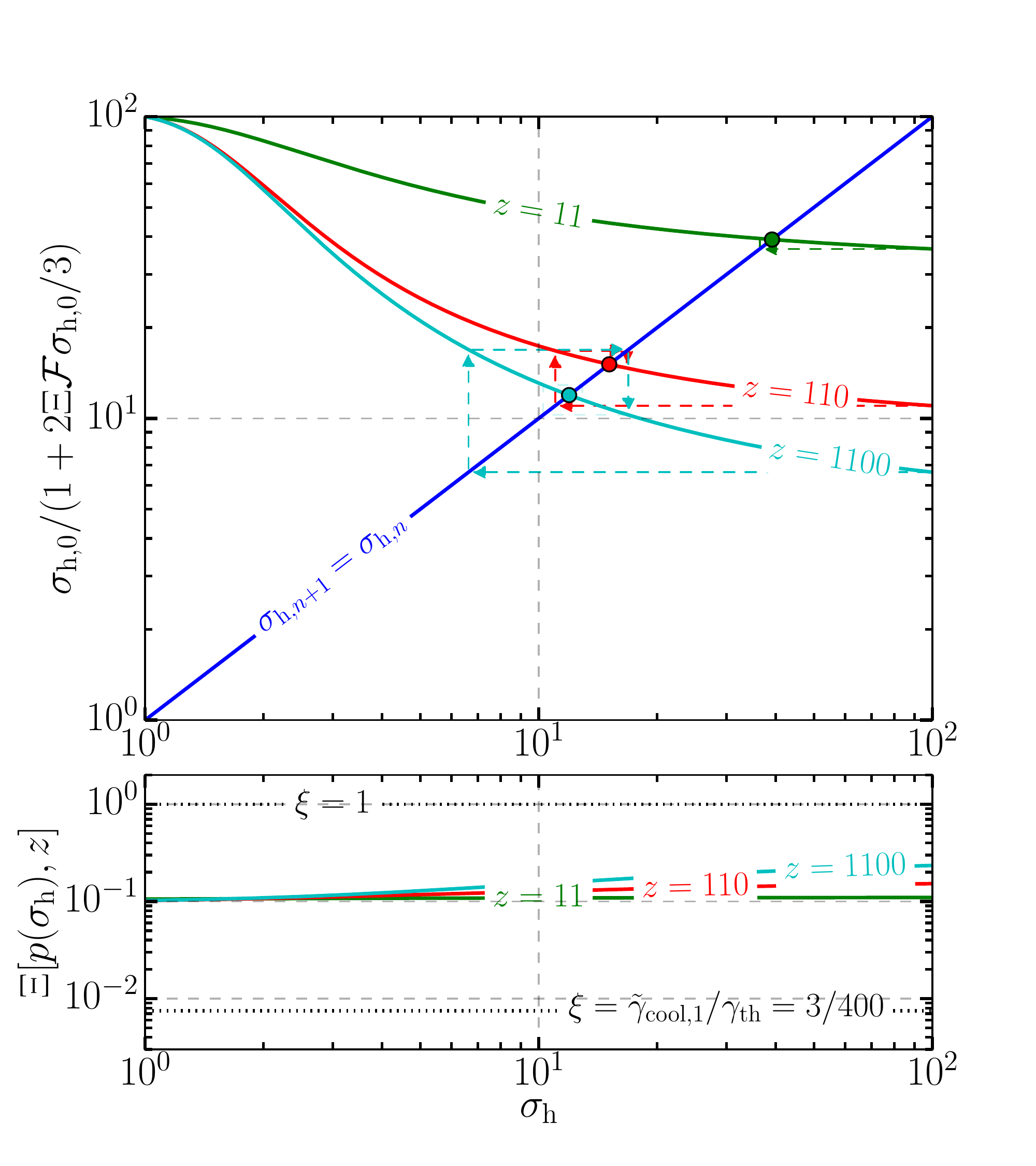}
    \end{subfigure}
    \begin{subfigure}{0.49\textwidth}
        \centering
        \includegraphics[width=\linewidth]{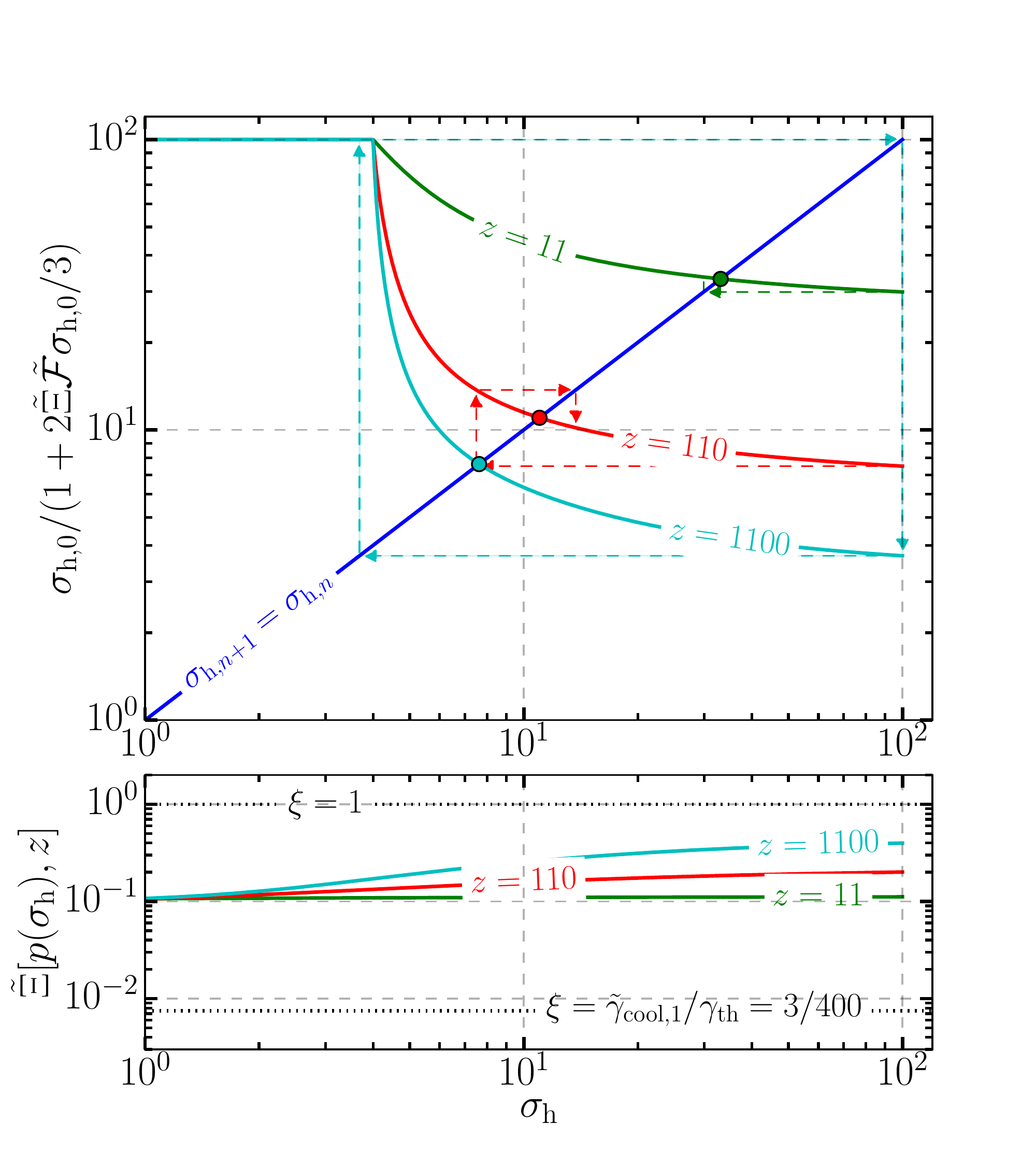}
    \end{subfigure}
    \caption{The same as Fig.~\ref{fig:scx_constoverlay}, but without {constant-$\xi$} contours of the functions~$h$ and~$\tilde{h}$. Instead, on the left (right), the functions~$H$~($\tilde{H}$) and~$\Xi$~($\tilde{\Xi}$) are plotted for several different values of~$z$. On both sides, the maximum {$\xi$-value} occurs at {high-$\sighgen$} and is sensitive to~$z$. Both radiation reaction and {high-$z$} are required for~$\xi$ to reach order unity. In contrast, the low value of~$\xi$, occurring at low-$\sighgen$, is virtually independent of~$z$ and is always around~$\xi = 0.1 \gg \fncmin$. Thus, order-unity efficiency requires both high~$z$ and high~$\sighgen$, but extremely low efficiency~($\xi \sim 10^{-2}$) is averted across the domain of all models.}
    \label{fig:scx_evolution}
\end{figure*}

\begin{figure*}
    \centering
    \begin{subfigure}{0.49\textwidth}
        \centering
        \includegraphics[width=\linewidth]{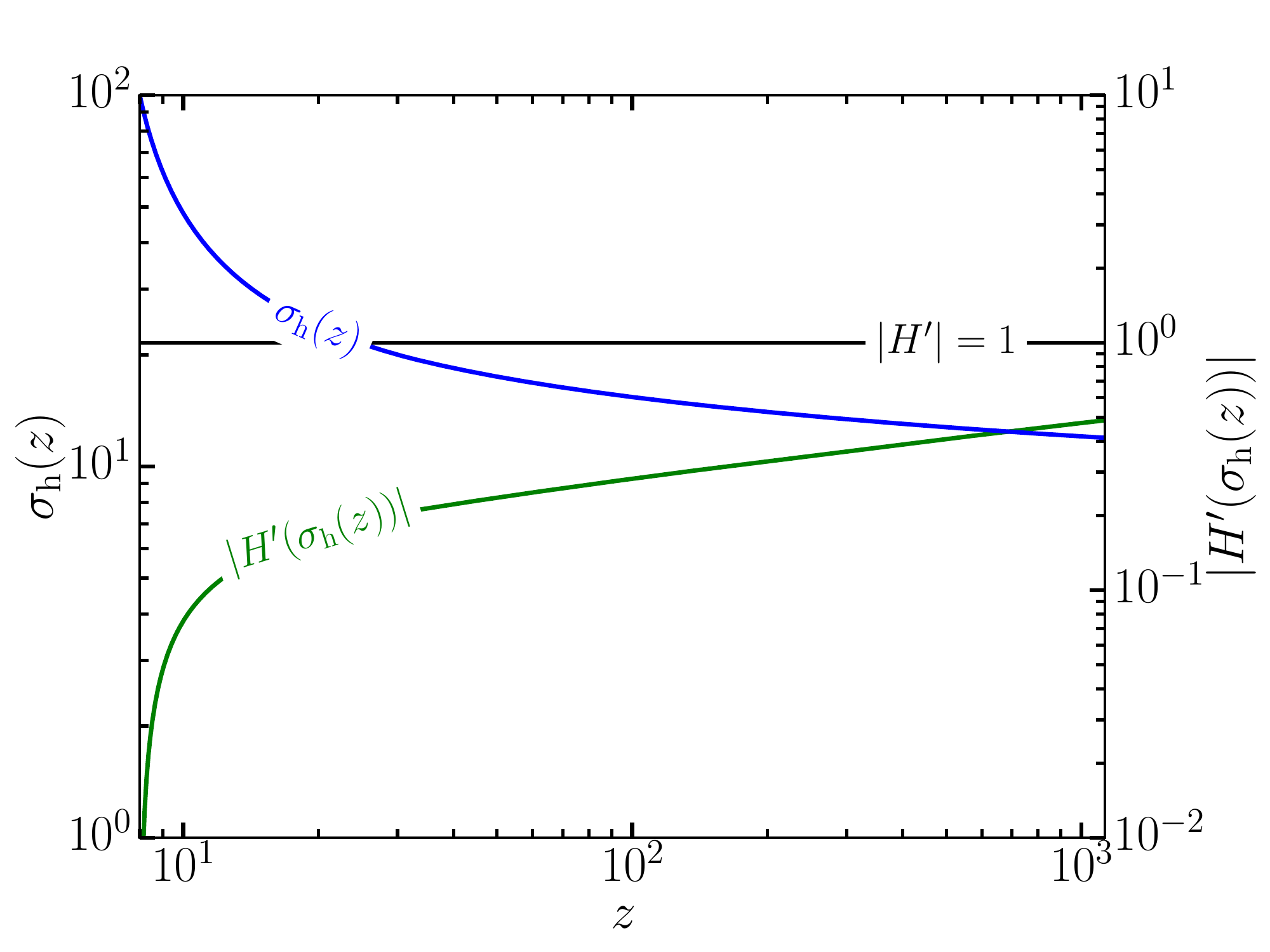}
    \end{subfigure}
    \begin{subfigure}{0.49\textwidth}
        \centering
        \includegraphics[width=\linewidth]{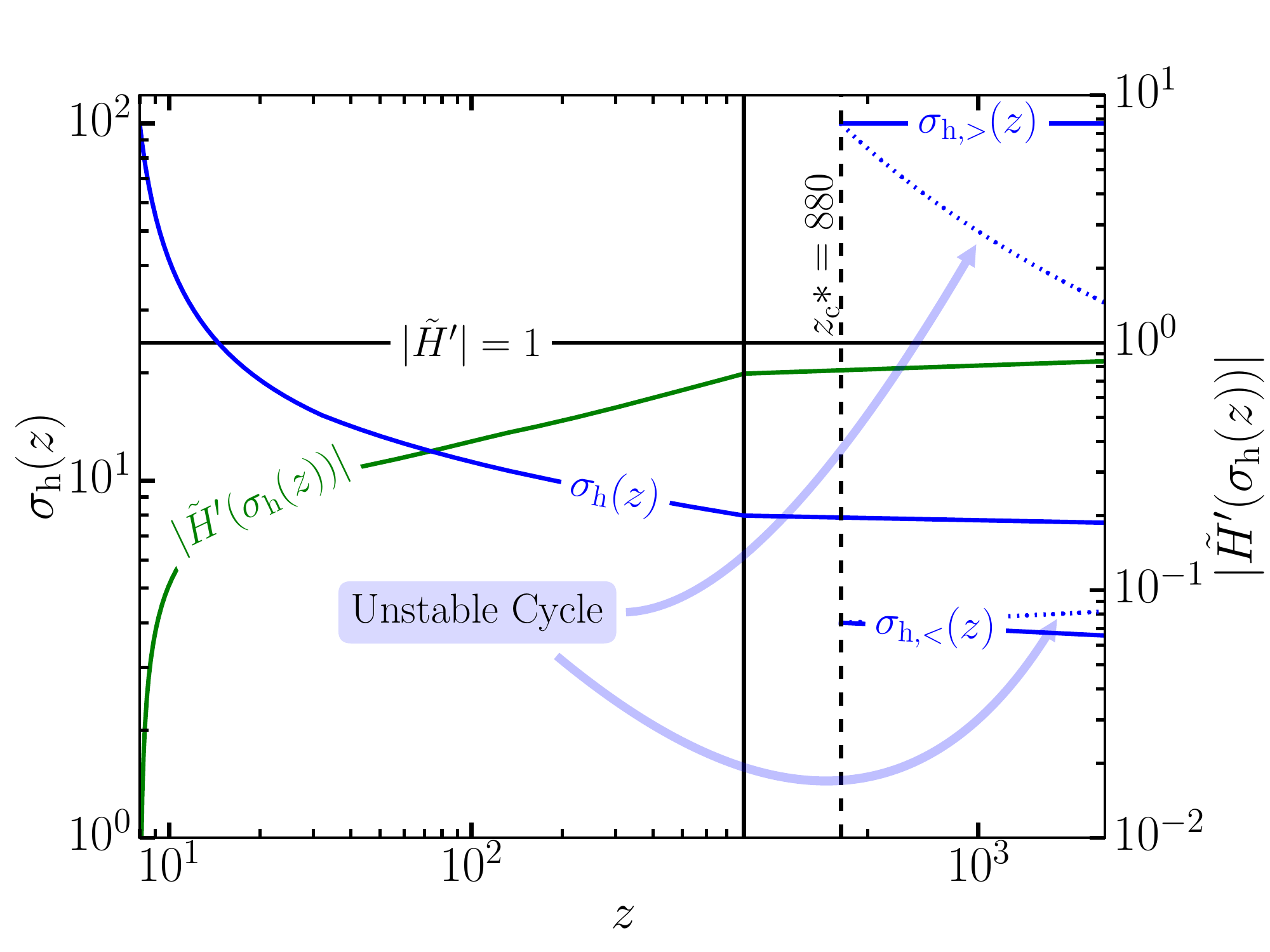}
    \end{subfigure}
    \caption{On the left (right), a stability analysis of the function~$H(\sighgen)$~[$\tilde{H}(\sighgen)$] for all~$z$ between~$8$ and~$1100$. For visual clarity on the right plot, the horizontal axis scale beyond~$z = 800$ is highly zoomed (and thus the kinks in curves crossing~$z = 800$ are not real). The model without radiation back reaction on the layer particles does not possess high enough efficiency for its fixed point~$\sighgen(z)$ to go unstable, or to develop a two-state swing cycle. In contrast, the model with radiation reaction displays limit-cycle behavi\spellor[.]There, pair feedback is so efficient that the system immediately starts, from the first readvection time, in a swing cycle whenever~$z > z_{\rm c}* = 880$. This is true even though the fixed point~$\sighgen(z)$ does not become locally unstable for any~$z < 1100$. Instead, an \textit{unstable} cycle appears and intercepts the flow on the iterated map, blocking all states that start from~$\sigh$ from ever reaching the fixed point. Thus, the model on the right illustrates our argument in Appendix~\ref{sec:hstability} that a locally unstable fixed point is a sufficient but not necessary condition for the system to globally asymptote to a limit cycle.}
    \label{fig:scx_stability}
\end{figure*}

Putting~$f_{\rm noesc} = 1$ as in section~\ref{sec:upstreamendens} (see also Appendix~\ref{sec:fnoesc}), the energy recapture efficiency is just~$\xi = f_{\rm noesc} \fnc = \fnc$. We denote the (new in this Appendix) self-consistent functional dependence of~$\xi$, through~$\fnc$, on~$\pk$ and~$z$ by writing
\begin{align}
    \xi = \Xi(\pk, z) = \fnc \, ,
    \label{eq:bigxidef}
\end{align}
where~$\fnc$ is evaluating according to equation~(\ref{eq:fncdef}) using equations~(\ref{eq:qinjexplicit}) and~{(\ref{eq:avgenninjeval})-(\ref{eq:avgenneval})}.
Note that~$\Xi$ is a function of~$p$ only through~$\pk$. This is because the power-law scaling of the upstream particle distributions~$\pgg$ is inherited strictly from the deep Klein-Nishina-regime layer particles:~$\pgg \simeq \pk + 1$.
Equation~(\ref{eq:bigxidef}) eliminates~$\xi$ as a free parameter from the problem -- just like~$\mathcal{F}$, it is entirely determined in terms of~$\pk$ (and, hence, in terms of~$\sighgen$) and~$z = \gf / \gkn$. This allows us to repeat our analysis from section~\ref{sec:upstreamendens}, except using our expression for~$\xi$ in~(\ref{eq:bigxidef}) instead of scanning across it as a free parameter. This we do in Figs.~\ref{fig:scx_constoverlay}-\ref{fig:scx_stability}.

Each of these figures presents results from two cases, one where radiation back reaction on the layer distribution of emitting particles is ignored [i.e.\ where~$\pt = \pk = p(\sighgen)$] and one where it is crudely taken into account [i.e.\ where~$\pt = p(\sighgen) + 1$ and~$\pk = p(\sighgen) - 0.5$]. In the former case, we \generaliz[e]equation~(\ref{eq:hdef}) to
\begin{align}
    \sighn{n+1} = H(\sighn{n}) \equiv \frac{\sigh}{1 + 2\, \Xi[p(\sighn{n}), z] \mathcal{F}[p(\sighn{n}), z] \sigh / 3} \, ,
    \label{eq:bighdef}
\end{align}
whereas in the latter we \generaliz[e]equation~(\ref{eq:htildedef}) to
\begin{align}
    \sighn{n+1} = \tilde{H}(\sighn{n}) \equiv \frac{\sigh}{1 + 2 \, \tilde{\Xi}[p(\sighn{n}), z] \tilde{\mathcal{F}}[p(\sighn{n}), z] \sigh / 3} \, .
    \label{eq:bightildedef}
\end{align}
Here, we have defined
\begin{align}
    \tilde{\Xi}(p, z) \equiv \Xi(p - 0.5, z) \, .
    \label{eq:bigxitildedef}
\end{align}

We note that, when evaluating~$\Xi$ and~$\tilde{\Xi}$ through equation~(\ref{eq:fncdef}), we replace the symbol~$\Qgg[n]$ in the expression for~$\Ngg[n]$ [equation~(\ref{eq:Ngeneval})] with~$\Qnorm[n] \gamma_{\rm min}^{-\pgg + 1} [1 - (4^{n+1/2} / z )^{\pgg - 1}]/(\pgg - 1)$ rather than adopt the cruder approximation~$\Qgg[n] \simeq \Qnorm[n] \gamma_{\rm min}^{-\pgg + 1} / (\pgg - 1)$ [see equation~(\ref{eq:qinjexplicit})]. This maintains continuity of~$\fnc$, ensuring that contributions from each successive generation turn on gradually with~$z$ (as they do in reality) rather than discretely. We have checked that this does not introduce significant error into our calculation, even though it is slightly inconsistent with the derivations of equations~{(\ref{eq:avgenninjeval})-(\ref{eq:avgenneval})} [particularly one of the steps in equation~(\ref{eq:nggsteadyexplicit})], which assume~$\Qgg[n] \simeq \Qnorm[n] \gamma_{\rm min}^{-\pgg + 1} / (\pgg - 1)$.

In Figs.~\ref{fig:scx_constoverlay}-\ref{fig:scx_stability}, we \analyz[e]more thoroughly the dependence of the solution~$\sighgen$ on~$z$ than in section~\ref{sec:upstreamendens}. There are two reasons for this. First, eliminating~$\xi$ as a free parameter renders~$z$ the only independent variable in the problem, and so examining~$z$-dependence is now easier. Second, unlike when we treated~$\xi$ as a free parameter,~$\Xi$ and~$\tilde{\Xi}$ both depend on~$z$, so examining~$z$-dependence is now more necessary. We conduct our analysis in the broad range of~$z = \gf / \gkn$ spanning from the minimum for pair production to occur in the upstream region,~$z = 8$, to the maximum such that all newborn particles are injected with Lorentz factors less than~$\gcti{2}$. In the latter case,~$z$ is determined by setting the \cutoff[]in the first generation's injected distribution~$\gfn[1]$ to~$\gcti{2}$, yielding~$z = \gf / \gkn = (4 \gfn[1]) / \gkn = 4 \gcti{2} / \gkn \simeq \zmax$ [equation~(\ref{eq:gct2})].

Fig.~\ref{fig:scx_constoverlay} shows how the solutions~$\sighgen = H(\sighgen)$ and~$\sighgen = \tilde{H}(\sighgen)$, where~$\xi$ is determined self-consistently, differ from the corresponding solutions~$\sighgen = h(\sighgen)$ and~$\sighgen = \tilde{h}(\sighgen)$, when~$\xi$ is taken to be constant. This is done for the extreme case~$z = \zmax$, which is most suitable for comparing with section~\ref{sec:upstreamendens}, where~$z$ is infinite. Additionally, Fig.~\ref{fig:scx_constoverlay} also displays the functions~$\Xi$ and~$\tilde{\Xi}$. One can see that, at small~$\sighgen$,~$\Xi$ and~$\tilde{\Xi}$ are small~[$\sim 0.1$, as predicted by equation~(\ref{eq:fnclim})]. Furthermore,~$\Xi$ and~$\tilde{\Xi}$ both increase with~$\sighgen$. This results from the power-law index~$p(\sighgen)$ of radiating layer particles becoming harder, which causes~$\fnc$ to be dominated by the high-energy particles. These particles have their radiative cooling somewhat suppressed by Klein-Nishina effects and, for high enough~$z$, may retain a large portion of the initially injected energy, substantially increasing~$\fnc$. Thus, as~$\sighgen$ grows, one observes the solution~$H(\sighgen)$~[$\tilde{H}(\sighgen)$] cross progressively larger {constant-$\xi$} contours of~$h(\sighgen)$~[$\tilde{h}(\sighgen)$]. In fact, when radiation reaction on the layer particles is taken into account,~$\tilde{\Xi}$ grows large enough at high~$z$ to initiate a dramatic 2-state swing cycle -- a very stark difference from when~$\xi$ is pessimistically estimated as~$f_{\rm nocool,min} \sim 0.01$.

In Fig.~\ref{fig:scx_evolution}, we illustrate how the picture changes at smaller~$z$. Here, the monotonic dependence of~$\xi$ on~$\sighgen$ from Fig.~\ref{fig:scx_constoverlay} remains but is weaker. In particular, the largest value that~$\xi$ reaches, which occurs at high-$\sighgen$, diminishes as one reduces~$z$. Thus, a high \cutoff[]in the layer particle energy distribution is needed to achieve order-unity efficiency. However, it is \textit{not} the case that a lower \cutoff[]causes~$\xi$ to plummet. As long as~$z \geq 8$, and consistent with equation~(\ref{eq:fnclim}), all models have appreciable efficiency, with~$\xi \sim 0.1 \gg \fncmin$. This implies an important conclusion: even for a radiating layer particle distribution that cuts off barely beyond~$z = 8$, establishing a universal steady state is still possible for finite~$\sigh$. The initial magnetization must only be greater than~$\sim 1 / \min(\xi) \sim 10$.

Finally, in Fig.~\ref{fig:scx_stability}, we present a complete stability analysis in~$z$ between~$8$ and~$\zmax$. This includes both how the fixed point solution~$\sighgen(z)$ varies with~$z$, and whether a two-state swing cycle appears, with respective high and low magnetizations~$\sighn{>}(z)$ and~$\sighn{<}(z)$. The model that neglects radiation reaction on the layer particles never develops oscillatory behavi\spellor[,]but always converges toward the fixed points~$\sighgen(z)$. (However, this could change if~$z$ is made even larger. In that case, layer particles with energies~$> 4 \gcti{2}$ spawn upstream pairs that retain nearly \textit{all} of their energy while traveling back toward the layer, increasing~$\xi$, and likely raising~$\, \abs{H'(\sighgen)}$ above~$1$ in Fig.~\ref{fig:scx_stability}.) However, in the more radiatively self-consistent model, a swing cycle develops at a critical~$z = z_{\rm c}^* \simeq 880$. Intriguingly, this is before the fixed point~$\sighgen(z)$ becomes unstable, which does not happen for any~$z < 1100$. Instead, an \textit{unstable} 2-cycle appears and intercepts the flow of the iterated map~$\sighn{n+1} = \tilde{H}(\sighn{n})$, blocking states that start at~$\sigh$ from reaching the fixed point.

While swing cycles appear to require quite extended particle distributions in our model~($z_{\rm c}* = 880$ is close to the upper limit~$z = 1100$ that we study), we do not think that this necessarily precludes their operation in all but the most extreme systems. Instead, we have found that the particular value of~$z$ where swing-cycles set in is highly sensitive to order-unity changes in~$\xi$, and, thus, this value could be either much higher (in which case swing cycles would be unlikely in reality) or much lower (in which case they may be quite common).

\subsection{Summary of self-consistent~$\fnc$}
In this discussion, we have seen how, by self-consistently determining~$\xi = \fnc$ (through the functions~$\Xi$ and~$\tilde{\Xi}$), one may achieve significantly higher efficiencies than a simple and pessimistic estimate,~$\xi = f_{\rm nocool,min} = 3/400$, would predict. Rather than radiating away most of the energy it receives from the reconnection layer, the newly created upstream plasma may catch and hold onto this energy, delivering an appreciable portion of it back to the layer. The reason is not that the fresh pairs radiate inefficiently. On the contrary, given a full readvection time, all of them would cool to energies~$\sim \gcti{1}$. Rather, the broad distribution of newborn pairs is constantly being replenished, due to injection from photon annihilation, at high energies, and this allows the typical energy of radiating particles, as they enter the layer, to be quite high.

Thus, at some finite large (but not excessively large) initial magnetization~$\sigh \ll 1 / \min(\xi) \sim 10$, the layer \textit{always} has a universal fixed point solution. Furthermore -- and depending on how significantly radiation back reaction modifies the distribution of layer particles -- energy recapture by the layer may reach order-unity efficiency when a long high-energy tail of layer particles extends beyond~$\gkn$. In this case, the reconnection layer may overshoot its fixed point steady state, and undergo late-time limit-cycle oscillations about this solution rather than converge toward it. 

\section{The particle escape factor~$\fne$}
\label{sec:fnoesc}
In this Appendix, we discuss a complementary channel, besides radiative cooling, through which newly created upstream matter may lose energy: particle escape. Escape occurs when newborn pairs stream a distance~$\sim L$ along the unreconnected magnetic field, vacating the system. Pairs can be born with essentially any pitch angle, so the typical escape time is~$\sim L/c$ (except for a few particles with very small pitch angles). Therefore, the escape factor~$f_{\rm noesc}$ is close to unity if the readvection time,~$\tread$, satisfies~$\tread  \simeq 10 L / \tau_{\gamma\gamma} c < L / c$, which requires~$\tau_{\gamma\gamma} > 10$.

Beyond this simple consideration, there are also other, more complicated kinetic effects that may influence the value of~$\fne$. Many of these effects, at the same time, also help to decide whether above-threshold photons radiated from the layer ever reach the upstream region in the first place, as assumed throughout this work [e.g.\ assumption~\ref{en:lmfpintermediate}]. These are, in principle, independent concepts:~$\fne$ pertains to pairs that have already been born into the upstream region; photon escape from the layer to the upstream region concerns the emission and propagation of radiation before pairs are ever produced. However, the same processes dictate both, and we discuss these parallel influences simultaneously. 

Consider first the shape of the unreconnected magnetic field lines at a transverse distance~$\, \abs{y} < 0.1 L$ from the layer. In this region, the unreconnected field is perturbed by the presence of large plasmoids, the largest of which may extend a distance~$\sim 0.1L$ into the upstream region \citep{uls10, sgp16}. Because the field is asymptotically uniform as~$\, \abs{y} \to \infty$, it is necessarily stronger (i.e.\ compressed), when~$\, \abs{y} \lesssim 0.1 L$, in regions above plasmoids. This may lead to a magnetic bottling effect, where particles produced at~$\, \abs{y} < 0.1L$ tend to be mirror-confined between large plasmoids (even when they might otherwise escape), pushing~$\fne$ closer to unity. Unlike the effects discussed below, this confinement mechanism only pertains to particles already born and, therefore, does not affect what fraction of photons escape the layer to the upstream region.

We move now to a separate issue: anisotropy in the distribution of radiating layer particles. The layer tends to drive bulk plasma motion and, through the kinetic beaming mechanism \citep[e.g.][]{cwu12, cwu13, mwu20}, collimated bunches of high-energy particles, into the~$\pm x$-directions (either to the left or to the right in Figs.~\ref{fig:knppdiagram} and~\ref{fig:detaileddiagram}). This biases the emitted photons (which are relativistically beamed along the directions of their emitters) into these same directions, and may, in turn, increase~$\fne$. Suppose, for example, that a given above-threshold photon travels at an angle~$\theta$ from the reconnection midplane. This reduces the total distance it propagates into the upstream region from~$\abs{y} \sim \lmfp$ to~$\abs{y} \sim \lmfp \sin \theta$. The readvection time for the produced pair is therefore reduced by the same factor:~$\sin \theta$. Meanwhile, the escape time is still~$\sim L/c$. Thus, the readvection time decreases relative to the escape time. This increases the fraction~$\fne$ of particles captured by the reconnection layer.

On the other hand, if the beaming of plasma motion (both on bulk and kinetic levels) is strong enough, the angle~$\theta$ discussed above could be so small that most photons do not cross the separatrices into the upstream plasma before annihilating. This would occur if the transverse propagation distance~$\, \abs{y} \sim \lmfp \sin \theta$ is smaller than the thickness~$\thickness$ of the layer radiation zone (discussed in more detail below). In such a scenario, the created pairs would not load the inflow plasma and thus the pair-regulation mechanism would be somewhat suppressed.

We next discuss how the thickness,~$\thickness$, of \radzone[s,]where most pair-producing photons are emitted (discussed in section~\ref{sec:knradreconn_mod} and Fig.~\ref{fig:detaileddiagram}), influences the escape of photons to the upstream region and~$\fne$. If most photons are radiated from inside large plasmoids, instead of from thin strip-like \radzone[s](of the kind argued for in section~\ref{sec:knradreconn_mod}), then the condition~$\tau_{\gamma\gamma} > 10$, although it maintains~$\tread < L / c$ and helps keep~$\fne$ near unity, also confines most of the above-threshold photons to within the separatrices: inside plasmoids exceeding~$\lmfp$ in size. This inhibits pair feedback. If, instead, the primary radiation sites are thin, even kinetic-scale, current sheets (e.g. with~$\thickness \ll \lmfp$, as we argue in section~\ref{sec:knradreconn_mod}) most photons travel to the upstream region and effectively load its plasma when they annihilate (modulo potential extreme beaming effects described above).\footnote{Note that, if~$\lmfp > 0.1 L$, it is irrelevant whether most of the pair-producing photons come from thin structures or from large round plasmoids. Either way, the photons escape to the upstream region.}

Finally, let us consider how reconnection current sheets between merging plasmoids play into this picture. These miniature reconnection layers are oriented perpendicularly with respect to the main reconnection current sheet and therefore bias the motion of particles and emitted photons in the transverse~$\pm y$-directions. The produced upstream pairs then have suppressed pitch angles (the~$\pm y$ direction is perpendicular to the unreconnected magnetic field), which inhibits particle escape. However, if~$\tau_{\gamma\gamma} > 10$, then not all radiation produced at merging-plasmoid current sheets escapes back into the inflow plasma: there will at least be some photon-confining plasmoids larger than~$\lmfp < L / 10$. Mergers between such plasmoids would not source significant self-regulating pair production. This, however, does not preclude merging-plasmoid reconnection from regulating itself (just on a smaller scale) in the same way that the main reconnection layer does as a whole.

In light of this discussion, there are clearly several details, influencing both~$\fne$ and the delivery of photons to the upstream region, that are beyond the scope of our present work to calculate quantitatively. These will require a future computational study in order to properly diagnose. Nevertheless, many of these effects (especially when~$\tau_{\gamma\gamma} > 10$) promote order-unity~$\fne$, and -- modulo extreme beaming near reconnection X-points -- allow photon escape to the upstream region from the primary large-scale current sheet.

\section{Global stability of the iterated map~$\xeqh$}
\label{sec:hstability}
In this section, the phrase \quoted{late-time}or \quoted{late times}refers to the limit~$\lim_{n \to \infty} t_{n}$ where~$t_n \equiv n \lmfp / \beta_{\rm rec} c$. Late times defined in this sense are not necessarily reached by the reconnection system before reconnection terminates.

Here, we argue that the condition deciding the local stability of the iterated magnetization map~$\sighn{n+1} = h(\sighn{n})$ from section~\ref{sec:upstreamendens} also reveals, in some cases, its global stability. In particular, we argue that local instability~$\, \abs{h'(\sighgen)} > 1$ implies that a system starting at magnetization~$\sigh$ asymptotically approaches a two-state swing-cycle. If, on the other hand, the fixed point is stable,~$\, \abs{h'(\sighgen)} < 1$, then the system may or may not converge toward it. Thus, local instability is sufficient, but not necessary, for global instability.

We focus here only on strictly decreasing functions~$h(x)$ [such that~$h(y) < h(x)$ if~$y > x$] that map into their own domain~$[1,\sigh]$ (i.e.~$h:[1,\sigh]\to[1,\sigh]$). Here, the upper end of the domain happens to coincide with the starting point of the map~$\sigh$. The particular functions~$h(x)$ and~$H(x)$ specified in equations~(\ref{eq:hdef}) and~(\ref{eq:bighdef}) fall into this class. [Note here that we are overloading the symbol~$h(x)$, designating with it a general class of functions and not necessarily the particular form in equation~(\ref{eq:hdef}).] The arguments in this section can be \generaliz[ed]to non-increasing functions~$h^* (x)$ such that~$h^*(y) \leq h^*(x)$ if~$y > x$~[into which category fall the functions~$\tilde{h}(x)$ and~$\tilde{H}(x)$ in equations~(\ref{eq:htildedef}) and~(\ref{eq:bightildedef})] without difficulty, but require additional edge cases (saturation of the inequalities) to be considered, and so we do not formally treat them here.

We begin with a few basic observations. First, if the map does not start on the unique fixed point~$\sighgen$, then each successive iteration lands on the opposite side of the fixed point from the preceding iteration. That is, if~$\sighn{n} < \sighgen$, then~$\sighn{n+1} > \sighgen$, and if~$\sighn{n} > \sighgen$, then~$\sighn{n+1} < \sighgen$. This is because
\begin{align}
    \sighn{n+1} = h(\sighn{n}) < h(\sighgen) = \sighgen \, ,
    \label{eq:parityhopping}
\end{align}
where the \quoted{$<$}follows when~$\sighn{n} > \sighgen$ by the strictly decreasing hypothesis. The same proof, but with a \quoted{$>$}sign, follows when~$\sighn{n} < \sighgen$. This precludes all non-fixed-point odd-period orbits. Hence, one cannot automatically infer the presence of chaos in the system using the 3-period theorem \citep{ly04}. As we now show, the dynamics are even more constrained: chaotic behavi\spellor[]is, in fact, completely precluded.

For our second observation, we note that if, for any starting index~$m$, one discovers that~$\sighn{m+2} < \sighn{m}$, then for all integers~$n \geq 1$, it is necessarily the case that~$\sighn{m + 2n} < \sighn{m + 2(n-1)}$. That is, even though each successive iteration bounces to the opposite side of the fixed point, every consecutive even (or odd) iteration moves strictly in one direction. This follows inductively because
\begin{align}
    \sighn{m + 3} &= h(\sighn{m + 2}) > h(\sighn{m}) = \sighn{m + 1} \notag \\
    &\Rightarrow \sighn{m + 4} = h(\sighn{m + 3}) < h(\sighn{m + 1}) = \sighn{m + 2}.
    \label{eq:monotonichh}
\end{align}
The first line follows from the strictly decreasing hypothesis and the starting assumption that~$\sighn{m + 2} < \sighn{m}$. The second line follows from the strictly decreasing hypothesis and the first line.

This is a very powerful constraint because it severely limits the potential late-time dynamics of the system. In effect, the system can only either approach the fixed point~$\sighgen = h(\sighgen)$ or a two-state swing-cycle. To see this, let us suppose that~$\sighn{2} < \sigh$. (Note that~$\sighn{2} = \sigh$ implies that the system begins in a two-cycle, and the condition~$\sighn{2} > \sigh$ is impossible because~$h(x)$ maps onto its own domain.) Then each successive~$\sighn{2n}$ marches resolutely toward smaller values. This continuing reduction in~$\sighn{2n}$ can only be terminated in one way. It must be the case that there exists some~$\sighn{\infty}$ such that~$\sighgen \leq \sighn{\infty} = \lim_{n \to \infty} \sighn{2n}$. If~$\sighn{\infty} = \sighgen$, then the system approaches the fixed point~$\sighgen$ in a late-time steady state. If~$\sighn{\infty} > \sighgen$, then the system approaches a two-cycle. Note that it is not possible for the system to approach any~$2n$-cycle for~$n > 1$ that is not also a two-cycle. If that were the case, then one of the~$\sighn{n}$'s on the cycle would be on the same side of the fixed point as another~$\sighn{n}$ of the same parity (even or odd), and the two would not be equal. This would contradict our result that~$\sighn{2n + 2} < \sighn{2n}$.

Thus, there are only two possible asymptotic behavi\spellor[s]of the map~$\sighn{n+1} = h(\sighn{n})$: either~$\sighn{n}$ approaches the fixed point~$\sighgen$ or the map converges to a two-state cycle, and one of the states has magnetization~$\sighn{>} > \sighgen$. It follows that the flow of the iterated map from~$\sigh$ to~$\sighgen$ is always intercepted by a two-cycle when the local instability criterion~$\abs{h'(\sighgen)} > 1$ is met. Otherwise, there would not be a suitable attractor [i.e.\ one consistent with both equations~(\ref{eq:parityhopping}) and~(\ref{eq:monotonichh})] to catch the strictly decreasing flow of the map~$\sighn{n+2} = h(h(\sighn{n}))$. Let us illustrate these remarks with an example.

\begin{figure}
    \centering
    \includegraphics[width=\columnwidth]{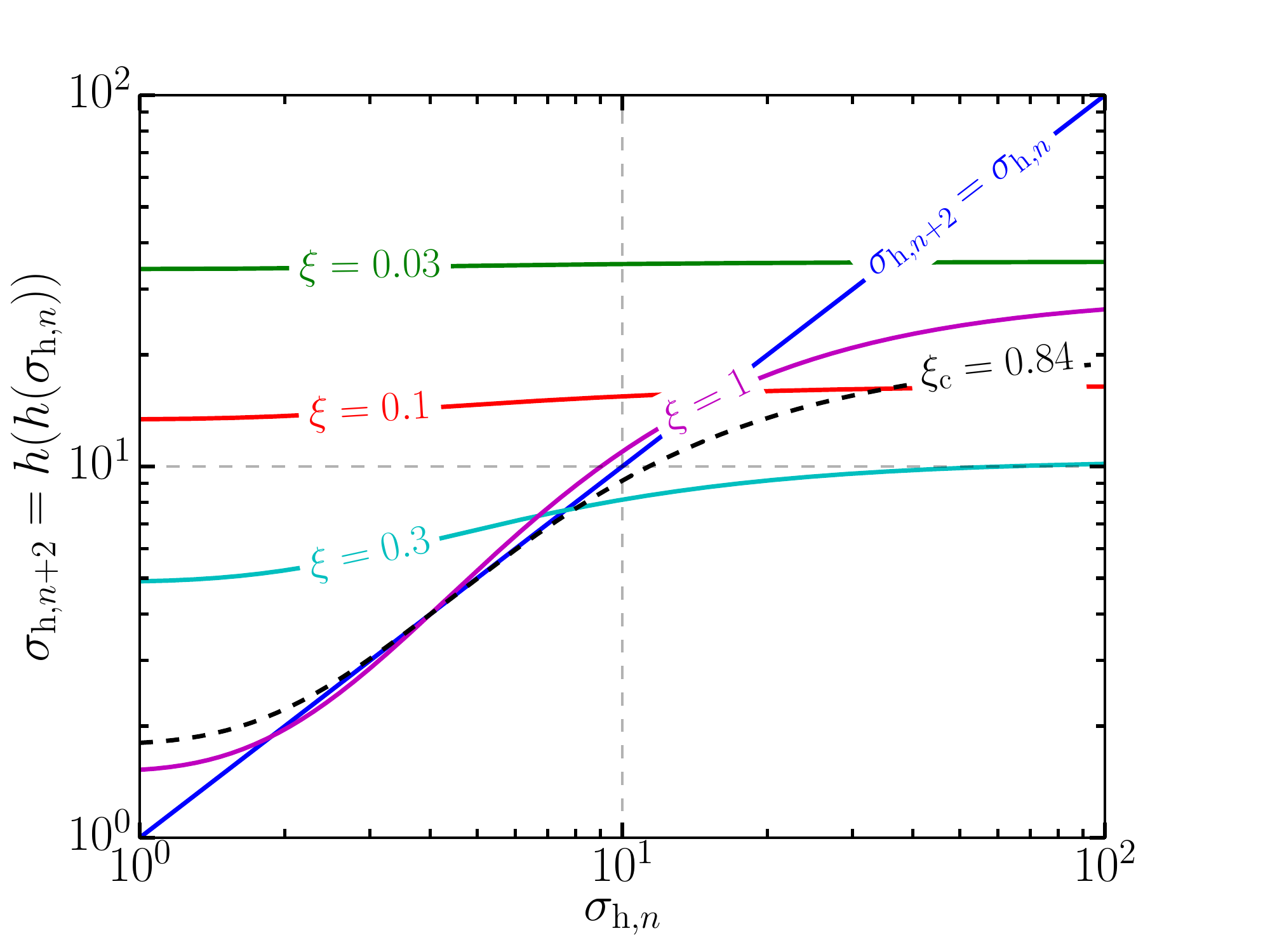}
    \caption{The twice-iterated map~$\sighn{n+2}=h(h(\sighn{n}))$ with~$h(x)$ as defined in equation~(\ref{eq:hdef}). The appearance of a stable two-cycle coincides with the critical value~$\xi_{\rm c} \simeq 0.84$ where the fixed point at~$\sighgen$ becomes unstable. This is consistent with the fact that there must be an attractor on the map~$\sighn{n+1} = h(\sighn{n})$ to intercept the flow~$\sighn{2n + 2} < \sighn{2n} \leq \sigh$ when the local instability criterion~$\abs{h'(\sighgen)} > 1$ is met. If locally unstable, the fixed point cannot be naked: it must be blocked from the initial state~$\sigh$ by a two-state cycle.}
    \label{fig:hh}
\end{figure}

We display the particular map~$\sighn{n+2}=h(h(\sighn{n}))$, where~$h$ is defined as in equation~(\ref{eq:hdef}), in Fig.~\ref{fig:hh}. For low~$\xi$ (in this case, lower than~$\xi_{\rm c} = 0.84$), all of the~$h(h(\sighn{n}))$ curves intersect the diagonal line only once and with a gentle slope~$\abs{\dif h(h(\sighn{n})) / \dif \sighn{n}} < 1$. These intersection points coincide with the stable fixed points~$\sighgen$ illustrated in Fig.~\ref{fig:sighunivnoradwithevolution}. The fixed point becomes unstable when the slope of the twice-iterated map becomes tangent to the diagonal at a critical value~$\xi_{\rm c} \simeq 0.84$. For~$\xi$ higher than this value, the map must intercept the diagonal at at least two other locations -- the values~$\sighn{<}$ and~$\sighn{>}$ corresponding to a two-cycle. This is demanded in order for~$h(h(x))$ to be at or below the diagonal as~$\sighn{n}$ approaches~$\sigh$ and to be at or above the diagonal as~$\sighn{n}$ approaches~$1$ (which itself is required by the fact that~$h$ maps into its own domain). 

We stress that this general behavi\spellor[]-- where a naked fixed point~$\sighgen$ is immediately concealed behind a two-cycle as soon as it goes unstable -- is demanded by equations~(\ref{eq:parityhopping}) and~(\ref{eq:monotonichh}). However, the inverse -- that all stable fixed points occur in isolation from two-cycles -- does not hold in general. The topological argument of the preceding paragraph, which was based on the fact that~$h(h(\sigh)) \leq \sigh$ and~$h(h(1)) \geq 1$, does not preclude the possibility that a stable fixed point could spontaneously become enshrouded by an even number of two-cycles with alternating stabilities. In this case, the function~$h(h(\sighn{n}))$ would simply intersect the diagonal an even number of times on either side of the fixed point. The first of these intersections, counted as one moves outward from the fixed point to the boundaries of the domain~($1$ and~$\sigh$) would be necessarily unstable (since~$\, \abs{\dif h(h(\sighgen)) / \dif \sighgen} > 1$), the next would be stable, the one after unstable, and so on until the outermost stable two-cycle. This would furthermore be consistent with equations~(\ref{eq:parityhopping}) and~(\ref{eq:monotonichh}), ensuring that a stable two-cycle attracts the late-time dynamics. Although such a state of affairs may seem at first rather unlikely, this is actually precisely what occurs for the map~$\tilde{H}$ defined in equation~(\ref{eq:bightildedef}). As shown in Fig.~\ref{fig:scx_stability}, the fixed point on that map becomes enclosed inside two 2-cycles without \textit{ever} becoming locally unstable.

All in all, if the fixed point~$\sighgen$ is unstable~($\, \abs{h'(\sighgen)} > 1$) then the system necessarily asymptotes to a two-cycle. On the other hand, if~$\sighgen$ is stable, then whether the system approaches it at late-times depends on whether a two-cycle is present to intercept the flow on the iterated map. If a stable two-cycle exists,~$\sighgen$ is never reached; otherwise, the system converges to~$\sighgen$. Fixed points and two-cycles are the only allowed late-time behavi\spellor[s]for monotonically decreasing maps.

\section{Exact solution to the Thomson-limit energy-advection equation}
\label{sec:fpexactsoln}
In this section, we solve equation~(\ref{eq:fppairst}) exactly subject to the initial condition~$\Ngg[n](\gamma, 0) = 0$. We evolve the solution for one readvection time~$\tread = \lmfp / \beta_{\rm rec} c$, which simulates the continual injection of particles as a parcel of plasma moves from~$\abs{y} \sim \lmfp$ to the layer. The solution follows identically for all generations, so we simplify our notation to
\begin{align}
    \frac{\partial}{\partial t} N(\gamma, t) + \frac{\partial}{\partial \gamma} \left[ \dot{\gamma_{\rm T}} N(\gamma, t) \right] = Q \delta(\gamma - \gkn) \, ,
    \label{eq:fppairstsimple}
\end{align}
and write~$-\dot{\gamma}_{\rm T} = \gamma / \tcoolt(\gamma) = c \gamma^2 / L \gcool \equiv \gamma^2 / \tau$, abbreviating~$\tau \equiv L \gcool / c$. (Note that~$\tau$ is a time and not an optical depth.)

We denote the Laplace transform in time with a tilde:
\begin{align}
    \tilde{N}(\gamma, s) \equiv \int_0^\infty N(\gamma, t) e^{-st} \dif t \, .
    \label{eq:ltdef}
\end{align}
Laplace-transforming equation~(\ref{eq:fppairstsimple}) gives
\begin{align}
    s \tilde{N} - \frac{1}{\tau} \frac{\partial}{\partial \gamma} \left(\gamma^2 \tilde{N} \right) = \frac{Q}{s} \delta(\gamma - \gkn) \, ,
    \label{eq:ltfppairstsimple}
\end{align}
where we used the initial condition~$N(\gamma, 0) = 0$. Expanding the~$\gamma$-derivative, rearranging, and multiplying through by the integrating factor
\begin{align}
    \mu(\gamma) = \gamma^2 e^{-s \tau (1 - 1/\gamma)}
    \label{eq:intfacdef}
\end{align}
gives
\begin{align}
    \frac{\partial}{\partial \gamma} \left[ \mu(\gamma) \tilde{N}(\gamma, s) \right] = -\frac{Q \tau}{s \gamma^2} \mu(\gamma) \delta(\gamma - \gkn) \, .
    \label{eq:intfaceqn}
\end{align}

Next, we integrate~(\ref{eq:intfaceqn}) from~$\gamma$ to some arbitrary~$\gamma_{\rm hi} > \gkn, \gamma$. Using~$\tilde{N}(\gamma_{\rm hi}, s) = 0$ (radiative cooling only populates energies lower than~$\gkn$) yields
\begin{align}
    \tilde{N}(\gamma, s) &= \frac{Q \tau}{s \gkn^2} \frac{\mu(\gkn)}{\mu(\gamma)} \Theta(\gkn - \gamma) \notag \\
    &= \frac{Q \tau}{s \gkn^2} e^{-s \tau (1/\gamma - 1/\gkn)} \Theta(\gkn - \gamma) \, ,
    \label{eq:solnlt}
\end{align}
where~$\Theta(x)$ is the Heaviside step function. Equation~(\ref{eq:solnlt}) is the Laplace transform of our sought solution:
\begin{align}
    N(\gamma, t) = \frac{Q \tau}{\gamma^2} \Theta \left[t - \tau \left( \frac{1}{\gamma} - \frac{1}{\gkn} \right) \right] \Theta(\gkn - \gamma) \, .
    \label{eq:soln}
\end{align}

Thus, constant injection at~$\gamma = \gkn$ develops into a power-law~$\gamma^{-2}$ extending into lower and lower energies with time. Putting~$t = \lmfp / \beta_{\rm rec} c \simeq 10 L / \tau_{\gamma\gamma} c$ allows the power-law to extend from~$\gkn$ all the way down to
\begin{align}
    \left(\frac{1}{\gkn} + \frac{1}{\gcti{1}}\right)^{-1} \simeq \gcti{1} \equiv \frac{3}{50} \gkn \, .
    \label{eq:loenlim}
\end{align}
Hence [cf.\ equation~(\ref{eq:nggsteadyt})],
\begin{align}
    N(\gamma, \tread) = Q \gcool \frac{L}{c} \begin{cases}
        \gamma^{-2} & \gcti{1} \leq \gamma < \gkn \\
        0 & \mathrm{otherwise}
    \end{cases} \, .
    \label{eq:steadystateresult}
\end{align}

\bsp	
\label{lastpage}
\end{document}